\documentclass[12pt,twoside]{article}   %For printing on two sides of page
\usepackage{amsmath,amssymb,amsthm}
\usepackage{graphicx}
\graphicspath{figure/}
\usepackage{subfigure}
\usepackage{caption}
\usepackage{float}
\usepackage{bm}
\usepackage{diagbox}

\usepackage{algorithm}
\usepackage{algorithmicx}
\usepackage{algpseudocode}
\usepackage{url}
\usepackage{indentfirst}
\usepackage{verbatim}
\usepackage{booktabs,multirow}
\usepackage{tikz}
\usepackage{cite}
\usetikzlibrary{spy}
\usepackage{booktabs}
\usepackage{multirow}
\usepackage{mathrsfs}

\newcommand{\x}		{\mathbf{x}}	
\newcommand{\y}		{\mathbf{y}}
\newcommand{\A}		{\mathbf{A}}
\newcommand{\B}     {\mathbf{B}}

\newcommand{\W}	 {\mathbf{W}}
\renewcommand{\P}	 {\mathbf{P}}
\newcommand{\omg}	 {\mathbf{\Omega}}

\newcommand{\Z}		{\mathbf{Z}}
\newcommand{\0}	     {\mathbf{0}}

\newcommand{\D}	 {\mathbf{D}}
\newcommand{\I}	     {\mathbf{I}}

\newcommand{\s}	 {\mathbf{s}}
\newcommand{\Rsf}{\mathsf{S}}
\newcommand{\g}	 {\mathbf{g}}
\newcommand{\h}	 {\mathbf{h}}
\newcommand{\ze}	 {\bm{\zeta}}

\newcommand{\R}	     {\mathbf{R}}

\newcommand{\diag}  {\mathsf{diag}}
\newcommand{\U}		{\mathbf{U}}	
\newcommand{\V}		{\mathbf{V}}
\newcommand{\Sig}		{\mathbf{\Sigma}}

\usepackage[super,sort,comma]{natbib}
\usepackage{fancyhdr}		%Gives headers and footers defined below
\usepackage[section]{placeins}   %
\usepackage{graphicx}

\makeatletter \renewcommand\@biblabel[1]{$^{#1}$} \makeatother
\newcommand{\cen}[1]{\begin{center} #1 \end{center}}

\usepackage[mathlines]{lineno}
\usepackage{hyperref}
\hypersetup{ colorlinks,
    citecolor=blue,
    filecolor=blue,
    linkcolor=blue,
    urlcolor=blue
}

\usepackage{xcolor}
\definecolor{gray}{rgb}{0.6,0.6,0.6}
\definecolor{red}{rgb}{0.85,0,0}
\definecolor{green}{rgb}{0,0.85,0}
\definecolor{blue}{rgb}{0,0,0.85}
\definecolor{beige}{rgb}{0.92,0.87,0.78}
\usepackage[all]{hypcap}    %causes link to figures to go to figure, not caption
\begin{document}

\cen{\sf {\Large {\bfseries Multi-layer Residual Sparsifying Transform (MARS) Model for Low-dose CT Image Reconstruction} \\  
\vspace*{10mm}
Xikai Yang$^1$, Yong Long$^1$, Saiprasad Ravishankar$^2$} \\
$^1$University of Michigan - Shanghai Jiao Tong University Joint Institute, \\
Shanghai Jiao Tong University, Shanghai 200240, China\\
$^2$Department of Computational Mathematics, Science and Engineering \\ and Department of Biomedical Engineering,\\
Michigan State University, East Lansing, MI 48824, USA\\
%\vspace{5mm}\\
%\vspace{5mm}
Version typeset \today\\
}

\pagenumbering{roman}
\setcounter{page}{1}
\pagestyle{plain}
Author to whom correspondence should be addressed. \\ Yong Long. E-mail: yong.long@sjtu.edu.cn
% note, probably best not to use a student's e-mail as it won't be valid for
% very long.

\vspace{-0.1in}
\begin{abstract}
	\vspace{-0.1in}
	\noindent {\bf Purpose:} Signal models based on sparse representations have received considerable attention in recent years. On the other hand, deep models consisting of a cascade of functional layers, commonly known as deep neural networks, have been highly successful for the task of object classification and have been recently introduced to image reconstruction. In this work, we develop a new image reconstruction approach based on a novel multi-layer model learned in an unsupervised manner by combining both sparse representations and deep models. The proposed framework extends the classical sparsifying transform model for images to a Multi-lAyer Residual Sparsifying transform (MARS) model, wherein the transform domain data are jointly sparsified over layers. We investigate the application of MARS models learned from limited regular-dose images for low-dose CT reconstruction using Penalized Weighted Least Squares (PWLS) optimization. \\
	{\bf Methods:} We propose new formulations for multi-layer transform learning and image reconstruction. We derive an efficient block coordinate descent algorithm to learn the transforms across layers, in an unsupervised manner from limited regular-dose images. The learned model is then incorporated into the low-dose image reconstruction phase. \\
	{\bf Results:} Low-dose CT experimental results with both the XCAT phantom and Mayo Clinic data show that the MARS model outperforms conventional methods such as FBP and PWLS methods based on the edge-preserving (EP) regularizer in terms of two numerical metrics (RMSE and SSIM) and noise suppression. Compared with the single-layer learned transform (ST) model, the MARS model performs better in maintaining some subtle details.\\
	{\bf Conclusions:} This work presents a novel data-driven regularization framework for CT image reconstruction that exploits learned multi-layer or cascaded residual sparsifying transforms. The image model is learned in an unsupervised manner from limited images. Our experimental results demonstrate the promising performance of the proposed multi-layer scheme over single-layer learned sparsifying transforms. Learned MARS models also offer better image quality than typical nonadaptive PWLS methods.  
	%\red{and are more robust to parameter choices.}\\
	
\end{abstract}

\newpage     %may or may not be needed

%\tableofcontents

\newpage

\setlength{\baselineskip}{0.7cm}      %double spacing		

\pagenumbering{arabic}
\setcounter{page}{1}
\pagestyle{fancy}
%\twocolumn

\section{Introduction}

Signal models exploiting sparsity have been shown to be useful in a variety of of imaging and image processing applications such as compression, restoration, denoising, reconstruction, etc. \cite{chen:08:pic,mairal:07:srf,elad:06:idv,zhang:17:tbd}
Natural signals can be modeled as sparse in a synthesis dictionary (i.e., represented as a linear combinations of a few dictionary atoms or columns) or in a sparsifying transform domain.
Transforms such as wavelets \cite{pati:93:omp} and the discrete cosine transform (DCT) are well-known to sparsify images. Synthesis dictionary learning \cite{aharon:06:ksa} and analysis dictionary learning \cite{rubinstein:13:aka} methods adapt such models to data and involve algorithms such as K-SVD \cite{rubinstein:13:aka}, the Chasing Butterflies approach \cite{Magoarou:15:cbi}, and some others.
The underlying dictionary learning problems are typically NP-hard and the corresponding algorithms often involve computationally expensive updates that limit their applicability to large-scale data. In contrast, the recently proposed sparsifying transform learning approaches \cite{ravishankar:13:lst} involve exact and highly efficient updates in the algorithms.
In particular, the transform model suggests that the signal is approximately sparse in a transformed domain. Furthermore, Ravishankar \textit{et al}\cite{Ravishankar:16:ddl,Wen:15:wsm,wen:19:vhd} demonstrated the applicability of adaptive sparsifying transforms for several applications such as image denoising and medical image reconstruction.

On the other hand, deep models with nested network structure popularly known as deep neural networks provide remarkable results for classification and regression across various fields\cite{lecun:15:dle}. 
Given a task-based loss function for network parameter estimation, algorithms based on gradient back-propagation sequentially reduce the error between a known target (ground truth) and the network prediction. Another approach from a few research groups combines deep network architectures with probabilistic models during learning, and this generative Bayesian model\cite{patel:16:apf} attains a superior performance during the inference process. 
Morever, the connections between sparse modeling and deep neural networks has also been exploited. For example, the multi-layer convolutional (synthesis) sparse coding model~\cite{papyan:17:cnn,Sulam:18:mcs} provides a new interpretation of convolutional
neural networks (CNNs), where the pursuit of sparse representation from a given input signal complies with the forward pass in a CNN. In the meantime, multi-layer sparsifying transforms make the most direct connection with CNNs in the model and enable sparsifying an input image successively over layers \cite{ravishankar:18:lml}, creating a rich and more complete sparsity model, whose learning in an unsupervised manner and from limited data also forms the core of this work.

One of the most important applications of such image models is for medical image reconstruction. In particular, an important problem in X-ray computed tomography (CT) is reducing the X-ray exposure to patients while maintaining good image reconstruction quality.
A conventional method for CT reconstruction is the analytical filtered back-projection (FBP) \cite{feldkamp:84:pcb}. However, image quality degrades severely for FBP when the radiation dose is reduced. In contrast, model-based image reconstruction (MBIR) exploits CT forward models and statistical models together with image priors to achieve often better image quality \cite{elbakri:02:sir}. 

A typical MBIR method for low-dose CT (LDCT) is the penalized weighted least squares (PWLS) approach. The cost function for PWLS includes a weighted quadratic data-fidelity term and a penalty term or regularizer capturing prior information or model of the object \cite{sauer:93:alu,thibault:06:arf,thibault:07:atd}. Recent works have shown promising LDCT reconstruction quality by incorporating data-driven models into the regularizer, where the models are learned from datasets of images or image patches. In particular, PWLS reconstruction with adaptive sparsifying transform-based regularization has shown promise for tomographic reconstruction \cite{pfister:14:mbi,pfister:14:trw,pfister:14:ast,zheng:18:pua,chun:17:esv}. Recent work has also shown that they may generalize better to unseen new data than supervised deep learning schemes \cite{ye:19:sld}. The adaptive transform-based image reconstruction algorithms can exploit a variety of image models \cite{pfister:14:mbi,zheng:18:pua,zhou:13:atf} learned in an unsupervised manner from limited training images, and involve efficient closed-form solutions for sparse coding.

In this work, we  propose a new formulation and algorithm for learning a multi-layer transform model \cite{ravishankar:18:lml}, where the transform domain residuals (the difference between transformed data and their sparse approximations) are successively sparsified over several layers. We refer to the model as the Multi-lAyer Residual Sparsifying transform (MARS) model. The transforms are learned over several layers from images to jointly minimize the transform domain residuals across layers, while enforcing sparsity conditions in each layer. Importantly, the filters beyond the first layer can help better exploit finer features (e.g., edges and correlations) in the residual maps. We investigate the performance of unsupervised learning of MARS models from limited data for LDCT reconstruction using PWLS. We propose efficient block coordinate descent algorithms for both learning and reconstruction.
Experimental results with the XCAT phantom and Mayo Clinic data illustrate that the learned MARS model 
outperforms conventional methods such as FBP and PWLS methods based on the non-adaptive edge-preserving (EP) regularizer in terms of two numerical metrics (RMSE and SSIM) and noise suppression. Compared with the recent learned single-layer transform model, the MARS model performs better in maintaining some subtle details.
%outperforms the conventional FBP as well as PWLS methods based on the non-adaptive edge-preserving (EP) regularizer and the recent learned single-layer transform model, especially for reducing noise and maintaining some subtle details. 

In the following sections, we will first study how to train our proposed model in detail in Section II, where we will discuss the corresponding problem formulations in Section II-A, followed by our algorithms in Section II-B. The experimental results with both the XCAT phantom and Mayo Clinic data are presented in Section III. Section IV presents a discussion of the proposed methods and results and concludes.

\section{Methods}
\subsection{Formulations for MARS Training and LDCT reconstruction}
\begin{figure}[!h]
	\centering
	\vspace{0.05in}
	\includegraphics[width=0.95\textwidth]{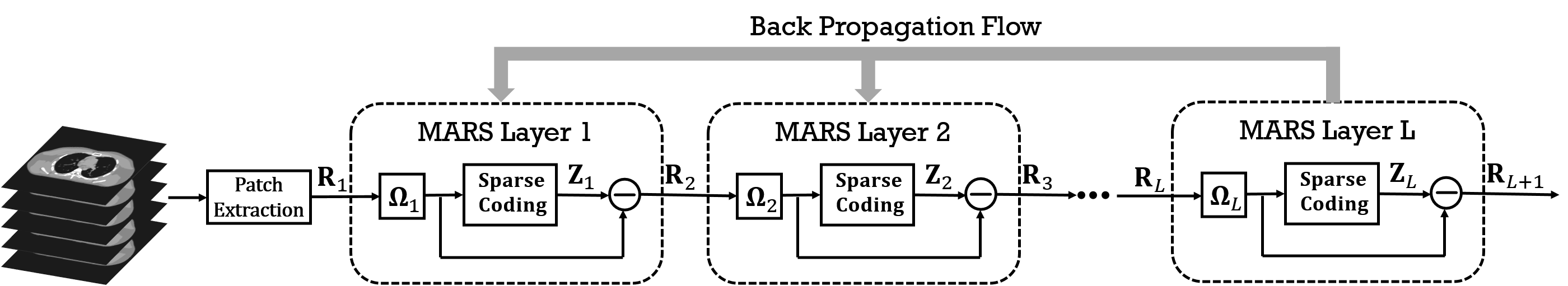}
	%	\vspace{-0.05in}
	\caption{MARS model with $L$ layers or modules. $\omg_l$ denotes the transform in the $l$th layer, which enables sparsifying the residual map arising from the $(l-1)$th module.}
	\label{fig:MRST_structure}
	%	\vspace{-0.27in}
\end{figure}

Here, we introduce the proposed general multi-layer transform learning framework and the formulation for LDCT image reconstruction. Fig.~\ref{fig:MRST_structure} illustrates the structure of our multi-layer residual sparsifying transform model, where $\omg_l$ denotes the transform in the $l$th layer. These transforms capture higher order image information by sparsifying the transform domain residual maps layer by layer.
The MARS learning cost and constraints are shown in Problem \eqref{eq:P0}, which is an extension of simple single-layer transform learning~\cite{ravishankar:13:lst,ravishankar:18:lml}.
\begin{equation}\label{eq:P0}
\begin{aligned}
& \quad \quad \quad \quad \min_{\{\omg_l,\Z_l\}}   \sum_{l=1}^L \bigg\{  \|\omg_l\R_l - \Z_l\|_F^2 + \eta_l^2\|\Z_l\|_0  \bigg\},
\\
& \quad \mathrm{s.t.} \quad \R_l = \omg_{l-1}\R_{l-1} - \Z_{l-1}, 2\leq l\leq L,\;\; \omg_l^T \omg_l  = \I, \forall l.
\end{aligned}
\tag{P0}
\end{equation}
Here, $\{\omg_l\in \mathbb{R}^{p \times p}\}$ and $\{\Z_l\in \mathbb{R}^{p \times N}\}$ denote the sets of learned transforms and sparse coefficient maps, respectively, for the $1 \leq l \leq L$ layers and ``F'' denotes the Frobenius norm. The total number of training patches is denoted by $N$. Parameter $\eta_l$ controls the maximum allowed sparsity level (computed using the $\ell_0$ ``norm'' penalty) for $\Z_l$. The residual maps $\{\R_l\in \mathbb{R}^{p \times N}\}$ are defined in recursive form over layers, with $\R_1$ denoting the input training data. We assume $\mathbf{R}_1$ to be a matrix, whose columns are (vectorized) patches drawn from image data sets. The unitary constraint for each $\omg_l$ enables closed-form solutions for the sparse coefficient and transform update steps in our algorithms.
The MARS model learned via \eqref{eq:P0} can then be used to construct a data-driven regularizer in PWLS as shown in Problem \eqref{eq:P1}.
\vspace{-0.05in}
\begin{equation}\label{eq:P1}	
\min_{\x \geq \0}  \frac{1}{2}\|\y - \A \x\|^2_{\W}  + \beta \Rsf(\x),   
\tag{P1}
\end{equation}
%\vspace{-0.10in}
\begin{equation*}\label{eq:Rx_DST}
\begin{aligned}
\quad \quad \quad &\Rsf(\x) \triangleq  \min_{\{\Z_l\}}  \sum_{l=1}^{L}  \bigg\{  \|\omg_l \R_l - \Z_l\|^2_F + \gamma_l^2\|\Z_l\|_0   \bigg\},
%\\ &\;   \mathrm{s.t.} \;  \textbf{r}_l^j =   \omg_{l-1} %\textbf{r}_{l-1}^j - \z_l^j, 2 \leq l \leq L \, ,  \textbf{r}_1^j = %\P^j\x, \, \omg_l^T \omg_l   = \I, \forall \, j, l \, .
\\ &\;   \mathrm{s.t.} \;  \R_l = \omg_{l-1}\R_{l-1} - \Z_{l-1}, 2 \leq l \leq L \, ,  \R_1^j = \P^j\x, \, \forall \, j.
\end{aligned}
%\vspace{-0.05in}
\end{equation*}
In particular, we reconstruct the image $\x \in \mathbb{R}^{N_p}$ from noisy sinogram data $\y \in \mathbb{R}^{N_d}$ by solving \eqref{eq:P1}, where $N_p$ denotes the number of pixels. $\A \in  \mathbb{R}^{  N_d  \times N_p}$ is the system matrix of the CT scan and $\W = \diag \{w_i\} \in \mathbb{R}^{N_d \times N_d}$ is the diagonal weighting matrix with elements being the estimated inverse variance of $y_i$. Operator $\P^j\in \mathbb{R}^{p\times N_p}$ extracts and vectorizes the $j$th patch of $\x$ as $\P^j\x$. Overlapping image patches are extracted with appropriate patch stride (1 pixel stride in our experiments). The $j$th columns of $\R_l$ and $\Z_l$ are denoted $\Z_l^j$ and $\R_l^j$, respectively. The non-negative parameters $\{\gamma_l\}$ control the sparsity of the coefficient maps in different layers, and $\beta>0$ captures the relative trade-off between the data-fidelity term and regularizer.
%\vspace{-0.05in}

\subsection{Algorithms for Learning and Reconstruction}\label{alg:MARS}
\begin{figure}[!h]
	\centering
	\vspace{0.05in}
	\includegraphics[width=0.9\textwidth]{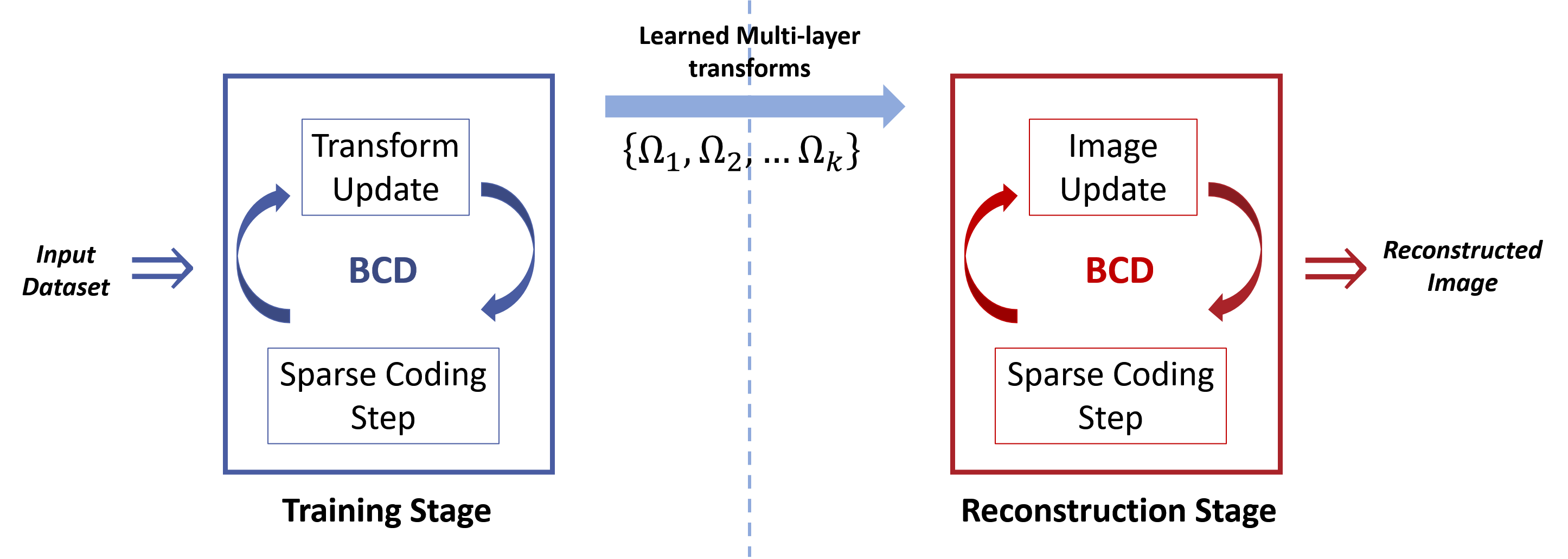}
	%	\vspace{-0.05in}
	\caption{Overview of algorithm scheme. Our approach involves a training stage and a reconstruction stage with block coordinate descent (BCD) algorithms being used in both stages.}
	\label{fig:MRST_scheme}
	%	\vspace{-0.27in}
\end{figure}

Fig.~\ref{fig:MRST_scheme} provides an overview of the proposed method. The whole algorithm is divided into two stages: training and reconstruction. In the training stage, we solve \eqref{eq:P0} using a block coordinate descent (BCD) method to learn a multi-layer sparsifying transform model in an unsupervised manner from (unpaired) regular-dose images. For the reconstruction stage, the prior information incorporated into learned transform would be designed into regularizer term, and iterative algorithm accomplishes the reconstruction for the CT image as we will show in the later section. 

\subsubsection{MARS Learning Algorithm}
We propose an exact block coordinate descent (BCD) algorithm for the nonconvex Problem \eqref{eq:P0} 
that cycles over updating $\Z_l$  (\emph{sparse coding step}) followed by updating the corresponding $\omg_l$ (\emph{transform update step}) for $1 \leq l \leq L$. The algorithmic details are shown in \textbf{Algorithm~\ref{alg: mrst_train}}. In each step, the remainder of the variables (that are not optimized) are kept fixed. 
The BCD algorithm provides a very efficient way to minimize the cost function and is shown to empirically work well with appropriate initialization. 
Recent works involving transform learning \cite{ravishankar:15:lst, ye:19:sld} have shown that such efficient alternating minimization or BCD algorithms can provably converge to the critical points of the underlying problems.
%The exact BCD algorithm ensures that the nonnegative objective in \eqref{eq:P0} is monotone decreasing over iterations and that it converges. 
In particular, we show that under the unitarity condition on the transforms, every subproblem in the block coordinate descent minimization approach can be solved exactly. We initialize the algorithm with the 2D DCT for $\omg_1$ and the identity matrices for $\{\omg_l\}_{l=2}^{L}$, respectively. The initial $\{\Z_l\}$ are all-zero matrices.

%\begin{equation}\label{eq:P0}
%\begin{aligned}
%& \quad \quad \quad \quad \min_{\{\omg_l,\Z_l\}}   \sum_{l=1}^L \bigg\{  \|\omg_l\R_l - \Z_l\|_F^2 + \eta_l^2\|\Z_l\|_0  \bigg\}
%\\
%& \quad \mathrm{s.t.} \quad \R_l = \omg_{l-1}\R_{l-1} - \Z_{l-1}, 2\leq l\leq L, \omg_l^T \omg_l  = \I, \forall l.
%\end{aligned}
%\tag{P0}
%\end{equation}
Since the residuals are defined recursively in \eqref{eq:P0}, for the sake of simplicity of the algorithmic description, we first define matrices $\B_p^q (p < q)$, which can be regarded as backpropagation matrices from the $q$th to $p$th layers.
\begin{equation}\label{eq:B_pq}
\begin{aligned}
&\B_p^q=\omg_{p+1}^T\Z_{p+1}+\omg_{p+1}^T\omg_{p+2}^T\Z_{p+2}+...+\omg_{p+1}^T\omg_{p+2}^T...\omg_q^T\Z_q
\\
&=\sum_{k=p+1}^q \bigg( \prod_{s=p+1}^k \omg_s^T \bigg)\Z_k.
\end{aligned} 
\end{equation}

\noindent(a) Sparse Coding Step for $\Z_l$

Here, we solve (P0) for $\Z_l$ with all other variables fixed. The corresponding nonconvex subproblem is as follows:
\begin{equation}\label{eq:sub_pro_Z}
\vspace{-0.05in}
\min_{\Z_l}   \sum_{i=l}^L \bigg\{  \|\omg_i\R_i - \Z_i\|_F^2 \bigg\}  + \eta_l^2\|\Z_l\|_0.
\end{equation}	

\noindent Using the definitions of the residual matrices and the backpropagation matrices $\B_p^q \, (p < q)$ along with the unitary property of the transforms allows us to rewrite \eqref{eq:sub_pro_Z} as:
\begin{equation}\label{eq:sub_derive_Z_1}
\vspace{-0.05in}
\begin{aligned}
\min_{\Z_l}\|\Z_l-\omg_l\R_l\|_F^2 + \sum_{i=l+1}^L \|\Z_l+\B_l^i-\omg_l\R_l\|_F^2 + \eta_l^2\|\Z_l\|_0.
\end{aligned} 
\end{equation}

\noindent We can now rewrite subproblem \eqref{eq:sub_derive_Z_1} as $\min_{\Z_l}(L-l+1)\times\|\Z_l - ( \omg_l\R_l - \frac{1}{L-l+1}\sum_{i=l+1}^L \B_l^i )\|_F^2 + \eta_{l}^{2}\|\Z_l\|_0$. This problem has a similar form as the single-transform sparse coding problem \cite{ravishankar:13:lst}, and the optimal solution $\hat{\Z}_l$ is obtained as in \eqref{eq:Z_l}, where $H_{\eta}(\cdot)$ denotes the \textit{hard-thresholding} operator that sets elements with magnitude less than the threshold $\eta$ to zero.
%\small{
\begin{equation}\label{eq:Z_l}
\begin{aligned}
\hat{\Z}_l=
%\\
&\begin{cases}
H_{\eta_l/\sqrt{L-l+1}} \bigg( \omg_l\R_l - \frac{1}{L-l+1}\sum_{i=l+1}^L \B_l^i \bigg),&  1\leq l \leq L-1,\\
H_{\eta_L} (\omg_L\R_L),& l=L.
\end{cases}
\end{aligned}
\end{equation}
%}
%when $1\leq l \leq L-1$:
%\begin{equation}\label{eq:Z_l}
%\hat{\Z}_l=H_{\eta_l/\sqrt{L-l+1}} \bigg( \omg_l\R_l - \frac{1}{L-l+1}\sum_{i=l+1}^L \B_l^i \bigg)
%\end{equation}
%
%when $l=L$:
%\begin{equation}\label{eq:Z_L}
%\hat{\Z}_L=H_{\eta_L} (\omg_L\R_L)
%\end{equation}
\noindent(b) Transform Update Step for $\omg_l$

Here, we fix $\{\Z_l\}$ and all $\omg_j$ (except the target $\omg_l$ in \eqref{eq:P0}) and solve the following subproblem:
\vspace{-0.05in}
\begin{equation}\label{eq:sub_pro_mOmega}
\min_{\omg_l}   \sum_{i=l}^L \bigg\{ \|\omg_i\R_i - \Z_i\|_F^2 \bigg\}
\quad \mathrm{s.t.} \quad  \omg_l^T \omg_l  = \I. 
\end{equation}

\noindent Similar to \eqref{eq:sub_derive_Z_1}, we rewrite \eqref{eq:sub_pro_mOmega} using the backpropagation matrices $\B_p^q\,(p < q)$ as follows:
\begin{equation}\label{eq:sub_derive_mOmega}
\vspace{-0.05in}
\begin{aligned}
&\min_{\omg_l: \omg_l^T \omg_l = \I}\|\omg_l\R_l - \Z_l\|_F^2 + \sum_{i=l+1}^L \|\omg_l\R_l - \Z_l - \B_l^i\|_F^2,
\\
\sim &\min_{\omg_l: \omg_l^T \omg_l = \I}(L-l+1)\times\bigg\|\omg_l\R_l - \Z_l - \frac{1}{L-l+1}\sum_{i=l+1}^L \B_l^i \bigg\|_F^2,
\end{aligned} 
\end{equation}

\noindent where the last relation (equality) holds up to an additive term that is independent of $\omg_l$. We can obtain a solution to \eqref{eq:sub_derive_mOmega} by exploiting the unitarity of $\omg_l$. First, denoting the full singular value decomposition (SVD) of the matrix $\mathbf{G}_l$ below by $\U_l\Sig_l\V_l^T$, the optimal solution to \eqref{eq:sub_derive_mOmega} is as \eqref{eq:mOmega_l_sol}.
\vspace{-0.10in}
\begin{equation}\label{eq:mOmega_l}
\mathbf{G}_l =
\begin{cases}
\R_l\bigg( \Z_l + \frac{1}{L-l+1} \sum_{i=l+1}^L \B_l^i \bigg)^T, & 1\leq l \leq L-1,\\
\R_L\Z_L^T, & l=L.
\end{cases}
\vspace{-0.05in}
\end{equation}

\vspace{-0.05in}
\begin{equation}\label{eq:mOmega_l_sol}
\hat{\omg_l}=\V_l\U_l^T
%\vspace{-0.05in}
\end{equation}

\begin{algorithm}[H]  
	\caption{MARS Learning Algorithm}\label{alg: mrst_train}
	\begin{algorithmic}[0]
		\State \textbf{Input:}
		training data $\R_1$, all-zero initial $\{\tilde{\Z}_l^{(0)}\}$, initial $\tilde{\omg}_1^{(0)} =$ 2D DCT, identity matrices for initial ${\{\tilde{\omg}_l^{(0)}\}}_{l=2}^L$, thresholds $\{\eta_l\}$, number of iterations $T$.			
		\State \textbf{Output:}  learned transforms $\{\tilde{\omg_l}^{(T)}\}$.
		\For {$t =1,2,\cdots,{T}$}		
		\For {$l =1,2,\cdots,{L}$}		
		\State \textbf{1)} Sparse Coding for $\tilde{\Z_l}^{(t)}$ via \eqref{eq:Z_l}.	
		
		\State \textbf{2)} Updating $\tilde{\omg_l}^{(t)}$ via \eqref{eq:mOmega_l_sol}.
		
		\EndFor	
		\EndFor	
	\end{algorithmic}
\end{algorithm}

\subsubsection{Image Reconstruction Algorithm}\label{sec:recon_alg}
The proposed PWLS-MARS algorithm for low-dose CT image reconstruction exploits the learned  model. We reconstruct the image by solving the PWLS problem \eqref{eq:P1}. We propose a block coordinate descent (BCD) algorithm for \eqref{eq:P1} that cycles over updating the image $\x$ and each of the sparse coefficient maps $\Z_l$ for $1 \leq l \leq L$.

\noindent(a) Image Update Step for $\x$
\vspace{-0.05in}

First, with the sparse coefficient maps $\{\Z_l\}$ fixed, we optimize for $\x$ in \eqref{eq:P1} by optimizing the following subproblem:
\vspace{-0.05in}
\begin{equation} \label{eq:sub_P1}
\min_{\x \geq \0} \frac{1}{2} \|\y - \A\x \|^2_{\W} + \beta \Rsf_2(\x),   
\end{equation}
\vspace{-0.05in}
where $\Rsf_2(\x) \triangleq   \sum_{l=1}^{L}  \bigg\{ \|\omg_l\R_l - \Z_l\|_F^2 \bigg\} $, with $\R_l=\omg_{l-1}\R_{l-1} - \Z_{l-1}$, $2 \leq l \leq L$, and $\R_1^j = \P^j \x$.	
We use the efficient relaxed linearized augmented Lagrangian method\cite{nien:16:rla} (relaxed LALM) to obtain the solution to \eqref{eq:sub_P1}. The algorithmic details are shown in \textbf{Algorithm~\ref{alg: mrst_recon}}. In each iteration of the relaxed LALM, we update the image $T_i$ times (corresponding to $T_i$ inner loops in \textbf{Algorithm~\ref{alg: mrst_recon}}).  We let matrix $\D_{\A}$ denote a diagonal majorizing matrix of $\A^T\W\A$ and precompute the Hessian matrix of  $\Rsf_2(\x)$ as $\D_{\Rsf_2}$ in \eqref{eq:D_R} to accelerate the algorithm, and the gradient of $\Rsf_2(\x)$ is shown in \eqref{eq:grad_R}, where $(\B_0^k)^j$ denotes the jth column of matrix $\B_0^k$. We decrease the parameter $\rho$ in Algorithm \ref{alg: mrst_recon} according to \eqref{eq:rho_decrease}~\cite{nien:16:rla}, where $r$ denotes the index of inner iterations and the relaxation parameter $\alpha\in [1,2)$ in \eqref{eq:rho_decrease}.
\vspace{-0.05in}
\begin{equation}\label{eq:grad_R}		
\nabla \Rsf_2(\x)= 2 \beta \sum_{j=1}^{N_p} (\P^j)^T \bigg\{ L\P^j\x - \sum_{k=1}^L (\B_0^k)^j \bigg\},
\end{equation}	
\vspace{-0.05in}
\begin{equation}\label{eq:D_R}		
\D_{\Rsf_2}  \triangleq  \nabla^2 \Rsf_2(\x) = 2L \beta \sum_{j=1}^{N_p}  (\P^j)^T\P^j, \\
\vspace{-0.05in}
\end{equation}
\begin{equation}\label{eq:rho_decrease}		
\rho_r(\alpha) =  
\begin{cases}
1, & r=0,\\
\frac{\pi}{\alpha(r+1)}\sqrt{1-(\frac{\pi}{2\alpha(r+1)})^2}, & \text{otherwise},
\end{cases}\\
%\vspace{-0.05in}
\end{equation}
%\begin{equation}\label{eq:D_A}		
%\D_{\A}  \triangleq \diag\{\A^T\W\A\textbf{1}\} \succeq \A^T\W\A  \\
%\end{equation}
\begin{algorithm}[H]  
	\caption{Image Reconstruction Algorithm}\label{alg: mrst_recon}
	\begin{algorithmic}[0]
		\State \textbf{Input:}
		initial image $\tilde{\x}^{(0)}$, all-zero initial $\{\tilde{\Z}_l^{(0)}\}$, pre-learned $\{\omg_l\}$, thresholds $\{\gamma_l\}$, \\
		$\alpha = 1.999$, $\D_{\A} $, $\D_{\Rsf_2}$, number of outer iterations $T_{O}$, number of inner iterations $T_i$.				
		\State \textbf{Output:}  reconstructed image $\tilde{\x}^{(T_{O})}$.
		\For {$t =0,1,2,\cdots,{T_{O}-1}$}		
		
		\State \textbf{1) Image Update}: With $\{\tilde{\Z}_l^{(t)}\}$ fixed,	
		
		\textbf{Initialization:} $\rho=1$, $\x^{(0)} = \tilde{\x}^{(t)}$, $\g^{(0)} = \ze^{(0)}  = \A^{T}\W(\A\x^{(0)}-\y) $ and $\h^{(0)} = \D_\A \x^{(0)} - \ze^{(0)}$.

		\For {$r =0,1,2,\cdots,T_i-1,$}				
		\begin{equation*}
		\left\{			
		\begin{aligned}
		\s^{(r+1)} &= \rho(\D_\A \x^{(r)} -\h^{(r)}) + (1-\rho)\g^{(r)} \\
		\x^{(r+1)} &= [\x^{(r)} - (\rho\D_\A+\D_{\Rsf_2})^{-1}(\s^{(r+1)} +\nabla \Rsf_2(\x^{(r)}))]_+ \\
		\ze^{(r+1)} & \triangleq  \A^T\W(\A\x^{(r+1)}-\y)   \\
		\g^{(r+1)} &= \frac{\rho}{\rho+1}(\alpha \ze^{(r+1)} + (1-\alpha)\g^{(r)}) +  \frac{1}{\rho+1}\g^{(r)}\\
		\h^{(r+1)}  &= \alpha(\D_{\A} \x^{(r+1)} -\ze^{(r+1)}) + (1-\alpha)\h^{(r)} 
		\end{aligned}
		\right.
		\end{equation*}  
		%		\State decreasing $\rho$ using \eqref{eq:rho}. % where $r = nM +m$.	
		\State decreasing $\rho$ using \eqref{eq:rho_decrease}. % where $r = nM +m$.		 	 		
		\EndFor	
		\State   $\tilde{\x}^{(t+1)} = \x^{(T_i)}$. 
		\State \textbf{2) Sparse Coding}: with $\tilde{\x}^{(t+1)}$ fixed, for each $1\leq l \leq L$, update $\tilde{\Z}_l^{(t+1)}$ sequentially by \eqref{eq:recon_formula_z_l}.
		\EndFor	
	\end{algorithmic}
\end{algorithm}
\noindent(b) Sparse Coding Step for Each $\Z_l$

Similar to the sparse coding step during transform learning, the solution of \eqref{eq:P1} with respect to each sparse coefficient map $\Z_l$ is shown in \eqref{eq:recon_formula_z_l}, and is the solution of \eqref{eq:recon_pro_z_l}.
\vspace{-0.10in}
\begin{equation}\label{eq:recon_pro_z_l}
\begin{aligned}
\min_{\Z_l} \sum_{i=l}^L \bigg\{  \|\omg_i\R_i - \Z_i\|_F^2 \bigg\}  + \gamma_l^2\|\Z_l\|_0, 
\\
\quad \mathrm{s.t.} \quad \R_i=\omg_{i-1}\R_{i-1} - \Z_{i-1},\;\; l \leq i \leq L,
\end{aligned}
\end{equation}		
\vspace{-0.05in}
\begin{equation}\label{eq:recon_formula_z_l}
\hat{\Z}_l=H_{\gamma_l/\sqrt{L-l+1}} \bigg\{ \omg_l\R_l - \frac{1}{L-l+1}\sum_{i=l+1}^L \B_l^i \bigg\}.
\end{equation}

%	\begin{equation}\label{eq:recon_formula_z_l}
%	\begin{aligned}
%	\hat{\Z}_l=
%	%\\
%	&\begin{cases}
%	H_{\gamma_l/\sqrt{L-l+1}} \bigg( \omg_l\R_l - \frac{1}{L-l+1}\sum_{i=l+1}^L \B_l^i \bigg),&  1\leq l \leq L-1,\\
%	H_{\gamma_L} (\omg_L\R_L),& l=L.
%	\end{cases}
%	\end{aligned}
%	\end{equation}

%\clearpage

\section{Experiments}
\vspace{-0.15in}
In this section, we evaluate the image reconstruction quality for the proposed PWLS-MARS algorithm and compare it with several conventional or related methods:

\noindent$\bullet$ \textbf{FBP}: conventional FBP method with a Hanning window.

\noindent$\bullet$ \textbf{PWLS-EP}~\cite{cho:15:rdf}: PWLS reconstruction combined with the edge-preserving regularizer $\mathsf{R}(\bf{x})=\sum_{j  =1}^{N_p} \sum_{k\in N_{j}}\kappa_{j} \kappa_{k} \phi(x_j - x_k)$, where $N_j$ denotes the set of neighborhood pixel indices, and $\kappa_{j}$ and $\kappa_{k}$ are the parameters that encourage uniform noise~\cite{cho:15:rdf}. We use $\phi(t) \triangleq \delta^2(\sqrt{1+\|t/\delta\|^2}-1)$ as the potential function. The relaxed OS-LALM~\cite{nien:16:rla} is the chosen optimizing approach for this PWLS cost function.

To compare the image quality quantitatively, we compute the root mean square error (RMSE) and the structural similarity index measure (SSIM)~\cite{xu:12:ldx,zhang:17:tbd}. The RMSE in Hounsfield units (HU) is computed between the ground truth image and reconstructed image as RMSE $= \sqrt{\Sigma_{i \in \text{ROI}}(\hat{x}_i-x^*_i)^2/{N_{\text{ROI}}}}$, where $\hat{x}_i$ and $x^*_i$ denote the pixel intensities of the reconstructed and ground truth images, respectively, and $N_{\text{ROI}}$ is the number of pixels in the region of interest (ROI). The ROI here was a circular (around center of image) region containing all the phantom tissues. We simulate the low-dose CT measurements using the ``Poisson + Gaussian" noisy model~\cite{ding:16:mmp}, i.e., $\hat{\bf{y}_i} = {\rm Poisson}\{I_0e^{-[\bf{Ax}]_i}\} + \mathcal{N}\{0, \sigma^2\}$, where $I_0$ is the incident X-ray intensity incorporating X-ray source illumination
and the detector gain, and $\sigma^2 = 5^2$ is the variance of electronic noise~\cite{ding:16:mmp}.

We conduct experiments with the XCAT phantom~\cite{Segars:08:rcs} and Mayo Clinic data~\cite{Mayo:16:data}, respectively.
Our first experiment uses the XCAT phantom data with a clean ground truth (reference) to demonstrate the performance of the MARS model over other schemes and illustrates the learned multi-layer filters.
%and algorithm convergence. 
In our second experiment, we investigate the performance of various methods on the Mayo Clinic data and provide a more detailed comparison between MARS and other methods. Lastly, we analyze the residual maps in the proposed model in different layers to better understand the MARS model.

\subsection{Parameter Selection}

For each MARS model, multiple parameters are tuned for the learning ($\{\eta_l, 1\leq l \leq L\}$) and reconstruction ($\beta, \{\gamma_l, 1\leq l \leq L\}$) stages. Even though the number of parameters here increases the difficulty of adjusting the model for optimal image quality, we can choose the values of the parameters with an empirical approach. The parameters $\{\eta_l\}$ during learning are to achieve a low sparsity of the sparse coefficient maps. Normally, we set $\{\eta_l\}$ to achieve $5-10\%$ sparsity for $\Z_l$. One clever method for selecting good sparsity penalty parameters is to set them in decreasing order over layers. This strategy is expected to work because the residual maps in subsequent layers always contain less (or finer) image information than the early layers. 
A similar approach works for adjusting parameters in the reconstruction stage.
In the reconstruction algorithm, we tune the parameters over ranges of values (decreasing over layers for $\gamma_l$) to achieve the best reconstruction quality (i.e., RMSE and SSIM). 

%\vspace{-0.15in}
\subsection{Results with the XCAT Phantom}
%\vspace{-0.10in}
\subsubsection{Behavior of the Learned MARS Models}
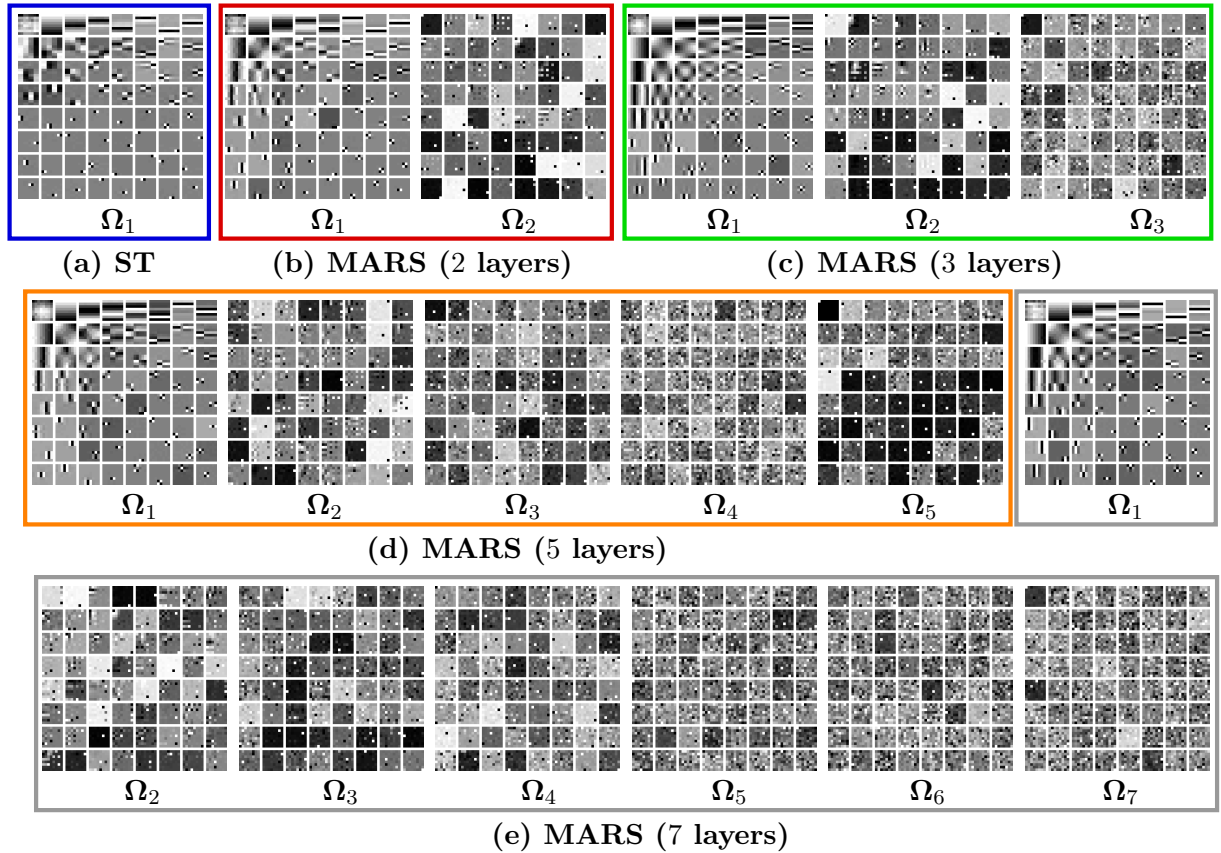
\begin{figure}[!h]
	\centering
	\begin{tabular}{ccc}
		\vspace{-0.15in}
		\hspace{-0.2in}
		\begin{tikzpicture}
		[spy using outlines={rectangle,green,magnification=2,size=6mm, connect spies}]
		\node {	\includegraphics[width=0.15\textwidth]{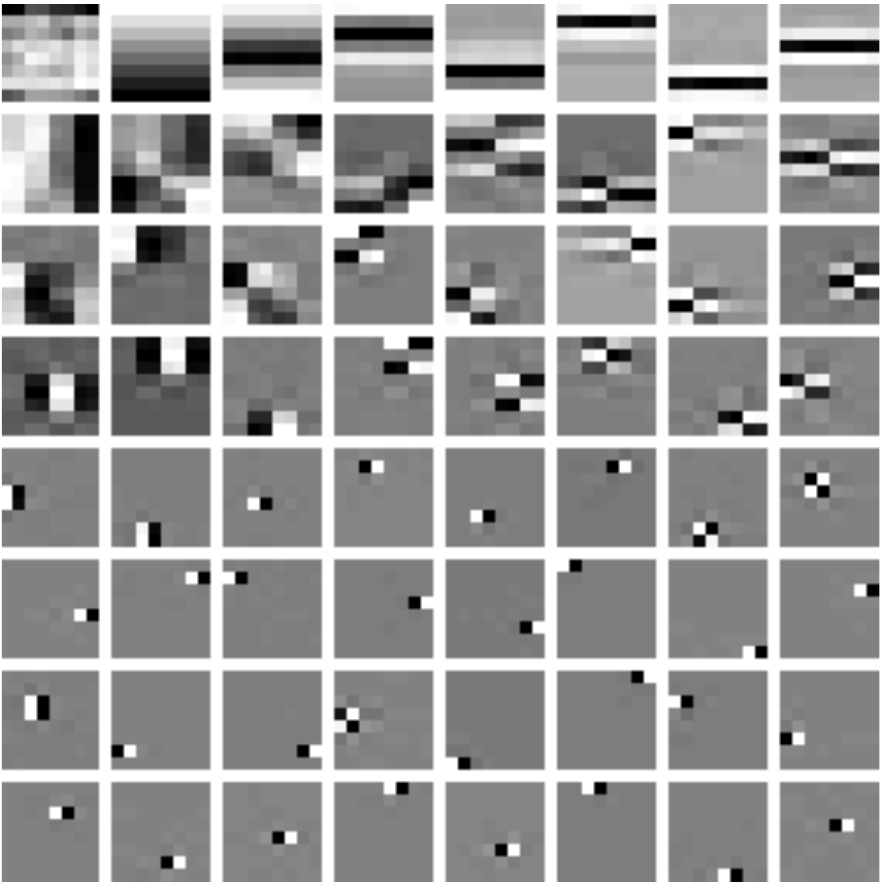}};
		\draw[blue, ultra thick] (-1.35,-1.75) rectangle (1.3,1.35);
		\put(-4,-46){ \color{black}{\bf \small{$\omg_1$}}}
		\end{tikzpicture}
		\put(-62,-12){ \color{black}{\bf \small{(a) ST}}}
		\begin{tikzpicture}
		[spy using outlines={rectangle,green,magnification=2,size=6mm, connect spies}]
		\node{\includegraphics[width=0.15\textwidth]{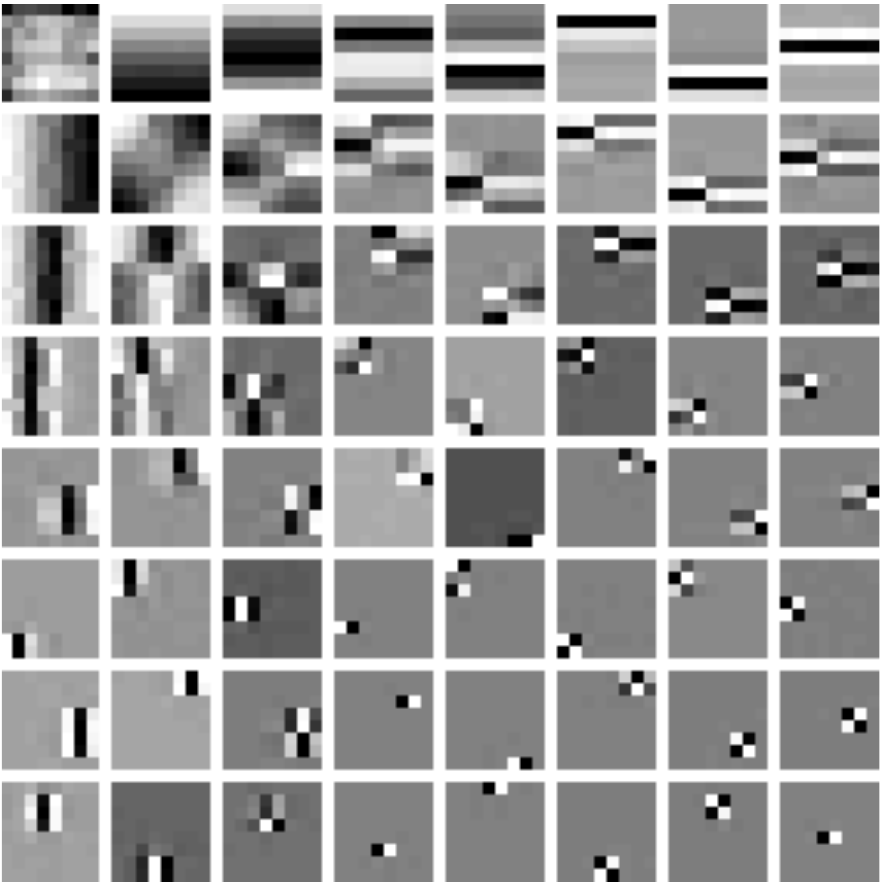}
			\includegraphics[width=0.15\textwidth]{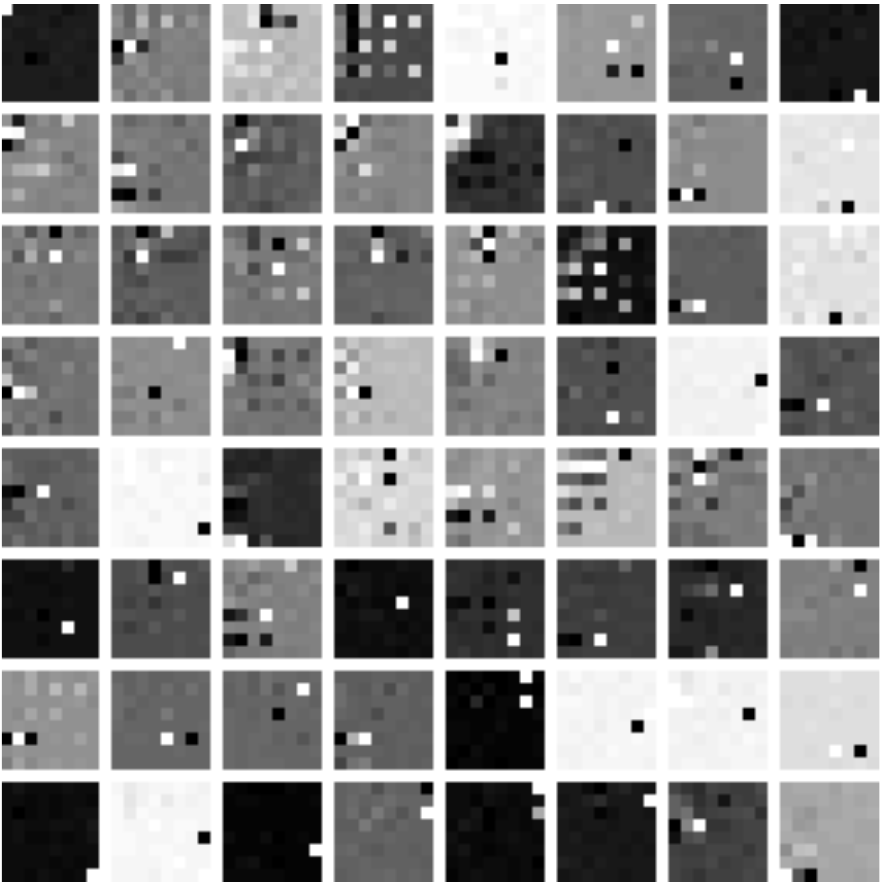}
		};
		\draw[red, ultra thick] (-2.60,-1.75) rectangle (2.6,1.35);
		\put(-40,-46){ \color{black}{\bf \small{$\omg_1$}}}
		\put(32,-46){ \color{black}{\bf \small{$\omg_2$}}}
		\end{tikzpicture}
		\put(-135,-12){ \color{black}{\bf \small{(b) MARS ($2$ layers)}}}
		\begin{tikzpicture}
		[spy using outlines={rectangle,green,magnification=2,size=6mm, connect spies}]
		\node{\includegraphics[width=0.15\textwidth]{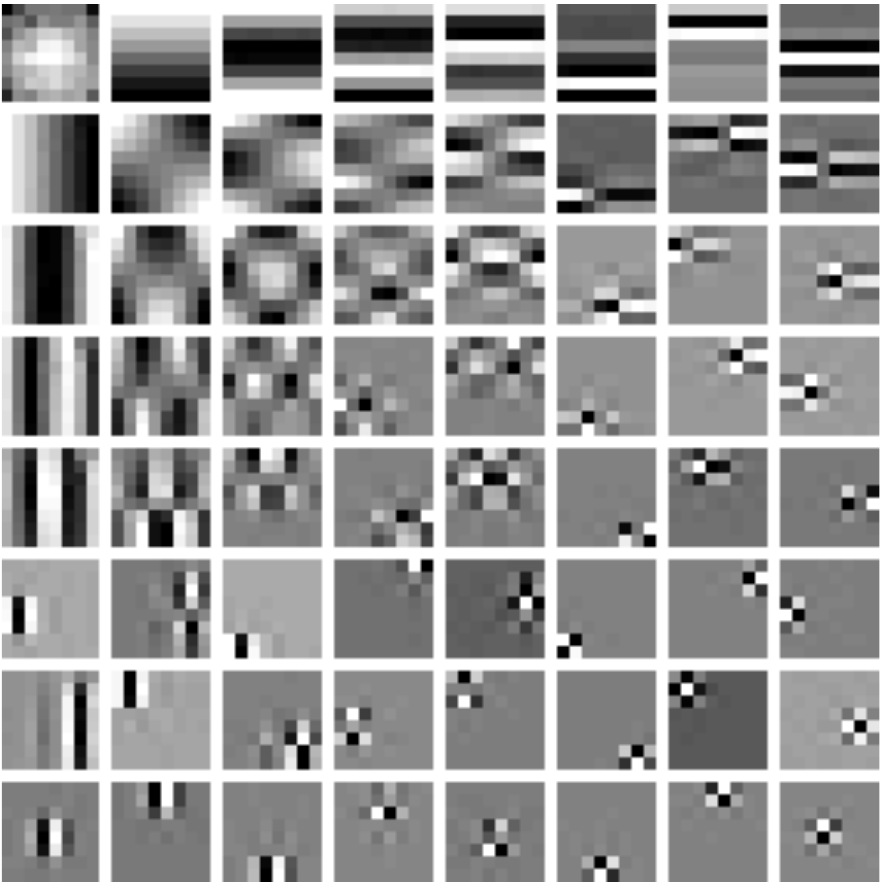}
			\includegraphics[width=0.15\textwidth]{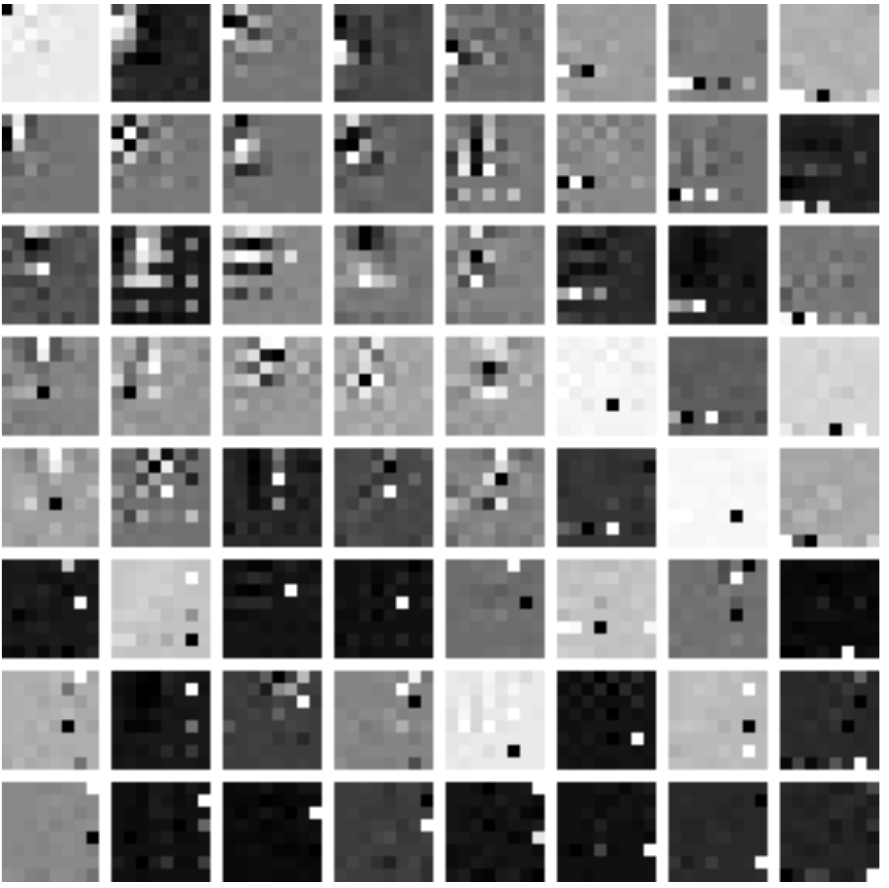}
			\includegraphics[width=0.15\textwidth]{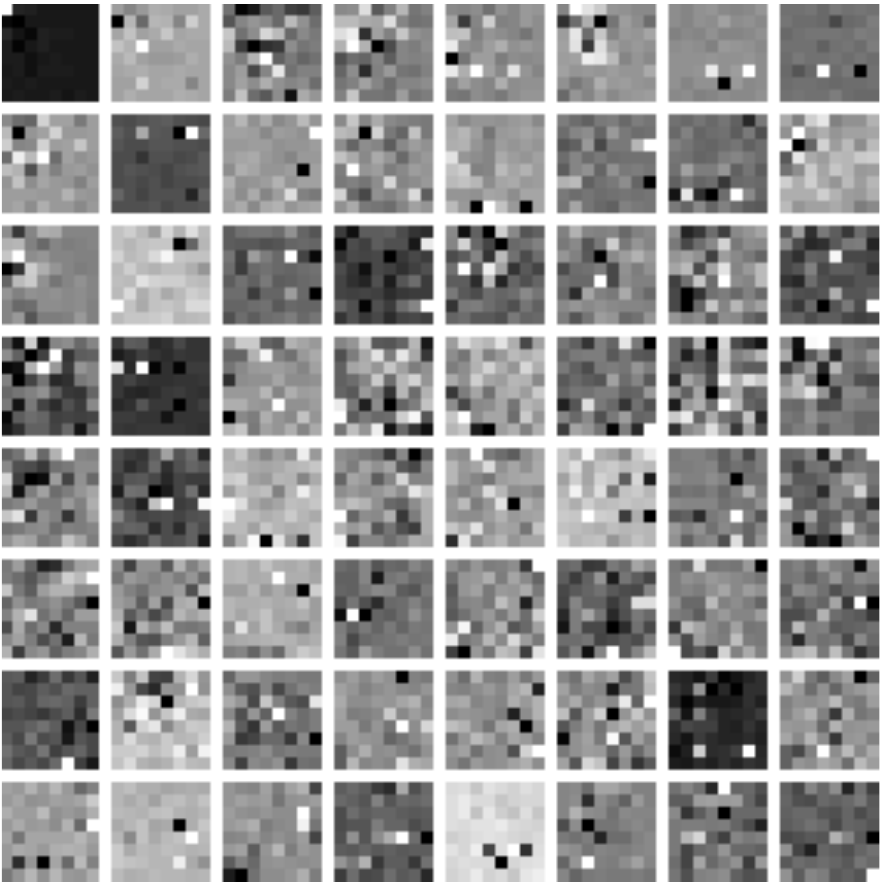}
		};
		\draw[green, ultra thick] (-3.90,-1.75) rectangle (3.9,1.35);
		\put(-80,-46){ \color{black}{\bf \small{$\omg_1$}}}
		\put(-5,-46){ \color{black}{\bf \small{$\omg_2$}}}
		\put(80,-46){ \color{black}{\bf \small{$\omg_3$}}}
		\end{tikzpicture}
		\put(-175,-12){ \color{black}{\bf \small{(c) MARS ($3$ layers)}}}
		\\
		\\
		\begin{tikzpicture}
		[spy using outlines={rectangle,green,magnification=2,size=6mm, connect spies}]
		\node{\includegraphics[width=0.15\textwidth]{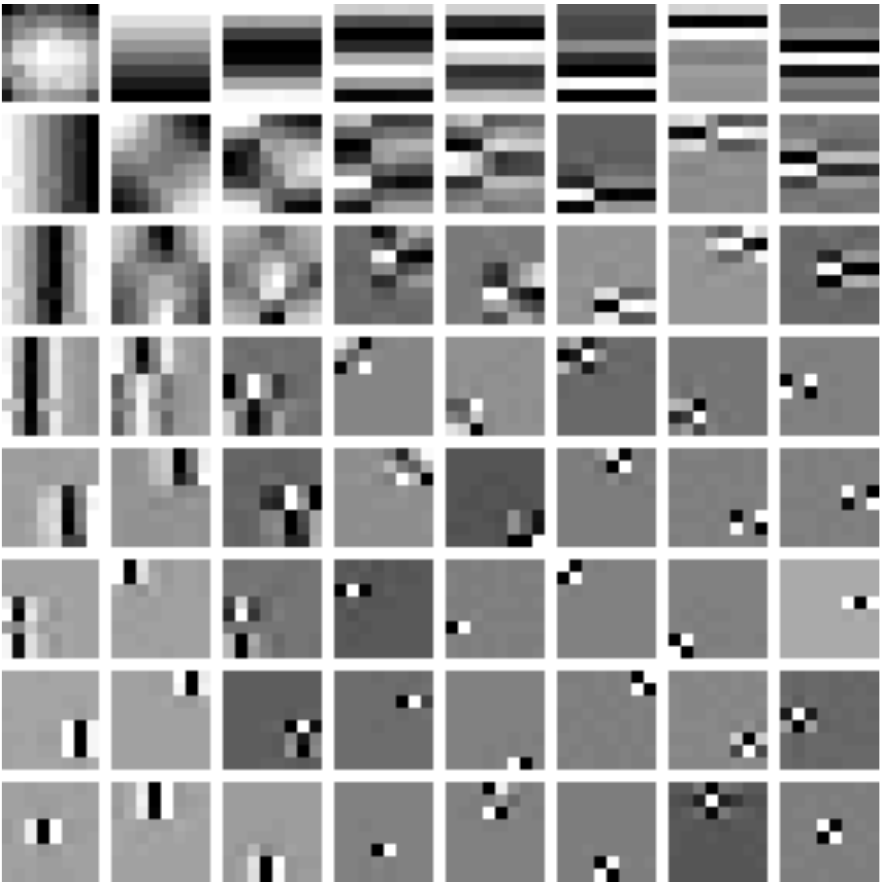}
			\includegraphics[width=0.15\textwidth]{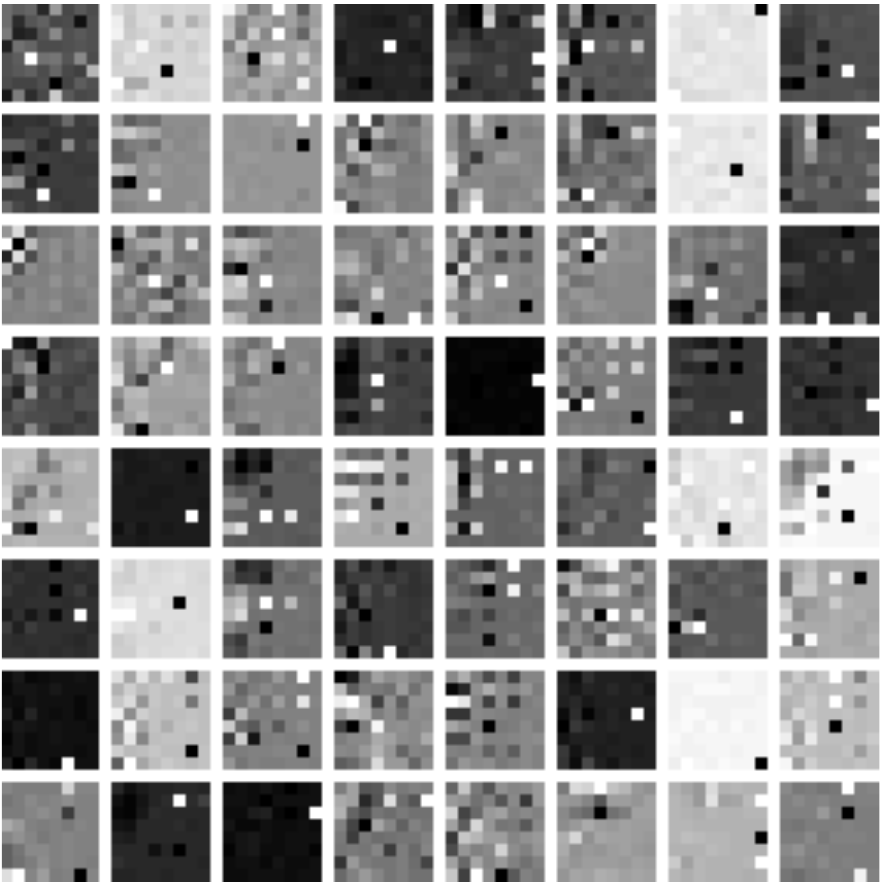}
			\includegraphics[width=0.15\textwidth]{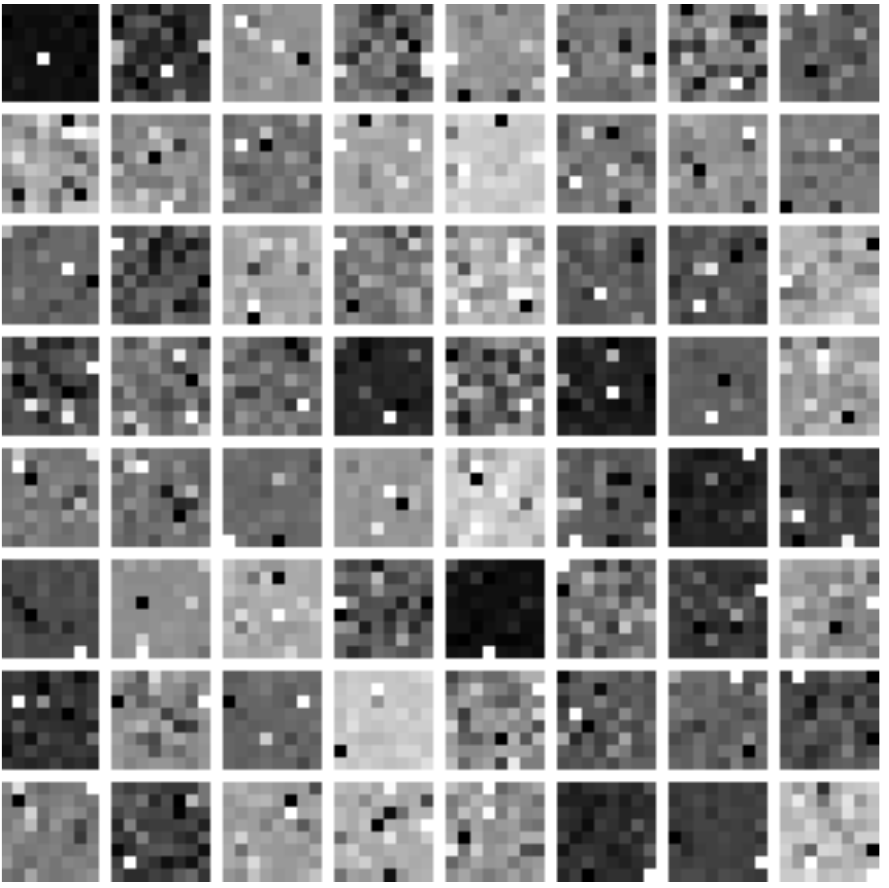}
			\includegraphics[width=0.15\textwidth]{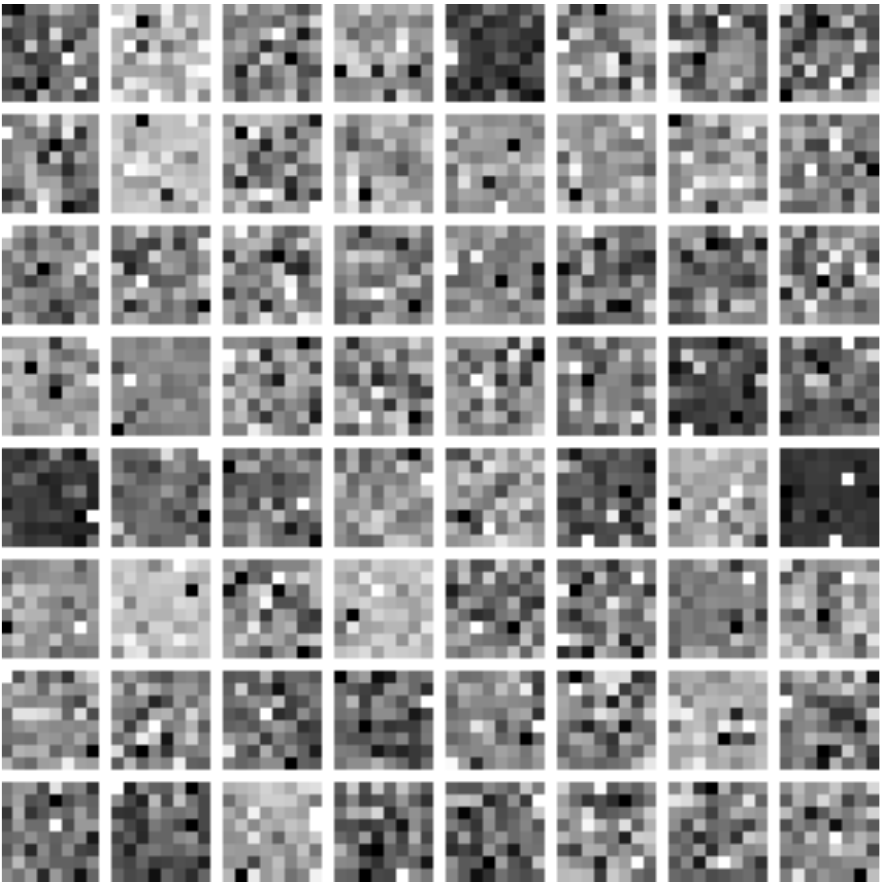}
			\includegraphics[width=0.15\textwidth]{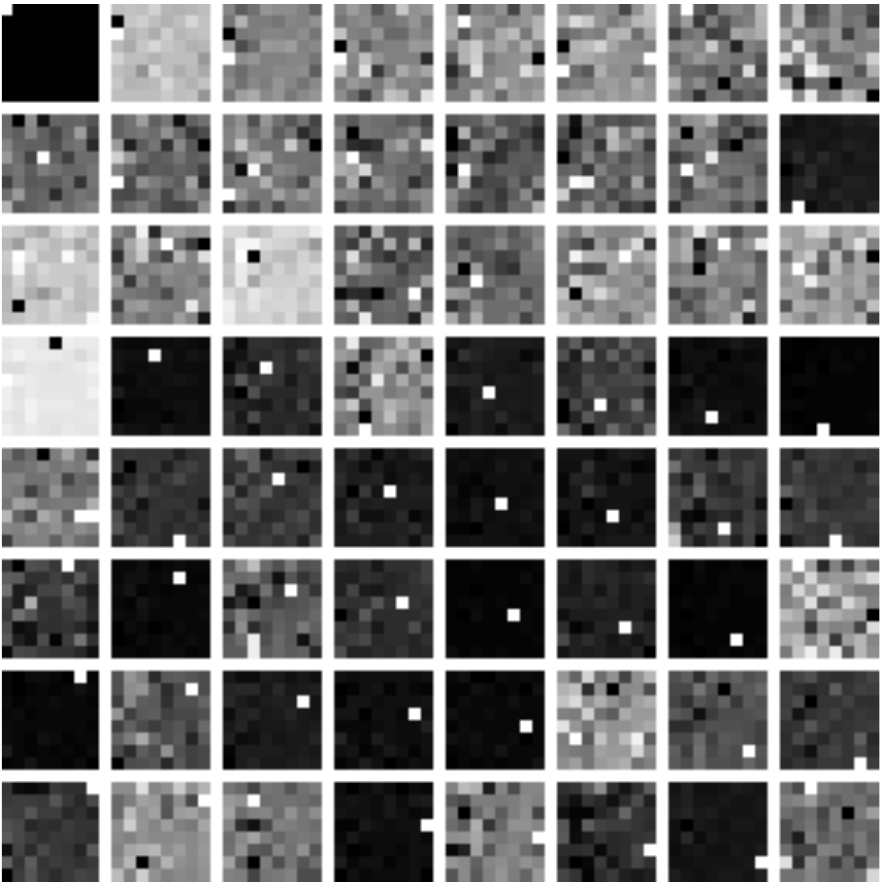}
		};
		\draw[orange, ultra thick] (-6.55,-1.75) rectangle (6.55,1.35);
		\put(-150,-46){ \color{black}{\bf \small{$\omg_1$}}}
		\put(-80,-46){ \color{black}{\bf \small{$\omg_2$}}}
		\put(-5,-46){ \color{black}{\bf \small{$\omg_3$}}}
		\put(70,-46){ \color{black}{\bf \small{$\omg_4$}}}
		\put(145,-46){ \color{black}{\bf \small{$\omg_5$}}}
		\end{tikzpicture}
		\put(-249,-12){ \color{black}{\bf \small{(d) MARS ($5$ layers)}}}
		\begin{tikzpicture}
		[spy using outlines={rectangle,green,magnification=2,size=6mm, connect spies}]
		\node {	\includegraphics[width=0.15\textwidth]{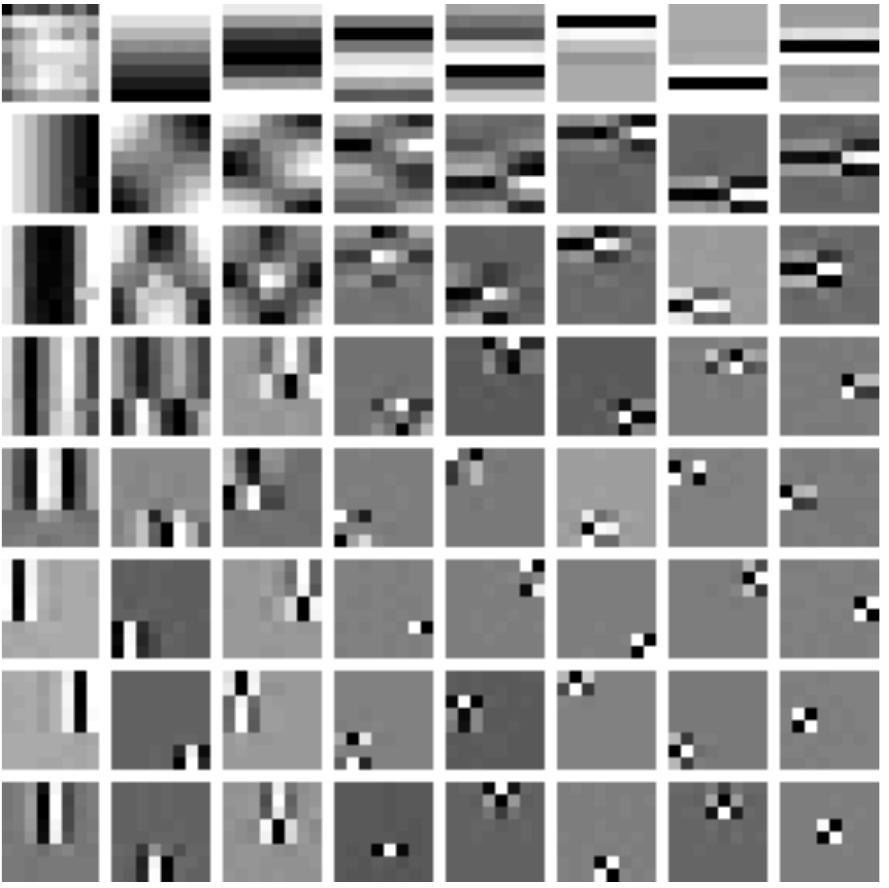}};
		\draw[gray, ultra thick] (-1.35,-1.75) rectangle (1.3,1.35);
		\put(-4,-46){ \color{black}{\bf \small{$\omg_1$}}}
		\end{tikzpicture}
		\vspace{0.05in}
		\\
		
		\begin{tikzpicture}
		[spy using outlines={rectangle,green,magnification=2,size=6mm, connect spies}]
		\node{\includegraphics[width=0.15\textwidth]{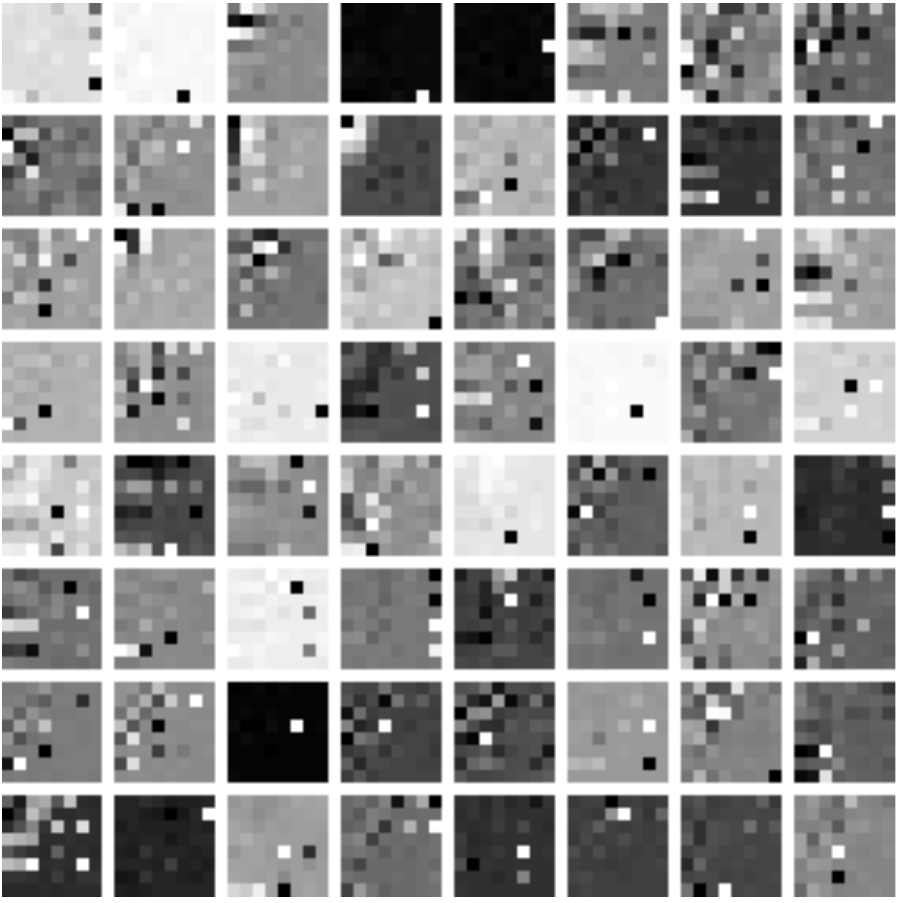}
			\includegraphics[width=0.15\textwidth]{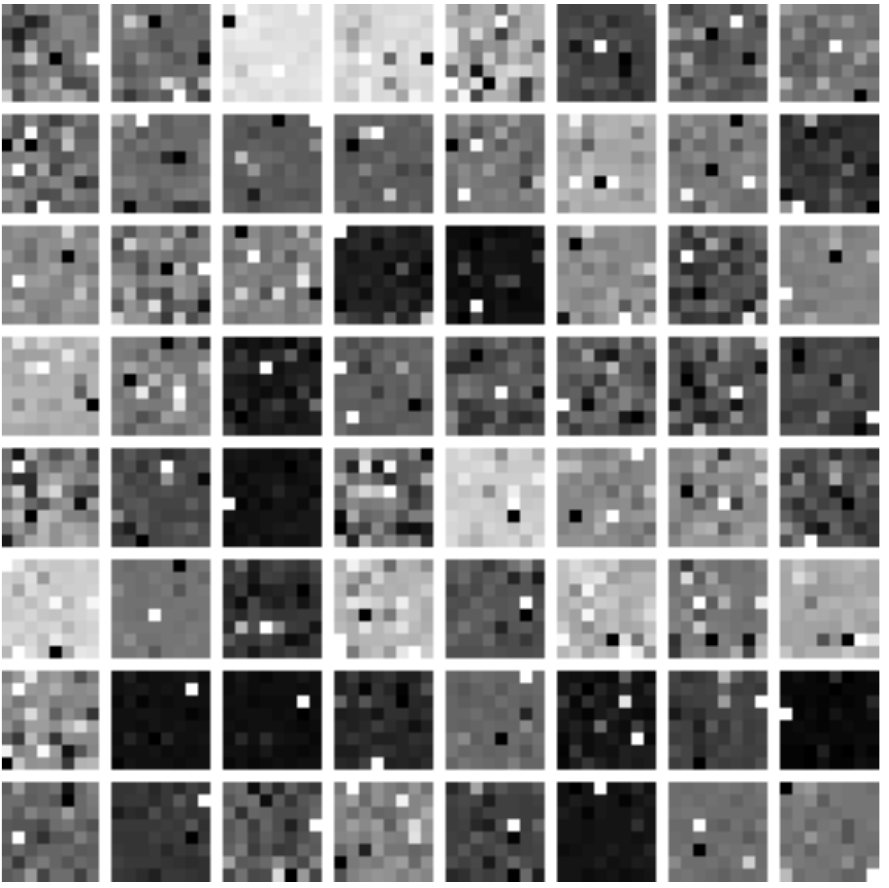}
			\includegraphics[width=0.15\textwidth]{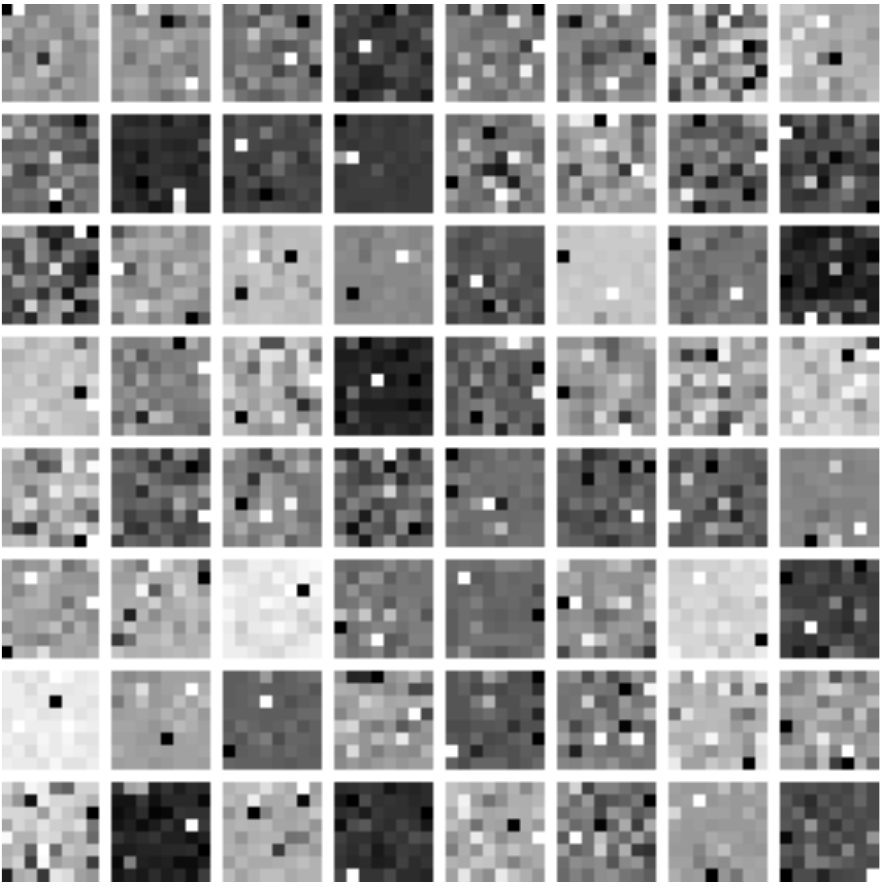}
			\includegraphics[width=0.15\textwidth]{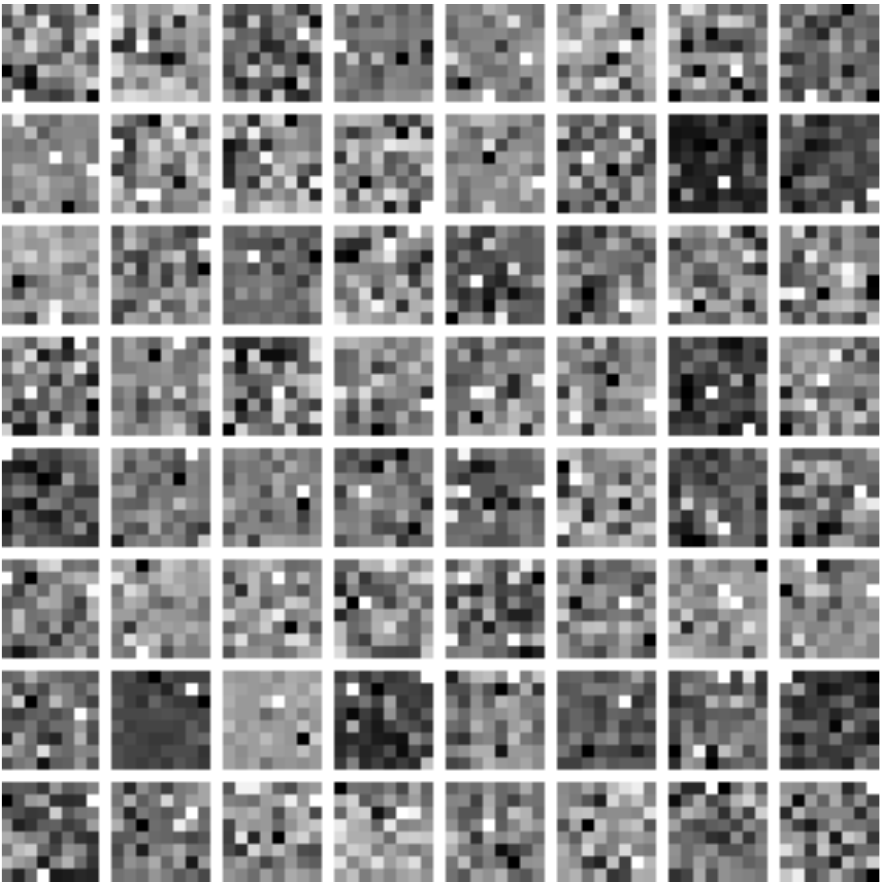}
			\includegraphics[width=0.15\textwidth]{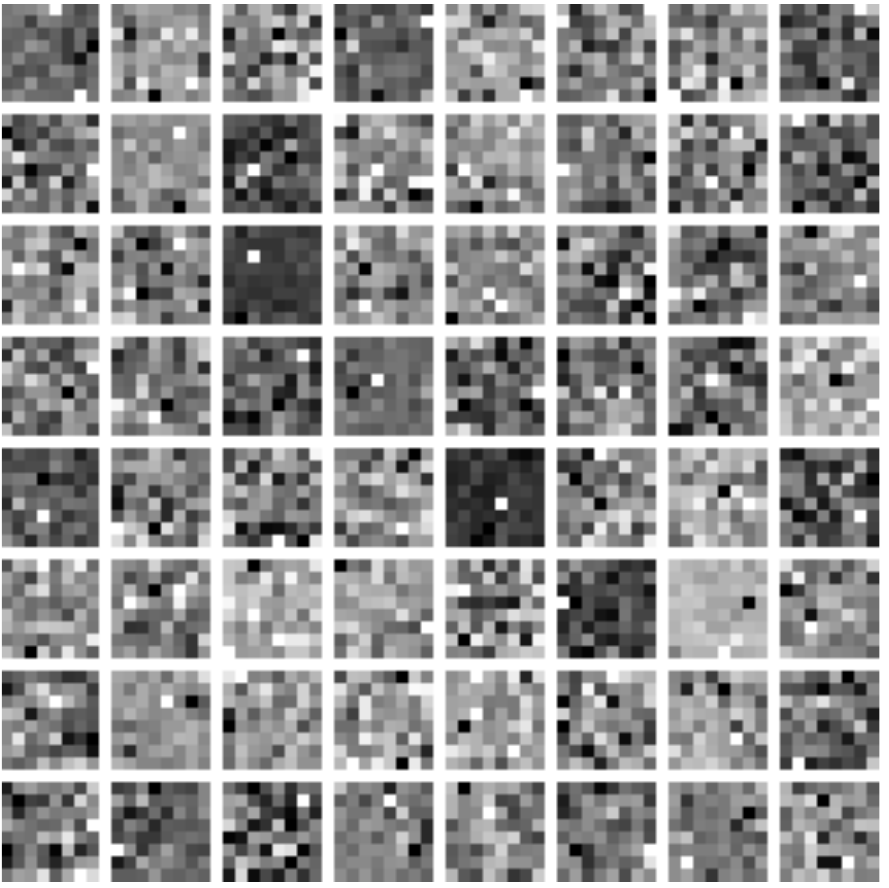}
			\includegraphics[width=0.15\textwidth]{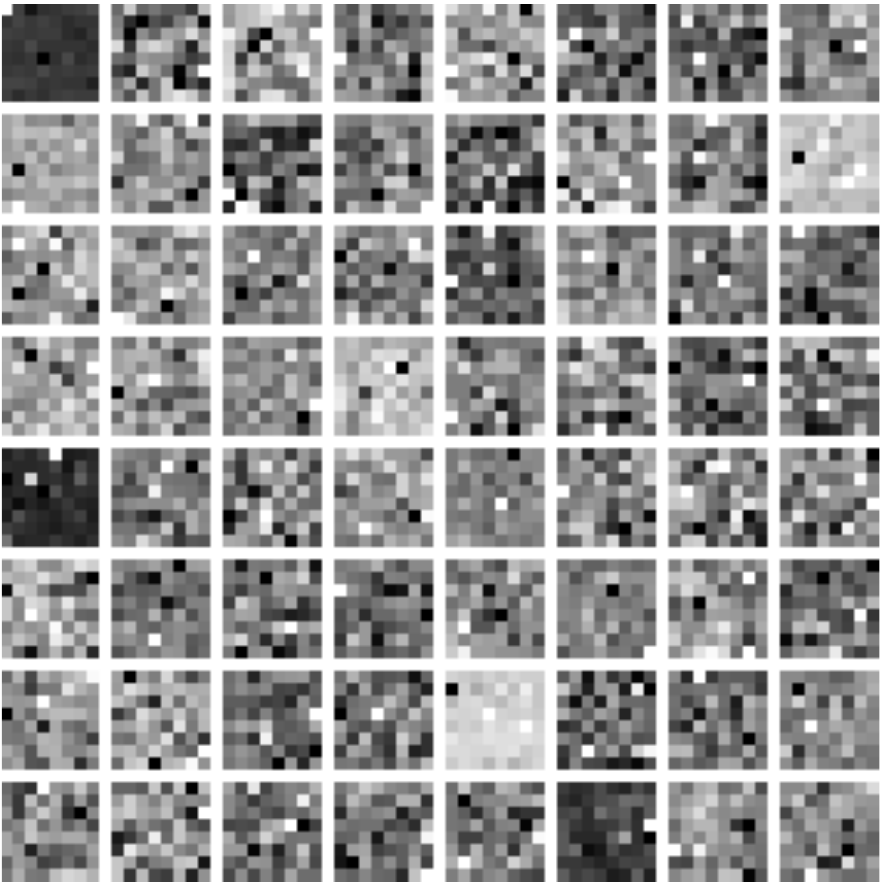}
		};
		\draw[gray, ultra thick] (-7.85,-1.75) rectangle (7.85,1.35);
		\put(-190,-46){ \color{black}{\bf \small{$\omg_2$}}}
		\put(-115,-46){ \color{black}{\bf \small{$\omg_3$}}}
		\put(-40,-46){ \color{black}{\bf \small{$\omg_4$}}}
		\put(32,-46){ \color{black}{\bf \small{$\omg_5$}}}
		\put(107,-46){ \color{black}{\bf \small{$\omg_6$}}}
		\put(180,-46){ \color{black}{\bf \small{$\omg_7$}}}
		\end{tikzpicture}
		\put(-280,-12){ \color{black}{\bf \small{(e) MARS ($7$ layers)}}}
	\end{tabular}
	\caption{Transforms learned from the XCAT phantom. Transform rows are shown as $8 \times 8$ patches. Beyond the first layer, the rows of the transforms sparsify across the residual channels (1D filters).}
	\label{fig:lear_XCAT_tran}
\end{figure}

%\vspace{-0.10in}
We pre-learn MARS models with different numbers of layers (depths) with $64 \times 64$ transforms. The models are learned from $8\times 8$ overlapping patches extracted from five $420\times 420$ XCAT phantom slices. The number of pixels $N_p$ and the number of overall training patches $N$ are about $1.7 \times 10^5$ and $8.5 \times 10^5$, respectively. 
%which is five times larger than the patch number in the testing phase ($N_p \approx 180$K)
The training slices are displayed in the supplement (Fig.~13). The patch stride is $1\times 1$. We choose $1$, $2$, $3$, $5$, and $7$ layers, respectively, during training, which corresponds to ST, MARS2, MARS3, MARS5, and MARS7 models. We initialize the MARS learning algorithm with the 2D DCT matrix for the transform in the first layer and identity matrices for transforms in deeper layers. For each model, we ran 1000 to 1500 iterations of the block coordinate descent training algorithm to ensure convergence. We choose $\eta=75$ for ST, $(\eta_1$, $\eta_2)$ $=$ $(80$, $60)$ for MARS2, $(\eta_1$, $\eta_2$, $\eta_3)$ $=$ $(90$, $80$, $60)$ for MARS3, $(\eta_1$, $\eta_2$, $\eta_3$, $\eta_4$, $\eta_5)$ $=$ $(120$, $120$, $120$, $110$, $110)$ for MARS5, $(\eta_1$, $\eta_2$, $\eta_3$, $\eta_4$, $\eta_5$, $\eta_6$, $\eta_7)$ $=$ $(120$, $120$, $120$, $110$, $110$, $80$, $60)$ for MARS7. Fig.~\ref{fig:lear_XCAT_tran} shows some of the learned transforms, with each transform matrix row displayed as a square patch for simplicity. The first layer transform in the models typically displays edge-like and gradient filters that sparsify the image. However, with more layers, finer level features are learned to sparsify transform-domain residuals in deeper layers. Nonetheless, the transforms in quite deep layers could potentially be more easily contaminated with noise in the training data, since the main image features are successively filtered out over layers.

\subsubsection{Simulation Framework and Visual Results}
\label{xcat_experiment}
\vspace{-0.1in}
We simulate low-dose CT measurements using $840\times 840$ XCAT phantom slices with $\Delta_x=\Delta_y=0.4883$ mm. The generated sinograms are of size $888\times 984$, obtained with GE 2D LightSpeed fan-beam geometry corresponding to a monoenergetic source with $I_0 = 1\times 10^4$ incident photons per ray and no scatter. For PWLS-EP, we ran $1000$ iterations of the relaxed LALM algorithm with the FBP reconstruction as initialization and regularization parameter $\beta = 2^{16}$. For the MARS model, we used the relaxed LALM algorithm for the image update step with $2$ inner iterations. We initialized PWLS-MARS schemes with the PWLS-EP reconstruction and used $T_O = 1500$ outer iterations for ST and all MARS schemes. 

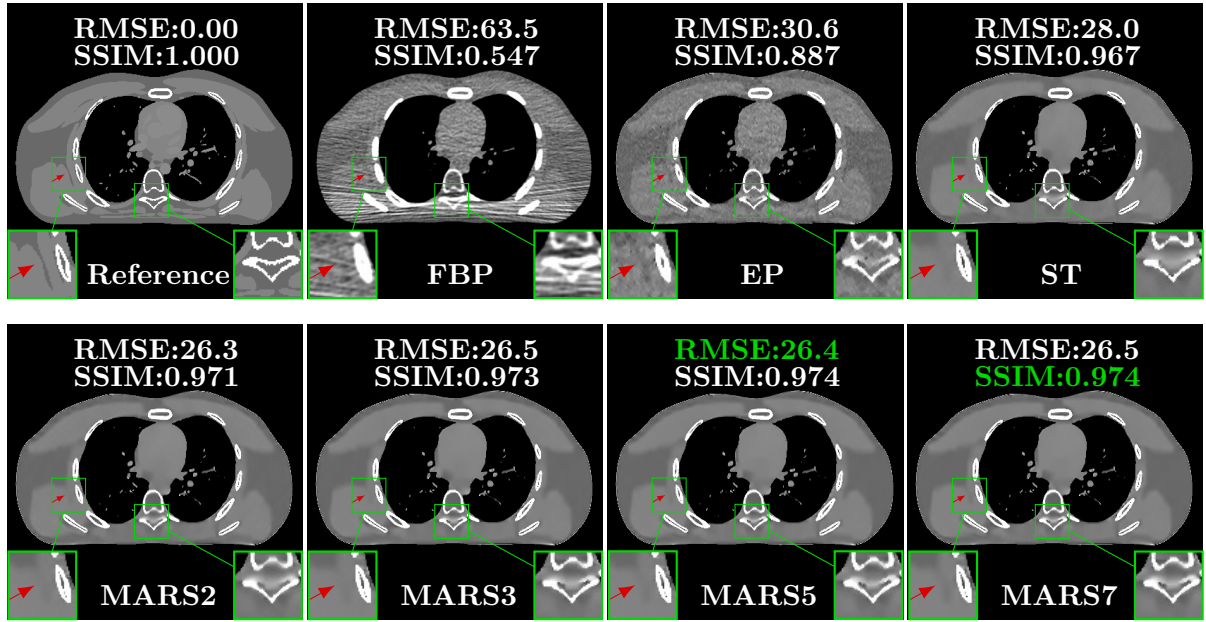
\begin{figure}[!h]
	\centering  
	\begin{tikzpicture}
	[spy using outlines={rectangle,green,magnification=2,size=9mm, connect spies}]
	\node {\includegraphics[width=0.24\textwidth]{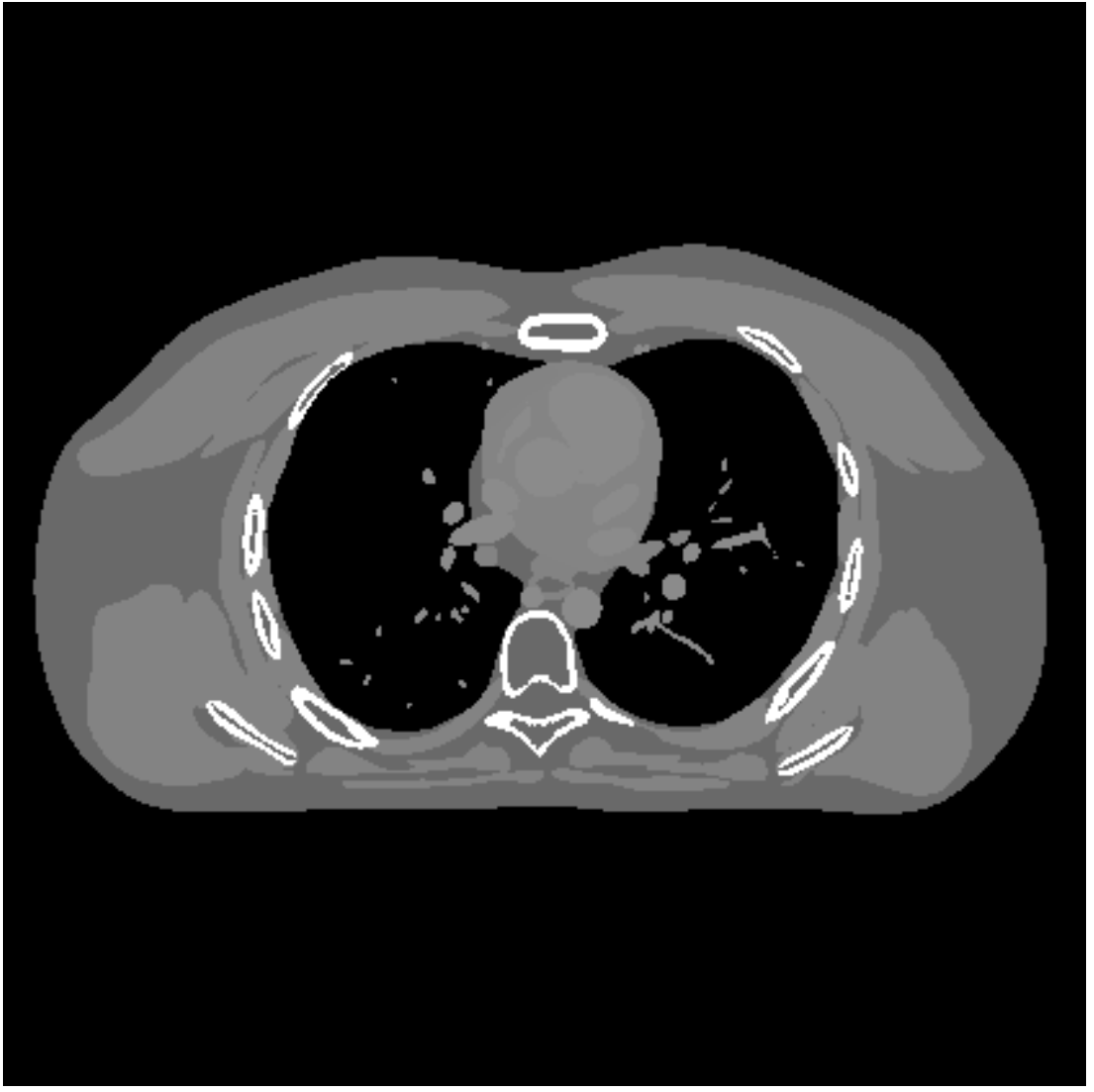}	};
	\spy on (-1.15,-0.3) in node [left] at (-1.05,-1.5);	
	\spy on (-0.05,-0.65) in node [left] at (1.95,-1.5);
	\draw[line width=1pt, ultra thin, -latex, red] (-1.38,-0.40) -- node[xshift=-0.05cm,yshift=-0.09cm] {\tiny{}} (-1.2,-0.3);	
	%\draw[red] (-1.95,0)--(1.95,0);
	%\draw[blue] (0,-1.95)--(0,1.95);
	\end{tikzpicture}
	\put(-95,103){ \color{white}{\bf \small{RMSE:0.00}}}
	\put(-95,93){ \color{white}{\bf \small{SSIM:1.000}}}
	\put(-90,10){ \color{white}{\bf \small{Reference}}} 
	\hspace{-0.15in}
	\begin{tikzpicture}
	[spy using outlines={rectangle,green,magnification=2,size=9mm, connect spies}]
	\node {\includegraphics[width=0.24\textwidth]{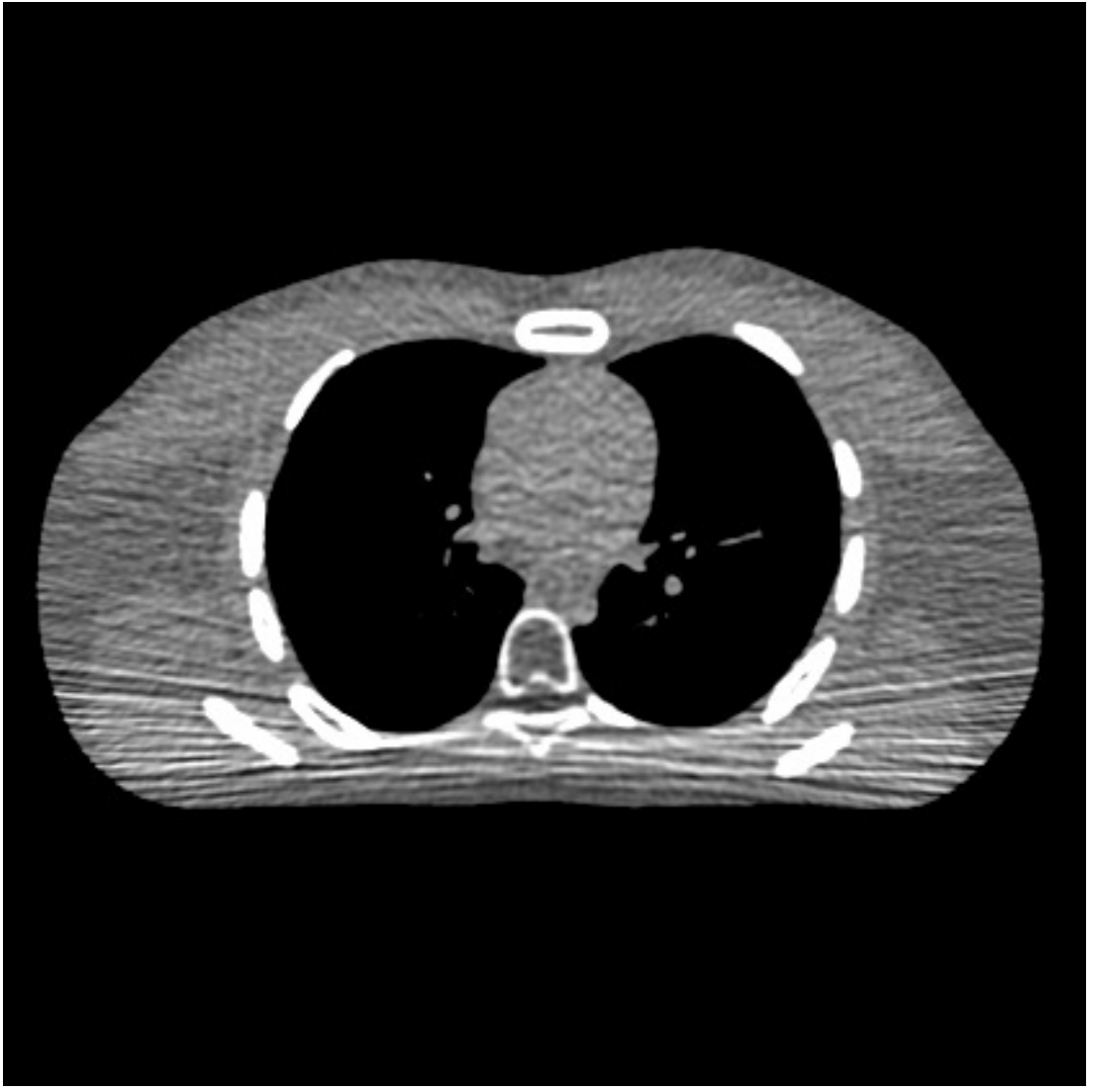}	};
	\spy on (-1.15,-0.3) in node [left] at (-1.05,-1.5);	
	\spy on (-0.05,-0.65) in node [left] at (1.95,-1.5);
	\draw[line width=1pt, ultra thin, -latex, red] (-1.38,-0.40) -- node[xshift=-0.05cm,yshift=-0.09cm] {\tiny{}} (-1.2,-0.3);		
	\end{tikzpicture}
	\put(-95,103){ \color{white}{\bf \small{RMSE:63.5}}}
	\put(-95,93){ \color{white}{\bf \small{SSIM:0.547}}}
	\put(-75,10){ \color{white}{\bf \small{FBP}}} 
	\hspace{-0.15in}
	\begin{tikzpicture}
	[spy using outlines={rectangle,green,magnification=2,size=9mm, connect spies}]
	\node {\includegraphics[width=0.24\textwidth]{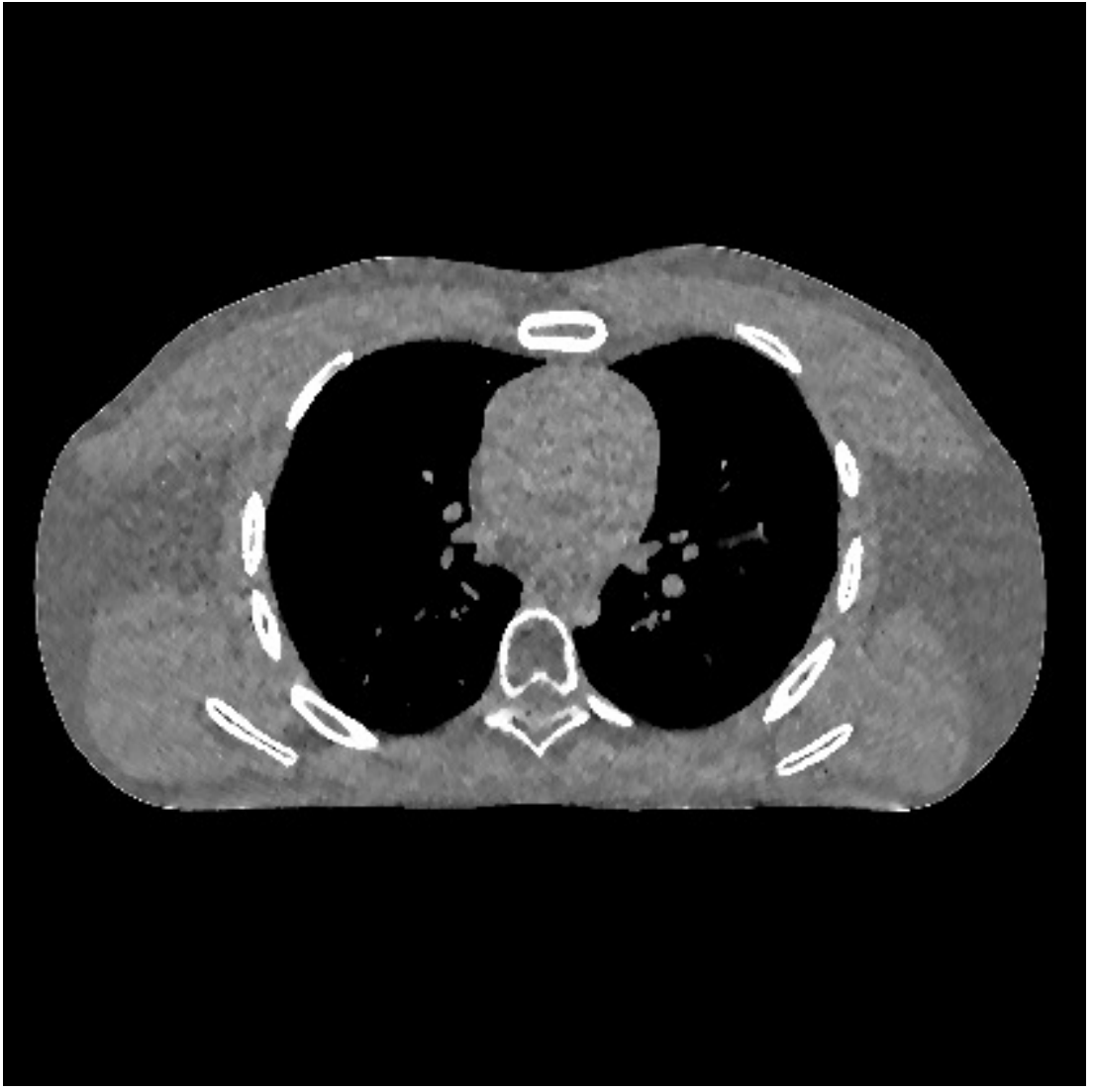}	};
	\spy on (-1.15,-0.3) in node [left] at (-1.05,-1.5);	
	\spy on (-0.05,-0.65) in node [left] at (1.95,-1.5);
	\draw[line width=1pt, ultra thin, -latex, red] (-1.38,-0.40) -- node[xshift=-0.05cm,yshift=-0.09cm] {\tiny{}} (-1.2,-0.3);		
	\end{tikzpicture}
	\put(-95,103){ \color{white}{\bf \small{RMSE:30.6}}}
	\put(-95,93){ \color{white}{\bf \small{SSIM:0.887}}}
	\put(-70,10){ \color{white}{\bf \small{EP}}}
	\hspace{-0.15in}	
	\begin{tikzpicture}
	[spy using outlines={rectangle,green,magnification=2,size=9mm, connect spies}]
	\node {\includegraphics[width=0.24\textwidth]{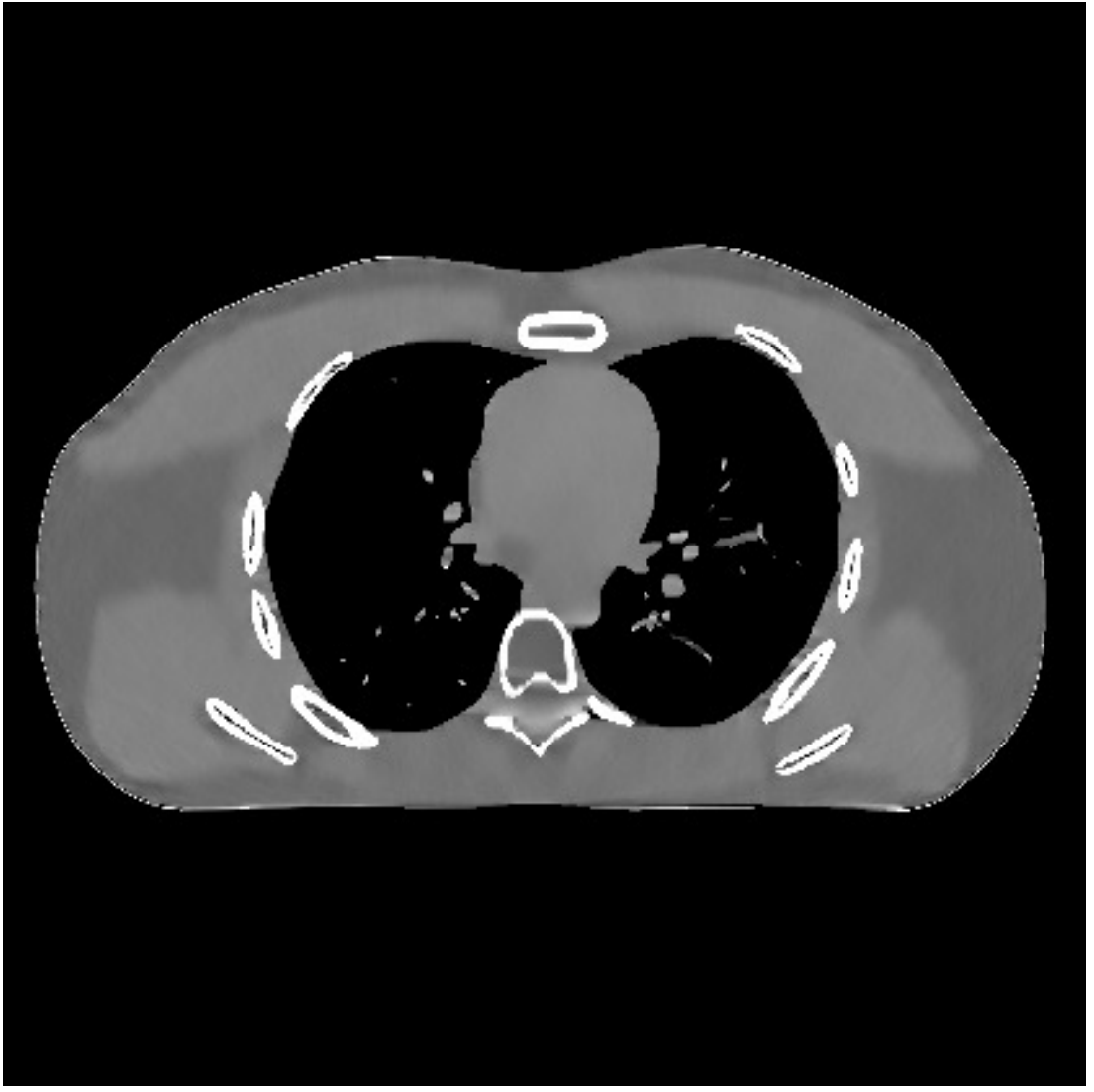}	};
	\spy on (-1.15,-0.3) in node [left] at (-1.05,-1.5);	
	\spy on (-0.05,-0.65) in node [left] at (1.95,-1.5);
	\draw[line width=1pt, ultra thin, -latex, red] (-1.38,-0.40) -- node[xshift=-0.05cm,yshift=-0.09cm] {\tiny{}} (-1.2,-0.3);		
	\end{tikzpicture}	
	\put(-95,103){ \color{white}{\bf \small{RMSE:28.0}}}
	\put(-95,93){ \color{white}{\bf \small{SSIM:0.967}}}
	\put(-70,10){ \color{white}{\bf \small{ST}}}
	\\
	\begin{tikzpicture}
	[spy using outlines={rectangle,green,magnification=2,size=9mm, connect spies}]
	\node {\includegraphics[width=0.24\textwidth]{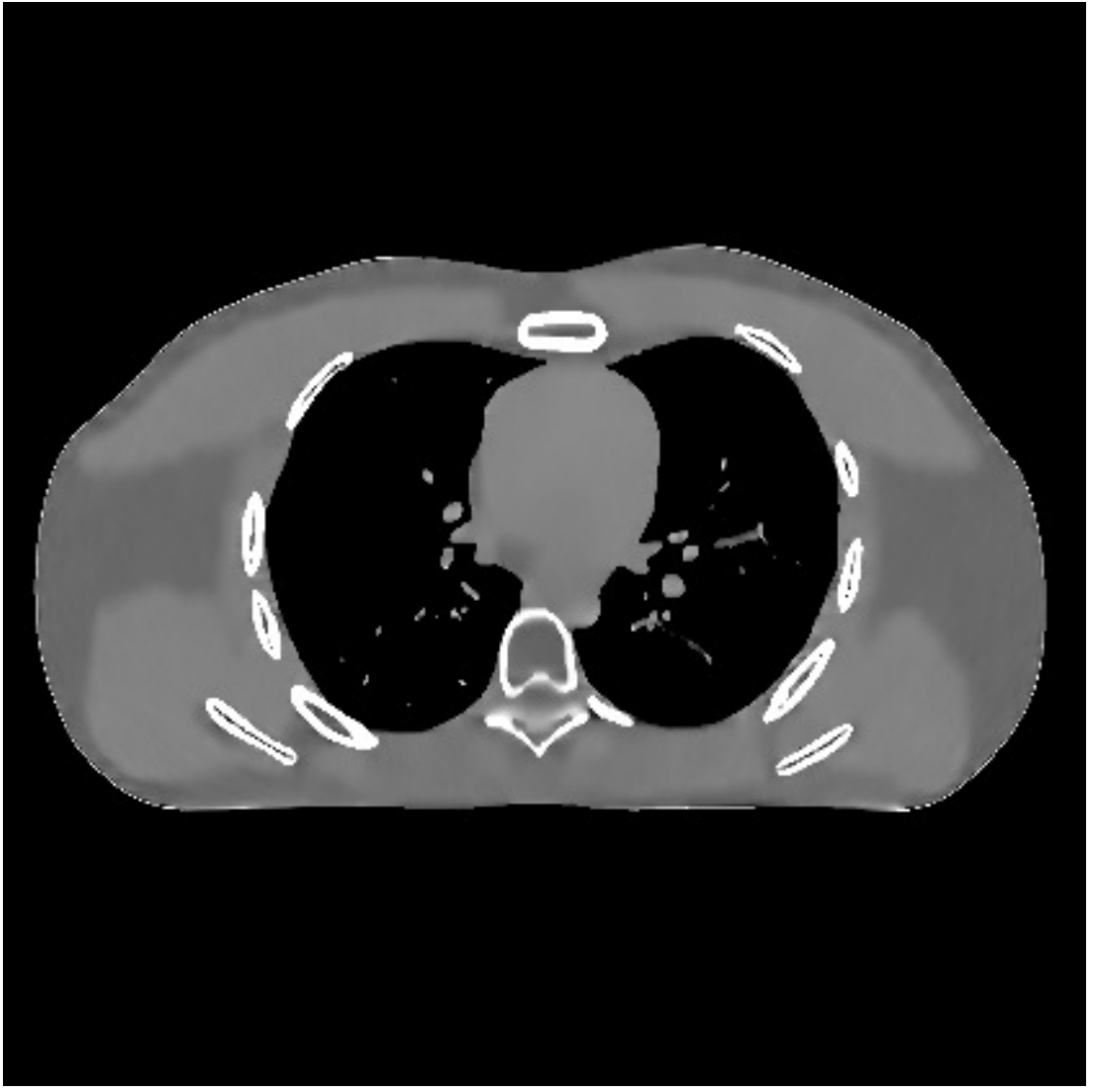}	};
	\spy on (-1.15,-0.3) in node [left] at (-1.05,-1.5);	
	\spy on (-0.05,-0.65) in node [left] at (1.95,-1.5);
	\draw[line width=1pt, ultra thin, -latex, red] (-1.38,-0.40) -- node[xshift=-0.05cm,yshift=-0.09cm] {\tiny{}} (-1.2,-0.3);		
	\end{tikzpicture}
	\put(-95,103){ \color{white}{\bf \small{RMSE:26.3}}}
	\put(-95,93){ \color{white}{\bf \small{SSIM:0.971}}}
	\put(-85,10){ \color{white}{\bf \small{MARS2}}}
	\hspace{-0.15in}
	\begin{tikzpicture}
	[spy using outlines={rectangle,green,magnification=2,size=9mm, connect spies}]
	\node {\includegraphics[width=0.24\textwidth]
		{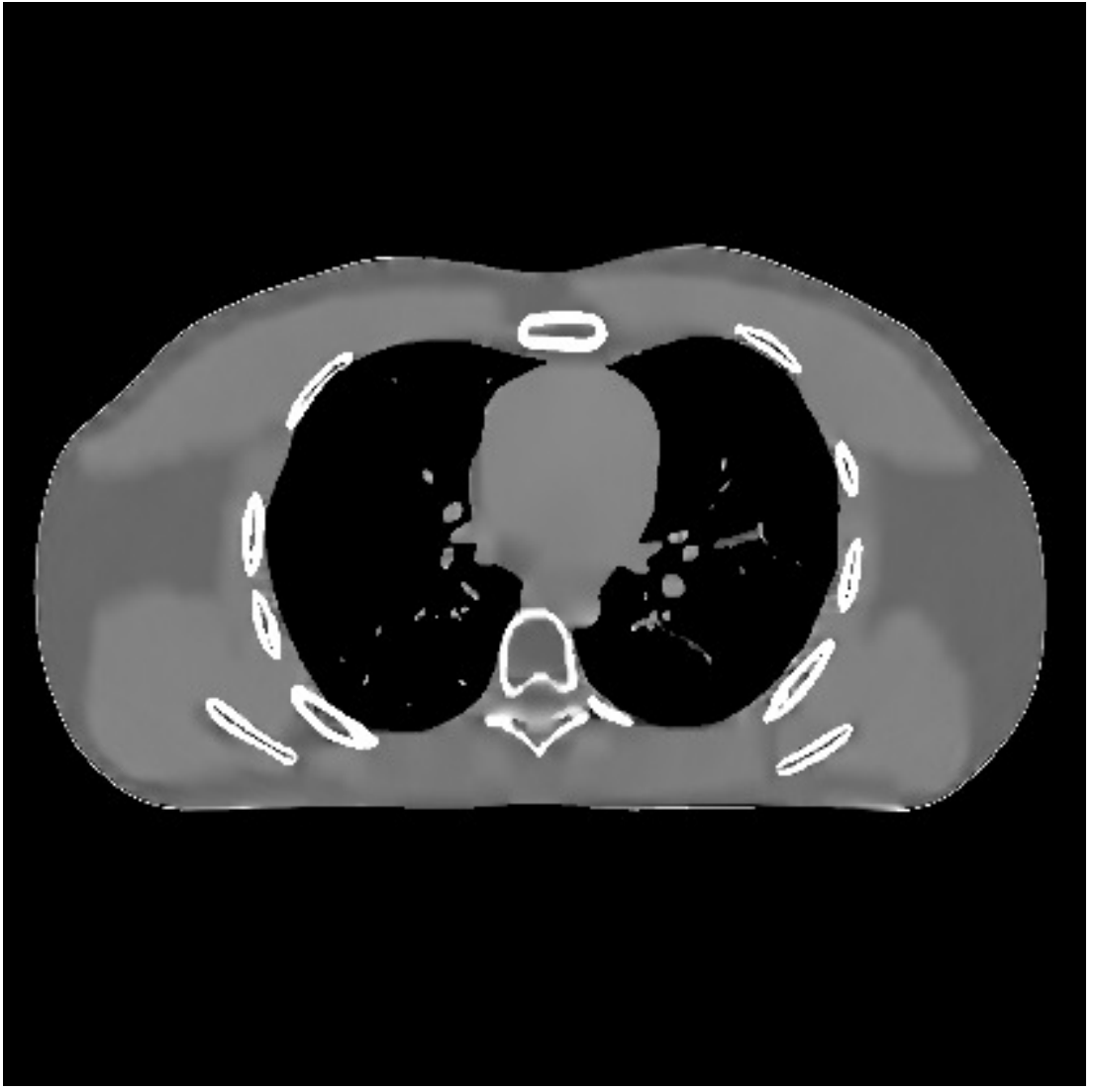}	};
	\spy on (-1.15,-0.3) in node [left] at (-1.05,-1.5);	
	\spy on (-0.05,-0.65) in node [left] at (1.95,-1.5);
	\draw[line width=1pt, ultra thin, -latex, red] (-1.38,-0.40) -- node[xshift=-0.05cm,yshift=-0.09cm] {\tiny{}} (-1.2,-0.3);		
	\end{tikzpicture}
	\put(-95,103){ \color{white}{\bf \small{RMSE:26.5}}}
	\put(-95,93){ \color{white}{\bf \small{SSIM:0.973}}}
	\put(-85,10){ \color{white}{\bf \small{MARS3}}}
	\hspace{-0.15in}
	\begin{tikzpicture}
	[spy using outlines={rectangle,green,magnification=2,size=9mm, connect spies}]
	\node {\includegraphics[width=0.24\textwidth]
		{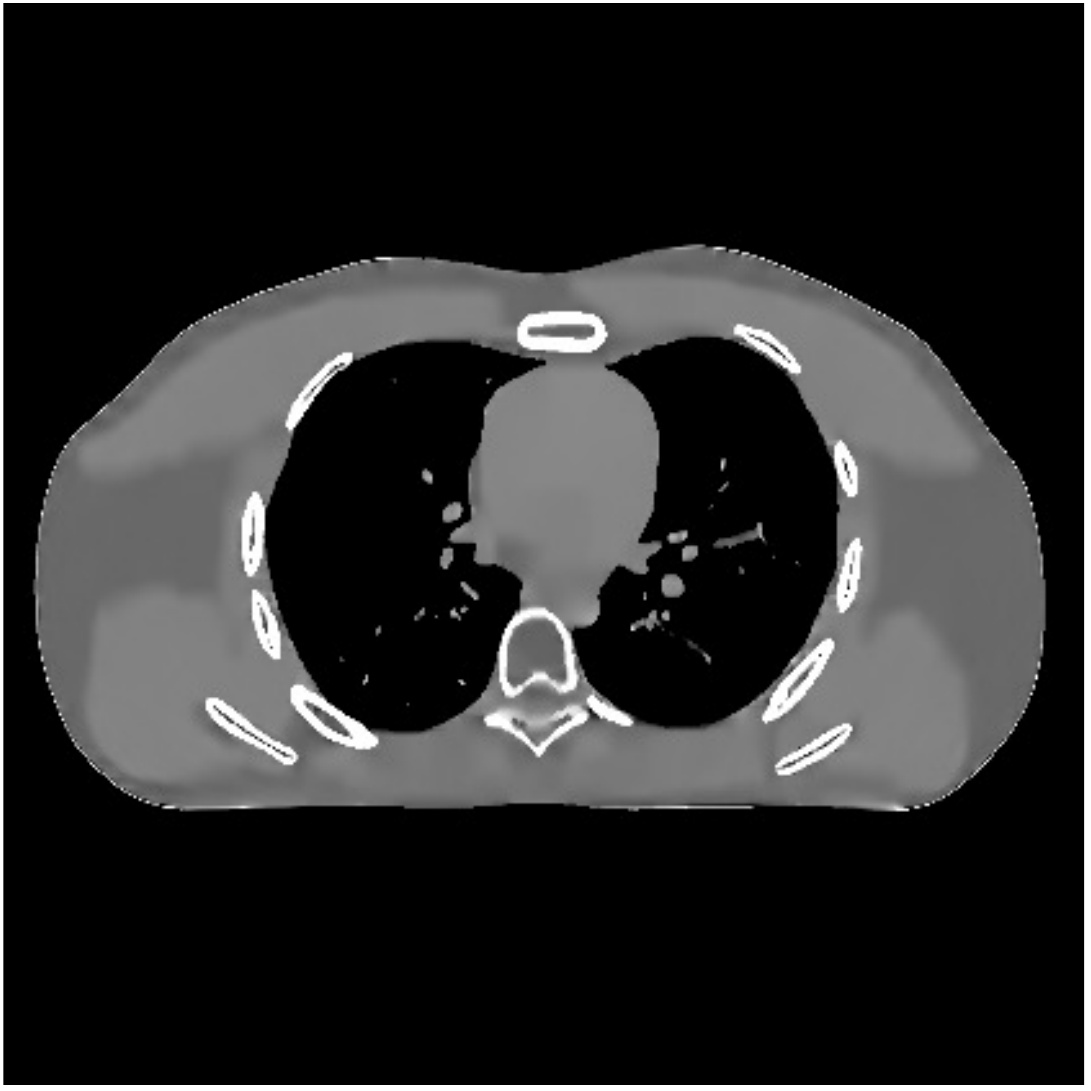}	};
	\spy on (-1.15,-0.3) in node [left] at (-1.05,-1.5);	
	\spy on (-0.05,-0.65) in node [left] at (1.95,-1.5);
	\draw[line width=1pt, ultra thin, -latex, red] (-1.38,-0.40) -- node[xshift=-0.05cm,yshift=-0.09cm] {\tiny{}} (-1.2,-0.3);		
	\end{tikzpicture}
	\put(-95,103){ \color{green}{\bf \small{RMSE:26.4}}}
	\put(-95,93){ \color{white}{\bf \small{SSIM:0.974}}}
	\put(-85,10){ \color{white}{\bf \small{MARS5}}}
	\hspace{-0.15in}
	\begin{tikzpicture}
	[spy using outlines={rectangle,green,magnification=2,size=9mm, connect spies}]
	\node {\includegraphics[width=0.24\textwidth]
		{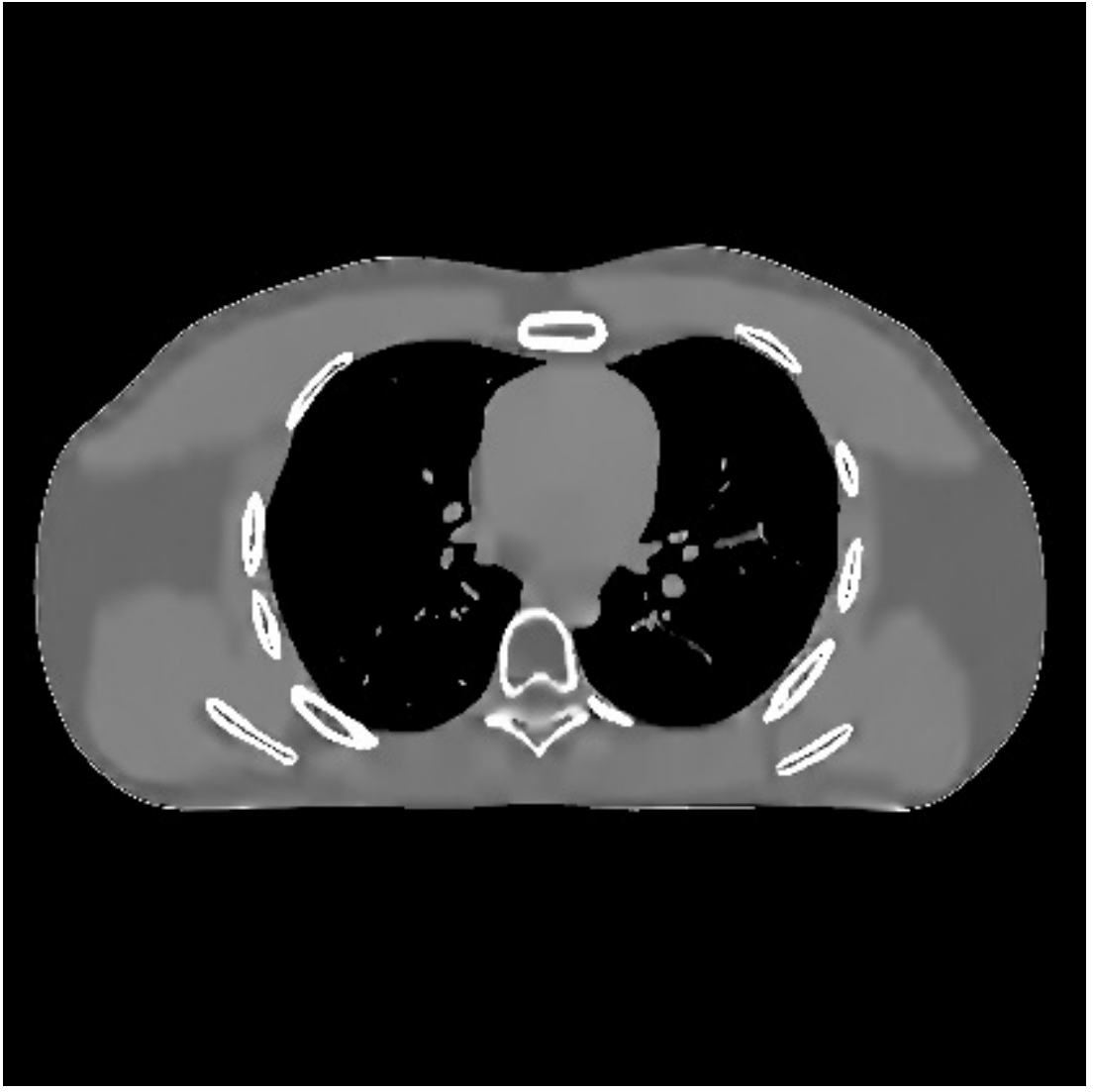}	};
	\spy on (-1.15,-0.3) in node [left] at (-1.05,-1.5);	
	\spy on (-0.05,-0.65) in node [left] at (1.95,-1.5);
	\draw[line width=1pt, ultra thin, -latex, red] (-1.38,-0.40) -- node[xshift=-0.05cm,yshift=-0.09cm] {\tiny{}} (-1.2,-0.3);		
	\end{tikzpicture}
	\put(-95,103){ \color{white}{\bf \small{RMSE:26.5}}}
	\put(-95,93){ \color{green}{\bf \small{SSIM:0.974}}}
	\put(-85,10){ \color{white}{\bf \small{MARS7}}}
	%	\vspace{-0.15in}、
	\caption{Comparison of reconstructions of slice $20$ of the XCAT phantom with FBP, PWLS-EP, PWLS-ST, PWLS-MARS2, PWLS-MARS3, PWLS-MARS5, and PWLS-MARS7, respectively, at incident photon intensity $I_0=1\times 10^{4}$. The display window is [800, 1200] HU.
	}
	\label{fig:recon_XCAT_slice20}
\end{figure}

\begin{figure}[!h]
	\centering  
	\begin{tikzpicture}
	[spy using outlines={rectangle,green,magnification=2,size=9mm, connect spies}]
	\node {\includegraphics[width=0.24\textwidth]{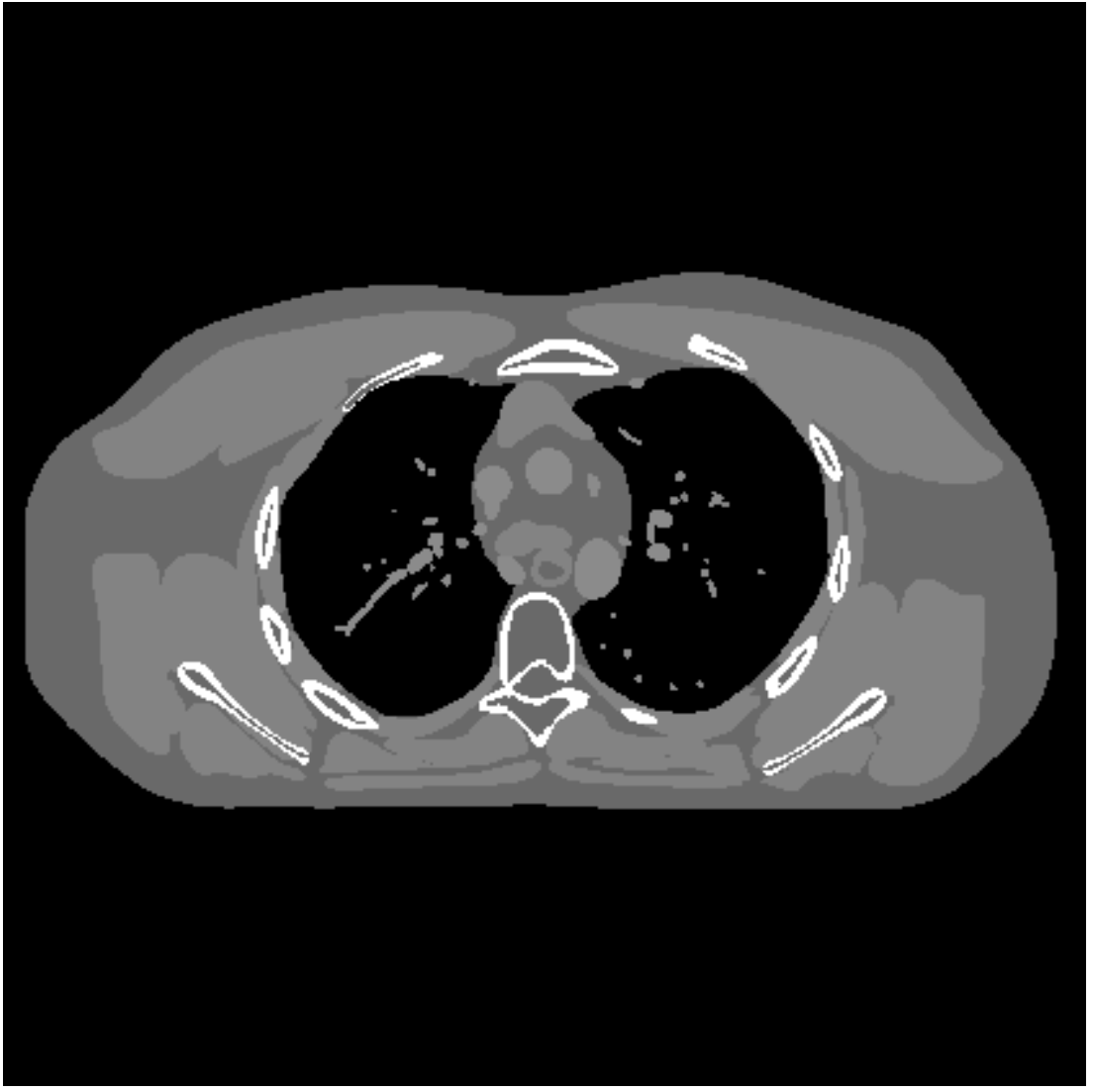}	};
	\spy on (0.00,0.30) in node [left] at (1.95,-1.5);	
	\spy on (-0.03,-0.60) in node [left] at (-1.05,-1.5);	
	%\draw[red] (-1.95,0)--(1.95,0);
	%\draw[blue] (0,-1.95)--(0,1.95);
	\end{tikzpicture}
	\put(-95,103){ \color{white}{\bf \small{RMSE:0.00}}}
	\put(-95,93){ \color{white}{\bf \small{SSIM:1.000}}}
	\put(-90,10){ \color{white}{\bf \small{Reference}}} 
	\hspace{-0.15in}
	\begin{tikzpicture}
	[spy using outlines={rectangle,green,magnification=2,size=9mm, connect spies}]
	\node {\includegraphics[width=0.24\textwidth]{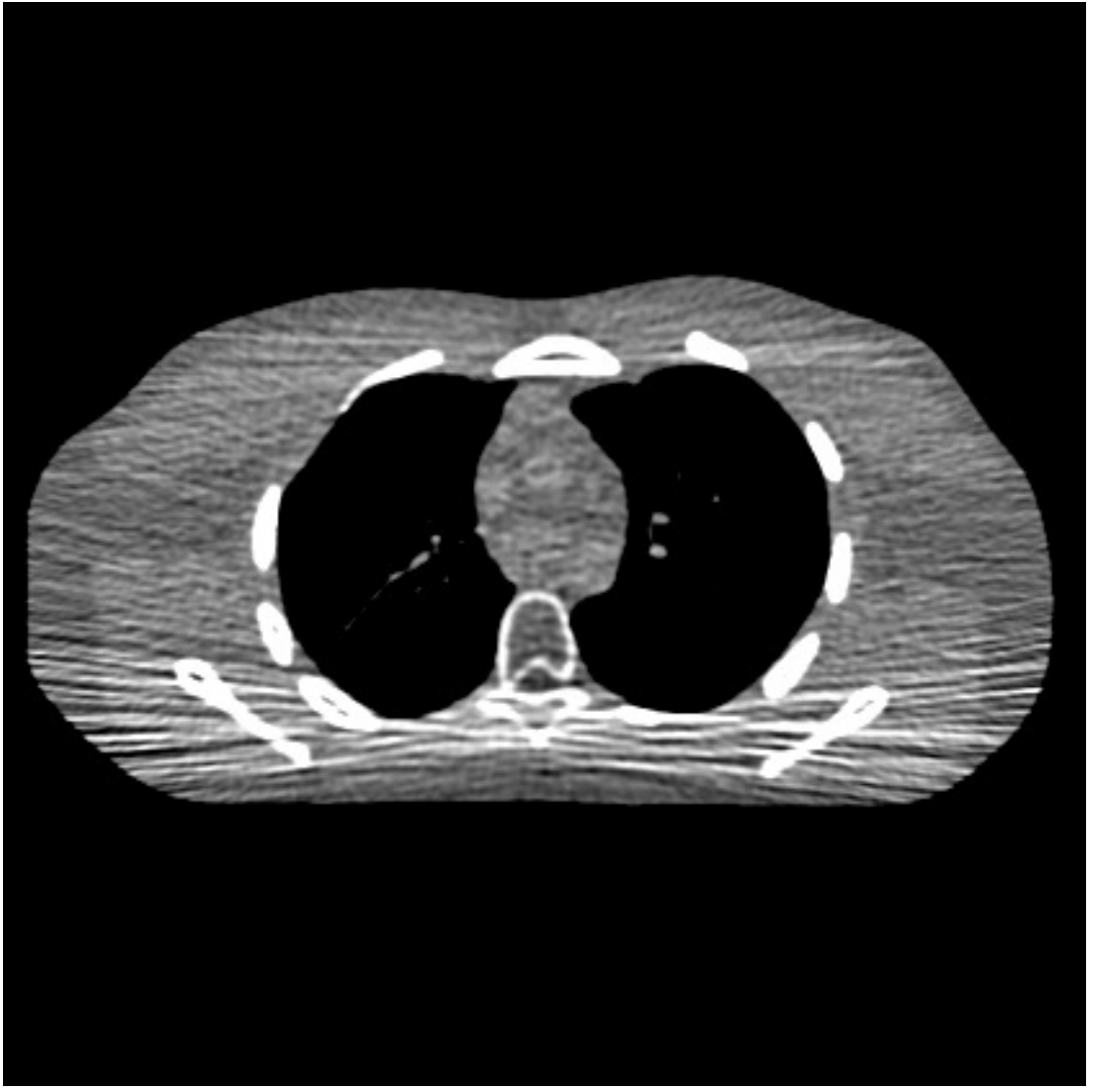}	};
	\spy on (0.00,0.30) in node [left] at (1.95,-1.5);	
	\spy on (-0.03,-0.60) in node [left] at (-1.05,-1.5);		
	\end{tikzpicture}
	\put(-95,103){ \color{white}{\bf \small{RMSE:67.7}}}
	\put(-95,93){ \color{white}{\bf \small{SSIM:0.539}}}
	\put(-75,10){ \color{white}{\bf \small{FBP}}} 
	\hspace{-0.15in}
	\begin{tikzpicture}
	[spy using outlines={rectangle,green,magnification=2,size=9mm, connect spies}]
	\node {\includegraphics[width=0.24\textwidth]{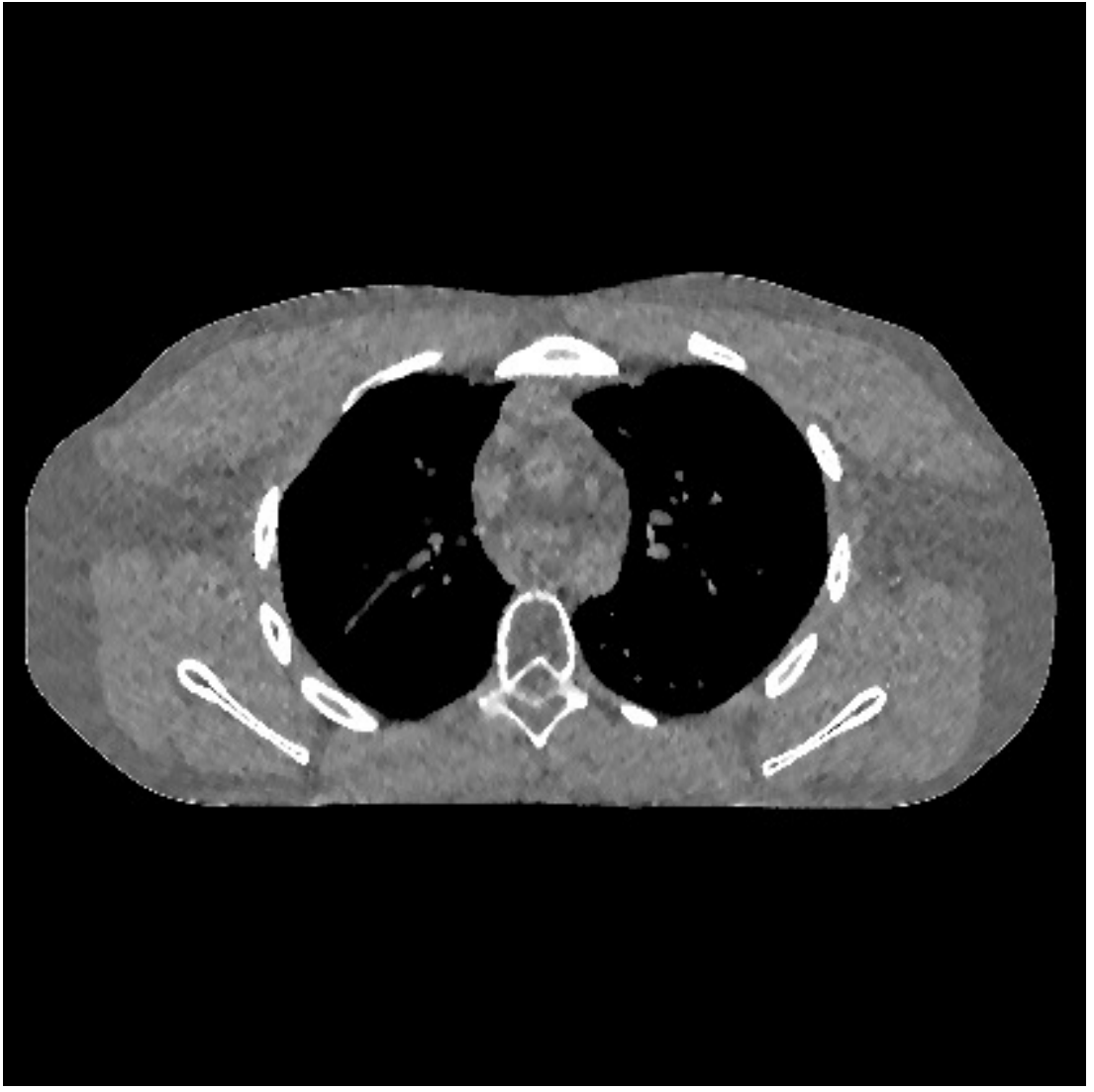}	};
	\spy on (0.00,0.30) in node [left] at (1.95,-1.5);	
	\spy on (-0.03,-0.60) in node [left] at (-1.05,-1.5);		
	\end{tikzpicture}
	\put(-95,103){ \color{white}{\bf \small{RMSE:33.0}}}
	\put(-95,93){ \color{white}{\bf \small{SSIM:0.894}}}
	\put(-70,10){ \color{white}{\bf \small{EP}}}
	\hspace{-0.15in}	
	\begin{tikzpicture}
	[spy using outlines={rectangle,green,magnification=2,size=9mm, connect spies}]
	\node {\includegraphics[width=0.24\textwidth]{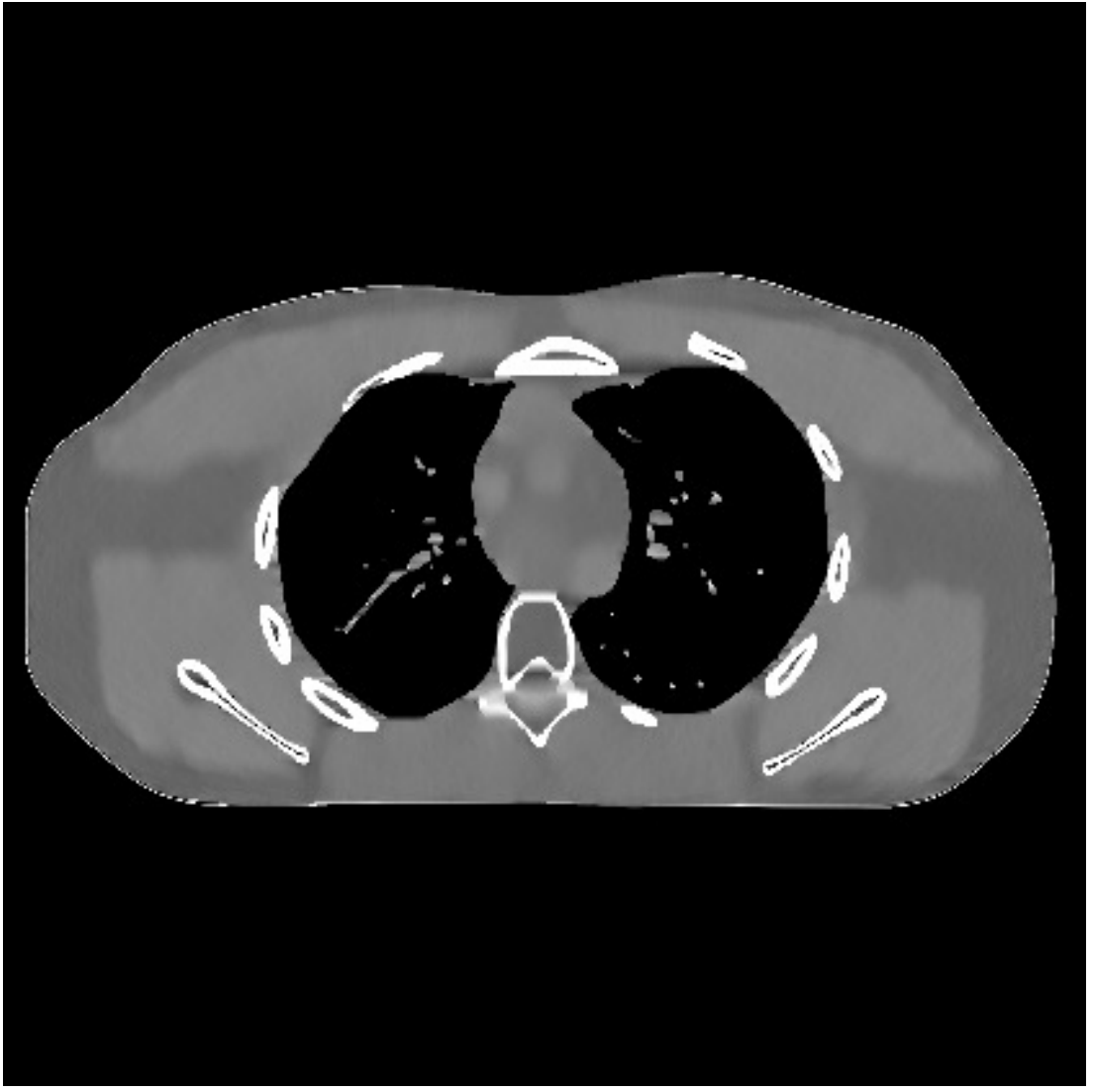}	};
	\spy on (0.00,0.30) in node [left] at (1.95,-1.5);	
	\spy on (-0.03,-0.60) in node [left] at (-1.05,-1.5);		
	\end{tikzpicture}	
	\put(-95,103){ \color{white}{\bf \small{RMSE:29.7}}}
	\put(-95,93){ \color{white}{\bf \small{SSIM:0.965}}}
	\put(-70,10){ \color{white}{\bf \small{ST}}}
	\\
	\begin{tikzpicture}
	[spy using outlines={rectangle,green,magnification=2,size=9mm, connect spies}]
	\node {\includegraphics[width=0.24\textwidth]{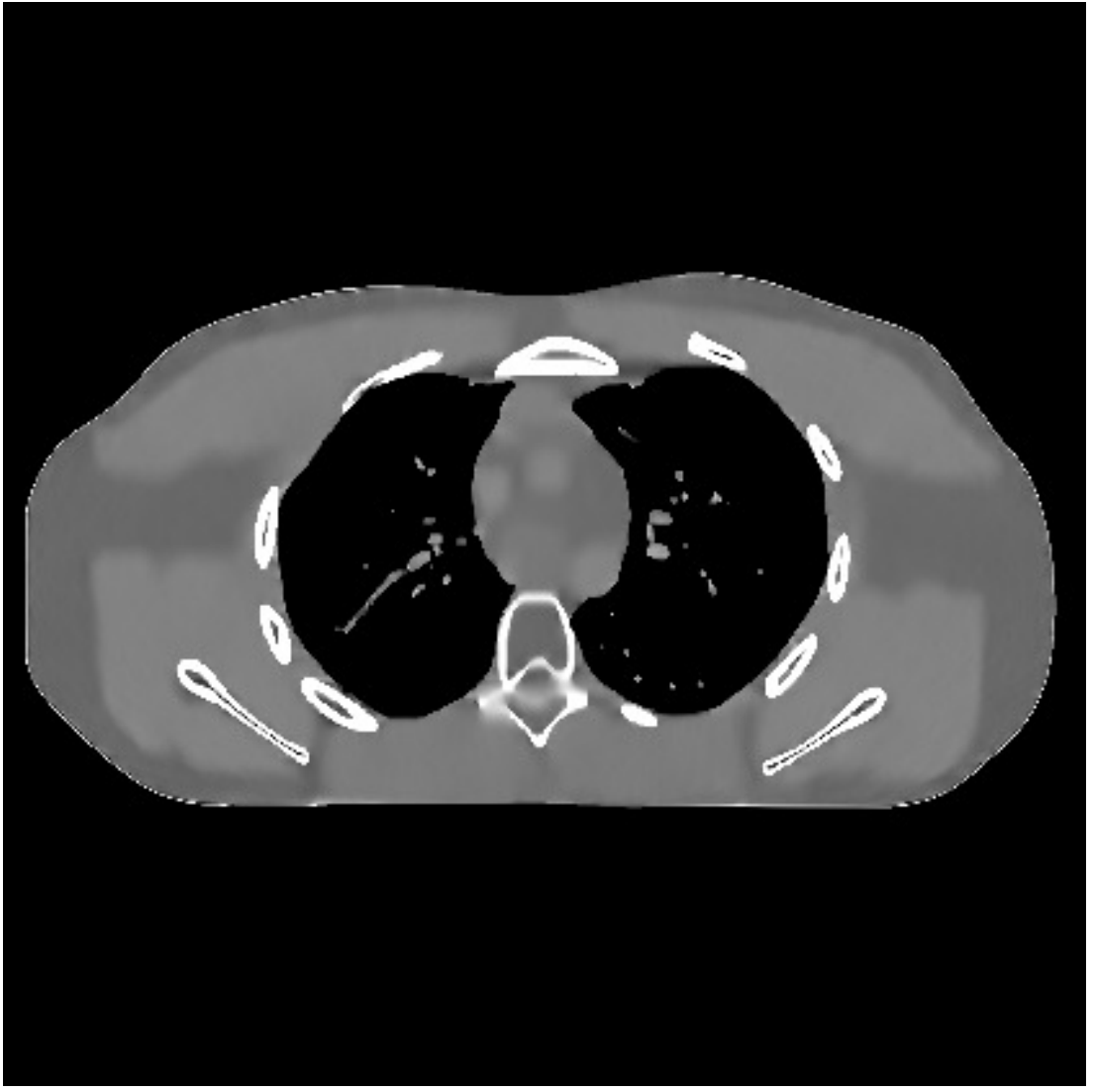}	};
	\spy on (0.00,0.30) in node [left] at (1.95,-1.5);	
	\spy on (-0.03,-0.60) in node [left] at (-1.05,-1.5);	
	\end{tikzpicture}
	\put(-95,103){ \color{white}{\bf \small{RMSE:28.5}}}
	\put(-95,93){ \color{white}{\bf \small{SSIM:0.969}}}
	\put(-85,10){ \color{white}{\bf \small{MARS2}}}
	\hspace{-0.15in}
	\begin{tikzpicture}
	[spy using outlines={rectangle,green,magnification=2,size=9mm, connect spies}]
	\node {\includegraphics[width=0.24\textwidth]
		{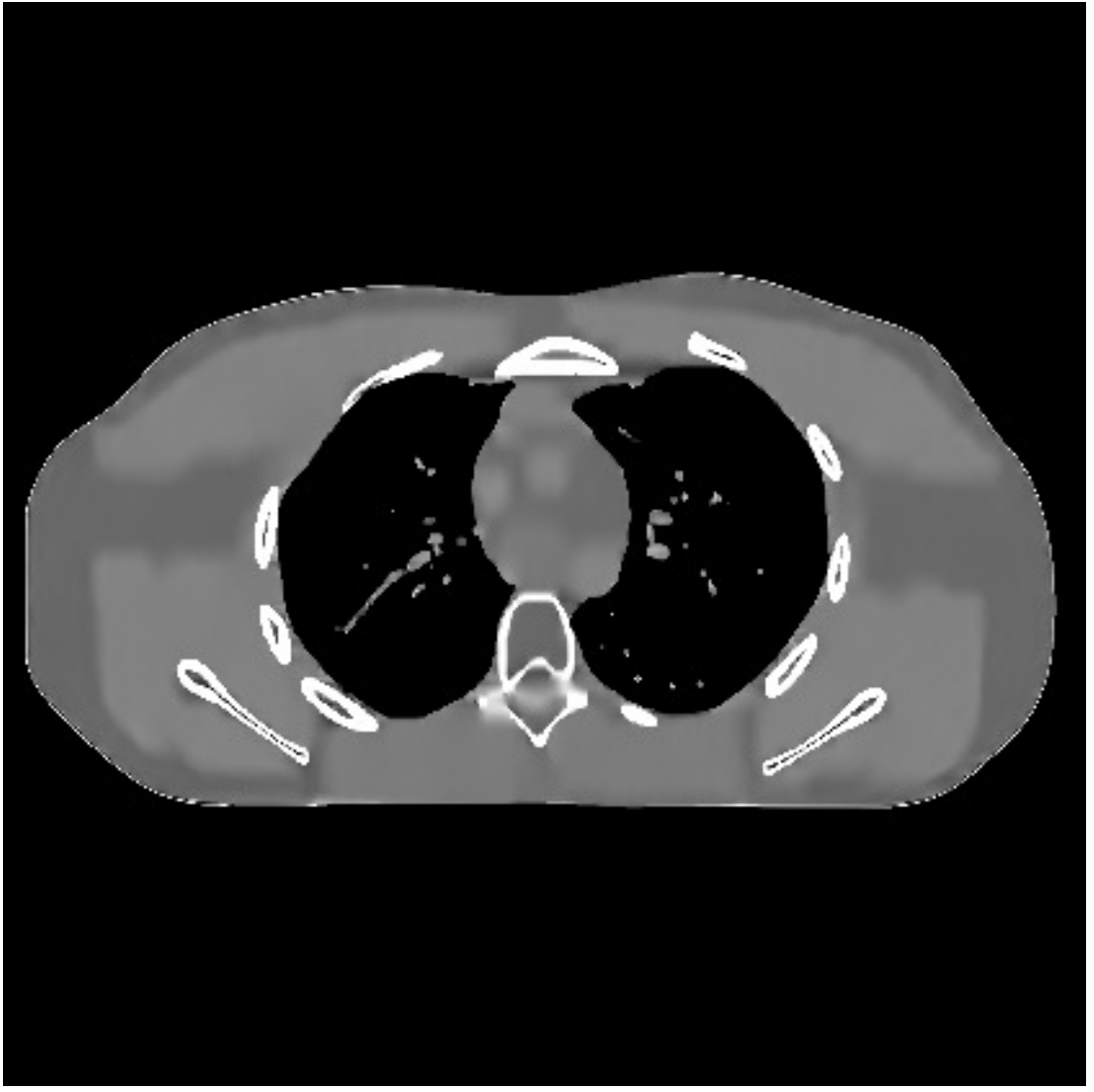}	};
	\spy on (0.00,0.30) in node [left] at (1.95,-1.5);	
	\spy on (-0.03,-0.60) in node [left] at (-1.05,-1.5);	
	\end{tikzpicture}
	\put(-95,103){ \color{green}{\bf \small{RMSE:28.3}}}
	\put(-95,93){ \color{white}{\bf \small{SSIM:0.971}}}
	\put(-85,10){ \color{white}{\bf \small{MARS3}}}
	\hspace{-0.15in}
	\begin{tikzpicture}
	[spy using outlines={rectangle,green,magnification=2,size=9mm, connect spies}]
	\node {\includegraphics[width=0.24\textwidth]
		{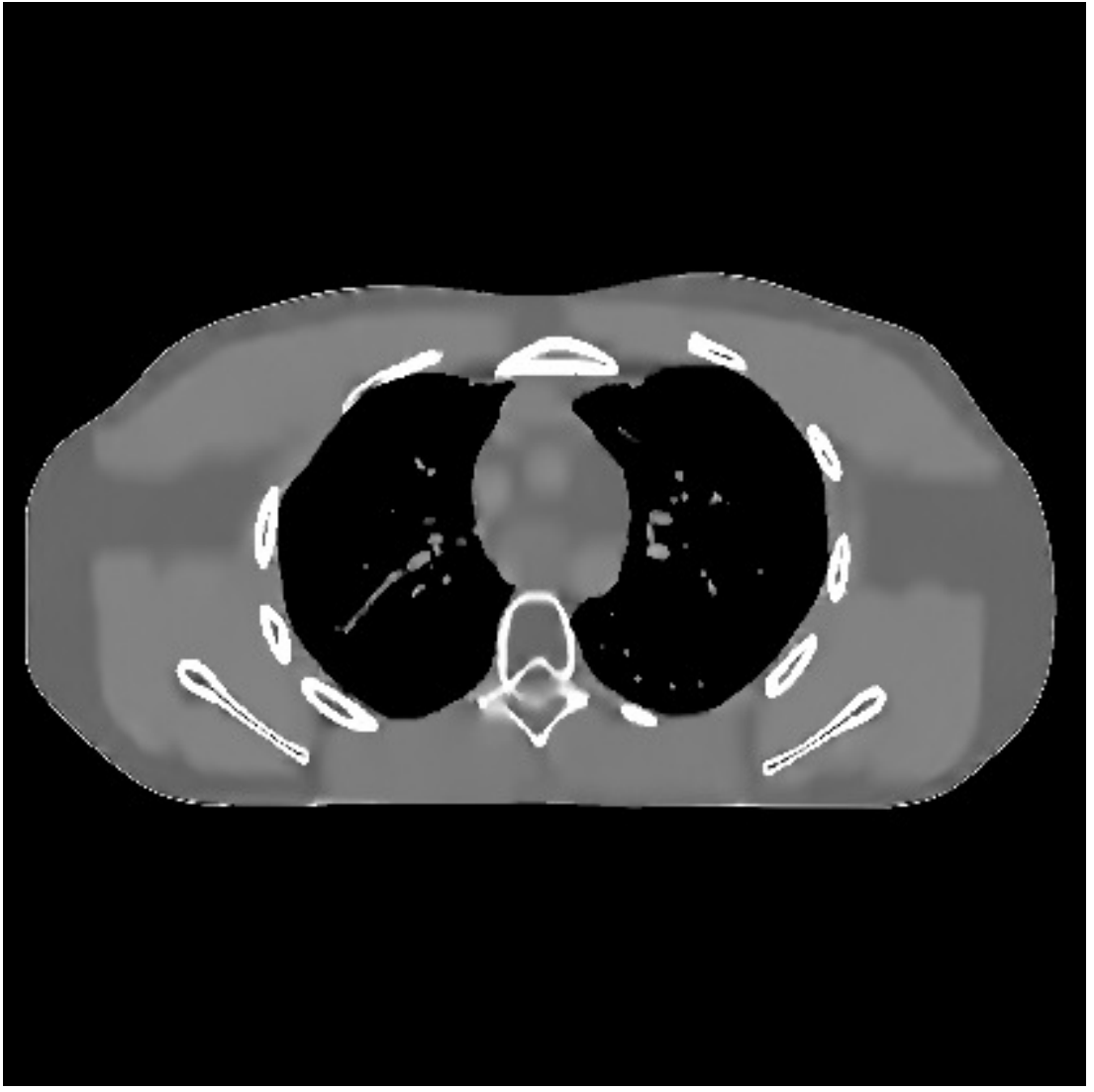}	};
	\spy on (0.00,0.30) in node [left] at (1.95,-1.5);	
	\spy on (-0.03,-0.60) in node [left] at (-1.05,-1.5);	
	\end{tikzpicture}
	\put(-95,103){ \color{green}{\bf \small{RMSE:28.3}}}
	\put(-95,93){ \color{green}{\bf \small{SSIM:0.972}}}
	\put(-85,10){ \color{white}{\bf \small{MARS5}}}
	\hspace{-0.15in}
	\begin{tikzpicture}
	[spy using outlines={rectangle,green,magnification=2,size=9mm, connect spies}]
	\node {\includegraphics[width=0.24\textwidth]
		{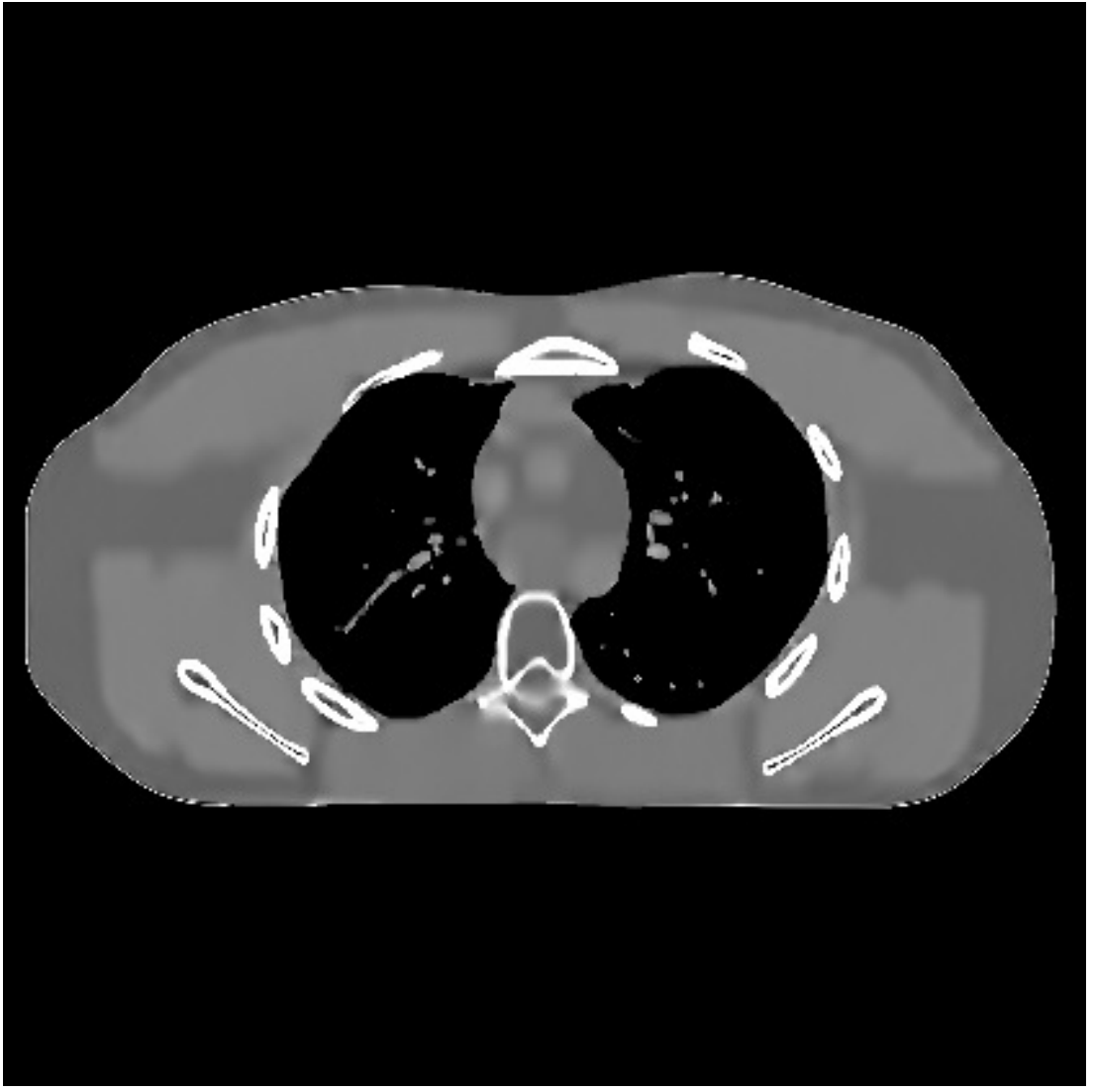}	};
	\spy on (0.00,0.30) in node [left] at (1.95,-1.5);	
	\spy on (-0.03,-0.60) in node [left] at (-1.05,-1.5);	
	\end{tikzpicture}
	\put(-95,103){ \color{white}{\bf \small{RMSE:28.4}}}
	\put(-95,93){ \color{green}{\bf \small{SSIM:0.972}}}
	\put(-85,10){ \color{white}{\bf \small{MARS7}}}
	%	\vspace{-0.15in}、
	\caption{Comparison of reconstructions of slice $60$ of the XCAT phantom with FBP, PWLS-EP, PWLS-ST, PWLS-MARS2, PWLS-MARS3, PWLS-MARS5, and PWLS-MARS7, respectively, at incident photon intensity $I_0=1\times 10^{4}$. 
		The display window is [800, 1200] HU.
	}
	\label{fig:recon_XCAT_slice60}
\end{figure}
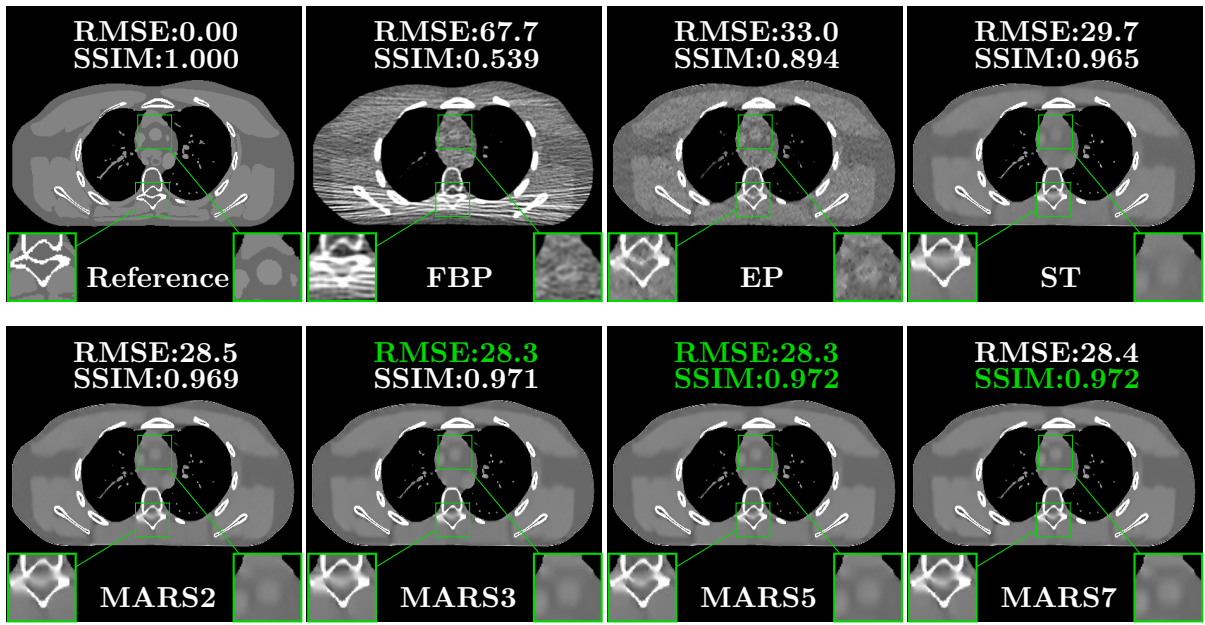

We firstly hand-tuned the reconstruction parameters ($\beta, \{\gamma_l, 1\leq l \leq L\}$) for one test slice and treated this set of parameters as the baseline. Similar to the PWLS-EP algorithm, we could determine the optimal (in terms of optimal RMSE) parameters for other testing slices by tuning the base parameters in a small range. However, we found that the change in reconstruction quality by picking a common set of parameters instead of slice-wise optimized parameters is quite small (only 0.2 HU in RMSE and without the loss of details). Therefore, the same set of parameters (baseline parameters) were used across testing cases and shown to be effective over the cases.
In particular, we selected slice 48 of the XCAT phantom as the case for parameter tuning and set the regularization parameters (after tuning over ranges of values) as $(\beta$, $\gamma)$ $=$ $(2\times10^5$, $20)$ for ST, $(\beta$, $\gamma_1$, $\gamma_2)$ $=$ $(9\times10^4$, $30$, $10)$ for MARS2, $(\beta$, $\gamma_1$, $\gamma_2$, $\gamma_3)$ $=$ $(9\times10^4$, $25$, $15$, $10)$ for MARS3, $(\beta$, $\gamma_1$, $\gamma_2$, $\gamma_3$, $\gamma_4$, $\gamma_5)$ $=$ $(9\times10^4$, $25$, $15$, $10$, $5$, $1)$ for MARS5, and $(\beta$, $\gamma_1$, $\gamma_2$, $\gamma_3$, $\gamma_4$, $\gamma_5$, $\gamma_6$, $\gamma_7)$ $=$ $(6\times10^4$, $30$, $25$, $20$, $15$, $10$, $5$, $1)$ for MARS7, respectively. In Fig. 14 in the supplement, we give the reconstructions for slice 48 of the XCAT phantom with various methods. Figs.~\ref{fig:recon_XCAT_slice20} and \ref{fig:recon_XCAT_slice60} here show the reconstructions for two independent test cases (slice 20 and 60 of the XCAT phantom). Both of them used the same set of parameters obtained for slice 48.
%Fig.~\ref{fig:recon_XCAT_slice60} displays the reconstruction results for slice 60 of the XCAT phantom. Additional XCAT phantom experimental results (Figs.~12 and 13) are shown in the supplementary document. 
The zoom-in regions give an explicit comparison between the multi-layer sparsifying transform models and other methods such as FBP, PWLS-EP, and PWLS-ST. PWLS-MARS achieves better noise reduction and higher contrast.

\subsection{Low-dose Experiments with Mayo Clinic Data}
\subsubsection{Study of Model Training}
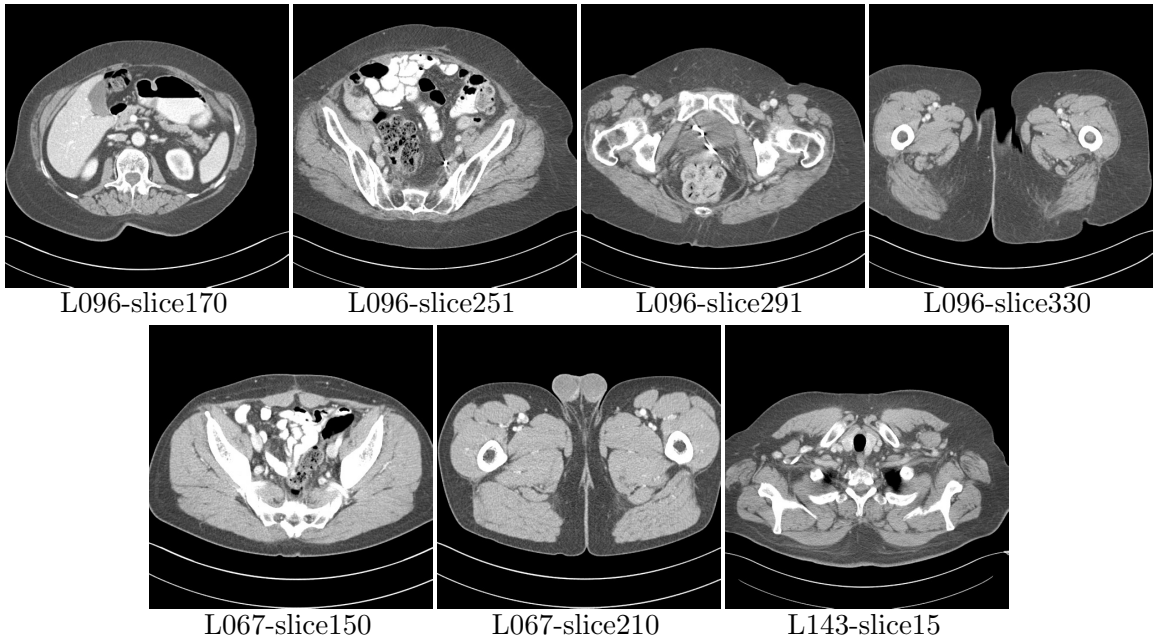
\begin{figure}[!h]
	\centering  
	%\vspace{-0.1in}
	\begin{tikzpicture}
	[spy using outlines={rectangle,green,magnification=2,size=7mm, connect spies}]
	\node {\includegraphics[width=0.23\textwidth]{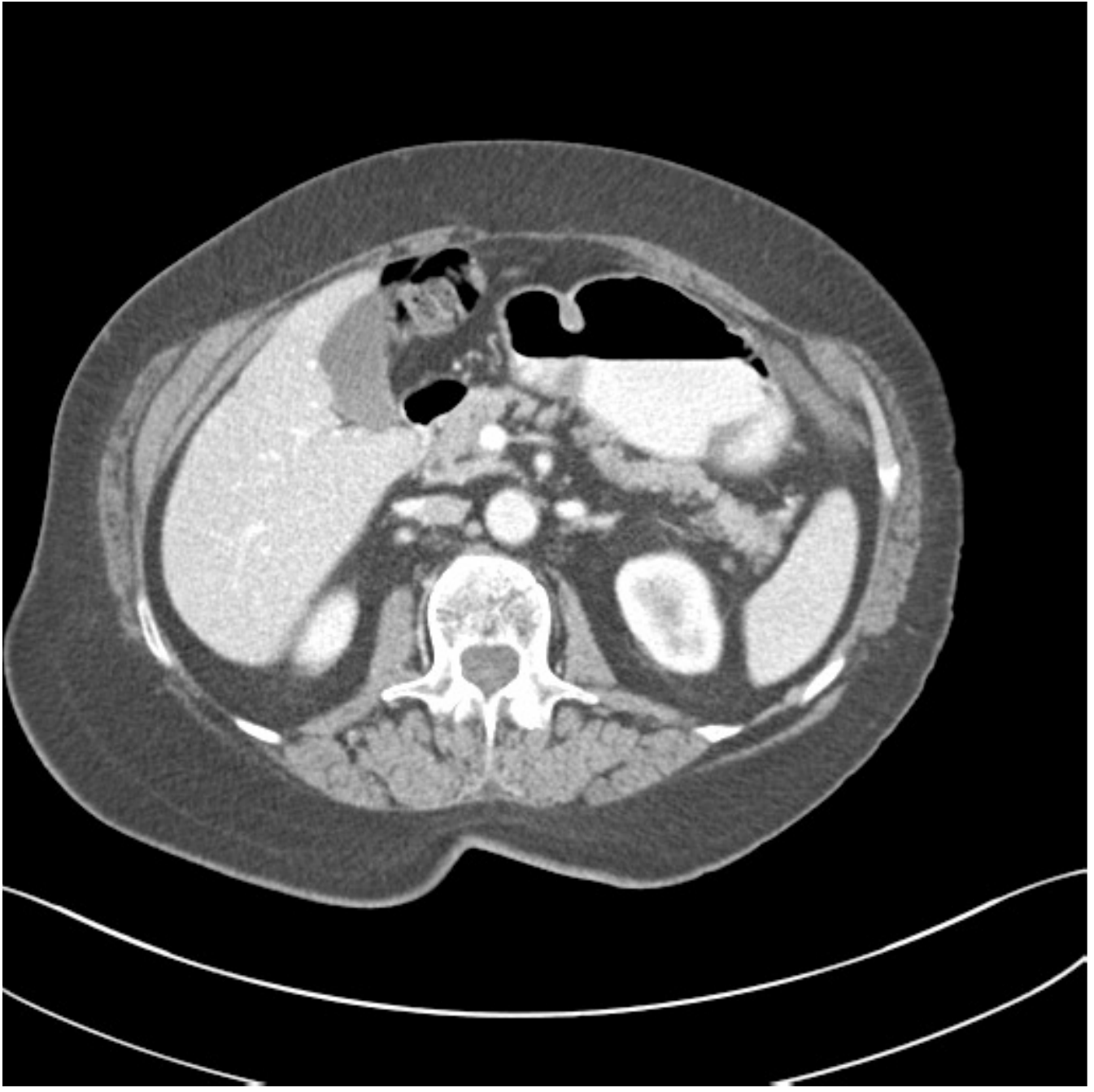}	};
	\end{tikzpicture}
	\put(-95,-5){ \color{black}{\bf \small{\rm{L096-slice170}}}} 
	\hspace{-0.15in}
	\begin{tikzpicture}
	[spy using outlines={rectangle,green,magnification=2,size=7mm, connect spies}]
	\node {\includegraphics[width=0.23\textwidth]{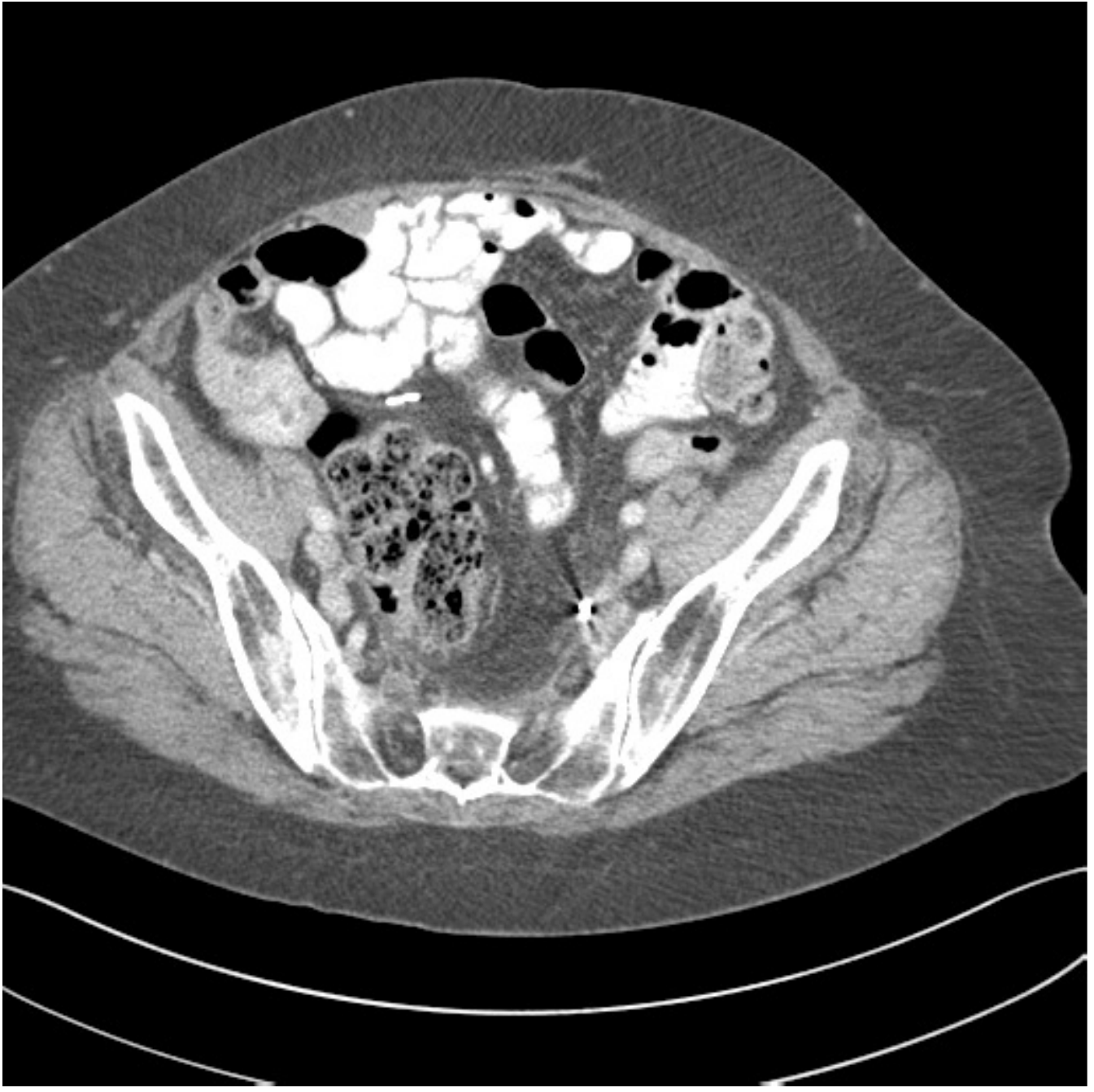}	};
	\end{tikzpicture}
	\put(-95,-5){ \color{black}{\bf \small{\rm{L096-slice251}}}}
	\hspace{-0.15in}
	\begin{tikzpicture}
	[spy using outlines={rectangle,green,magnification=2,size=7mm, connect spies}]
	\node {\includegraphics[width=0.23\textwidth]{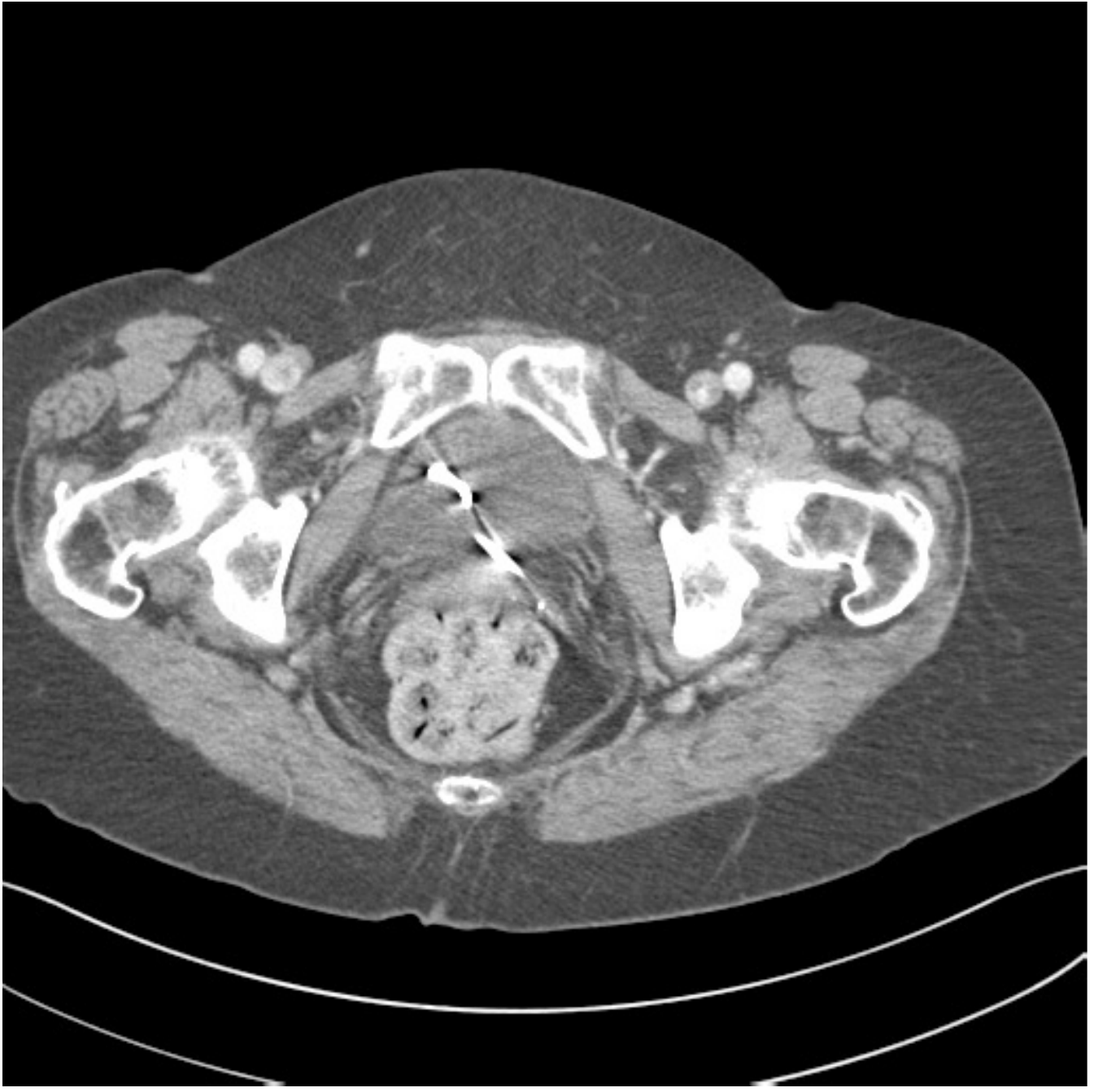}	};
	\end{tikzpicture}
	\put(-95,-5){ \color{black}{\bf \small{\rm{L096-slice291}}}} 
	\hspace{-0.15in}	
	\begin{tikzpicture}
	[spy using outlines={rectangle,green,magnification=2,size=7mm, connect spies}]
	\node {\includegraphics[width=0.23\textwidth]{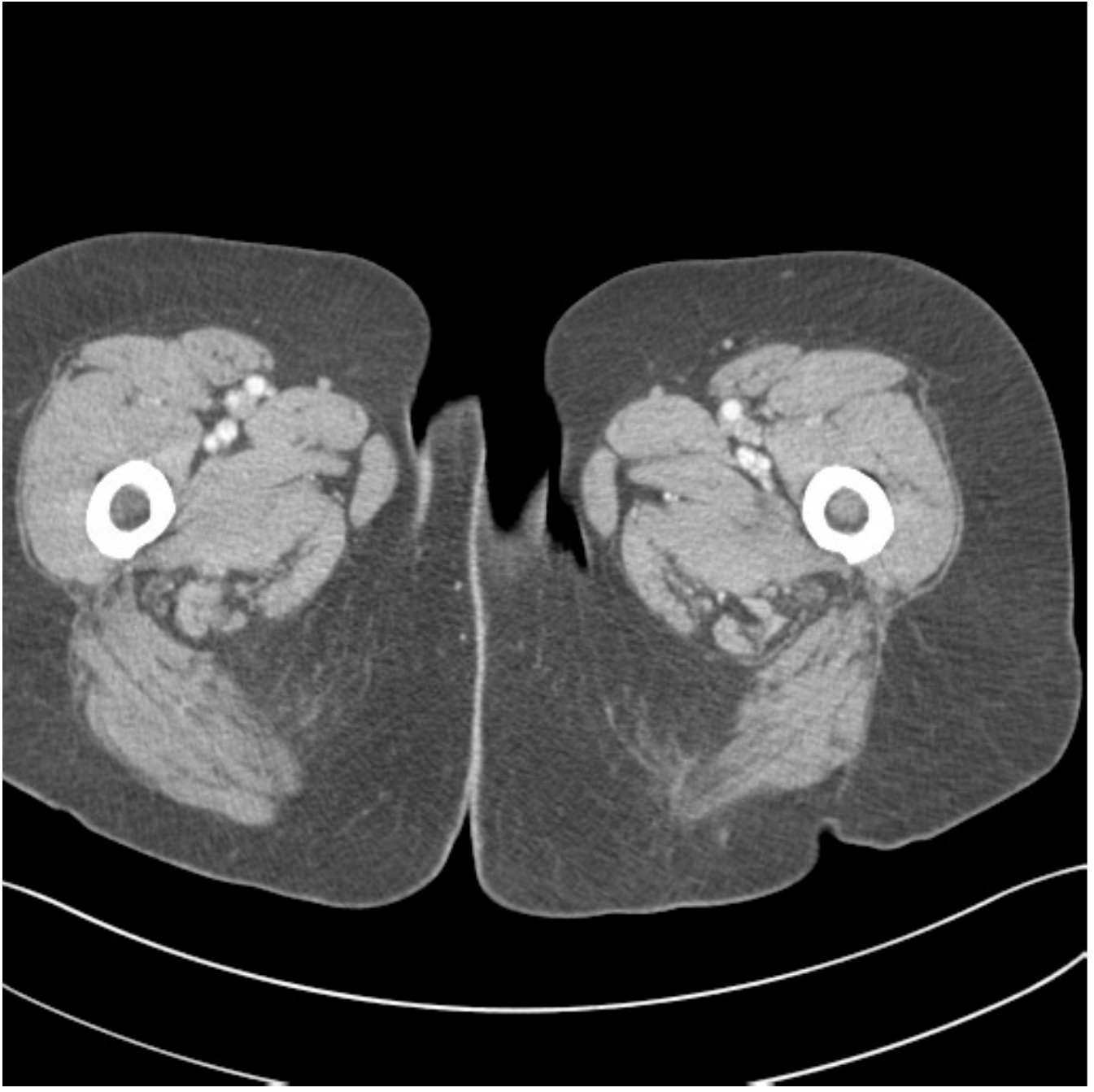}	};
	\end{tikzpicture}	
	\put(-95,-5){ \color{black}{\bf \small{\rm{L096-slice330}}}} 
	\\
	\begin{tikzpicture}
	[spy using outlines={rectangle,green,magnification=2,size=7mm, connect spies}]
	\node {\includegraphics[width=0.23\textwidth]{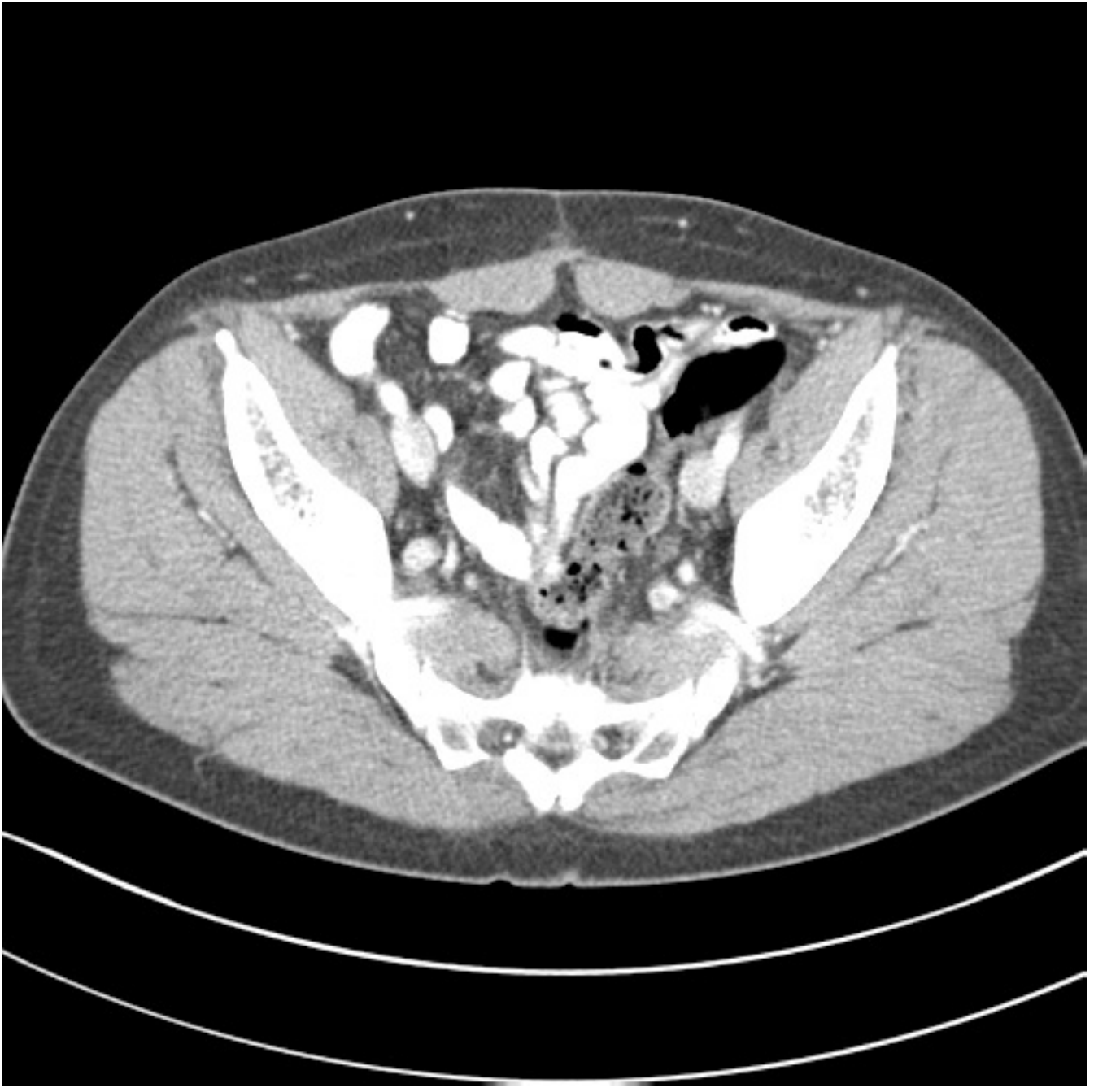}	};	
	\end{tikzpicture}
	\put(-95,-5){ \color{black}{\bf \small{\rm{L067-slice150}}}} 
	\hspace{-0.15in}
	\begin{tikzpicture}
	[spy using outlines={rectangle,green,magnification=2,size=7mm, connect spies}]
	\node {\includegraphics[width=0.23\textwidth]{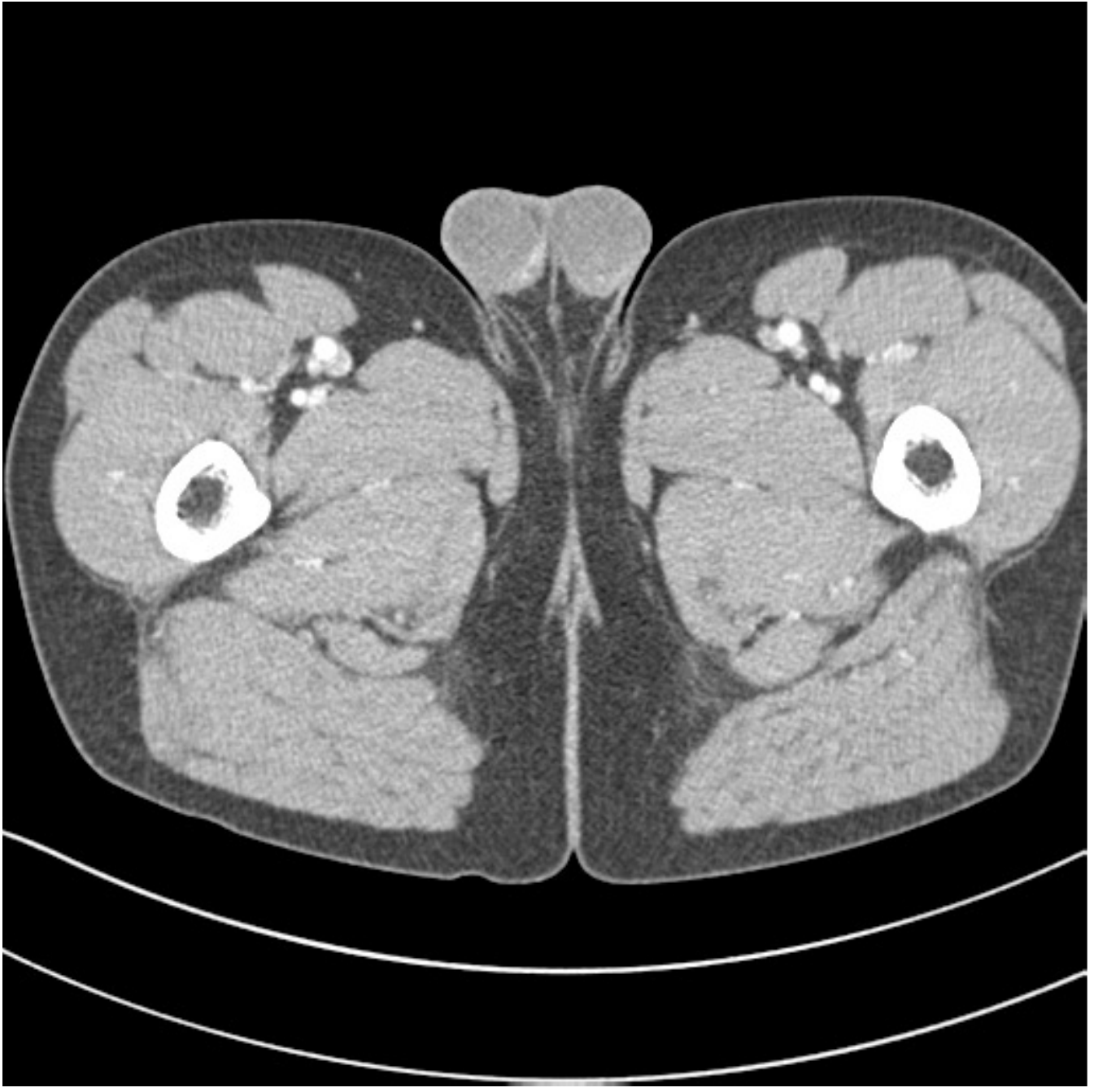}	};
	\end{tikzpicture}
	\put(-95,-5){ \color{black}{\bf \small{\rm{L067-slice210}}}} 
	\hspace{-0.15in}
	\begin{tikzpicture}
	[spy using outlines={rectangle,green,magnification=2,size=7mm, connect spies}]
	\node {\includegraphics[width=0.23\textwidth]{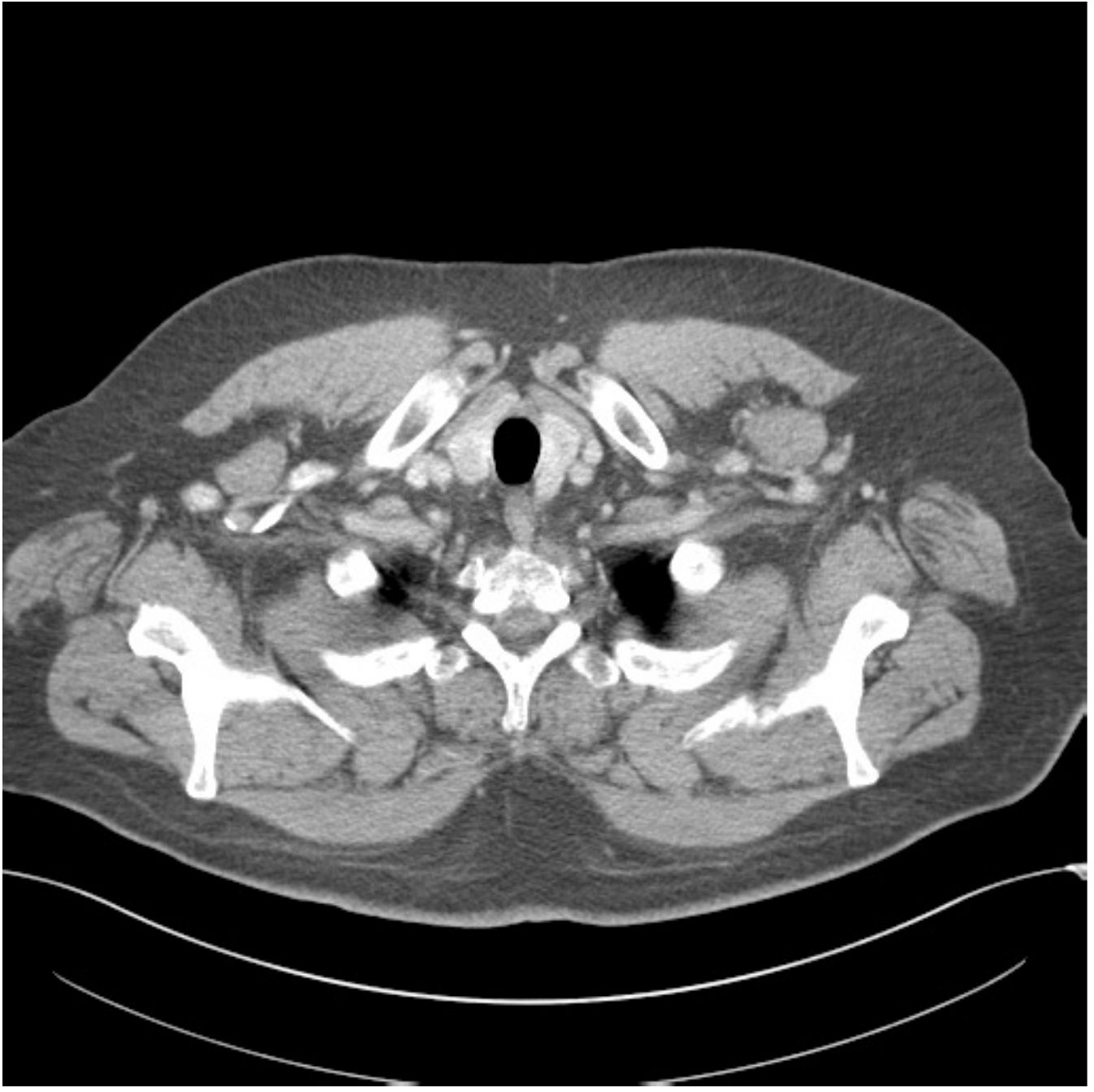}	};
	\end{tikzpicture}
	\put(-92,-5){ \color{black}{\bf \small{\rm{L143-slice15}}}} 
	\caption{Seven regular-dose slices for training the MARS model. The first row displays four slices of patient L096 and the second row shows three training slices from patients L067 and L143, respectively.}
	\label{fig:train_mayo_dataset}
	\vspace{0.2in}
\end{figure}

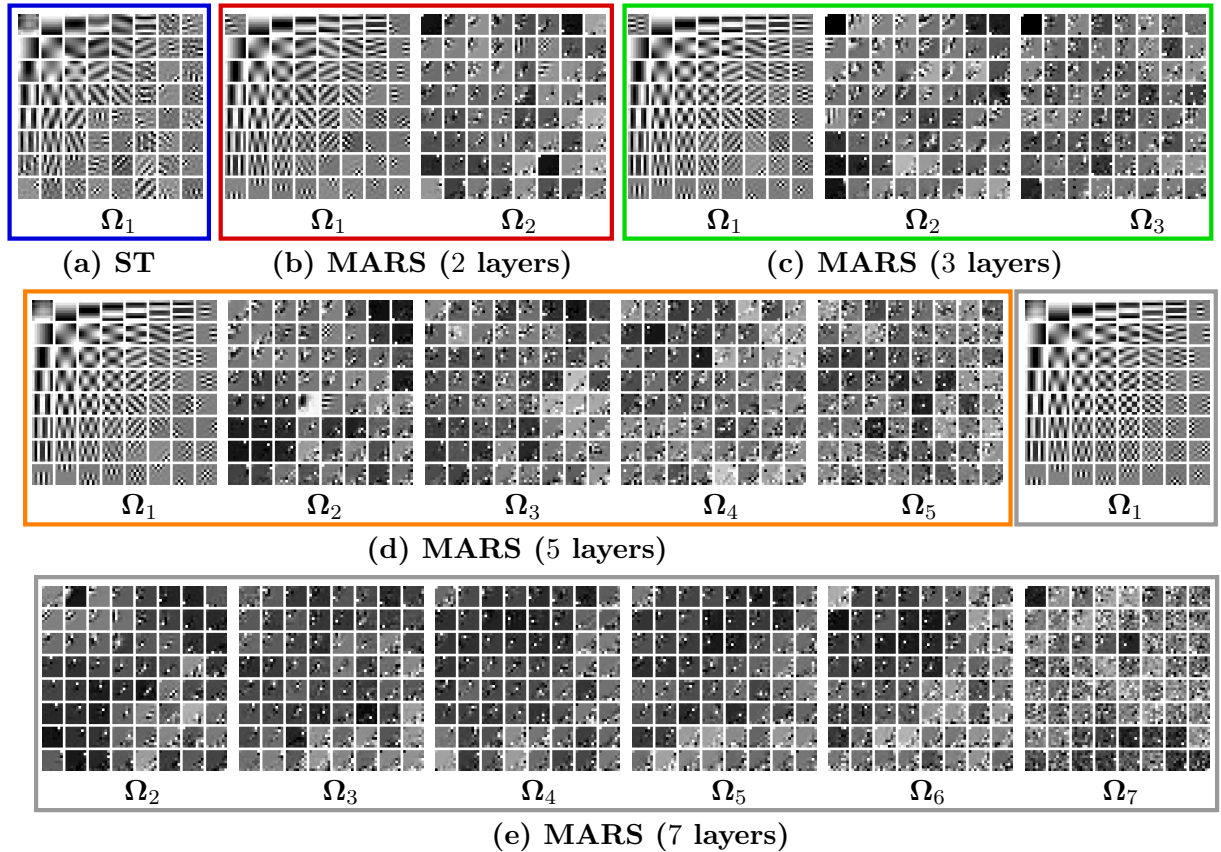
\begin{figure}[!h]
	\centering
	\begin{tabular}{ccc}
		\vspace{-0.15in}
		\hspace{-0.2in}
		\begin{tikzpicture}
		[spy using outlines={rectangle,green,magnification=2,size=6mm, connect spies}]
		\node {	\includegraphics[width=0.15\textwidth]{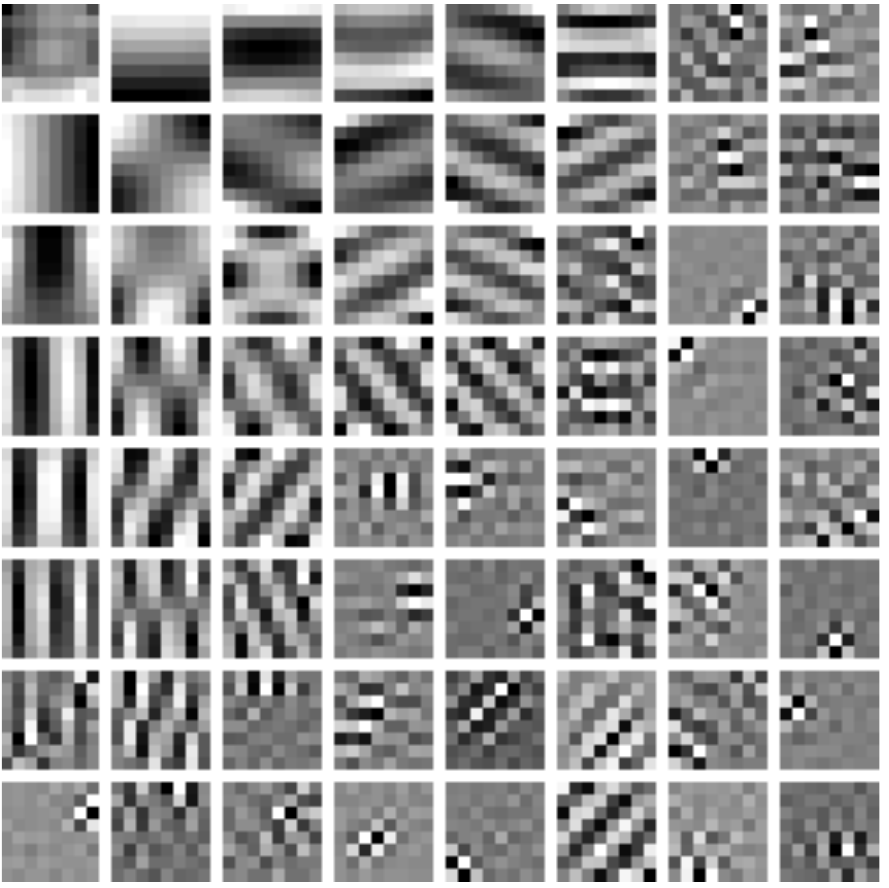}};
		\draw[blue, ultra thick] (-1.35,-1.75) rectangle (1.3,1.35);
		\put(-4,-46){ \color{black}{\bf \small{$\omg_1$}}}
		\end{tikzpicture}
		\put(-62,-12){ \color{black}{\bf \small{(a) ST}}}
		\begin{tikzpicture}
		[spy using outlines={rectangle,green,magnification=2,size=6mm, connect spies}]
		\node{\includegraphics[width=0.15\textwidth]{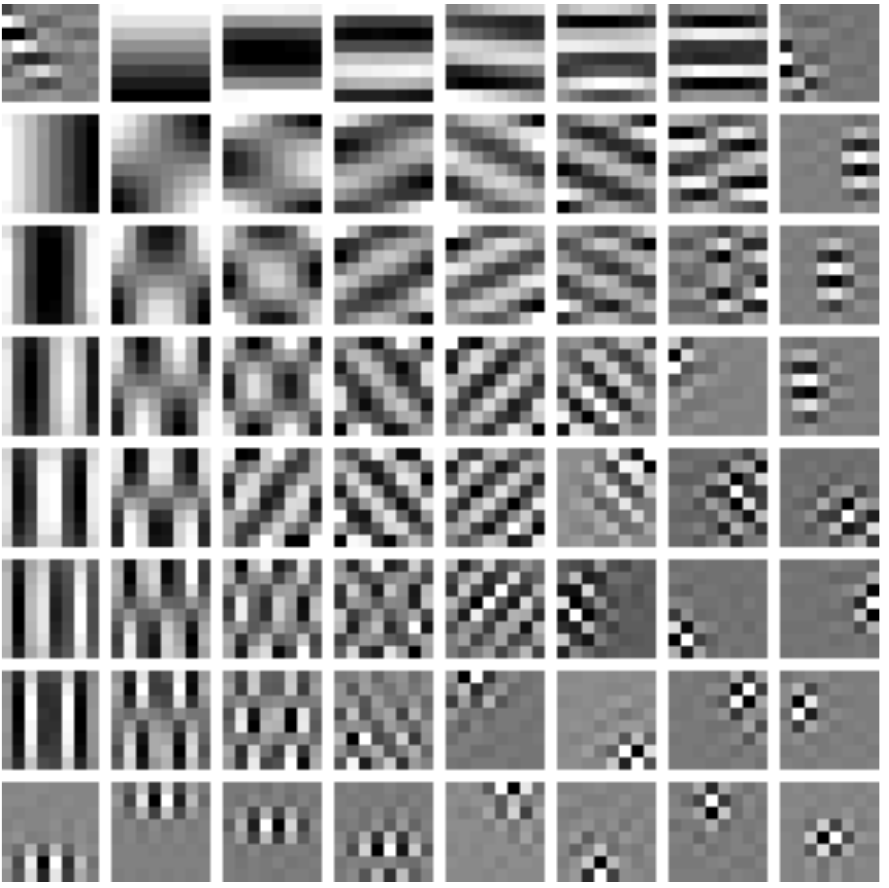}
			\includegraphics[width=0.15\textwidth]{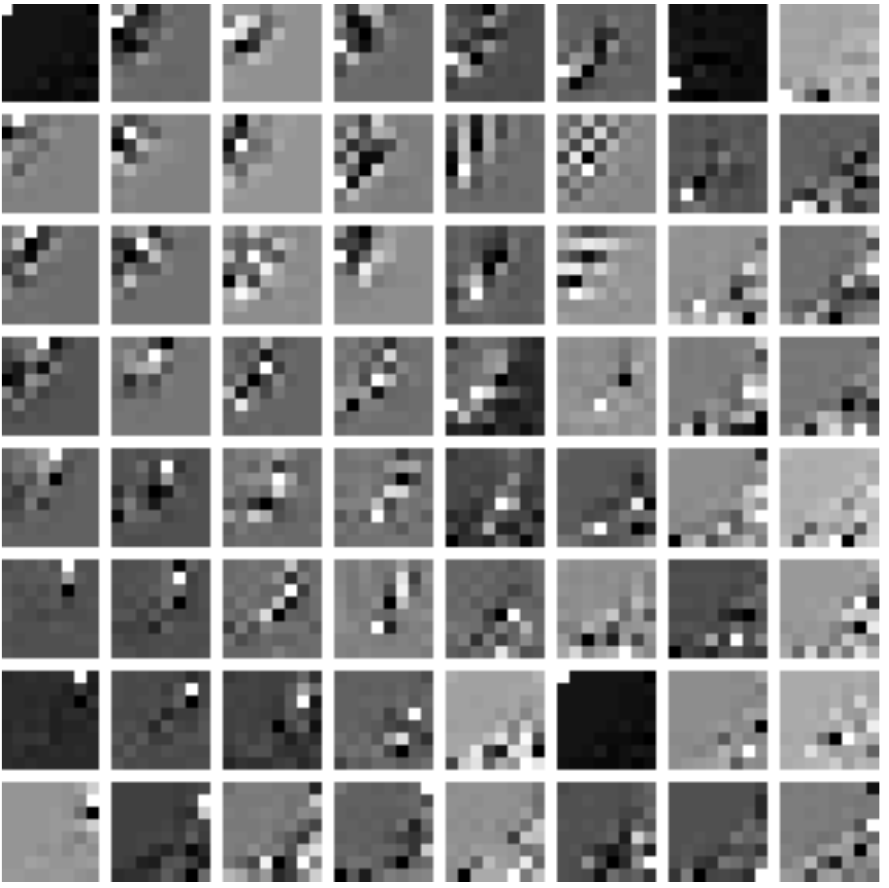}
		};
		\draw[red, ultra thick] (-2.60,-1.75) rectangle (2.6,1.35);
		\put(-40,-46){ \color{black}{\bf \small{$\omg_1$}}}
		\put(32,-46){ \color{black}{\bf \small{$\omg_2$}}}
		\end{tikzpicture}
		\put(-135,-12){ \color{black}{\bf \small{(b) MARS ($2$ layers)}}}
		\begin{tikzpicture}
		[spy using outlines={rectangle,green,magnification=2,size=6mm, connect spies}]
		\node{\includegraphics[width=0.15\textwidth]{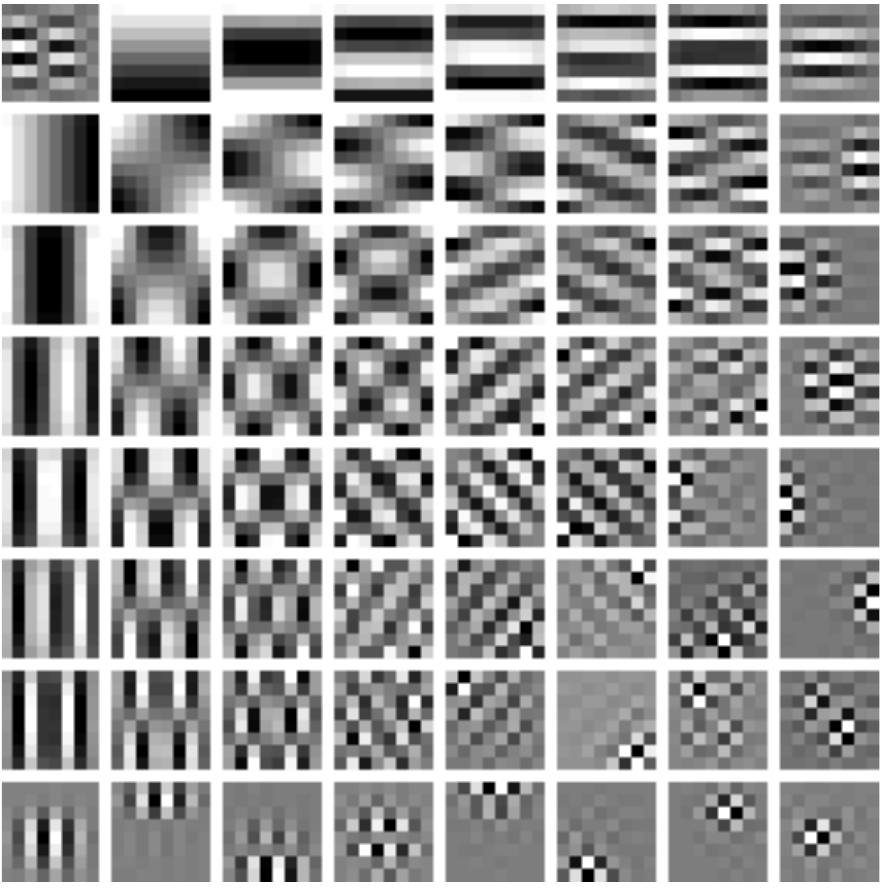}
			\includegraphics[width=0.15\textwidth]{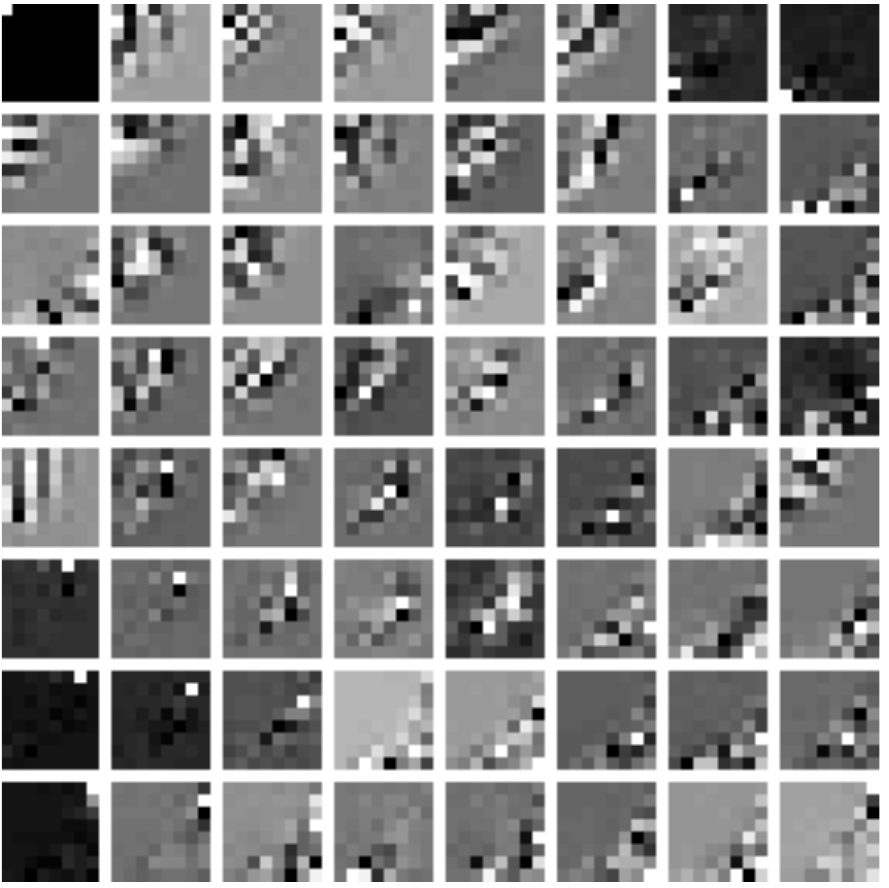}
			\includegraphics[width=0.15\textwidth]{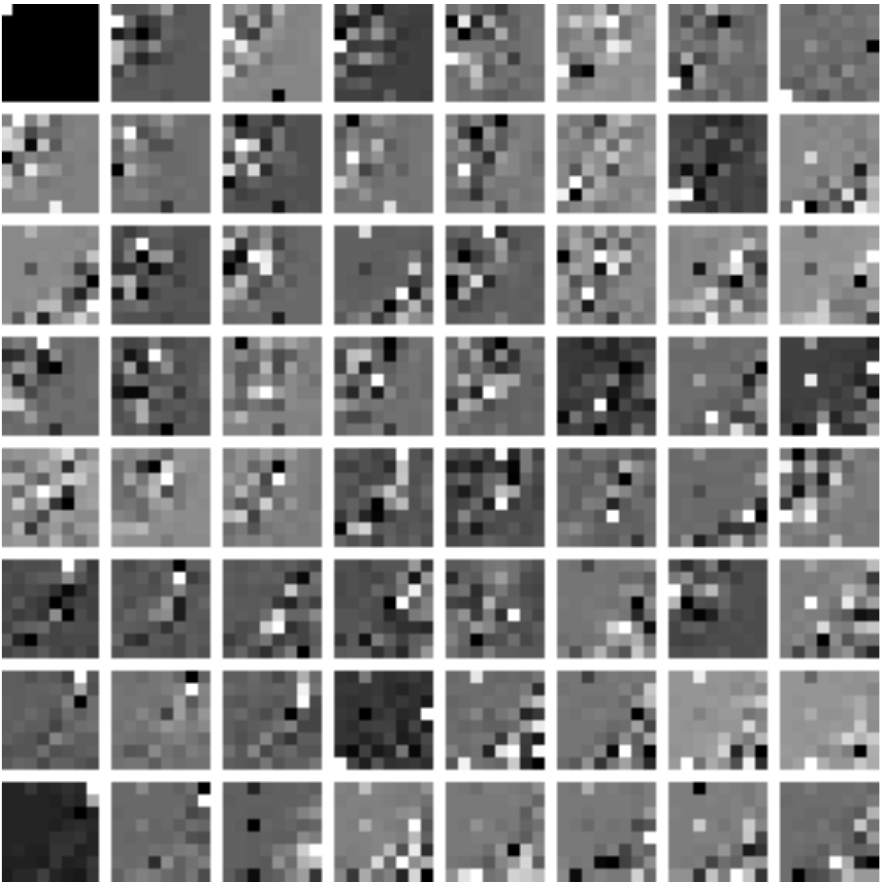}
		};
		\draw[green, ultra thick] (-3.90,-1.75) rectangle (3.9,1.35);
		\put(-80,-46){ \color{black}{\bf \small{$\omg_1$}}}
		\put(-5,-46){ \color{black}{\bf \small{$\omg_2$}}}
		\put(80,-46){ \color{black}{\bf \small{$\omg_3$}}}
		\end{tikzpicture}
		\put(-175,-12){ \color{black}{\bf \small{(c) MARS ($3$ layers)}}}
		\\
		\\
		\begin{tikzpicture}
		[spy using outlines={rectangle,green,magnification=2,size=6mm, connect spies}]
		\node{\includegraphics[width=0.15\textwidth]{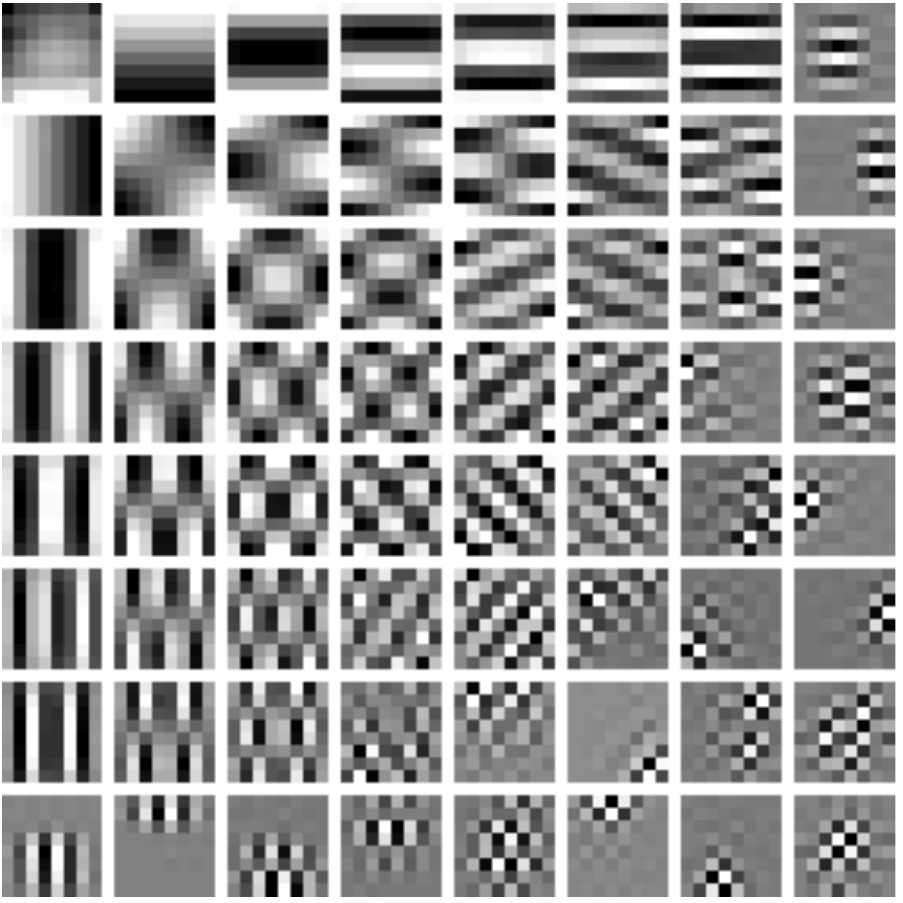}
			\includegraphics[width=0.15\textwidth]{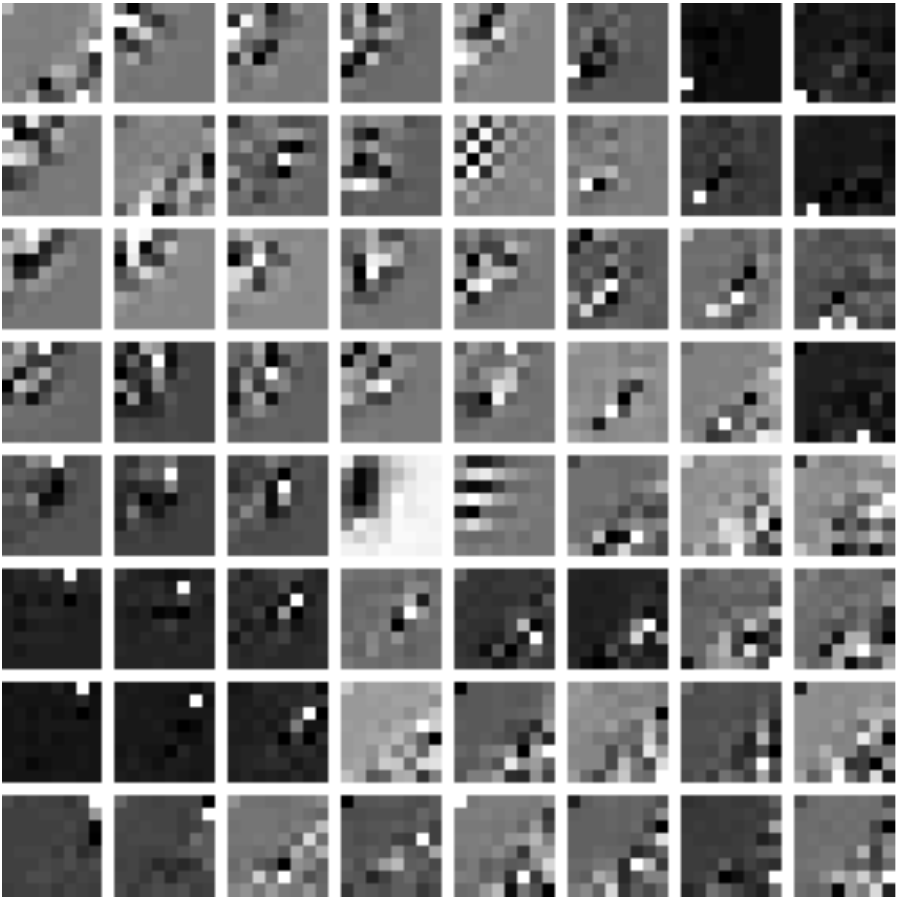}
			\includegraphics[width=0.15\textwidth]{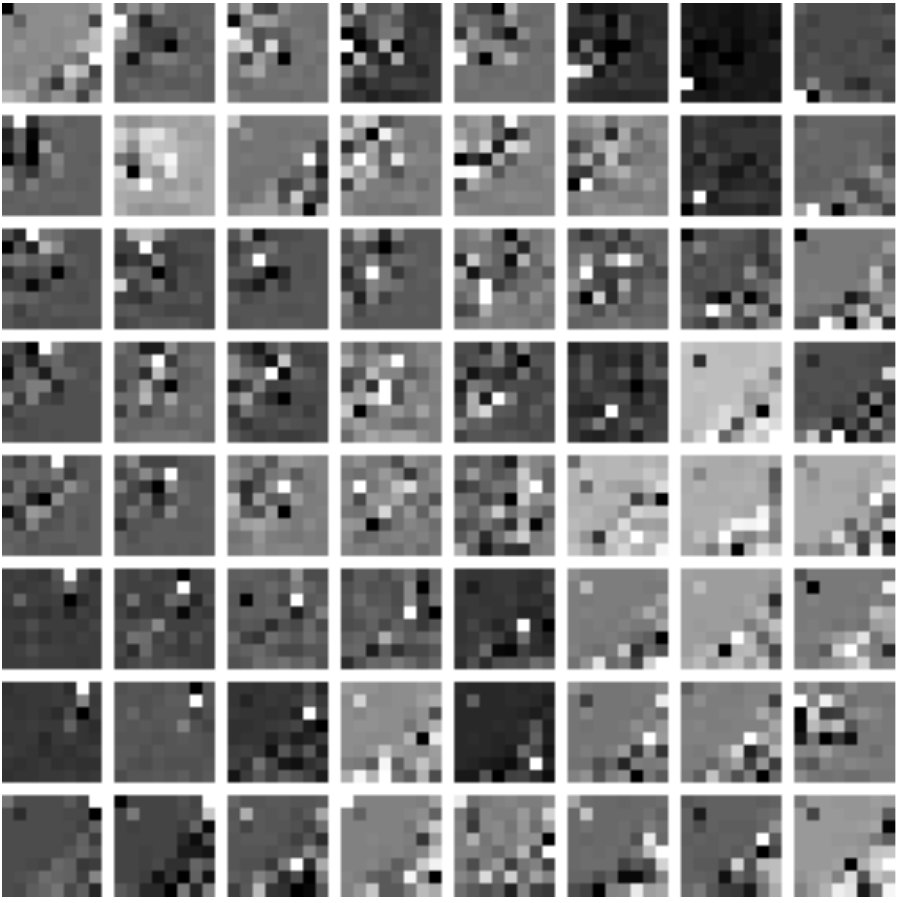}
			\includegraphics[width=0.15\textwidth]{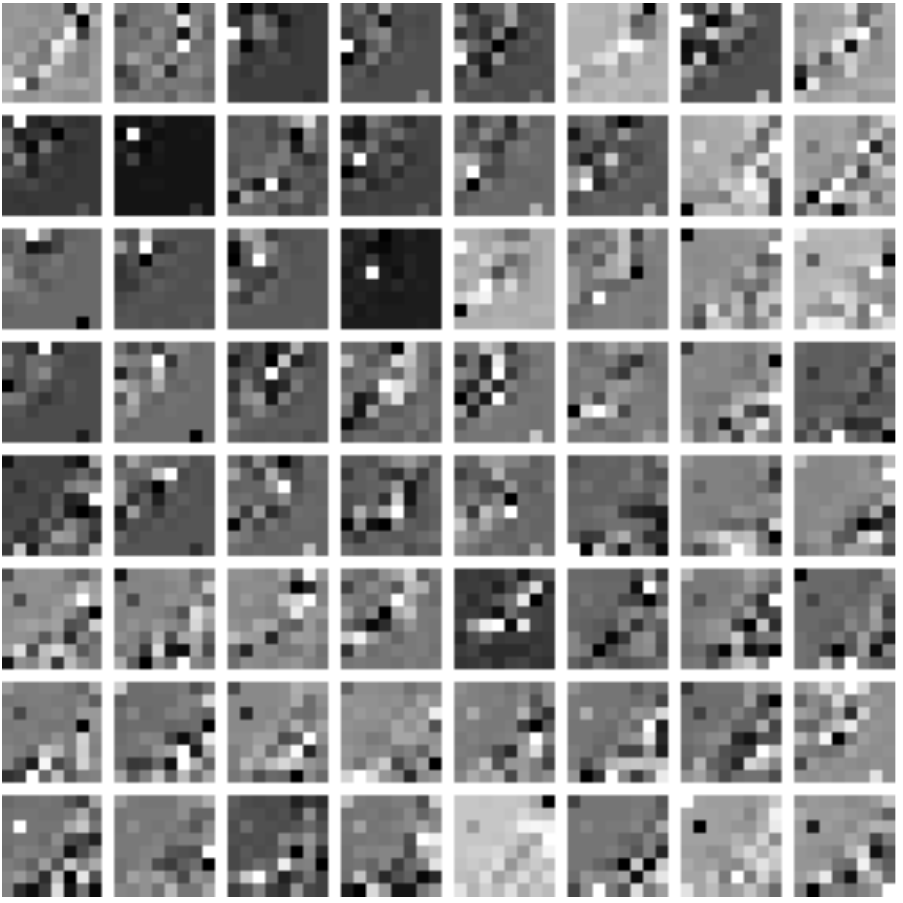}
			\includegraphics[width=0.15\textwidth]{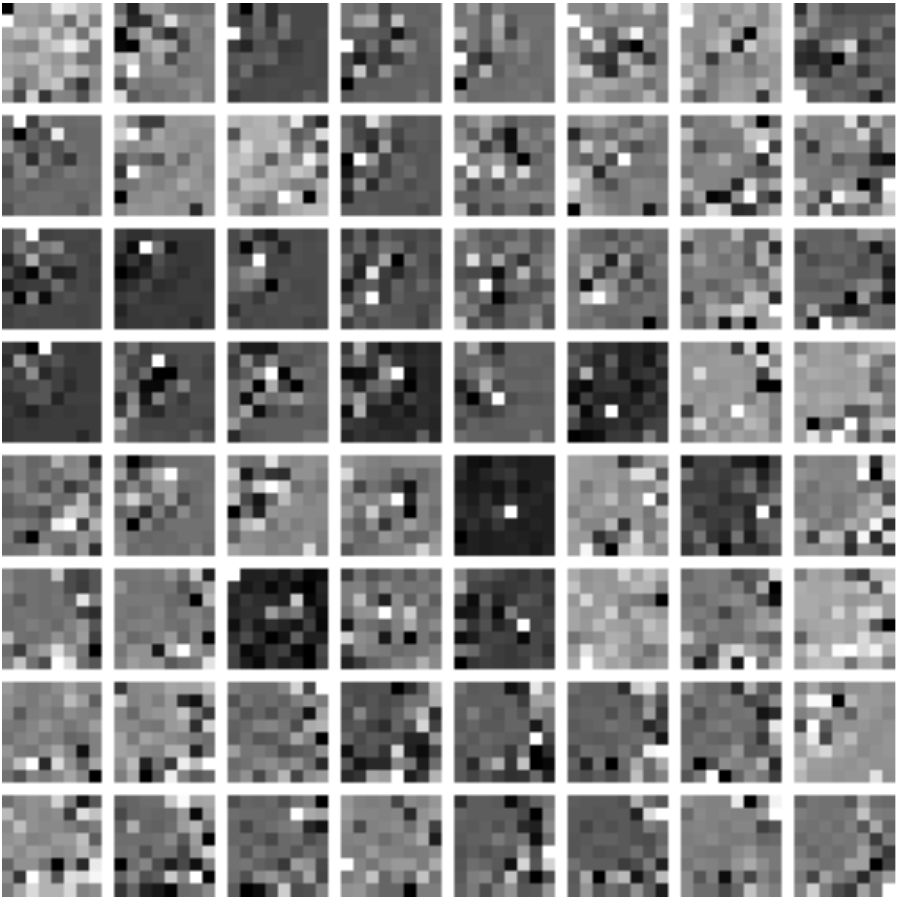}
		};
		\draw[orange, ultra thick] (-6.55,-1.75) rectangle (6.55,1.35);
		\put(-150,-46){ \color{black}{\bf \small{$\omg_1$}}}
		\put(-80,-46){ \color{black}{\bf \small{$\omg_2$}}}
		\put(-5,-46){ \color{black}{\bf \small{$\omg_3$}}}
		\put(70,-46){ \color{black}{\bf \small{$\omg_4$}}}
		\put(145,-46){ \color{black}{\bf \small{$\omg_5$}}}
		\end{tikzpicture}
		\put(-249,-12){ \color{black}{\bf \small{(d) MARS ($5$ layers)}}}
		\begin{tikzpicture}
		[spy using outlines={rectangle,green,magnification=2,size=6mm, connect spies}]
		\node {	\includegraphics[width=0.15\textwidth]{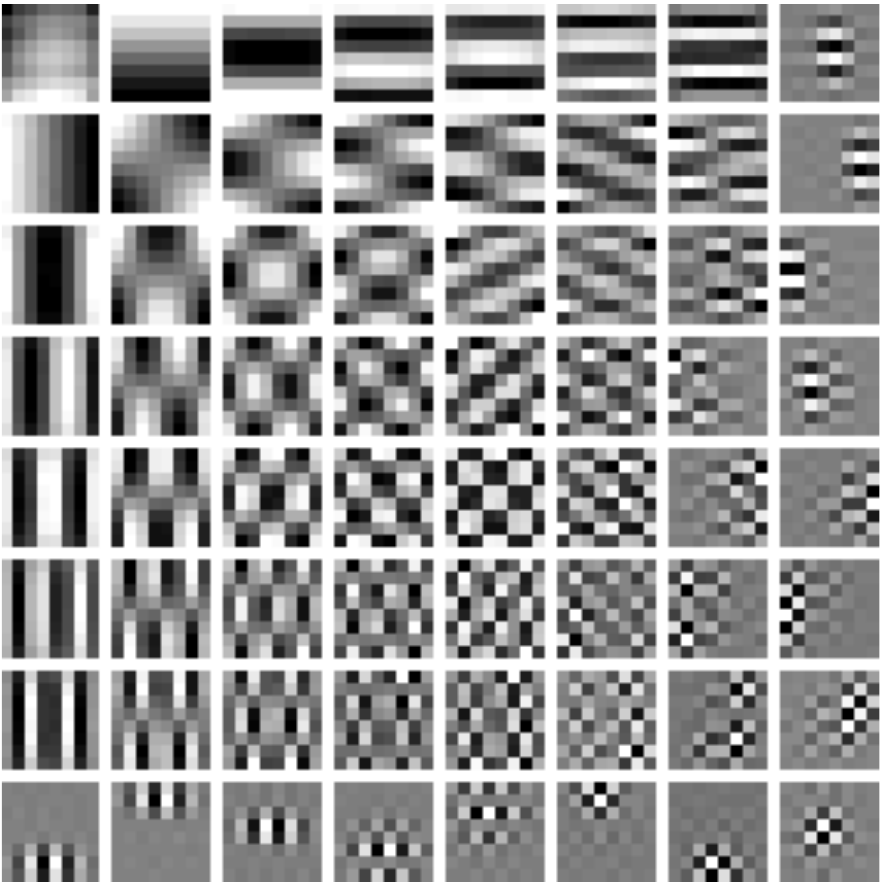}};
		\draw[gray, ultra thick] (-1.35,-1.75) rectangle (1.3,1.35);
		\put(-4,-46){ \color{black}{\bf \small{$\omg_1$}}}
		\end{tikzpicture}
		\vspace{0.05in}
		\\
		
		\begin{tikzpicture}
		[spy using outlines={rectangle,green,magnification=2,size=6mm, connect spies}]
		\node{\includegraphics[width=0.15\textwidth]{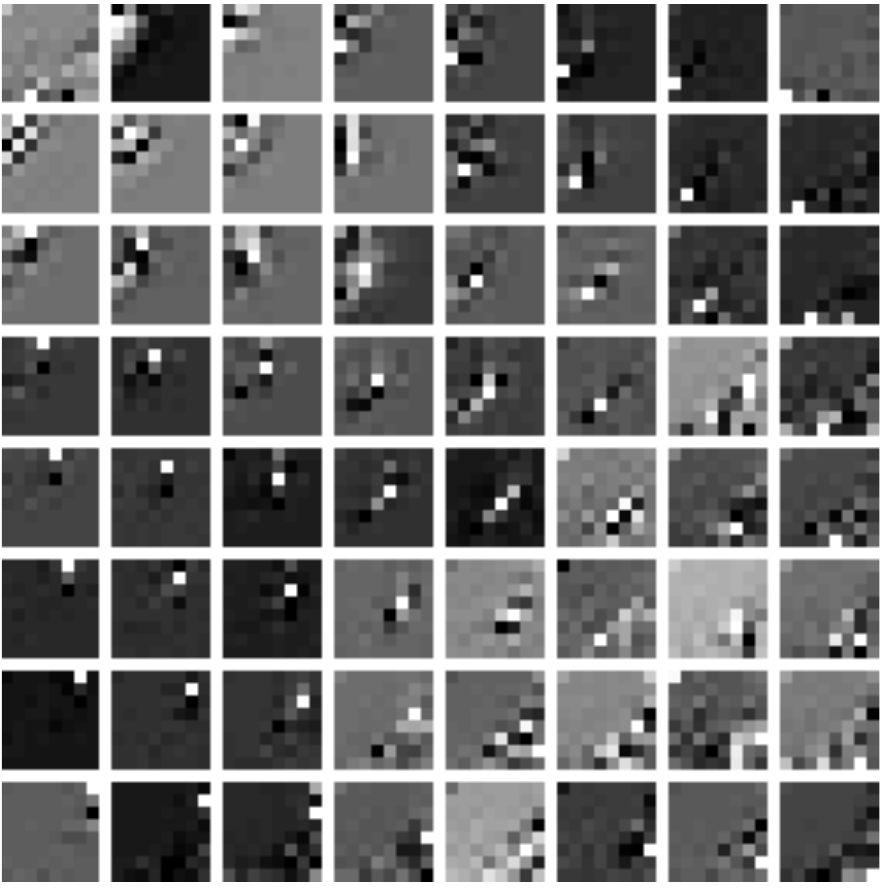}
			\includegraphics[width=0.15\textwidth]{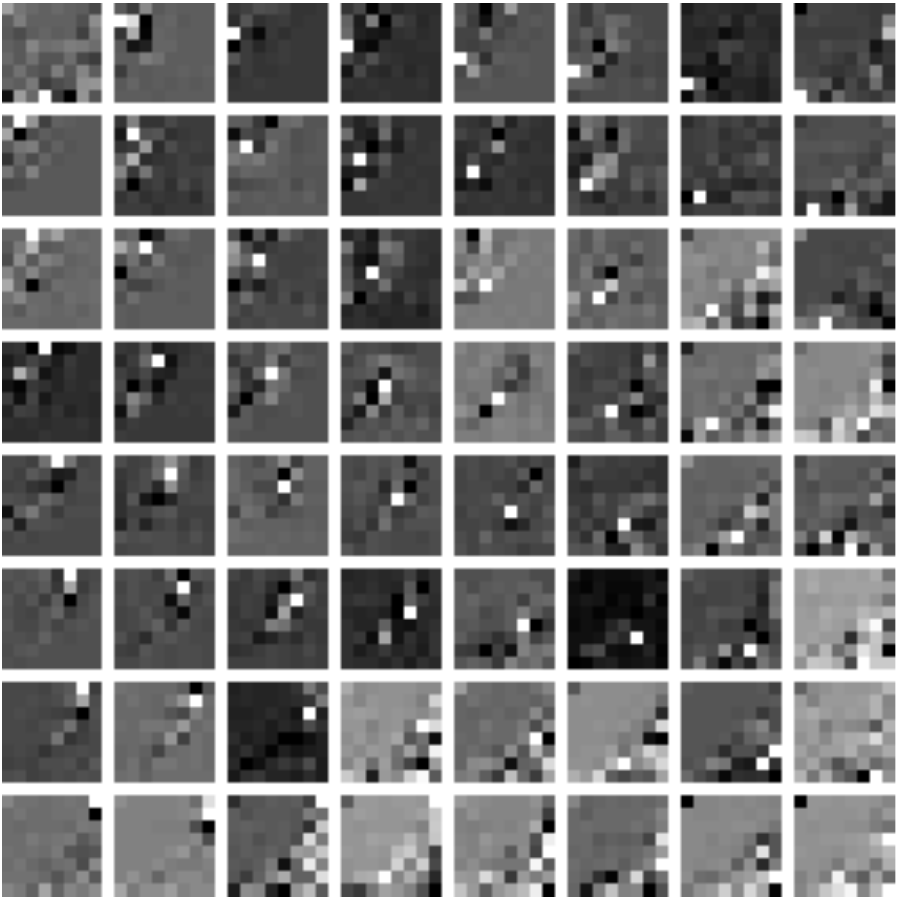}
			\includegraphics[width=0.15\textwidth]{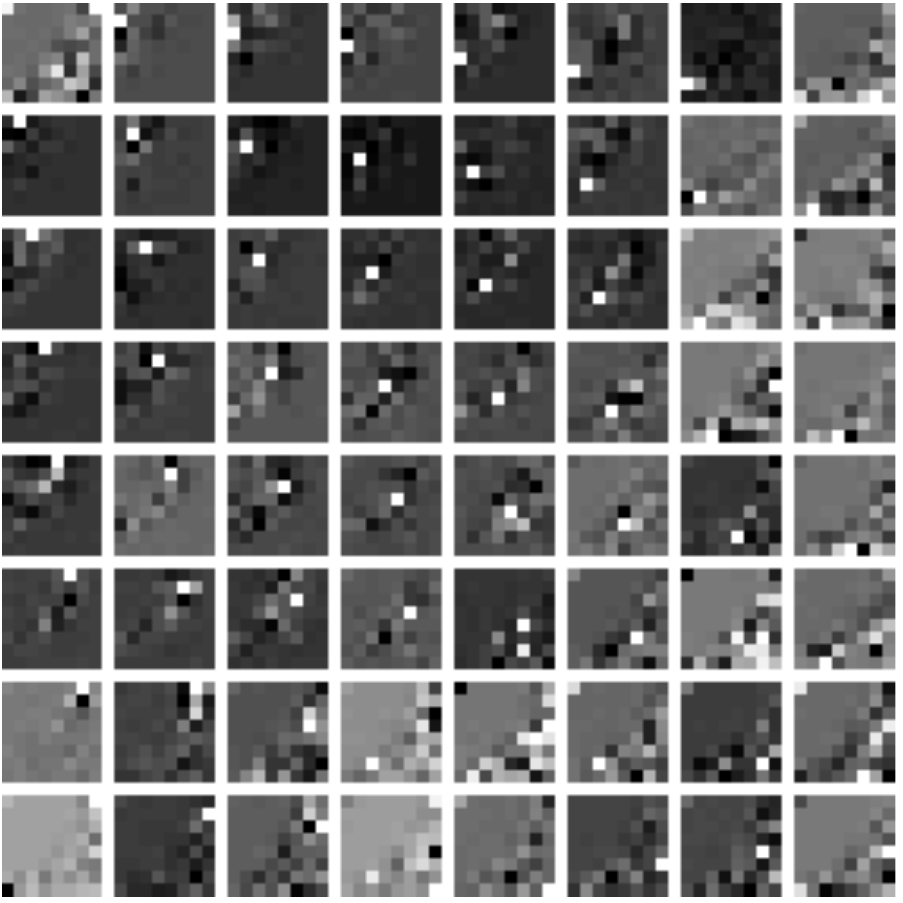}
			\includegraphics[width=0.15\textwidth]{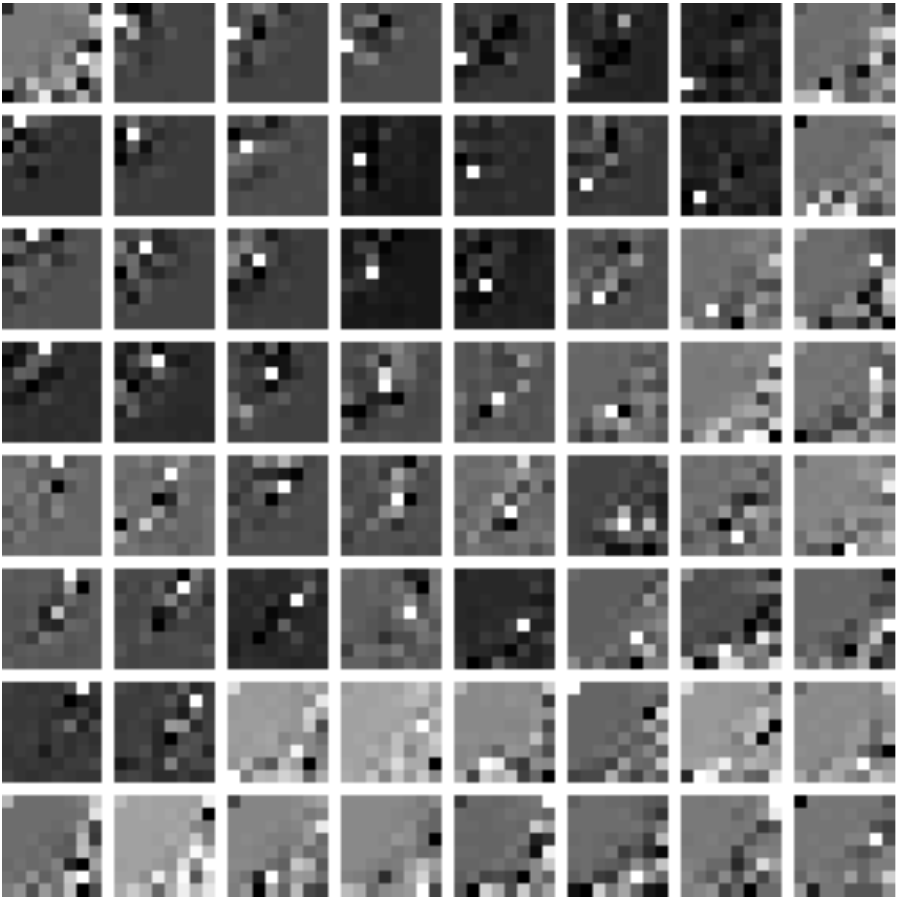}
			\includegraphics[width=0.15\textwidth]{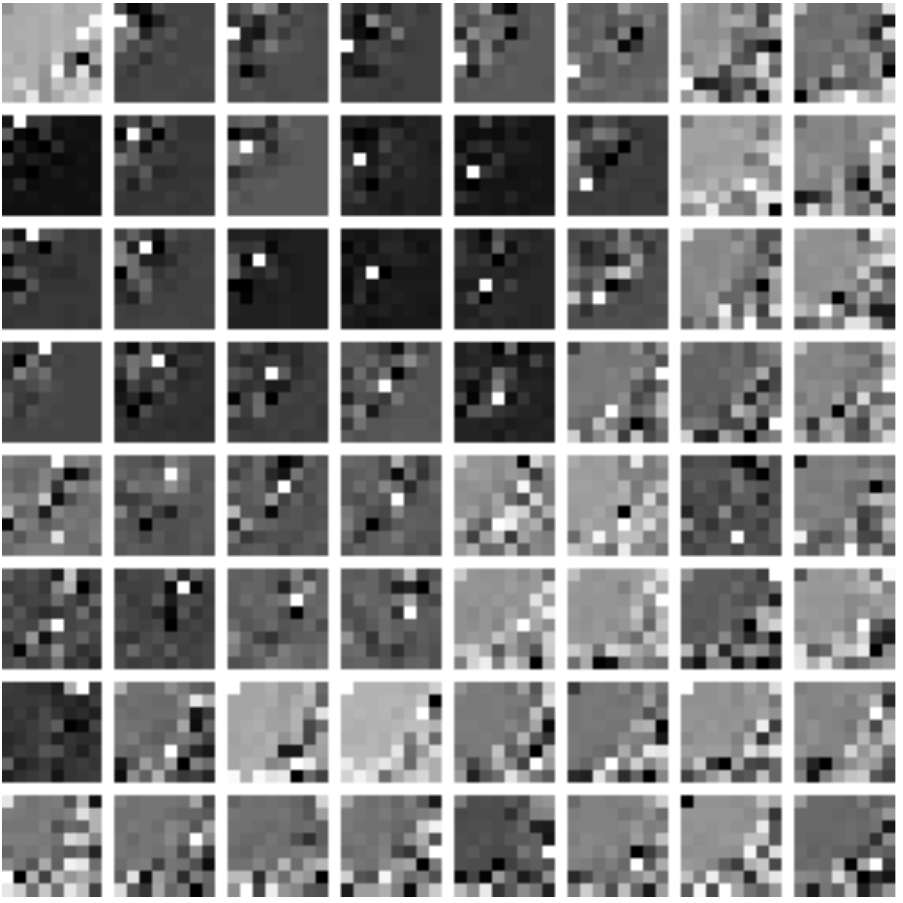}
			\includegraphics[width=0.15\textwidth]{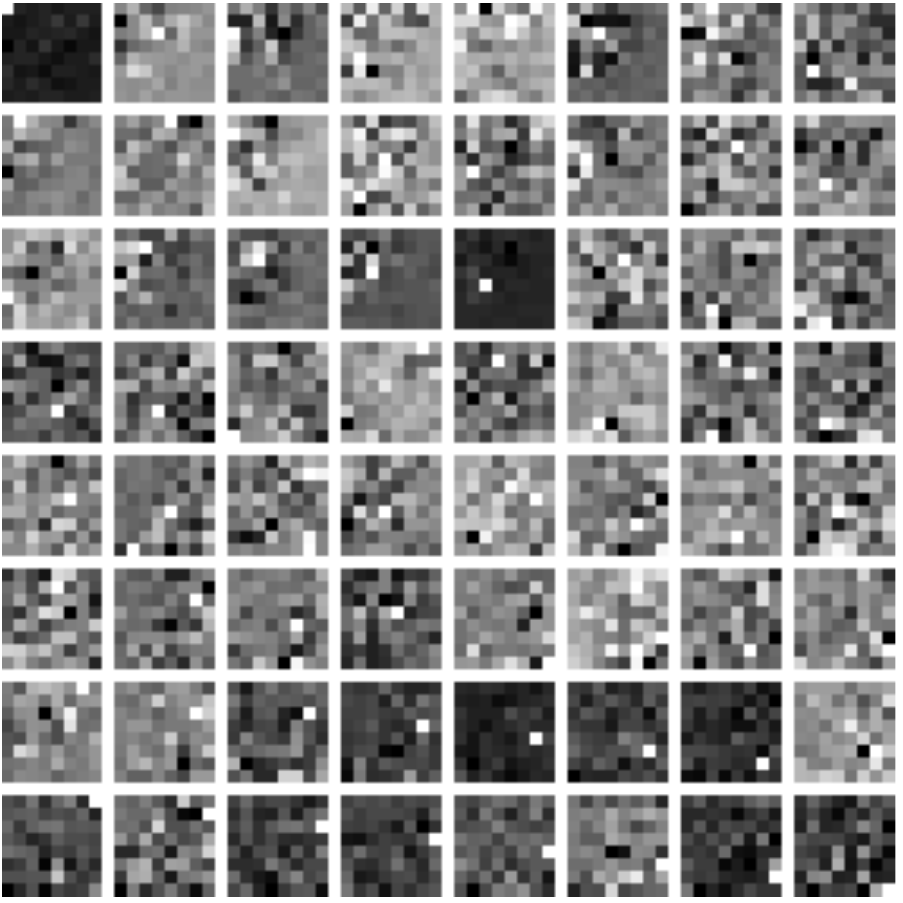}
		};
		\draw[gray, ultra thick] (-7.85,-1.75) rectangle (7.85,1.35);
		\put(-190,-46){ \color{black}{\bf \small{$\omg_2$}}}
		\put(-115,-46){ \color{black}{\bf \small{$\omg_3$}}}
		\put(-40,-46){ \color{black}{\bf \small{$\omg_4$}}}
		\put(32,-46){ \color{black}{\bf \small{$\omg_5$}}}
		\put(107,-46){ \color{black}{\bf \small{$\omg_6$}}}
		\put(180,-46){ \color{black}{\bf \small{$\omg_7$}}}
		\end{tikzpicture}
		\put(-280,-12){ \color{black}{\bf \small{(e) MARS ($7$ layers)}}}
	\end{tabular}
	\caption{Transforms learned from Mayo Clinic data. Beyond the first layer, the rows of the transforms are shown as (square) 2D patches and sparsify transform-domain residuals.}
	\vspace{0.1in}
	\label{fig:lear_mayo_tran}
\end{figure}

First, we study transform training based on Mayo Clinic data. As shown in Fig.~\ref{fig:train_mayo_dataset}, seven $512\times 512$ slices obtained at regular dose from three patients are used for transform learning. The number of pixels $N_p\approx 2.6 \times 10^5$.
Similar to the phantom experiments, $8 \times 8$ overlapping patches are extracted with a $1\times 1$ patch stride. The number of overall training patches $N$ is about $1.8 \times 10^6$.
%, while the patch number in the testing is about $255$K.
We set $\eta=100$ for ST, $(\eta_1$, $\eta_2)$ $=$ $(80$, $60)$ for MARS2, $(\eta_1$, $\eta_2$, $\eta_3)$ $=$ $(60$, $60$, $40)$ for MARS3, $(\eta_1$, $\eta_2$, $\eta_3$, $\eta_4$, $\eta_5)$ $=$ $(100$, $100$, $80$, $80$, $60)$ for MARS5, $(\eta_1$, $\eta_2$, $\eta_3$, $\eta_4$, $\eta_5$, $\eta_6$, $\eta_7)$ $=$ $(150$, $140$, $130$, $120$, $110$, $100$, $90)$ for MARS7. The iteration number $T=1000$ in \textbf{Algorithm~\ref{alg: mrst_train}.} Fig.~\ref{fig:lear_mayo_tran} illustrates the learned transforms obtained with Mayo Clinic data. Different from the XCAT phantom case, these transforms up to MARS5 display more complex features and structures. The rich features of the MARS models better sparsify the training images over layers compared to the single-layer model (ST).

\subsubsection{Simulation Framework, Reconstruction Results, and Comparisons}
The synthesized low-dose clinical measurements are simulated from regular-dose images at a resolution of $\Delta_x=\Delta_y=0.9766$ mm with a fan-beam CT geometry corresponding to a monoenergetic source at incident photon intensity $I_0=1\times 10^{4}$. The sinograms are of size $736\times 1152$. The width of each detector column is $1.2858$ mm, the source to detector distance is $1085.6$ mm, and the source to rotation center distance is $595$ mm. We reconstruct images of size $512\times 512$ with the pixel size being $0.69$ mm $\times$ $0.69$ mm.

We conducted experiments on one test slice used for parameter tuning (L067-slice 120) and four independent test slices (L109-slice 90, L192-slice90, L333-slice140, L506-slice 100) of the Mayo Clinic data.  
For PWLS-EP, we ran $1000$ iterations using relaxed OS-LALM and set regularization parameter $\beta = 2^{15.5}$.
We used the same $T_O = 1500$ as the phantom experiments for \textbf{Algorithm~\ref{alg: mrst_recon}}. 
%We set the regularization parameters as TABLE~\ref{tab:Mayo_para}.
%We set the regularization parameters as $(\beta$, $\gamma)$ $=$ $(2.5\times10^4$, $20)$ for ST, $(\beta$, $\gamma_1$, $\gamma_2)$ $=$ $(1.8\times10^4$, $30$, $10)$ for MARS2, $(\beta$, $\gamma_1$, $\gamma_2$, $\gamma_3)$ $=$ $(1.8\times10^4$, $30$, $12$, $10)$ for MARS3, $(\beta$, $\gamma_1$, $\gamma_2$, $\gamma_3$, $\gamma_4$, $\gamma_5)$ $=$ $(1.8\times10^4$, $25$, $15$, $10$, $5$, $1)$ for MARS5, and $(\beta$, $\gamma_1$, $\gamma_2$, $\gamma_3$, $\gamma_4$, $\gamma_5$, $\gamma_6$, $\gamma_7)$ $=$ $(3.5\times10^4$, $18$, $15$, $12$, $9$, $5$, $3$, $1)$ for MARS7, respectively.
The process of selecting a general set of reconstruction parameters ($\beta, \{\gamma_l, 1\leq l \leq L\}$) for the Mayo Clinic test slices is identical to that for the XCAT phantom in Section \ref{xcat_experiment} The selected regularization parameter $\beta$ and the parameters $\gamma_l$ that control the sparsity of the coefficient maps are $(\beta, \gamma)$ $=$ $(2.5\times 10^4, 30)$ for ST, $(\beta$, $\gamma_1$, $\gamma_2)$ $=$ $(1.8\times 10^4$, $30$, $10)$ for MARS2, $(\beta$, $\gamma_1$, $\gamma_2$, $\gamma_3)$ $=$ $(1.8\times 10^4$, $30$, $12$, $10)$ for MARS3, $(\beta$, $\gamma_1$, $\gamma_2$, $\gamma_3$, $\gamma_4$, $\gamma_5)$ $=$ $(1.6\times 10^4$, $30$, $20$, $10$, $7$, $5)$ for MARS5, and $(\beta$, $\gamma_1$, $\gamma_2$, $\gamma_3$, $\gamma_4$, $\gamma_5$, $\gamma_6$, $\gamma_7)$ $=$ $(3.5\times 10^4$, $20$, $17$, $14$, $11$, $7$, $4$, $1)$ for MARS7, respectively. 

\begin{figure}[H]
	\centering  
	\begin{tikzpicture}
	[spy using outlines={rectangle,green,magnification=2,size=7mm, connect spies}]
	\node {\includegraphics[width=0.24\textwidth]{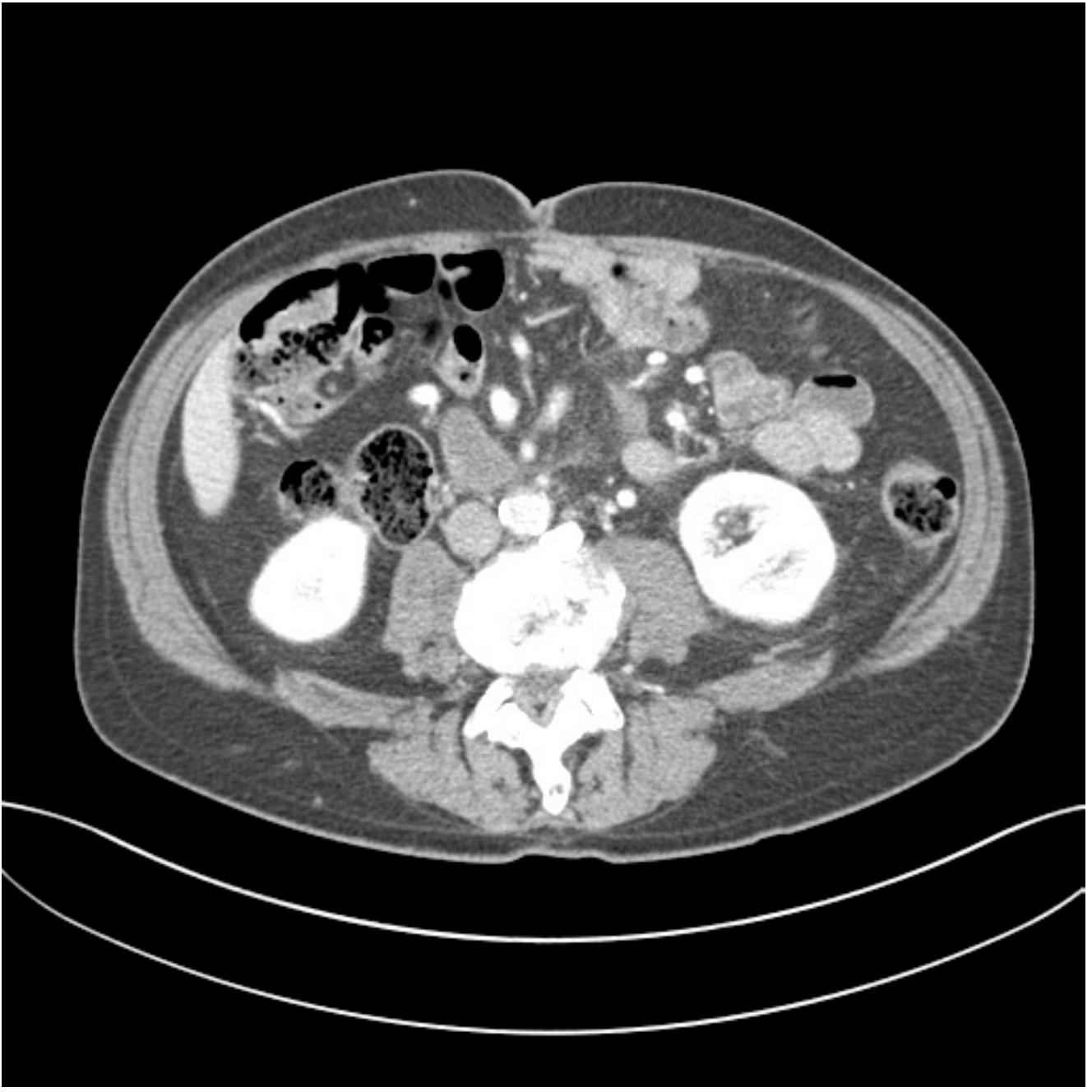}	};
	\spy [green, draw, height = 0.8cm, width = 0.8cm, magnification = 2.5,
	connect spies] on (-0.05,0.40) in node [left] at (-1.16,1.57);
	\spy [green, draw, height = 0.8cm, width = 0.8cm, magnification = 2.2,
	connect spies] on (-0.37,-0.70) in node [left] at (-1.17,-1.56);
	\spy [green, draw, height = 0.7cm, width = 0.9cm, magnification = 2.0,
	connect spies] on (0.45,-0.75) in node [left] at (1.95,-1.60);
	\end{tikzpicture}
	%	\put(-95,20){ \color{white}{\bf \small{RMSE:0.00}}}
	%	\put(-95,10){ \color{white}{\bf \small{SSIM:1.000}}}
	\put(-90,8){ \color{white}{\bf \small{Reference}}} 
	\hspace{-0.15in}
	\begin{tikzpicture}
	[spy using outlines={rectangle,green,magnification=2,size=7mm, connect spies}]
	\node {\includegraphics[width=0.24\textwidth]{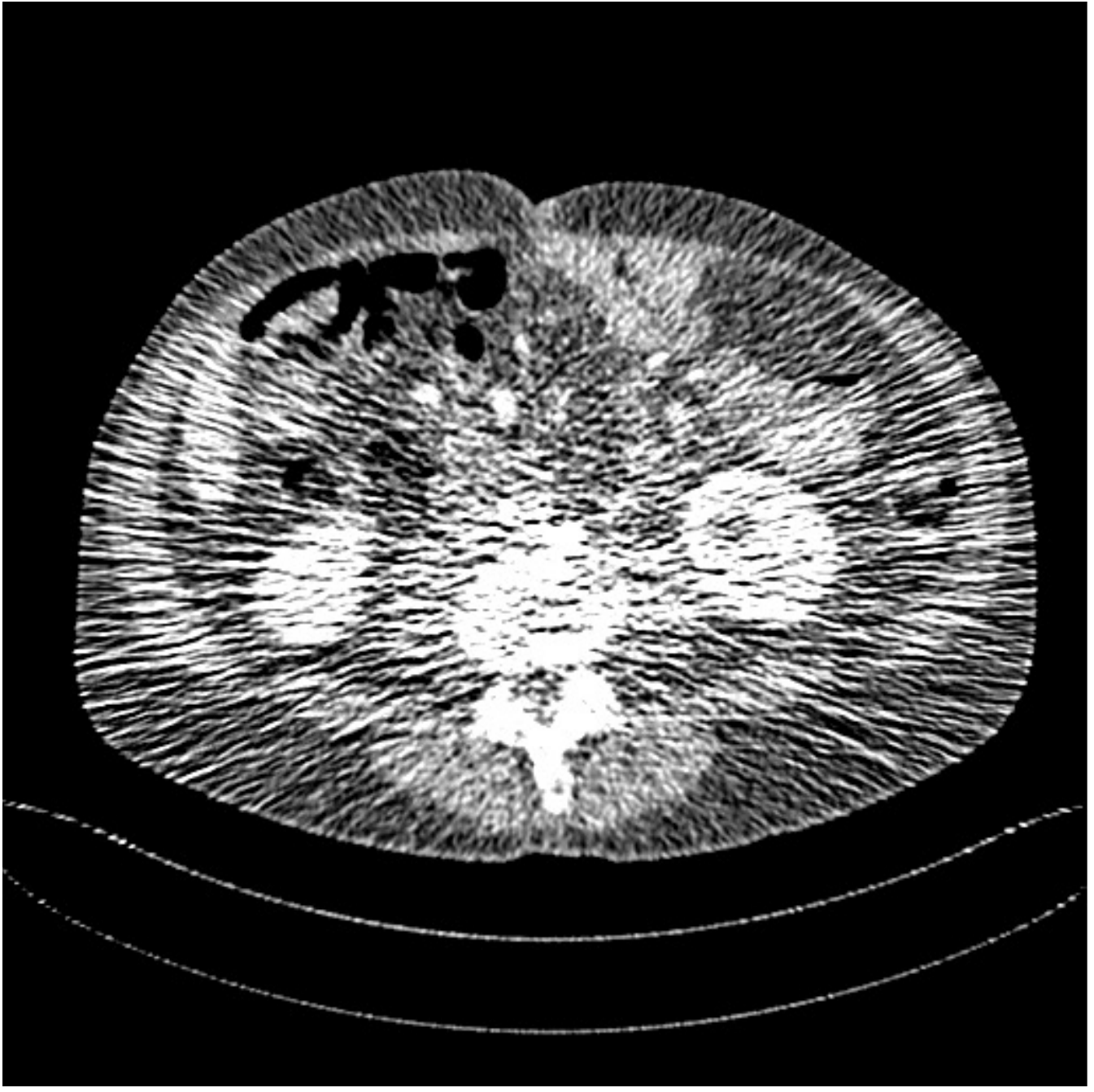}	};
	\spy [green, draw, height = 0.8cm, width = 0.8cm, magnification = 2.5,
	connect spies] on (-0.05,0.40) in node [left] at (-1.16,1.57);
	\spy [green, draw, height = 0.8cm, width = 0.8cm, magnification = 2.2,
	connect spies] on (-0.37,-0.70) in node [left] at (-1.17,-1.56);
	\spy [green, draw, height = 0.7cm, width = 0.9cm, magnification = 2.0,
	connect spies] on (0.45,-0.75) in node [left] at (1.95,-1.60);
	\end{tikzpicture}
	%	\put(-95,20){ \color{white}{\bf \small{RMSE:101.1}}}
	%	\put(-95,10){ \color{white}{\bf \small{SSIM:0.359}}}
	\put(-75,8){ \color{white}{\bf \small{FBP}}} 
	\hspace{-0.15in}
	\begin{tikzpicture}
	[spy using outlines={rectangle,green,magnification=2,size=7mm, connect spies}]
	\node {\includegraphics[width=0.24\textwidth]{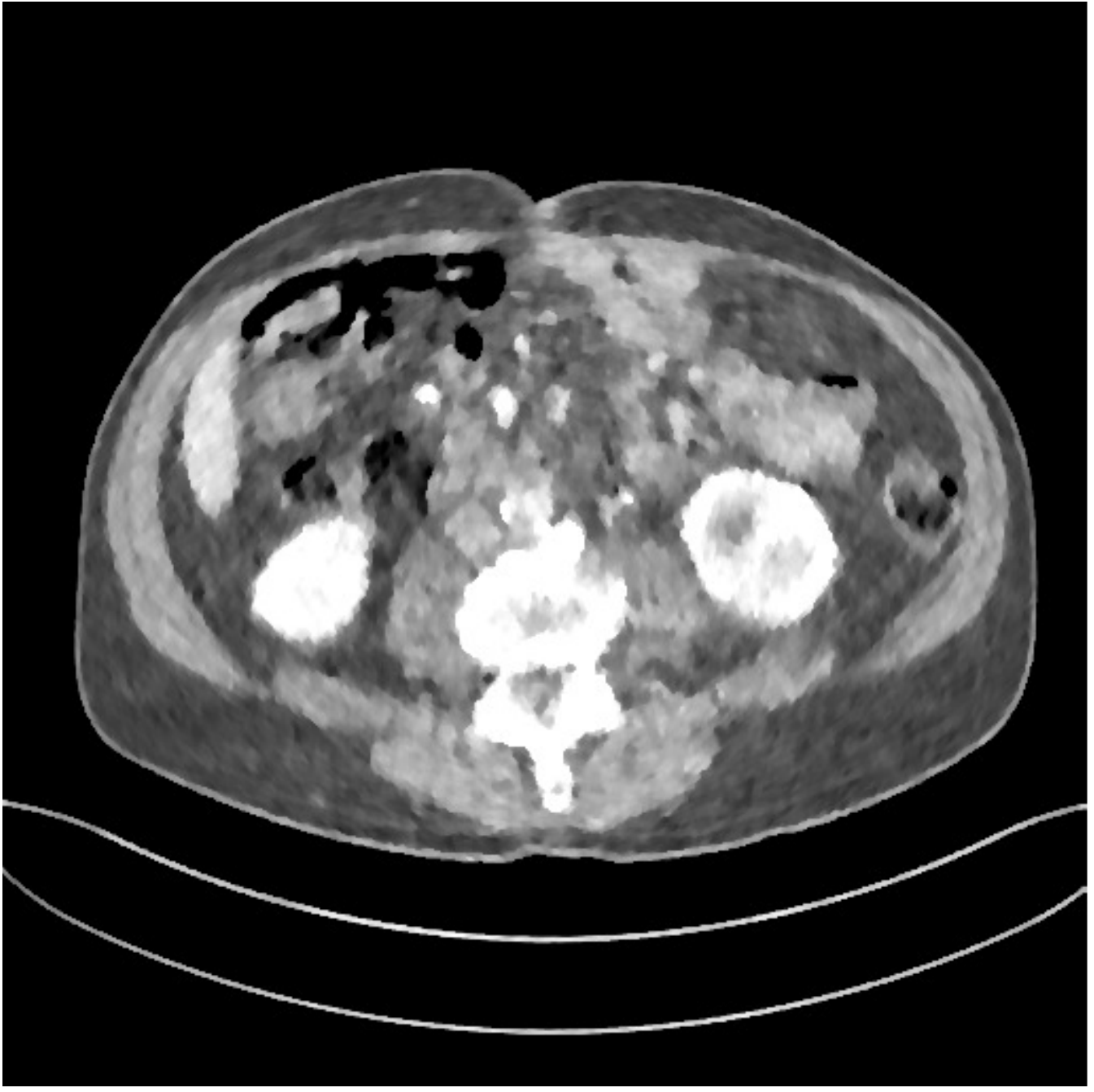}	};
	\spy [green, draw, height = 0.8cm, width = 0.8cm, magnification = 2.5,
	connect spies] on (-0.05,0.40) in node [left] at (-1.16,1.57);
	\spy [green, draw, height = 0.8cm, width = 0.8cm, magnification = 2.2,
	connect spies] on (-0.37,-0.70) in node [left] at (-1.17,-1.56);
	\spy [green, draw, height = 0.7cm, width = 0.9cm, magnification = 2.0,
	connect spies] on (0.45,-0.75) in node [left] at (1.95,-1.60);
	\end{tikzpicture}
	%	\put(-95,20){ \color{white}{\bf \small{RMSE:34.2}}}
	%	\put(-95,10){ \color{white}{\bf \small{SSIM:0.728}}}
	\put(-70,8){ \color{white}{\bf \small{EP}}}
	\hspace{-0.15in}	
	\begin{tikzpicture}
	[spy using outlines={rectangle,green,magnification=2,size=7mm, connect spies}]
	\node {\includegraphics[width=0.24\textwidth]{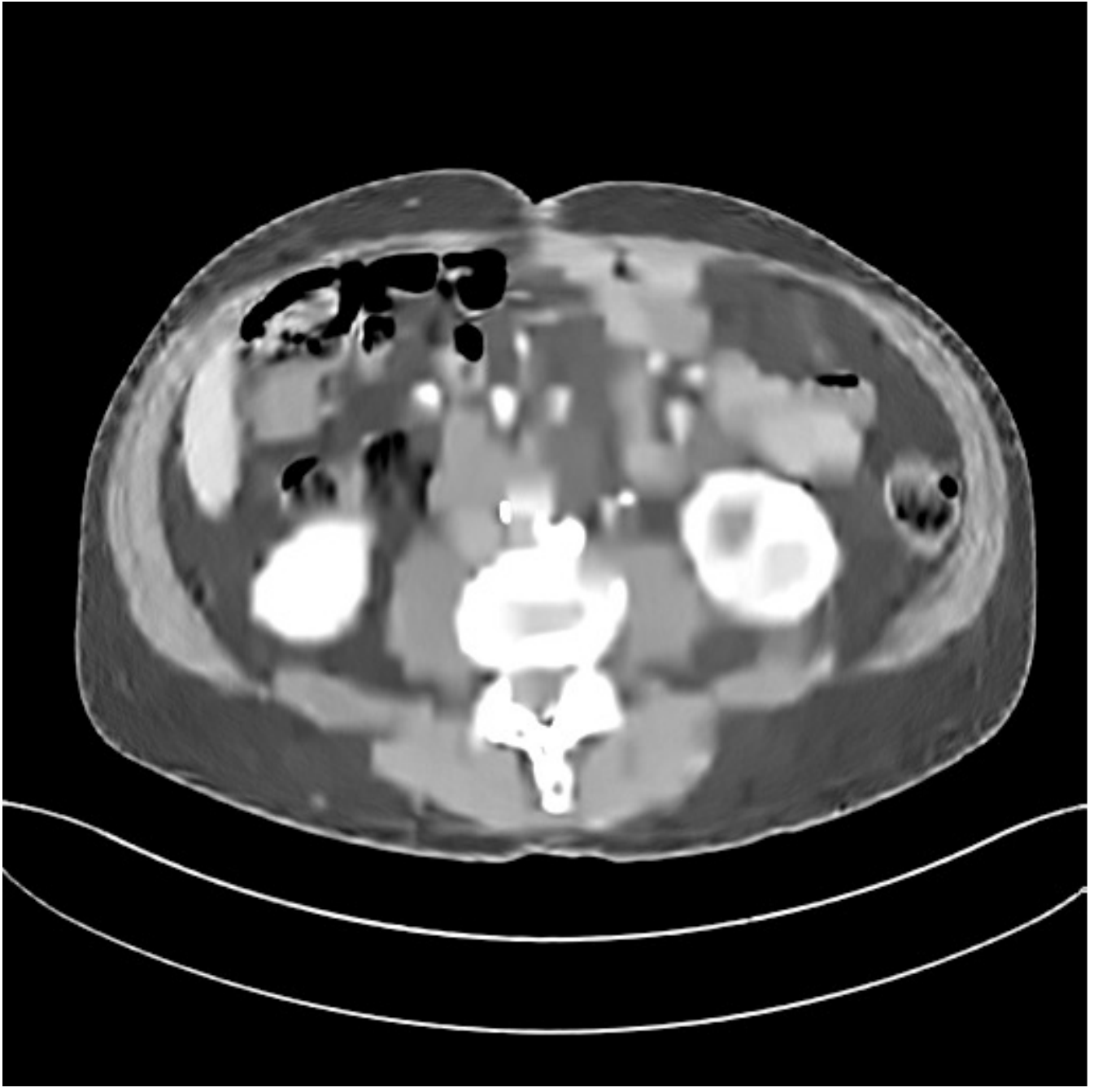}	};
	\spy [green, draw, height = 0.8cm, width = 0.8cm, magnification = 2.5,
	connect spies] on (-0.05,0.40) in node [left] at (-1.16,1.57);
	\spy [green, draw, height = 0.8cm, width = 0.8cm, magnification = 2.2,
	connect spies] on (-0.37,-0.70) in node [left] at (-1.17,-1.56);
	\spy [green, draw, height = 0.7cm, width = 0.9cm, magnification = 2.0,
	connect spies] on (0.45,-0.75) in node [left] at (1.95,-1.60);
	\end{tikzpicture}	
	%	\put(-95,20){ \color{white}{\bf \small{RMSE:30.3}}}
	%	\put(-95,10){ \color{white}{\bf \small{SSIM:0.724}}}
	\put(-70,8){ \color{white}{\bf \small{ST}}}
	\\
	\vspace{-0.10in}
	\begin{tikzpicture}
	[spy using outlines={rectangle,green,magnification=2,size=7mm, connect spies}]
	\node {\includegraphics[width=0.24\textwidth]{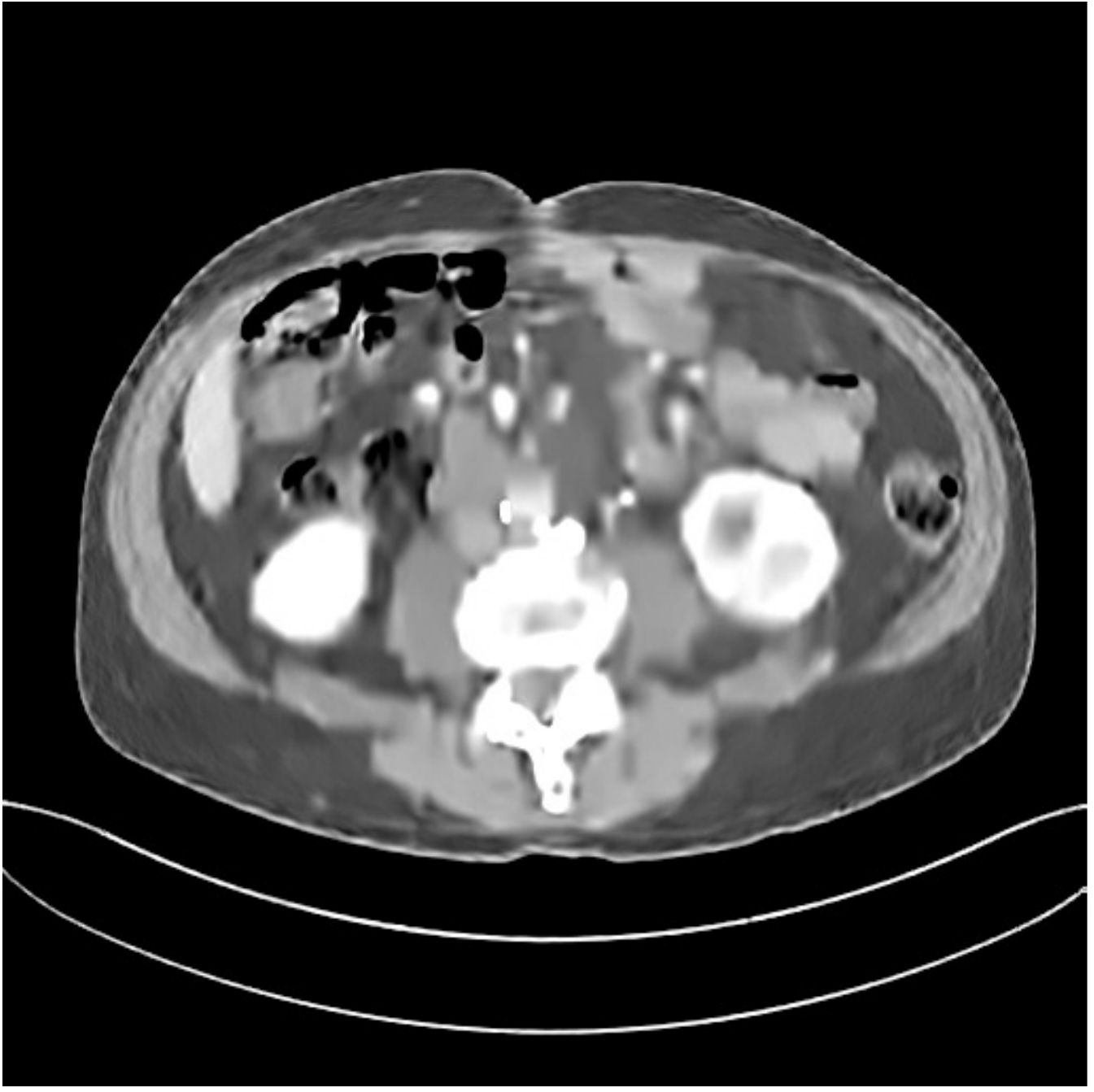}	};
	\spy [green, draw, height = 0.8cm, width = 0.8cm, magnification = 2.5,
	connect spies] on (-0.05,0.40) in node [left] at (-1.16,1.57);
	\spy [green, draw, height = 0.8cm, width = 0.8cm, magnification = 2.2,
	connect spies] on (-0.37,-0.70) in node [left] at (-1.17,-1.56);
	\spy [green, draw, height = 0.7cm, width = 0.9cm, magnification = 2.0,
	connect spies] on (0.45,-0.75) in node [left] at (1.95,-1.60);
	\end{tikzpicture}
	%	\put(-95,20){ \color{white}{\bf \small{RMSE:29.1}}}
	%	\put(-95,10){ \color{white}{\bf \small{SSIM:0.731}}}
	\put(-83,8){ \color{white}{\bf \small{MARS2}}}
	\hspace{-0.15in}
	\begin{tikzpicture}
	[spy using outlines={rectangle,green,magnification=2,size=7mm, connect spies}]
	\node {\includegraphics[width=0.24\textwidth]
		{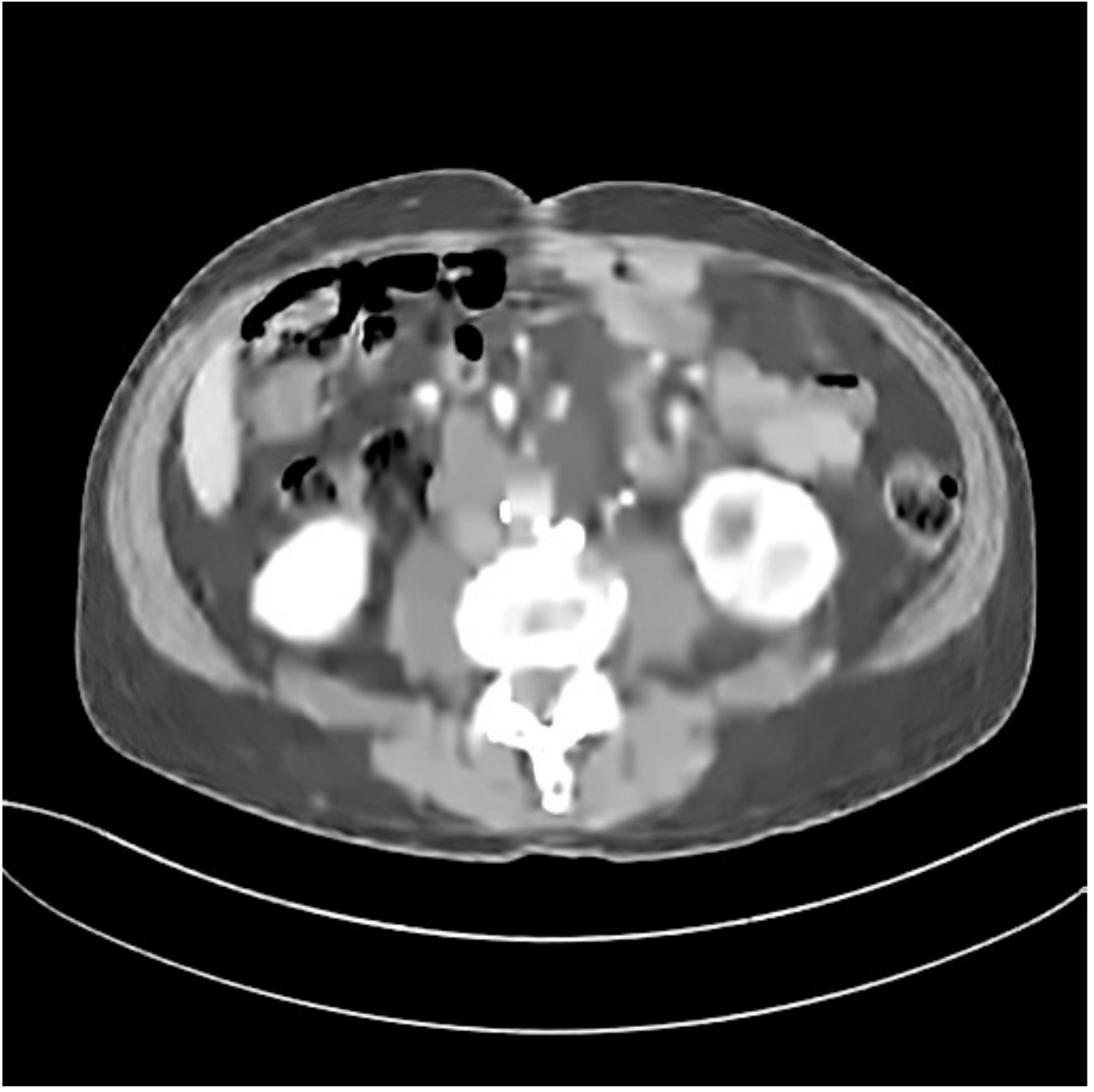}	};
	\spy [green, draw, height = 0.8cm, width = 0.8cm, magnification = 2.5,
	connect spies] on (-0.05,0.40) in node [left] at (-1.16,1.57);
	\spy [green, draw, height = 0.8cm, width = 0.8cm, magnification = 2.2,
	connect spies] on (-0.37,-0.70) in node [left] at (-1.17,-1.56);
	\spy [green, draw, height = 0.7cm, width = 0.9cm, magnification = 2.0,
	connect spies] on (0.45,-0.75) in node [left] at (1.95,-1.60);
	\end{tikzpicture}
	%	\put(-95,20){ \color{white}{\bf \small{RMSE:29.8}}}
	%	\put(-95,10){ \color{white}{\bf \small{SSIM:0.731}}}
	\put(-83,8){ \color{white}{\bf \small{MARS3}}}
	\hspace{-0.15in}
	\begin{tikzpicture}
	[spy using outlines={rectangle,green,magnification=2,size=7mm, connect spies}]
	\node {\includegraphics[width=0.24\textwidth]
		{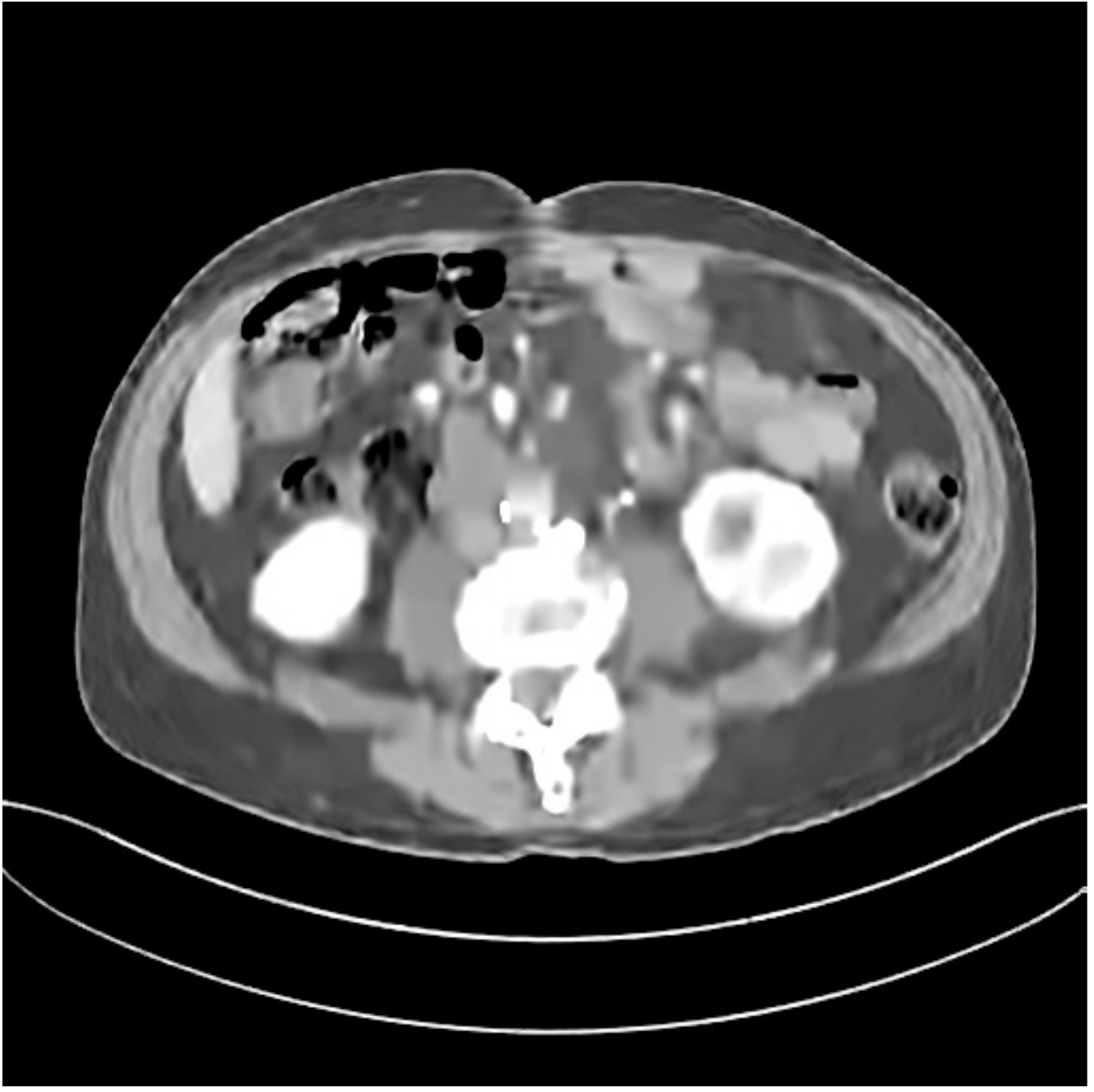}	};
	\spy [green, draw, height = 0.8cm, width = 0.8cm, magnification = 2.5,
	connect spies] on (-0.05,0.40) in node [left] at (-1.16,1.57);
	\spy [green, draw, height = 0.8cm, width = 0.8cm, magnification = 2.2,
	connect spies] on (-0.37,-0.70) in node [left] at (-1.17,-1.56);
	\spy [green, draw, height = 0.7cm, width = 0.9cm, magnification = 2.0,
	connect spies] on (0.45,-0.75) in node [left] at (1.95,-1.60);
	\end{tikzpicture}
	%	\put(-95,20){ \color{white}{\bf \small{RMSE:30.5}}}
	%	\put(-95,10){ \color{white}{\bf \small{SSIM:0.732}}}
	\put(-83,8){ \color{white}{\bf \small{MARS5}}}
	\hspace{-0.15in}
	\begin{tikzpicture}
	[spy using outlines={rectangle,green,magnification=2,size=7mm, connect spies}]
	\node {\includegraphics[width=0.24\textwidth]
		{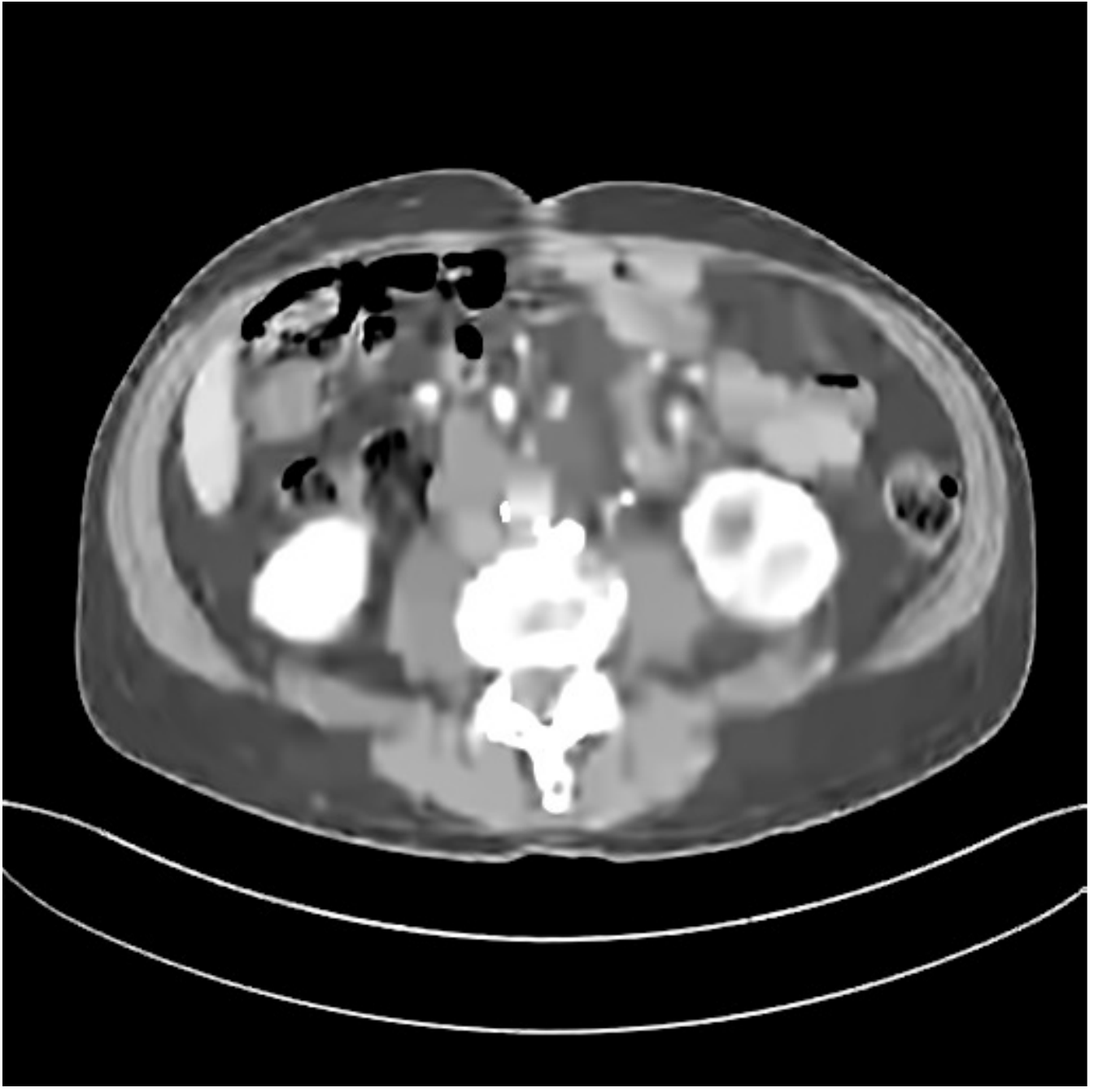}	};
	%		{figure/MayoRecon_L333_slice130_3mm_1e4_3e4beta_25_18_15_12_10_5_1gamma_learn150_140_130_120_110_100_90_tranSize_all_64_traindata_v3_iter1500_init20delta_NUFFT}	};
	\spy [green, draw, height = 0.8cm, width = 0.8cm, magnification = 2.5,
	connect spies] on (-0.05,0.40) in node [left] at (-1.16,1.57);
	\spy [green, draw, height = 0.8cm, width = 0.8cm, magnification = 2.2,
	connect spies] on (-0.37,-0.70) in node [left] at (-1.17,-1.56);
	\spy [green, draw, height = 0.7cm, width = 0.9cm, magnification = 2.0,
	connect spies] on (0.45,-0.75) in node [left] at (1.95,-1.60);
	\end{tikzpicture}
	%	\put(-95,20){ \color{white}{\bf \small{RMSE:31.4}}}
	%	\put(-95,10){ \color{white}{\bf \small{SSIM:0.731}}}
	\put(-83,8){ \color{white}{\bf \small{MARS7}}}
	\caption{Reconstructions of slice 90 of patient L109 at incident photon intensity $I_0=1\times 10^4$. The first row shows the reference image and reconstructions with FBP, PWLS-EP, and PWLS-ST, respectively, and the second row shows the results with MARS models with $2$, $3$, $5$, and $7$ layers, respectively. The display window is [800, 1200] HU.}
	\label{fig:recon_mayo_L109}
\end{figure}

\begin{figure}[H]
	\centering  
	\begin{tikzpicture}
	[spy using outlines={rectangle,green,magnification=2,size=7mm, connect spies}]
	\node {\includegraphics[width=0.24\textwidth]{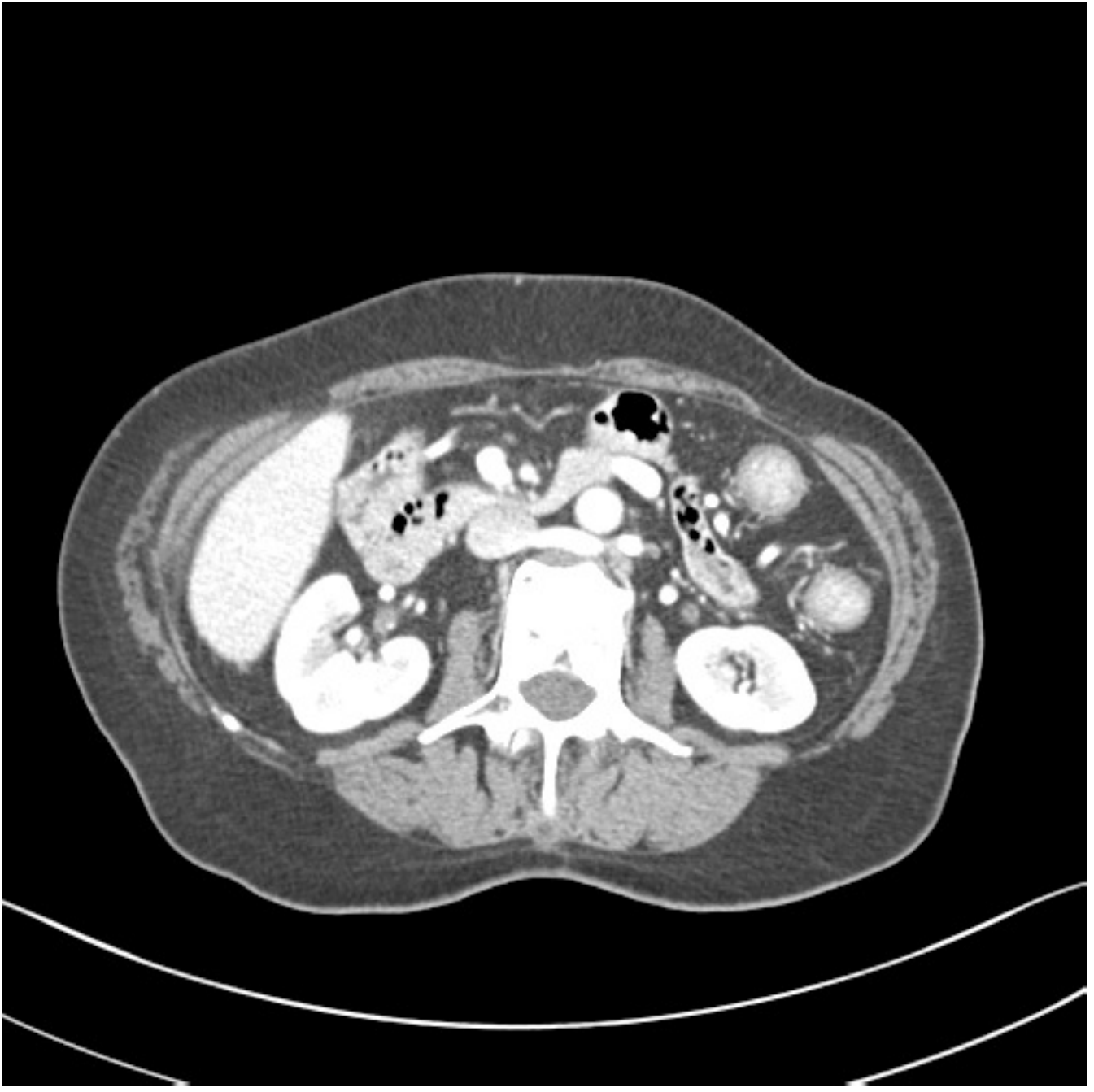}	};
	\spy [green, draw, height = 1.1cm, width = 1.1cm, magnification = 2,
	connect spies] on (-1.2,0.20) in node [left] at (-0.88,1.42);
	%	\spy [green, draw, height = 0.8cm, width = 0.7cm, magnification = 2.2,
	%	connect spies] on (0.18,-0.9) in node [left] at (1.95,-1.55);
	\end{tikzpicture}
	%		\put(-88,102){ \color{white}{\bf \small{RMSE:0.00}}}
	%		\put(-88,92){ \color{white}{\bf \small{SSIM:1.000}}}
	\put(-90,8){ \color{white}{\bf \small{Reference}}} 
	\hspace{-0.15in}
	\begin{tikzpicture}
	[spy using outlines={rectangle,green,magnification=2,size=7mm, connect spies}]
	\node {\includegraphics[width=0.24\textwidth]{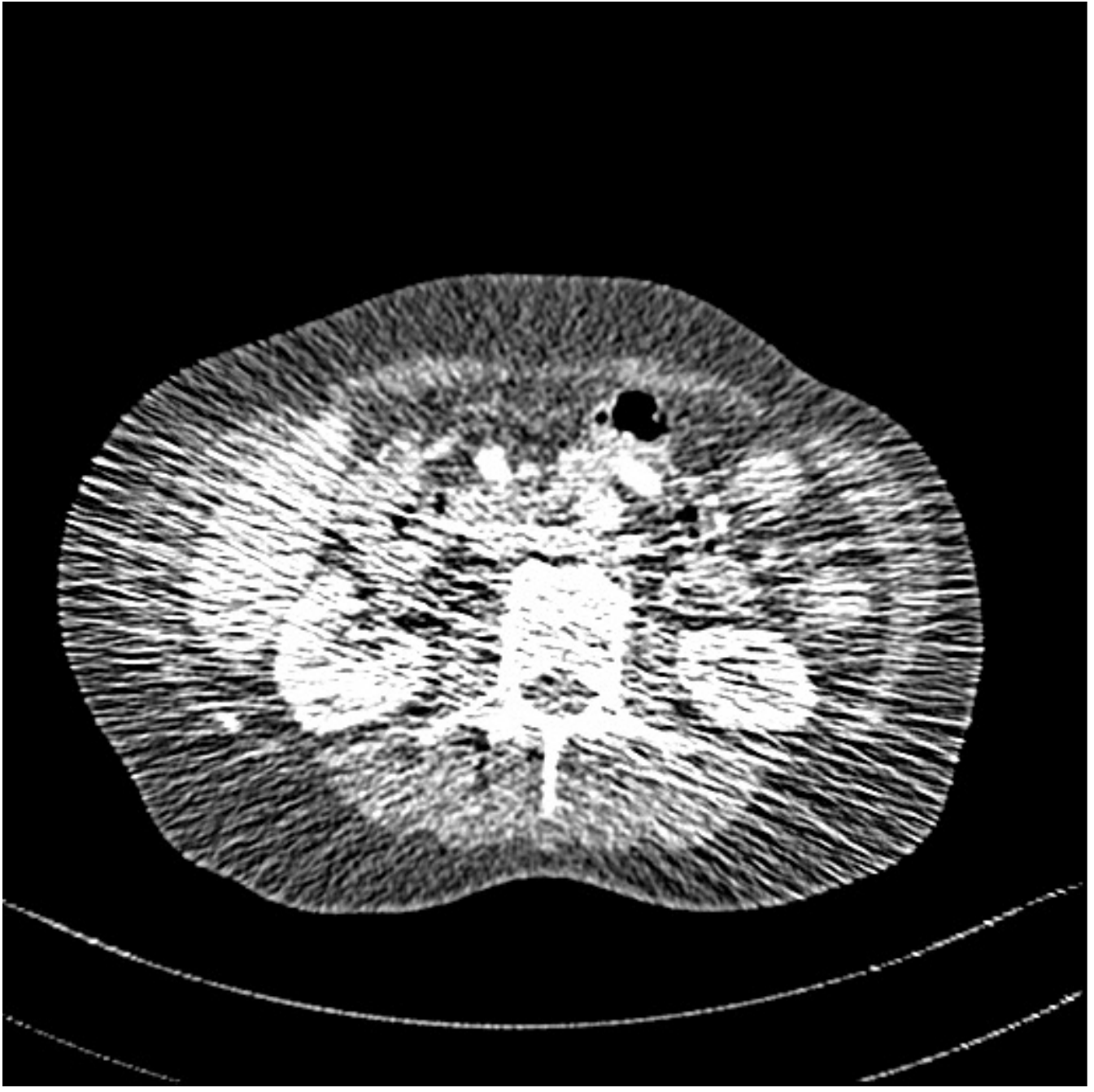}	};
	\spy [green, draw, height = 1.1cm, width = 1.1cm, magnification = 2,
	connect spies] on (-1.2,0.20) in node [left] at (-0.88,1.42);
	%	\spy [green, draw, height = 0.8cm, width = 0.7cm, magnification = 2.2,
	%	connect spies] on (0.18,-0.9) in node [left] at (1.95,-1.55);
	\end{tikzpicture}
	%		\put(-88,102){ \color{white}{\bf \small{RMSE:93.7}}}
	%		\put(-88,92){ \color{white}{\bf \small{SSIM:0.350}}}
	\put(-75,8){ \color{white}{\bf \small{FBP}}} 
	\hspace{-0.15in}
	\begin{tikzpicture}
	[spy using outlines={rectangle,green,magnification=2,size=7mm, connect spies}]
	\node {\includegraphics[width=0.24\textwidth]{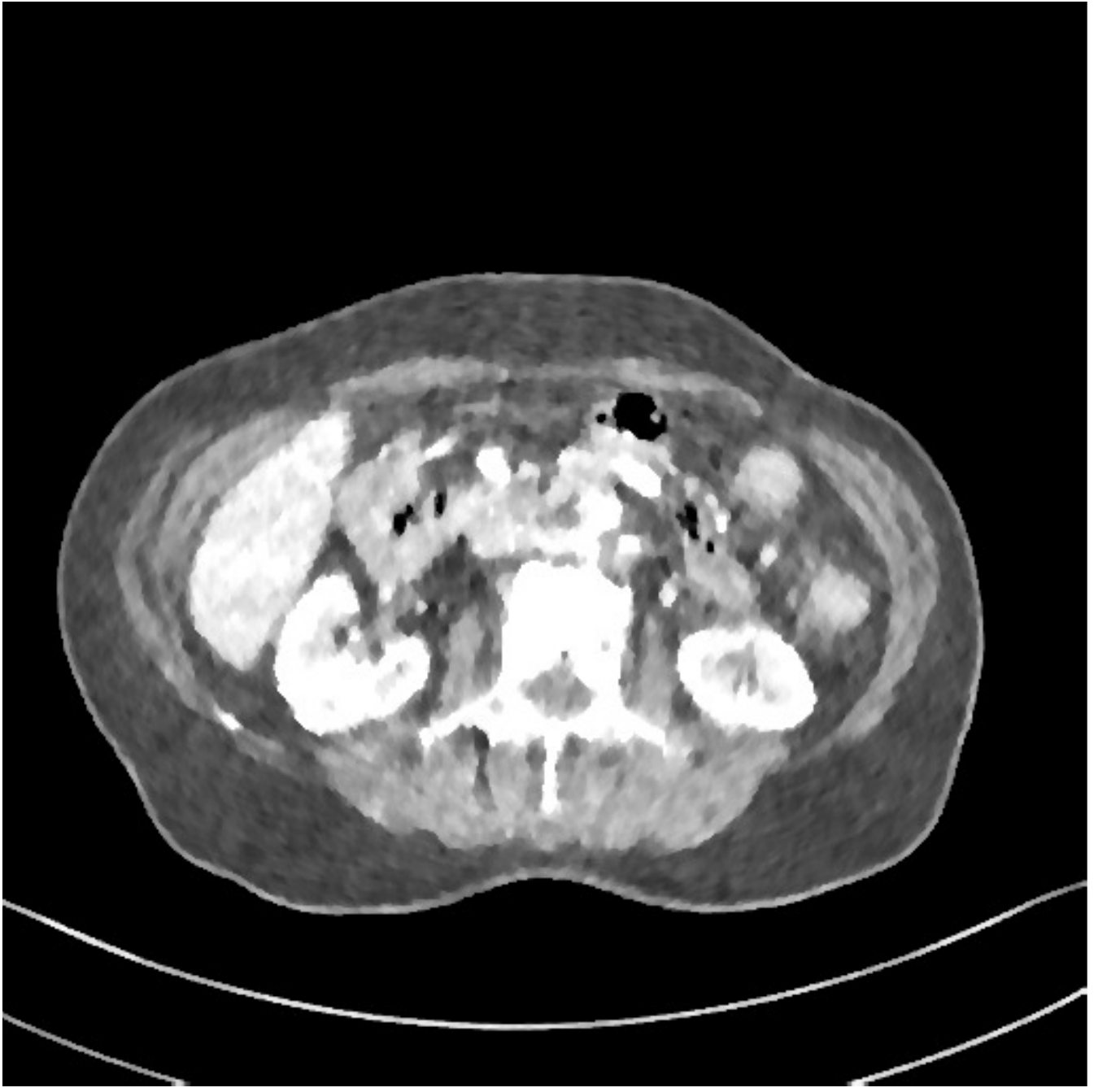}	};
	\spy [green, draw, height = 1.1cm, width = 1.1cm, magnification = 2,
	connect spies] on (-1.2,0.20) in node [left] at (-0.88,1.42);
	%	\spy [green, draw, height = 0.8cm, width = 0.7cm, magnification = 2.2,
	%	connect spies] on (0.18,-0.9) in node [left] at (1.95,-1.55);
	\end{tikzpicture}
	%		\put(-88,102){ \color{white}{\bf \small{RMSE:31.5}}}
	%		\put(-88,92){ \color{white}{\bf \small{SSIM:0.747}}}
	\put(-70,8){ \color{white}{\bf \small{EP}}}
	\hspace{-0.15in}	
	\begin{tikzpicture}
	[spy using outlines={rectangle,green,magnification=2,size=7mm, connect spies}]
	\node {\includegraphics[width=0.24\textwidth]{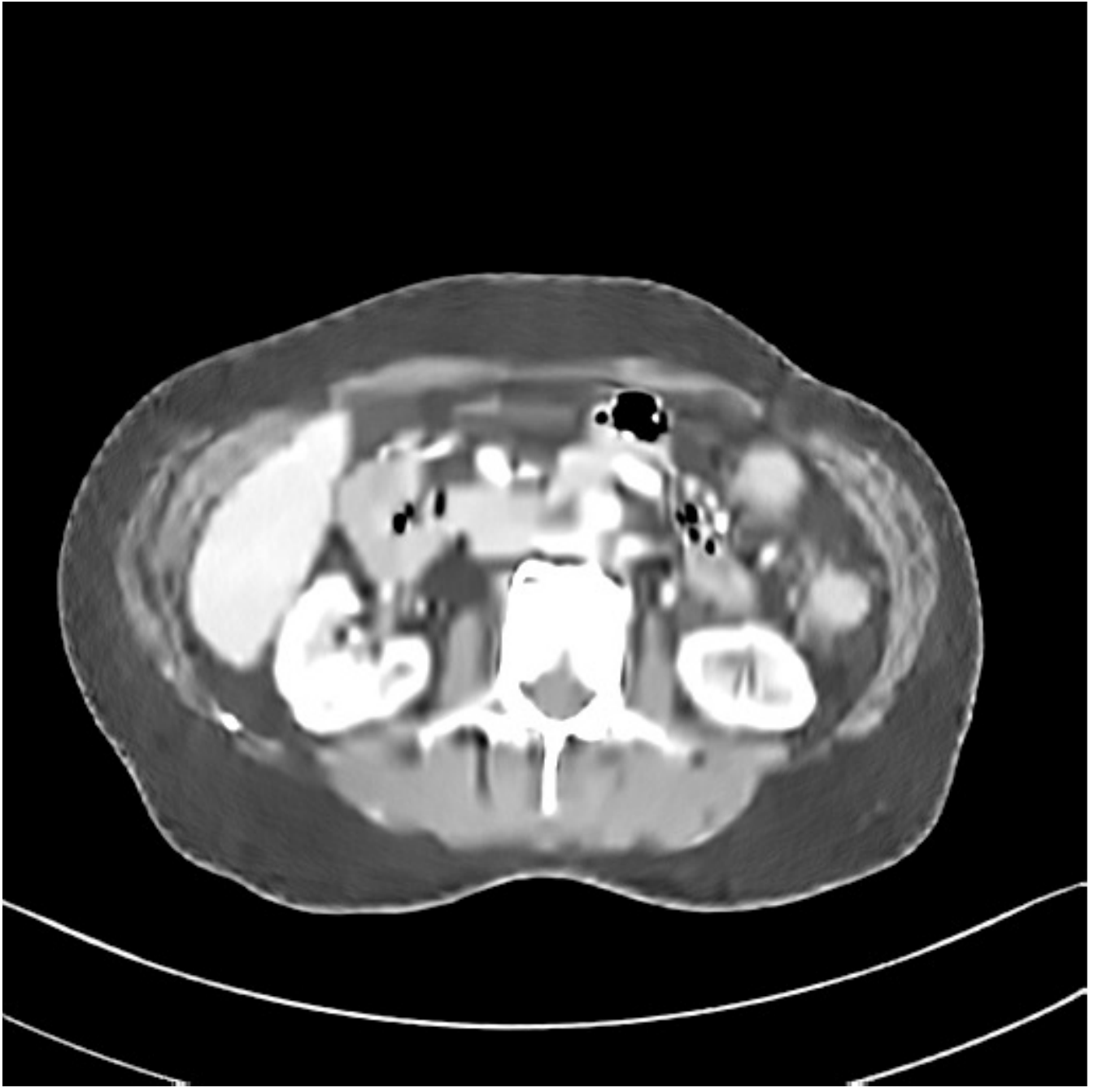}	};
	\spy [green, draw, height = 1.1cm, width = 1.1cm, magnification = 2,
	connect spies] on (-1.2,0.20) in node [left] at (-0.88,1.42);
	%	\spy [green, draw, height = 0.8cm, width = 0.7cm, magnification = 2.2,
	%	connect spies] on (0.18,-0.9) in node [left] at (1.95,-1.55);
	\end{tikzpicture}	
	%		\put(-88,102){ \color{white}{\bf \small{RMSE:26.3}}}
	%		\put(-88,92){ \color{white}{\bf \small{SSIM:0.737}}}
	\put(-70,8){ \color{white}{\bf \small{ST}}}
	\\
	\vspace{-0.10in}
	\begin{tikzpicture}
	[spy using outlines={rectangle,green,magnification=2,size=7mm, connect spies}]
	\node {\includegraphics[width=0.24\textwidth]{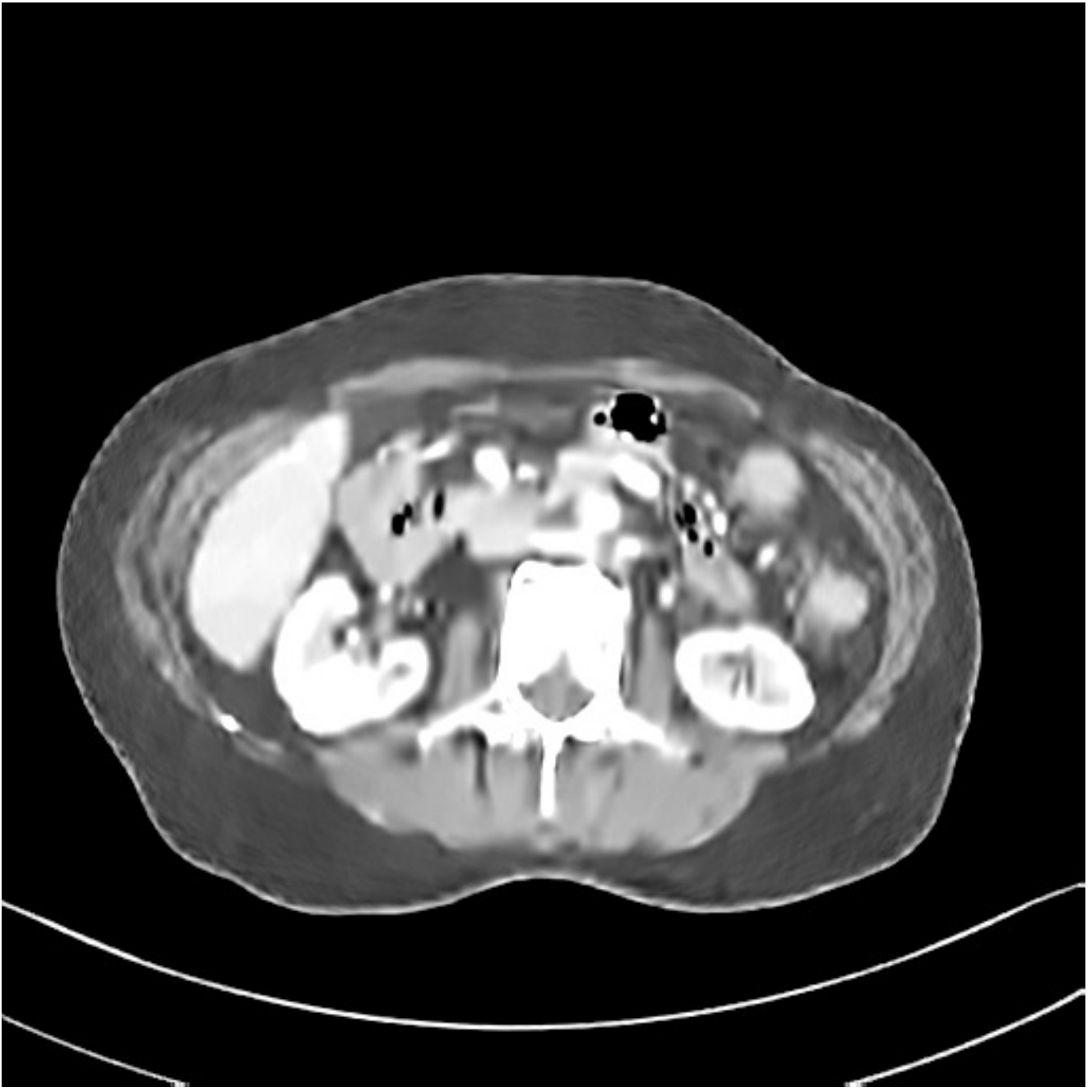}	};
	\spy [green, draw, height = 1.1cm, width = 1.1cm, magnification = 2,
	connect spies] on (-1.2,0.20) in node [left] at (-0.88,1.42);
	%	\spy [green, draw, height = 0.8cm, width = 0.7cm, magnification = 2.2,
	%	connect spies] on (0.18,-0.9) in node [left] at (1.95,-1.55);
	\end{tikzpicture}
	%		\put(-88,102){ \color{white}{\bf \small{RMSE:25.3}}}
	%		\put(-88,92){ \color{white}{\bf \small{SSIM:0.744}}}
	\put(-83,8){ \color{white}{\bf \small{MARS2}}}
	\hspace{-0.15in}
	\begin{tikzpicture}
	[spy using outlines={rectangle,green,magnification=2,size=7mm, connect spies}]
	\node {\includegraphics[width=0.24\textwidth]
		{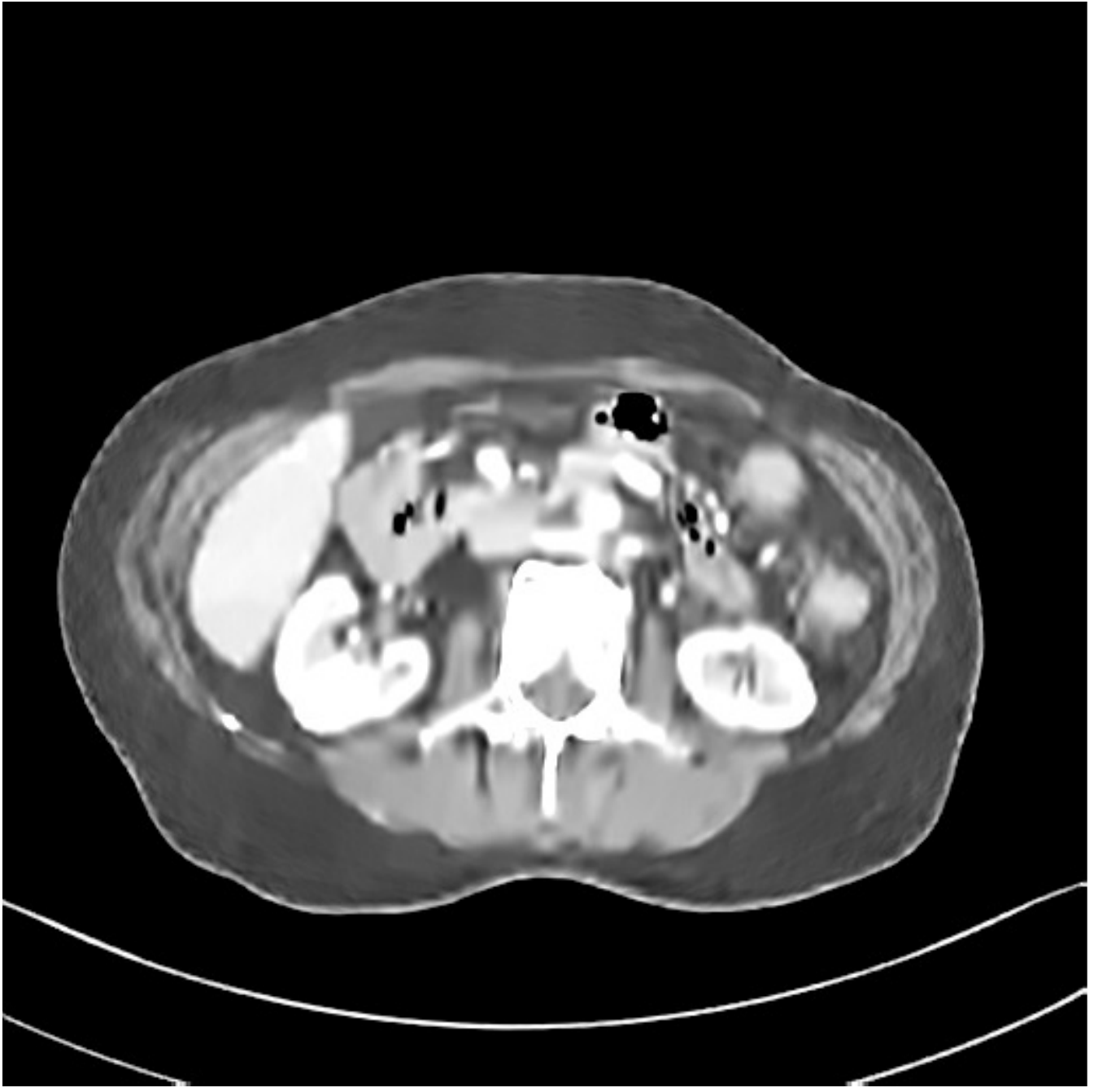}	};
	\spy [green, draw, height = 1.1cm, width = 1.1cm, magnification = 2,
	connect spies] on (-1.2,0.20) in node [left] at (-0.88,1.42);
	%	\spy [green, draw, height = 0.8cm, width = 0.7cm, magnification = 2.2,
	%	connect spies] on (0.18,-0.9) in node [left] at (1.95,-1.55);
	\end{tikzpicture}
	%		\put(-88,102){ \color{white}{\bf \small{RMSE:24.9}}}
	%		\put(-88,92){ \color{white}{\bf \small{SSIM:0.750}}}
	\put(-83,8){ \color{white}{\bf \small{MARS3}}}
	\hspace{-0.15in}
	\begin{tikzpicture}
	[spy using outlines={rectangle,green,magnification=2,size=7mm, connect spies}]
	\node {\includegraphics[width=0.24\textwidth]
		{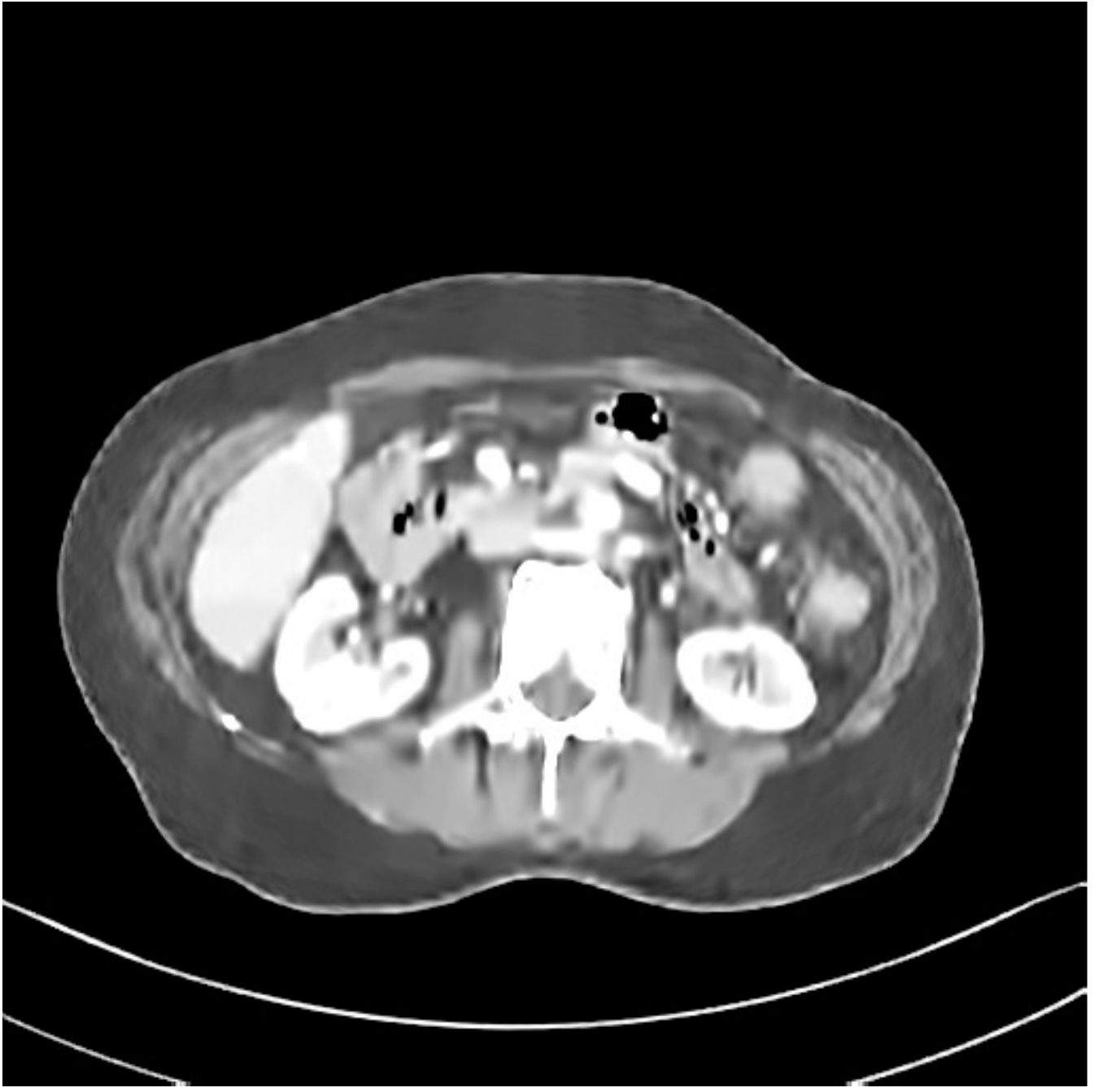}	};
	\spy [green, draw, height = 1.1cm, width = 1.1cm, magnification = 2,
	connect spies] on (-1.2,0.20) in node [left] at (-0.88,1.42);
	%	\spy [green, draw, height = 0.8cm, width = 0.7cm, magnification = 2.2,
	%	connect spies] on (0.18,-0.9) in node [left] at (1.95,-1.55);
	\end{tikzpicture}
	%		\put(-88,102){ \color{green}{\bf \small{RMSE:24.6}}}
	%		\put(-88,92){ \color{white}{\bf \small{SSIM:0.765}}}
	\put(-83,8){ \color{white}{\bf \small{MARS5}}}
	\hspace{-0.15in}
	\begin{tikzpicture}
	[spy using outlines={rectangle,green,magnification=2,size=7mm, connect spies}]
	\node {\includegraphics[width=0.24\textwidth]
		{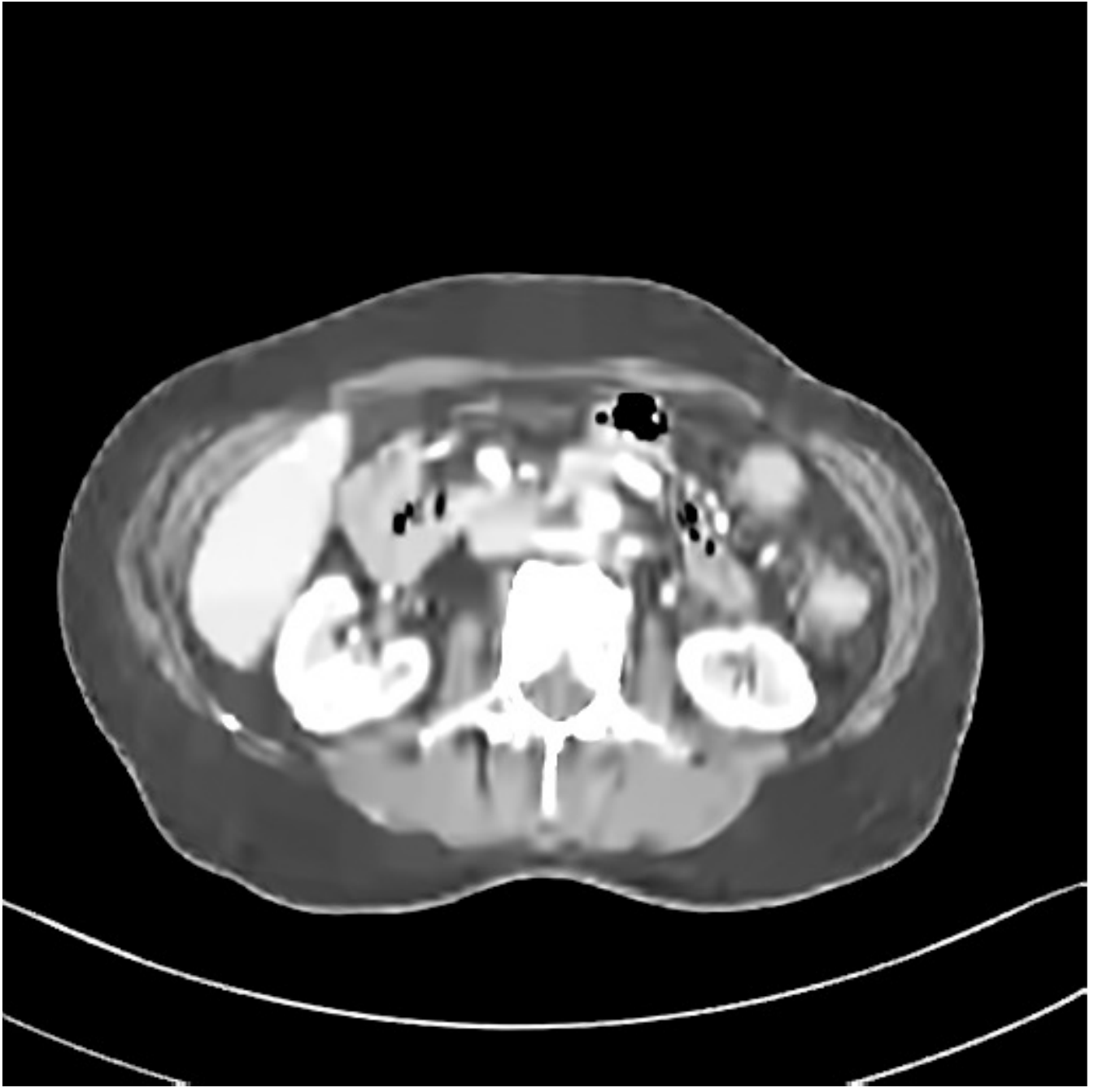}	};
	%		{figure/MayoRecon_L333_slice130_3mm_1e4_3e4beta_25_18_15_12_10_5_1gamma_learn150_140_130_120_110_100_90_tranSize_all_64_traindata_v3_iter1500_init20delta_NUFFT}	};
	\spy [green, draw, height = 1.1cm, width = 1.1cm, magnification = 2,
	connect spies] on (-1.2,0.20) in node [left] at (-0.88,1.42);
	%	\spy [green, draw, height = 0.8cm, width = 0.7cm, magnification = 2.2,
	%	connect spies] on (0.18,-0.9) in node [left] at (1.95,-1.55);
	\end{tikzpicture}
	%		\put(-88,102){ \color{white}{\bf \small{RMSE:24.9}}}
	%		\put(-88,92){ \color{green}{\bf \small{SSIM:0.781}}}
	\put(-83,8){ \color{white}{\bf \small{MARS7}}}
	\caption{Reconstructions of slice 90 of patient L192 at incident photon intensity $I_0=1\times 10^4$. The first row shows the reference image and reconstructions with FBP, PWLS-EP, and PWLS-ST, respectively, and the second row shows the results with MARS models with $2$, $3$, $5$, and $7$ layers, respectively. The display window is [800, 1200] HU.}
	\label{fig:recon_mayo_L192}
\end{figure}

\begin{figure}[H]
	\centering  
	\begin{tikzpicture}
	[spy using outlines={rectangle,green,magnification=2,size=7mm, connect spies}]
	\node {\includegraphics[width=0.24\textwidth]{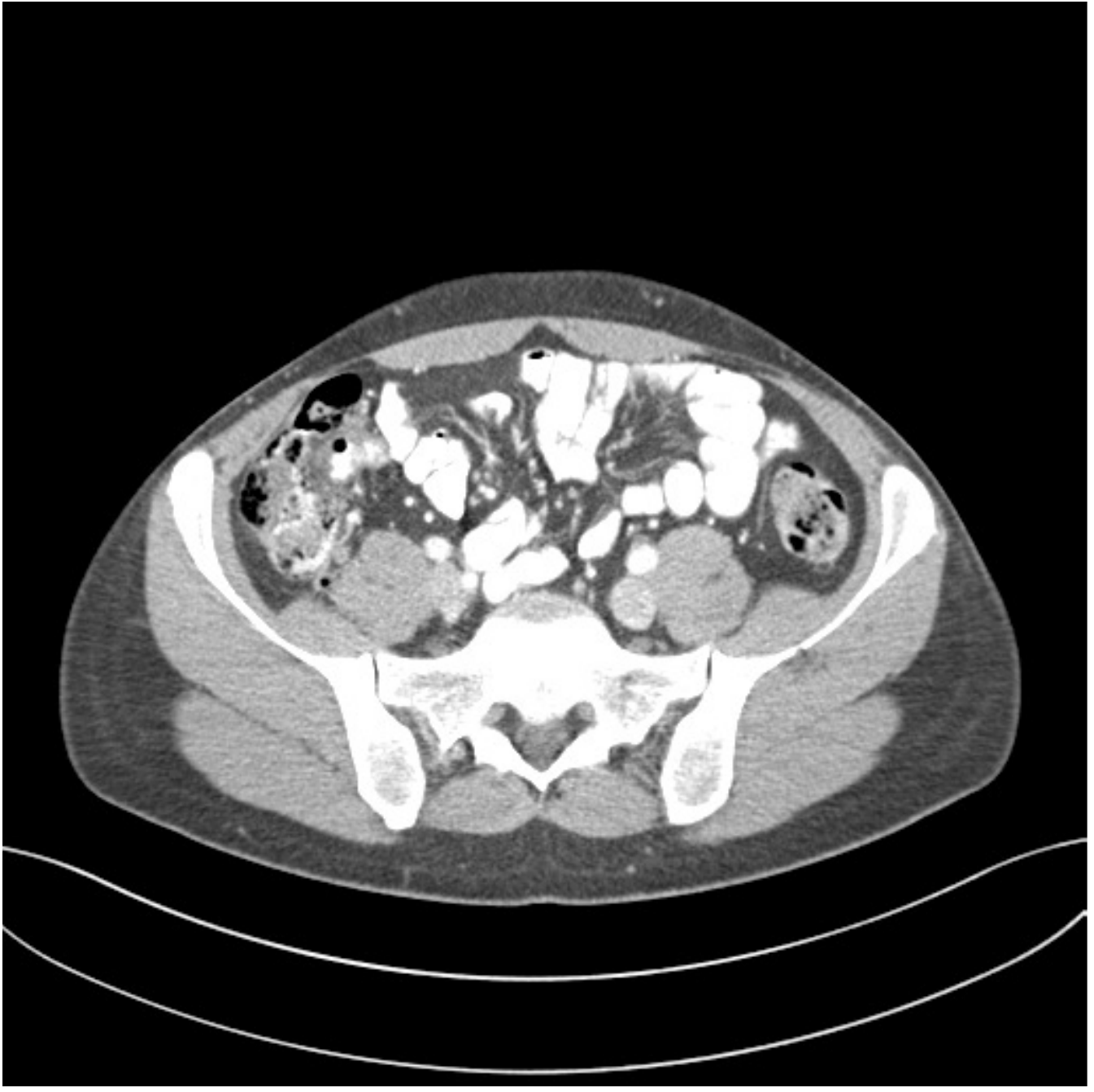}	};
	\spy [green, draw, height = 1.0cm, width = 1.0cm, magnification = 2.5,
	connect spies] on (-0.3,0.45) in node [left] at (1.95,1.46);
	\spy [green, draw, height = 0.8cm, width = 0.8cm, magnification = 3,
	connect spies] on (-0.5,0.10) in node [left] at (-1.17,-1.56);
	\end{tikzpicture}
	%		\put(-105,102){ \color{white}{\bf \small{RMSE:0.00}}}
	%		\put(-105,92){ \color{white}{\bf \small{SSIM:1.000}}}
	\put(-90,8){ \color{white}{\bf \small{Reference}}} 
	\hspace{-0.15in}
	\begin{tikzpicture}
	[spy using outlines={rectangle,green,magnification=2,size=7mm, connect spies}]
	\node {\includegraphics[width=0.24\textwidth]{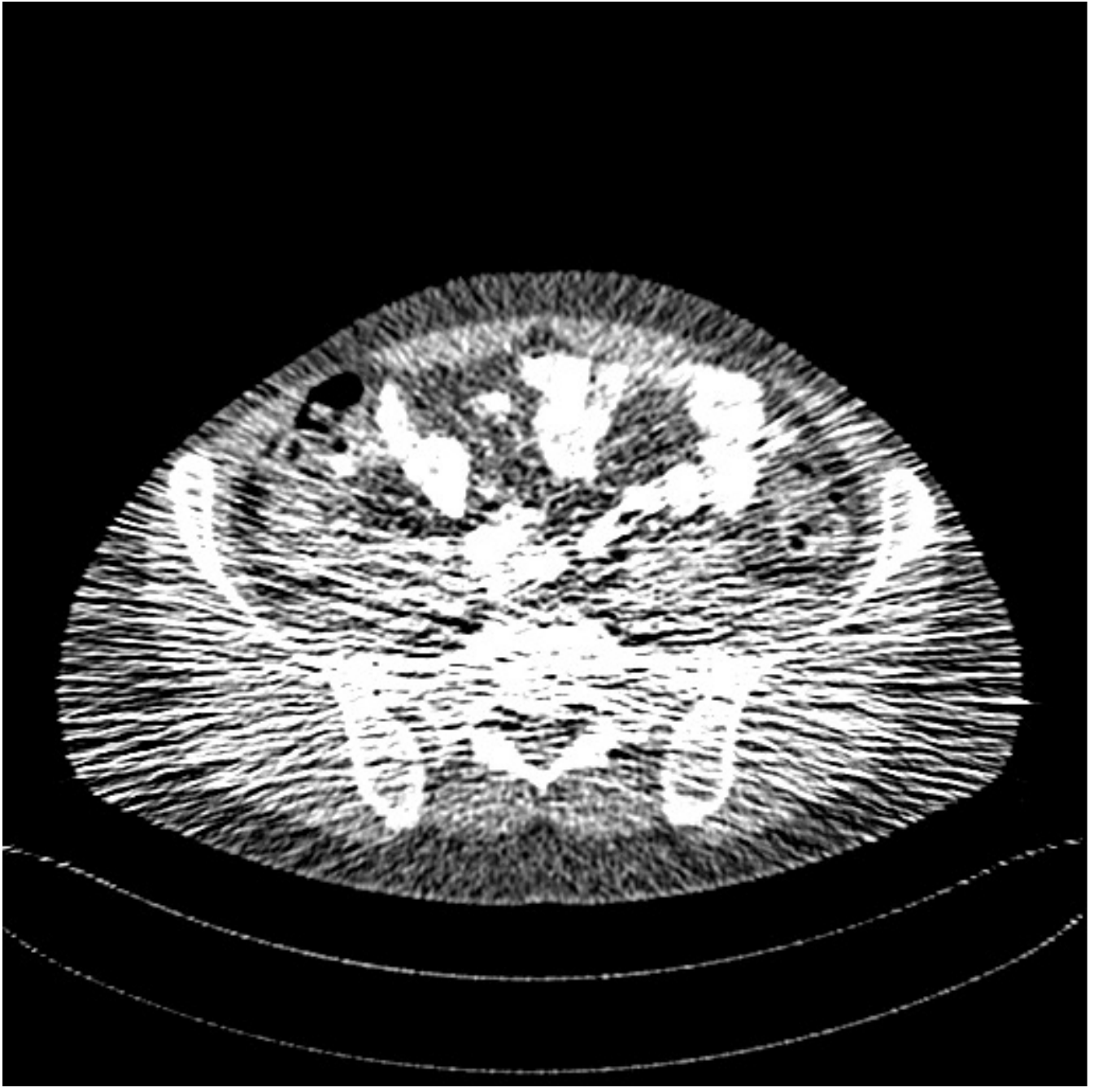}	};
	\spy [green, draw, height = 1.0cm, width = 1.0cm, magnification = 2.5,
	connect spies] on (-0.3,0.45) in node [left] at (1.95,1.46);
	\spy [green, draw, height = 0.8cm, width = 0.8cm, magnification = 3,
	connect spies] on (-0.5,0.10) in node [left] at (-1.17,-1.56);
	\end{tikzpicture}
	%		\put(-105,102){ \color{white}{\bf \small{RMSE:113.1}}}
	%		\put(-105,92){ \color{white}{\bf \small{SSIM:0.358}}}
	\put(-75,8){ \color{white}{\bf \small{FBP}}} 
	\hspace{-0.15in}
	\begin{tikzpicture}
	[spy using outlines={rectangle,green,magnification=2,size=7mm, connect spies}]
	\node {\includegraphics[width=0.24\textwidth]{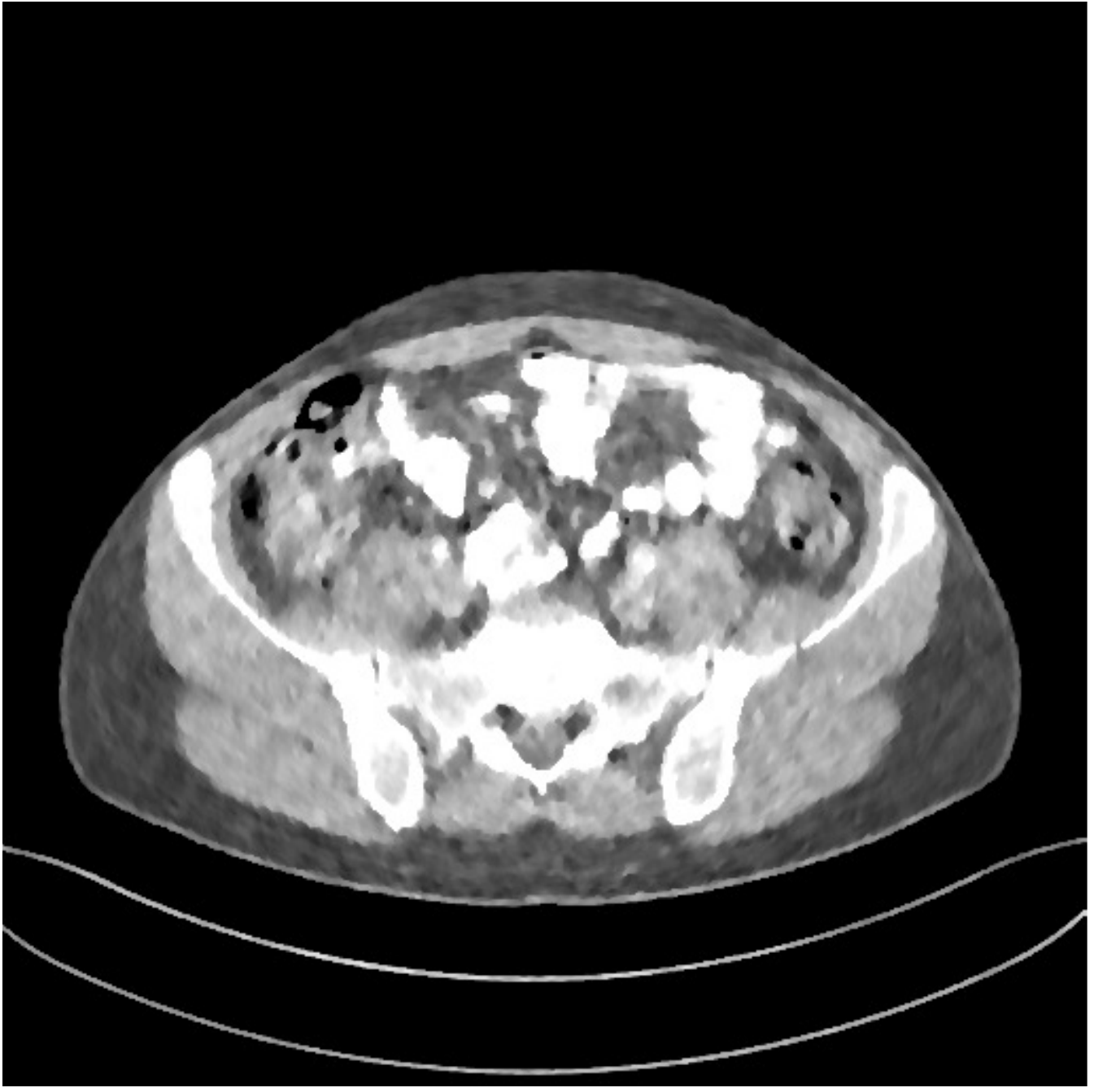}	};
	\spy [green, draw, height = 1.0cm, width = 1.0cm, magnification = 2.5,
	connect spies] on (-0.3,0.45) in node [left] at (1.95,1.46);
	\spy [green, draw, height = 0.8cm, width = 0.8cm, magnification = 3,
	connect spies] on (-0.5,0.10) in node [left] at (-1.17,-1.56);
	\end{tikzpicture}
	%		\put(-105,102){ \color{white}{\bf \small{RMSE:36.3}}}
	%		\put(-105,92){ \color{white}{\bf \small{SSIM:0.758}}}
	\put(-70,8){ \color{white}{\bf \small{EP}}}
	\hspace{-0.15in}	
	\begin{tikzpicture}
	[spy using outlines={rectangle,green,magnification=2,size=7mm, connect spies}]
	\node {\includegraphics[width=0.24\textwidth]{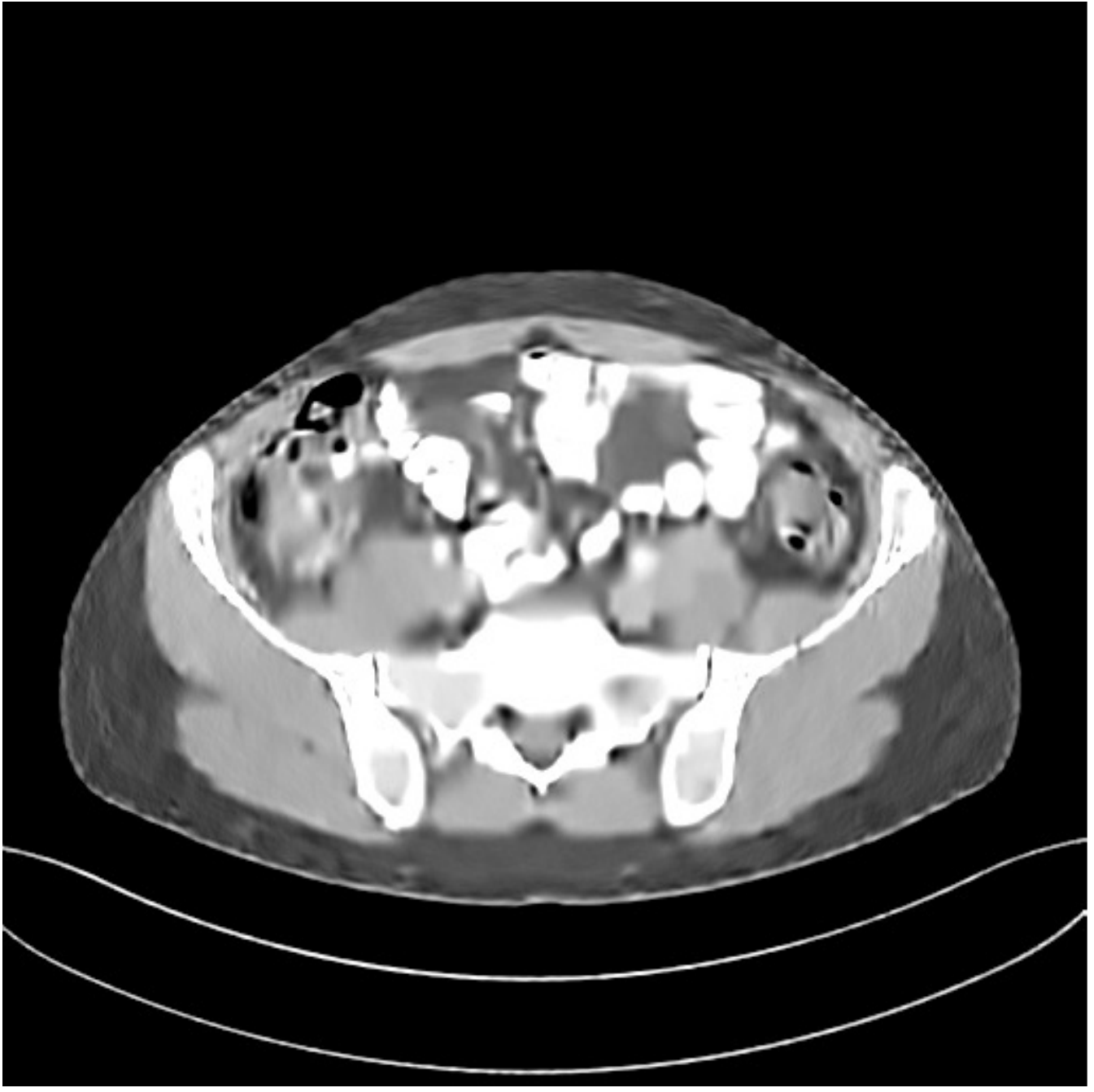}	};
	\spy [green, draw, height = 1.0cm, width = 1.0cm, magnification = 2.5,
	connect spies] on (-0.3,0.45) in node [left] at (1.95,1.46);
	\spy [green, draw, height = 0.8cm, width = 0.8cm, magnification = 3,
	connect spies] on (-0.5,0.10) in node [left] at (-1.17,-1.56);
	\end{tikzpicture}	
	%		\put(-105,102){ \color{white}{\bf \small{RMSE:29.7}}}
	%		\put(-105,92){ \color{white}{\bf \small{SSIM:0.739}}}
	\put(-70,8){ \color{white}{\bf \small{ST}}}
	\\
	\vspace{-0.10in}
	\begin{tikzpicture}
	[spy using outlines={rectangle,green,magnification=2,size=7mm, connect spies}]
	\node {\includegraphics[width=0.24\textwidth]{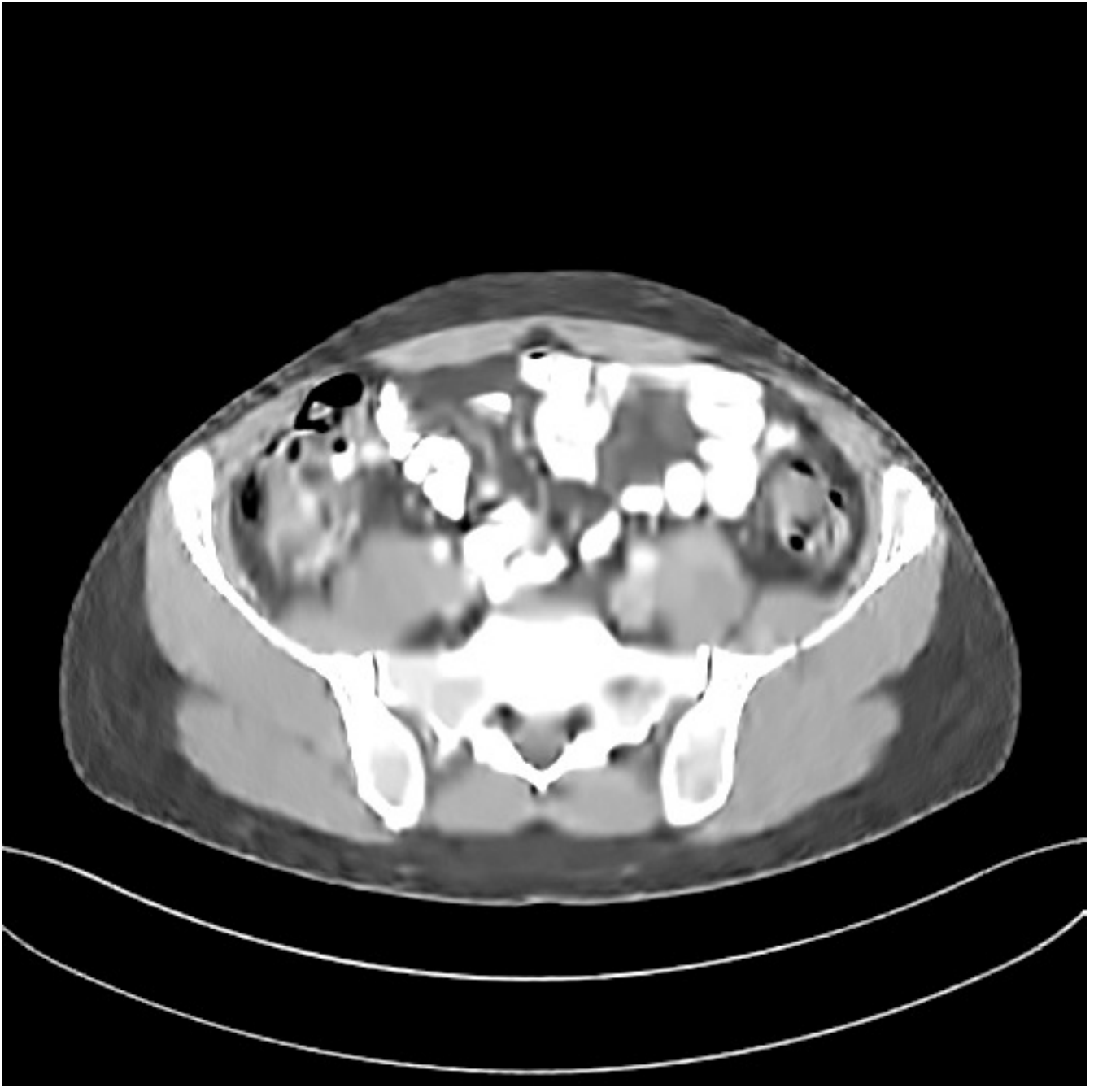}	};
	\spy [green, draw, height = 1.0cm, width = 1.0cm, magnification = 2.5,
	connect spies] on (-0.3,0.45) in node [left] at (1.95,1.46);
	\spy [green, draw, height = 0.8cm, width = 0.8cm, magnification = 3,
	connect spies] on (-0.5,0.10) in node [left] at (-1.17,-1.56);
	\end{tikzpicture}
	%		\put(-105,102){ \color{white}{\bf \small{RMSE:28.5}}}
	%		\put(-105,92){ \color{white}{\bf \small{SSIM:0.744}}}
	\put(-83,8){ \color{white}{\bf \small{MARS2}}}
	\hspace{-0.15in}
	\begin{tikzpicture}
	[spy using outlines={rectangle,green,magnification=2,size=7mm, connect spies}]
	\node {\includegraphics[width=0.24\textwidth]
		{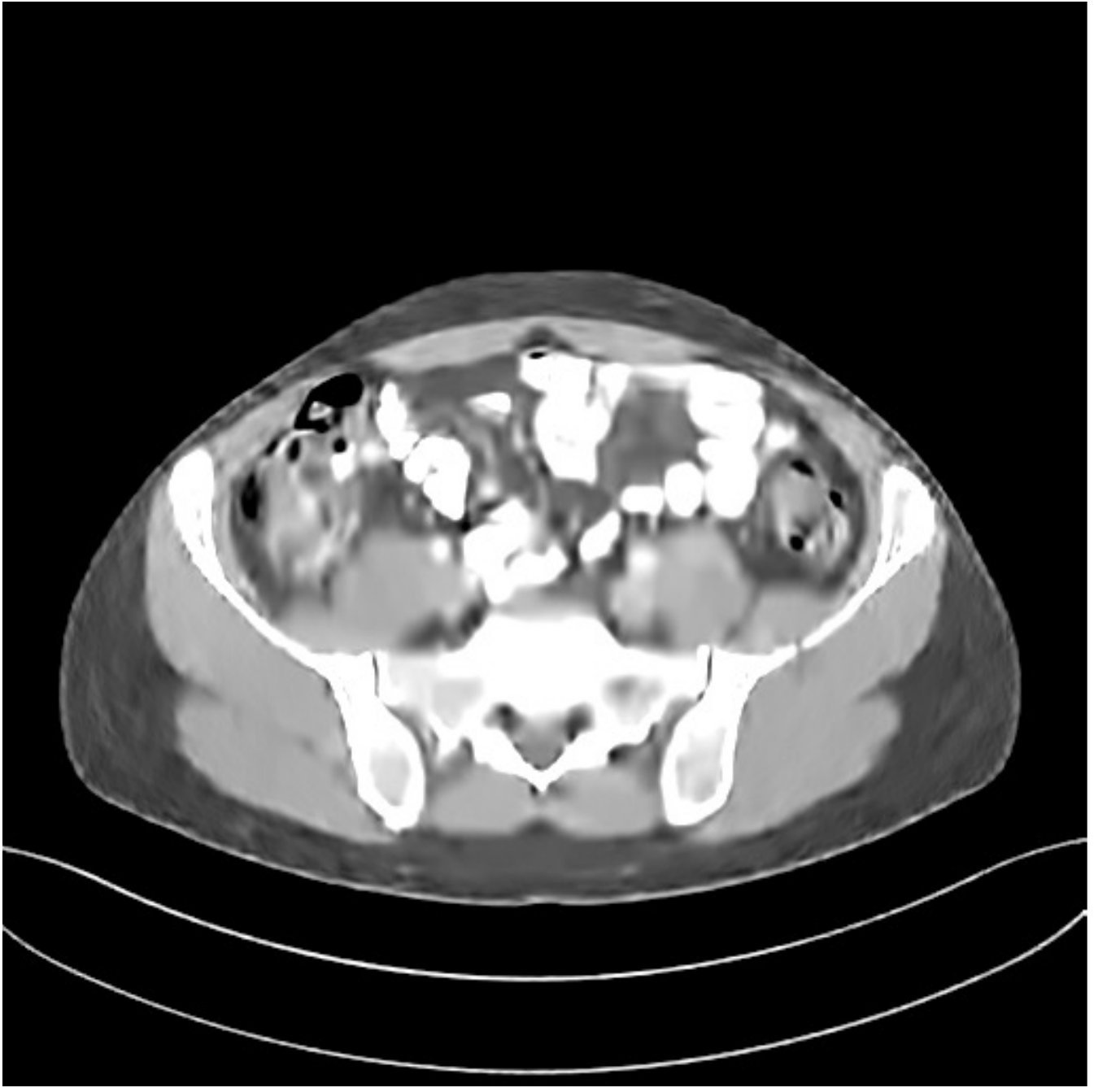}	};
	\spy [green, draw, height = 1.0cm, width = 1.0cm, magnification = 2.5,
	connect spies] on (-0.3,0.45) in node [left] at (1.95,1.46);
	\spy [green, draw, height = 0.8cm, width = 0.8cm, magnification = 3,
	connect spies] on (-0.5,0.10) in node [left] at (-1.17,-1.56);
	\end{tikzpicture}
	%		\put(-105,102){ \color{white}{\bf \small{RMSE:28.3}}}
	%		\put(-105,92){ \color{white}{\bf \small{SSIM:0.750}}}
	\put(-83,8){ \color{white}{\bf \small{MARS3}}}
	\hspace{-0.15in}
	\begin{tikzpicture}
	[spy using outlines={rectangle,green,magnification=2,size=7mm, connect spies}]
	\node {\includegraphics[width=0.24\textwidth]
		{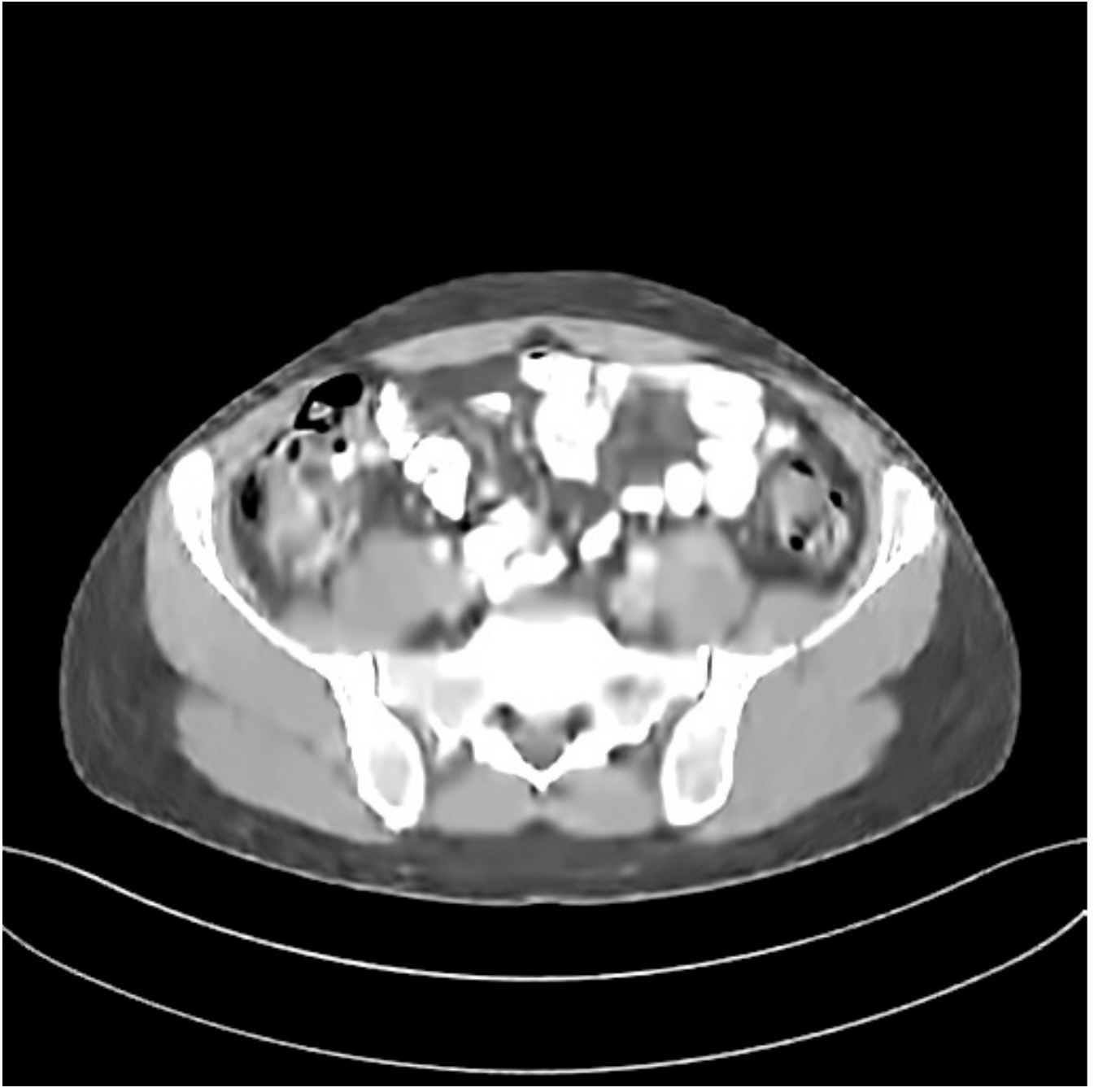}	};
	\spy [green, draw, height = 1.0cm, width = 1.0cm, magnification = 2.5,
	connect spies] on (-0.3,0.45) in node [left] at (1.95,1.46);
	\spy [green, draw, height = 0.8cm, width = 0.8cm, magnification = 3,
	connect spies] on (-0.5,0.10) in node [left] at (-1.17,-1.56);
	\end{tikzpicture}
	%		\put(-105,102){ \color{green}{\bf \small{RMSE:28.1}}}
	%		\put(-105,92){ \color{white}{\bf \small{SSIM:0.766}}}
	\put(-83,8){ \color{white}{\bf \small{MARS5}}}
	\hspace{-0.15in}
	\begin{tikzpicture}
	[spy using outlines={rectangle,green,magnification=2,size=7mm, connect spies}]
	\node {\includegraphics[width=0.24\textwidth]
		{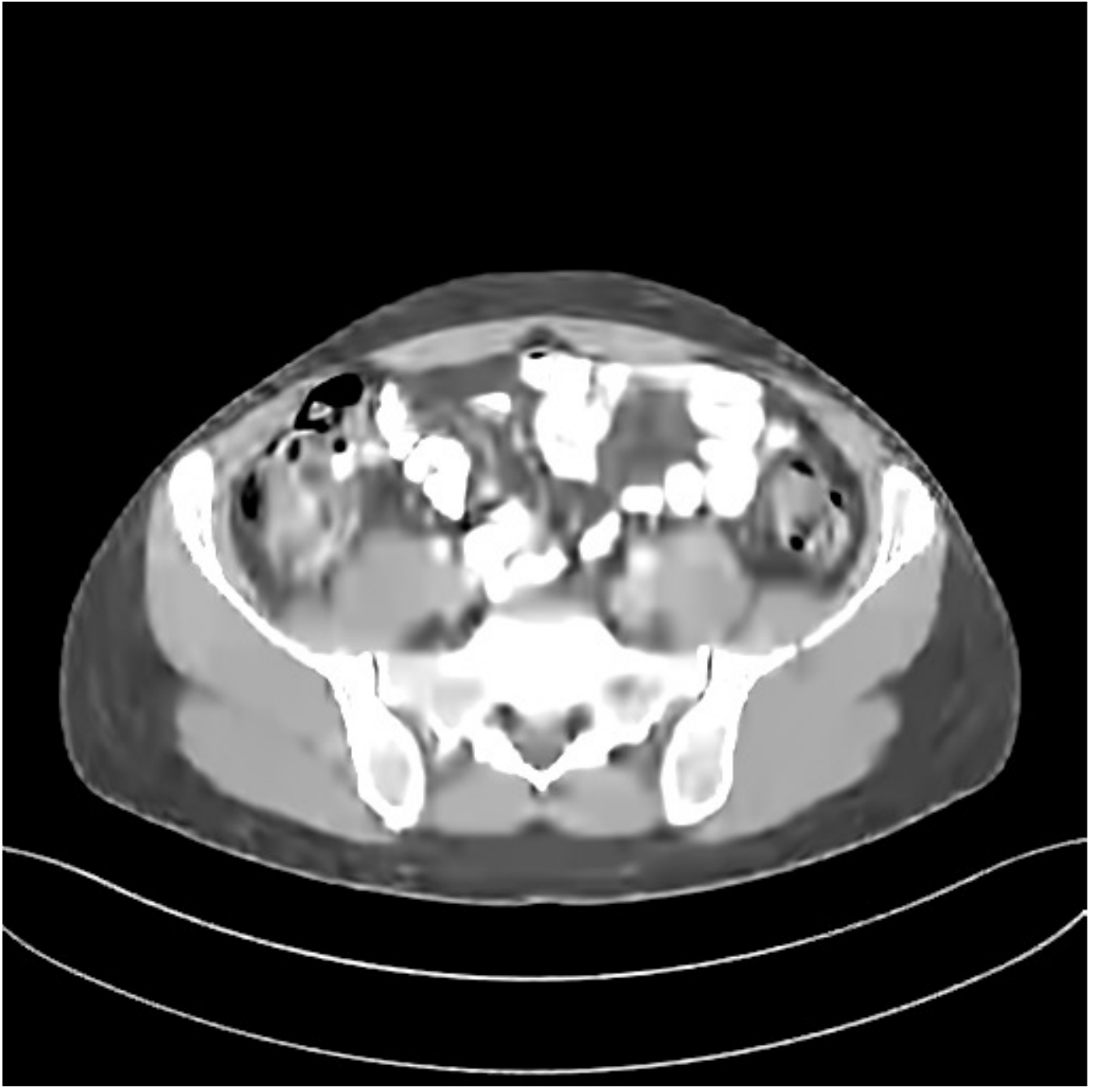}	};
	\spy [green, draw, height = 1.0cm, width = 1.0cm, magnification = 2.5,
	connect spies] on (-0.3,0.45) in node [left] at (1.95,1.46);
	\spy [green, draw, height = 0.8cm, width = 0.8cm, magnification = 3,
	connect spies] on (-0.5,0.10) in node [left] at (-1.17,-1.56);
	\end{tikzpicture}
	%		\put(-105,102){ \color{white}{\bf \small{RMSE:28.4}}}
	%		\put(-105,92){ \color{green}{\bf \small{SSIM:0.786}}}
	\put(-83,8){ \color{white}{\bf \small{MARS7}}}
	\caption{Reconstructions of slice 140 of patient L333 at incident photon intensity $I_0=1\times 10^4$. The first row shows the reference image and reconstructions with FBP, PWLS-EP, and PWLS-ST, respectively, and the second row shows the results with MARS models with $2$, $3$, $5$, and $7$ layers, respectively. The display window is [800, 1200] HU.}
	\label{fig:recon_mayo_L333}
\end{figure}

\begin{figure}[!h]
	\centering  
	%\vspace{-0.1in}
	\begin{tikzpicture}
	[spy using outlines={rectangle,green,magnification=2,size=7mm, connect spies}]
	\node {\includegraphics[width=0.24\textwidth]{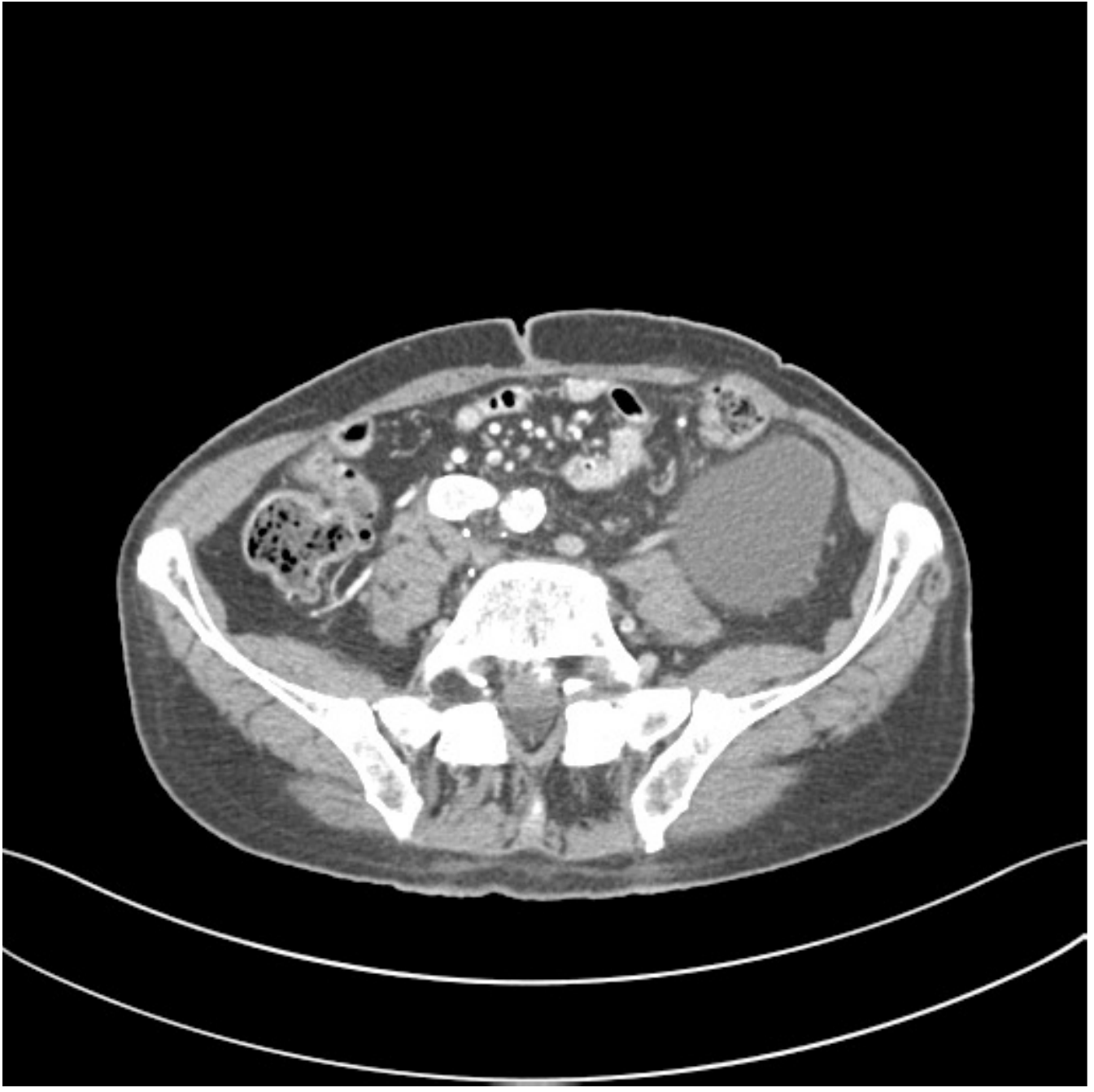}	};
	\spy [green, draw, height = 1.2cm, width = 1.2cm, magnification = 3.0,
	connect spies] on (-0.60,0.40) in node [left] at (-0.76,1.37);
	\spy [green, draw, height = 0.8cm, width = 0.8cm, magnification = 2.5,
	connect spies] on (-0.25,-0.00) in node [left] at (-1.16,-1.55);
	\spy [green, draw, height = 0.8cm, width = 0.8cm, magnification = 3.0,
	connect spies] on (0.30,0.10) in node [left] at (1.95,-1.56);	
	\spy [green, draw, height = 1.2cm, width = 1.2cm, magnification = 2.6,
	connect spies] on (-0.07,0.45) in node [left] at (1.95,1.38);
	\draw[line width=1pt, ultra thin, -latex, red] (-0.15,-0.10) -- node[xshift=-0.05cm,yshift=-0.09cm] {\tiny{}} (-0.25,-0.00);
	\end{tikzpicture}
	%	\put(-95,105){ \color{white}{\bf \small{RMSE:0.00}}}
	%	\put(-95,93){ \color{white}{\bf \small{SSIM:1.000}}}
	\put(-90,8){ \color{white}{\bf \small{Reference}}} 
	\put(-110,105){ \color{green}{\bf \small{1}}} 
	\put(-20,105){ \color{blue}{\bf \small{2}}} 
	\put(-110,18){ \color{cyan}{\bf \small{3}}} 
	\put(-20,18){ \color{magenta}{\bf \small{4}}}
	\hspace{-0.15in}
	\begin{tikzpicture}
	[spy using outlines={rectangle,green,magnification=2,size=7mm, connect spies}]
	\node {\includegraphics[width=0.24\textwidth]{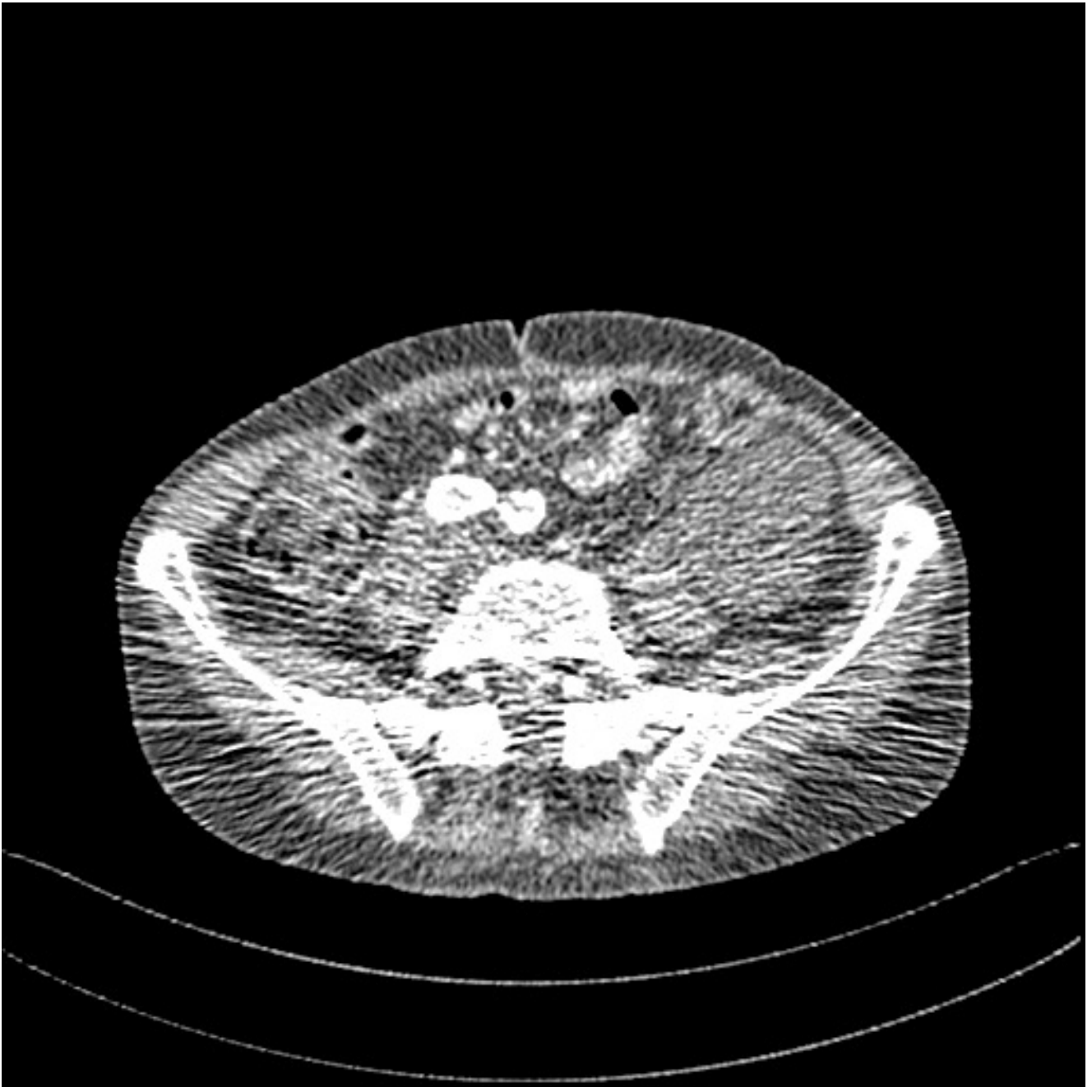}	};
	\spy [green, draw, height = 1.2cm, width = 1.2cm, magnification = 3.0,
	connect spies] on (-0.60,0.40) in node [left] at (-0.76,1.37);
	\spy [green, draw, height = 0.8cm, width = 0.8cm, magnification = 2.5,
	connect spies] on (-0.25,-0.00) in node [left] at (-1.16,-1.55);
	\spy [green, draw, height = 0.8cm, width = 0.8cm, magnification = 3.0,
	connect spies] on (0.30,0.10) in node [left] at (1.95,-1.56);	
	\spy [green, draw, height = 1.2cm, width = 1.2cm, magnification = 2.6,
	connect spies] on (-0.07,0.45) in node [left] at (1.95,1.38);
	\draw[line width=1pt, ultra thin, -latex, red] (-0.15,-0.10) -- node[xshift=-0.05cm,yshift=-0.09cm] {\tiny{}} (-0.25,-0.00);
	\end{tikzpicture}
	%	\put(-95,105){ \color{white}{\bf \small{RMSE:56.8}}}
	%	\put(-95,93){ \color{white}{\bf \small{SSIM:0.489}}}
	\put(-75,8){ \color{white}{\bf \small{FBP}}} 
	\hspace{-0.15in}
	\begin{tikzpicture}
	[spy using outlines={rectangle,green,magnification=2,size=7mm, connect spies}]
	\node {\includegraphics[width=0.24\textwidth]{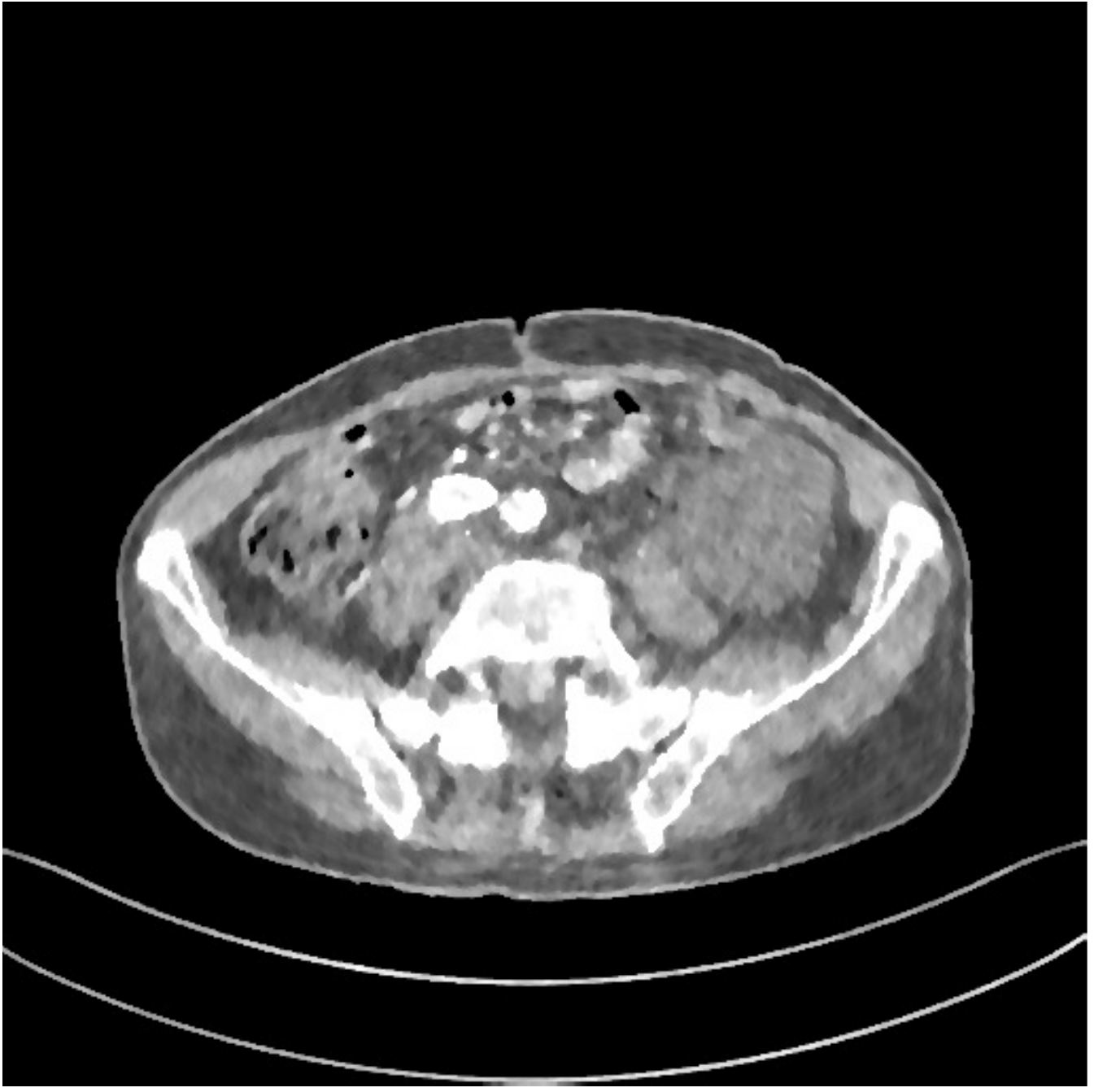}	};
	\spy [green, draw, height = 1.2cm, width = 1.2cm, magnification = 3.0,
	connect spies] on (-0.60,0.40) in node [left] at (-0.76,1.37);
	\spy [green, draw, height = 0.8cm, width = 0.8cm, magnification = 2.5,
	connect spies] on (-0.25,-0.00) in node [left] at (-1.16,-1.55);
	\spy [green, draw, height = 0.8cm, width = 0.8cm, magnification = 3.0,
	connect spies] on (0.30,0.10) in node [left] at (1.95,-1.56);	
	\spy [green, draw, height = 1.2cm, width = 1.2cm, magnification = 2.6,
	connect spies] on (-0.07,0.45) in node [left] at (1.95,1.38);
	\draw[line width=1pt, ultra thin, -latex, red] (-0.15,-0.10) -- node[xshift=-0.05cm,yshift=-0.09cm] {\tiny{}} (-0.25,-0.00);
	\end{tikzpicture}
	%	\put(-95,105){ \color{white}{\bf \small{RMSE:32.6}}}
	%	\put(-95,93){ \color{white}{\bf \small{SSIM:0.780}}}
	\put(-70,8){ \color{white}{\bf \small{EP}}}
	\hspace{-0.15in}	
	\begin{tikzpicture}
	[spy using outlines={rectangle,green,magnification=2,size=7mm, connect spies}]
	\node {\includegraphics[width=0.24\textwidth]{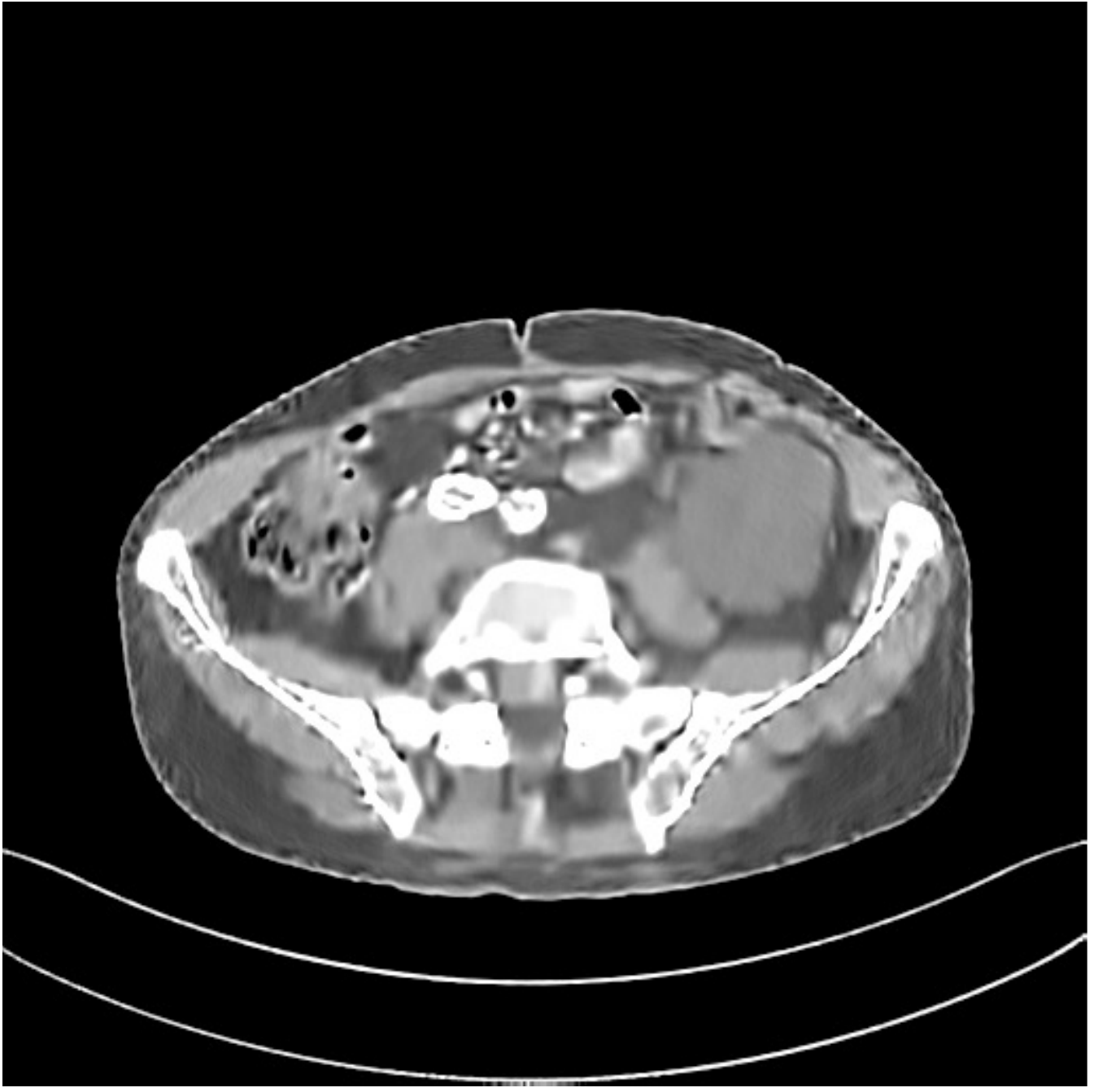}	};
	\spy [green, draw, height = 1.2cm, width = 1.2cm, magnification = 3.0,
	connect spies] on (-0.60,0.40) in node [left] at (-0.76,1.37);
	\spy [green, draw, height = 0.8cm, width = 0.8cm, magnification = 2.5,
	connect spies] on (-0.25,-0.00) in node [left] at (-1.16,-1.55);
	\spy [green, draw, height = 0.8cm, width = 0.8cm, magnification = 3.0,
	connect spies] on (0.30,0.10) in node [left] at (1.95,-1.56);	
	\spy [green, draw, height = 1.2cm, width = 1.2cm, magnification = 2.6,
	connect spies] on (-0.07,0.45) in node [left] at (1.95,1.38);
	\draw[line width=1pt, ultra thin, -latex, red] (-0.15,-0.10) -- node[xshift=-0.05cm,yshift=-0.09cm] {\tiny{}} (-0.25,-0.00);
	\end{tikzpicture}	
	%	\put(-95,105){ \color{white}{\bf \small{RMSE:27.1}}}
	%	\put(-95,93){ \color{white}{\bf \small{SSIM:0.790}}}
	\put(-70,8){ \color{white}{\bf \small{ST}}}
	\\
	\vspace{-0.1in}
	\begin{tikzpicture}
	[spy using outlines={rectangle,green,magnification=2,size=7mm, connect spies}]
	\node {\includegraphics[width=0.24\textwidth]{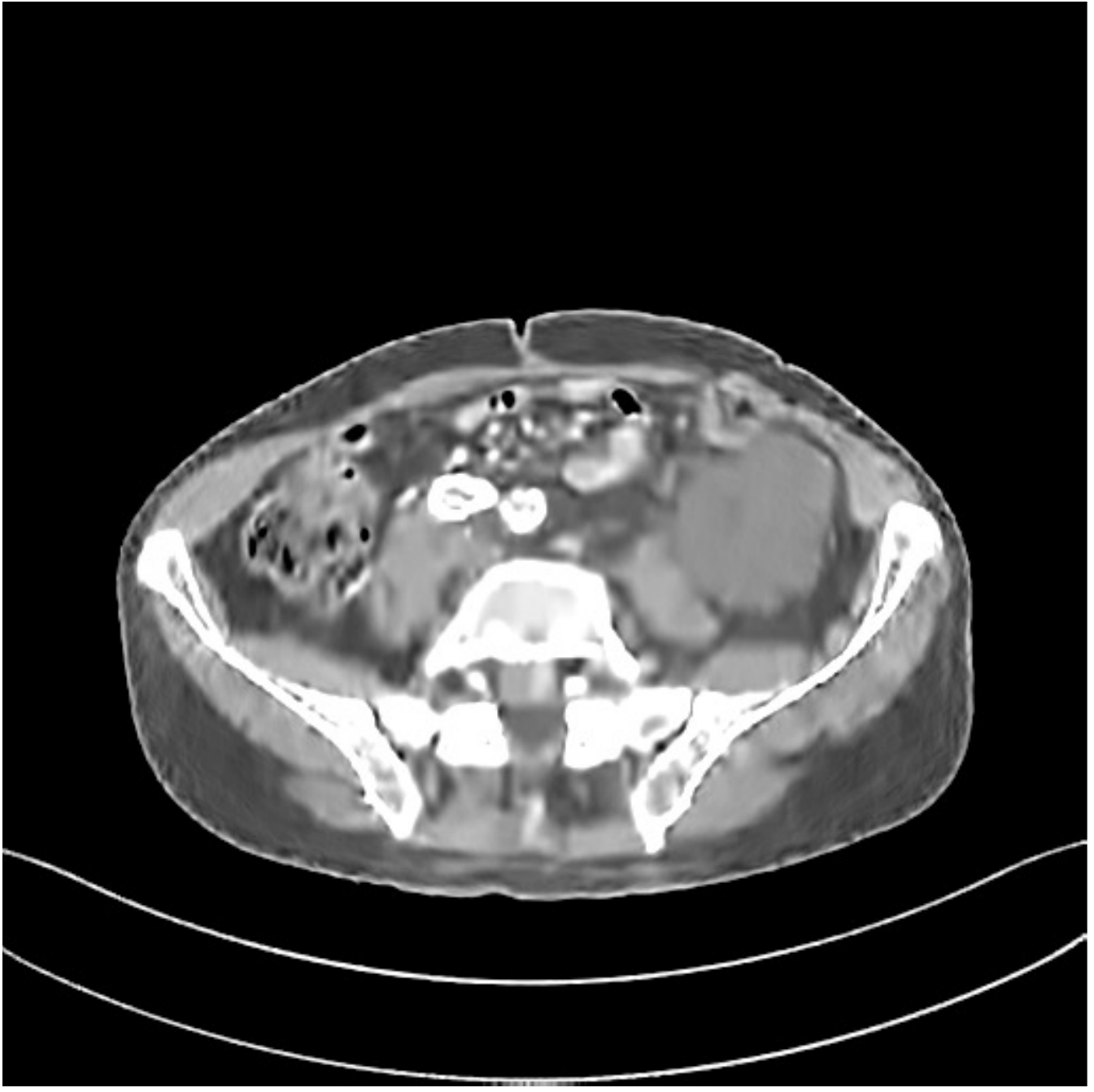}	};
	\spy [green, draw, height = 1.2cm, width = 1.2cm, magnification = 3.0,
	connect spies] on (-0.60,0.40) in node [left] at (-0.76,1.37);
	\spy [green, draw, height = 0.8cm, width = 0.8cm, magnification = 2.5,
	connect spies] on (-0.25,-0.00) in node [left] at (-1.16,-1.55);
	\spy [green, draw, height = 0.8cm, width = 0.8cm, magnification = 3.0,
	connect spies] on (0.30,0.10) in node [left] at (1.95,-1.56);	
	\spy [green, draw, height = 1.2cm, width = 1.2cm, magnification = 2.6,
	connect spies] on (-0.07,0.45) in node [left] at (1.95,1.38);
	\draw[line width=1pt, ultra thin, -latex, red] (-0.15,-0.10) -- node[xshift=-0.05cm,yshift=-0.09cm] {\tiny{}} (-0.25,-0.00);
	\end{tikzpicture}
	%	\put(-95,105){ \color{white}{\bf \small{RMSE:26.0}}}
	%	\put(-95,93){ \color{white}{\bf \small{SSIM:0.791}}}
	\put(-83,8){ \color{white}{\bf \small{MARS2}}}
	\hspace{-0.15in}
	\begin{tikzpicture}
	[spy using outlines={rectangle,green,magnification=2,size=7mm, connect spies}]
	\node {\includegraphics[width=0.24\textwidth]
		{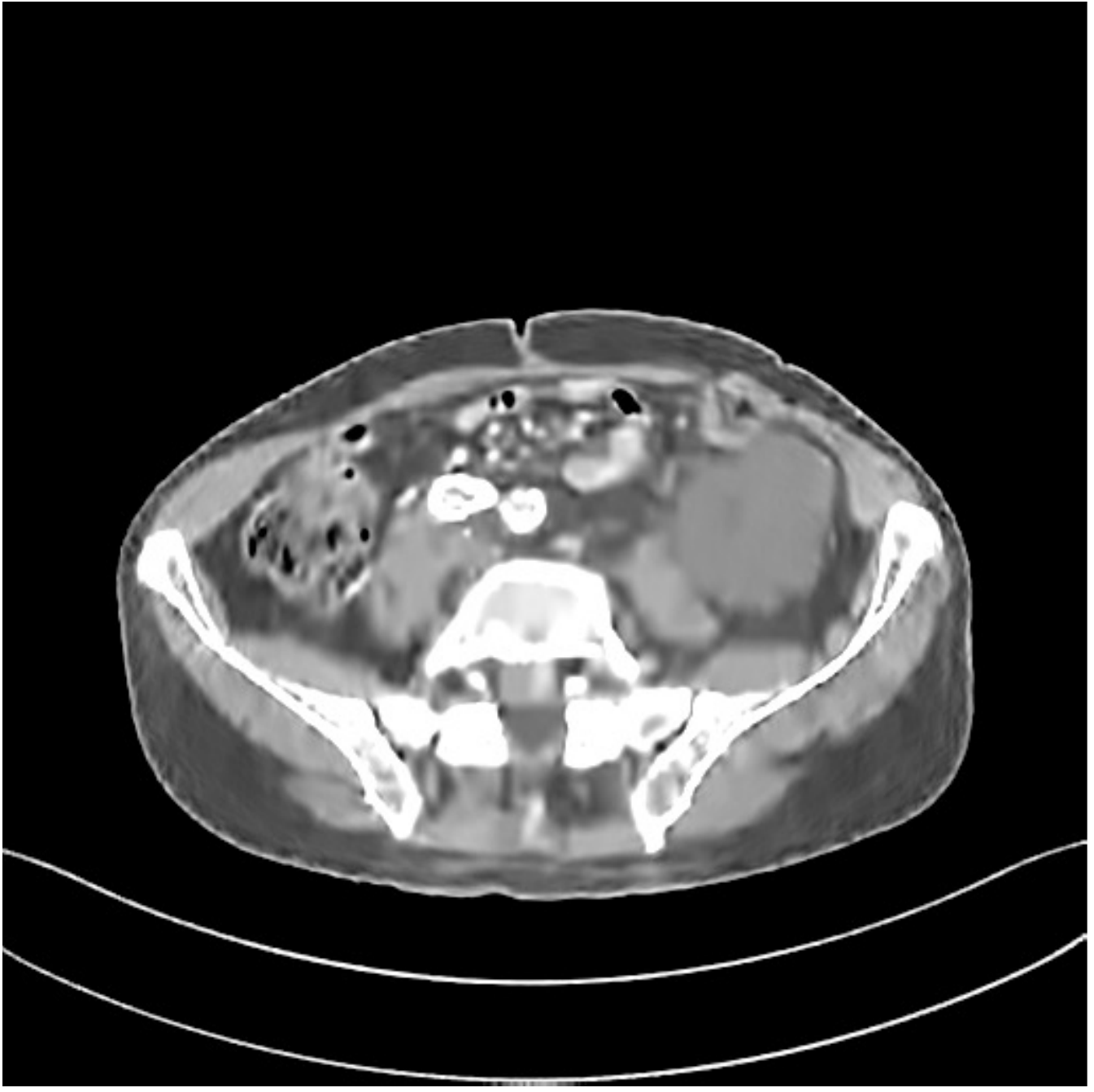}	};
	\spy [green, draw, height = 1.2cm, width = 1.2cm, magnification = 3.0,
	connect spies] on (-0.60,0.40) in node [left] at (-0.76,1.37);
	\spy [green, draw, height = 0.8cm, width = 0.8cm, magnification = 2.5,
	connect spies] on (-0.25,-0.00) in node [left] at (-1.16,-1.55);
	\spy [green, draw, height = 0.8cm, width = 0.8cm, magnification = 3.0,
	connect spies] on (0.30,0.10) in node [left] at (1.95,-1.56);	
	\spy [green, draw, height = 1.2cm, width = 1.2cm, magnification = 2.6,
	connect spies] on (-0.07,0.45) in node [left] at (1.95,1.38);
	\draw[line width=1pt, ultra thin, -latex, red] (-0.15,-0.10) -- node[xshift=-0.05cm,yshift=-0.09cm] {\tiny{}} (-0.25,-0.00);
	\end{tikzpicture}
	%	\put(-95,105){ \color{white}{\bf \small{RMSE:26.2}}}
	%	\put(-95,93){ \color{white}{\bf \small{SSIM:0.793}}}
	\put(-83,8){ \color{white}{\bf \small{MARS3}}}
	\hspace{-0.15in}
	\begin{tikzpicture}
	[spy using outlines={rectangle,green,magnification=2,size=7mm, connect spies}]
	\node {\includegraphics[width=0.24\textwidth]
		{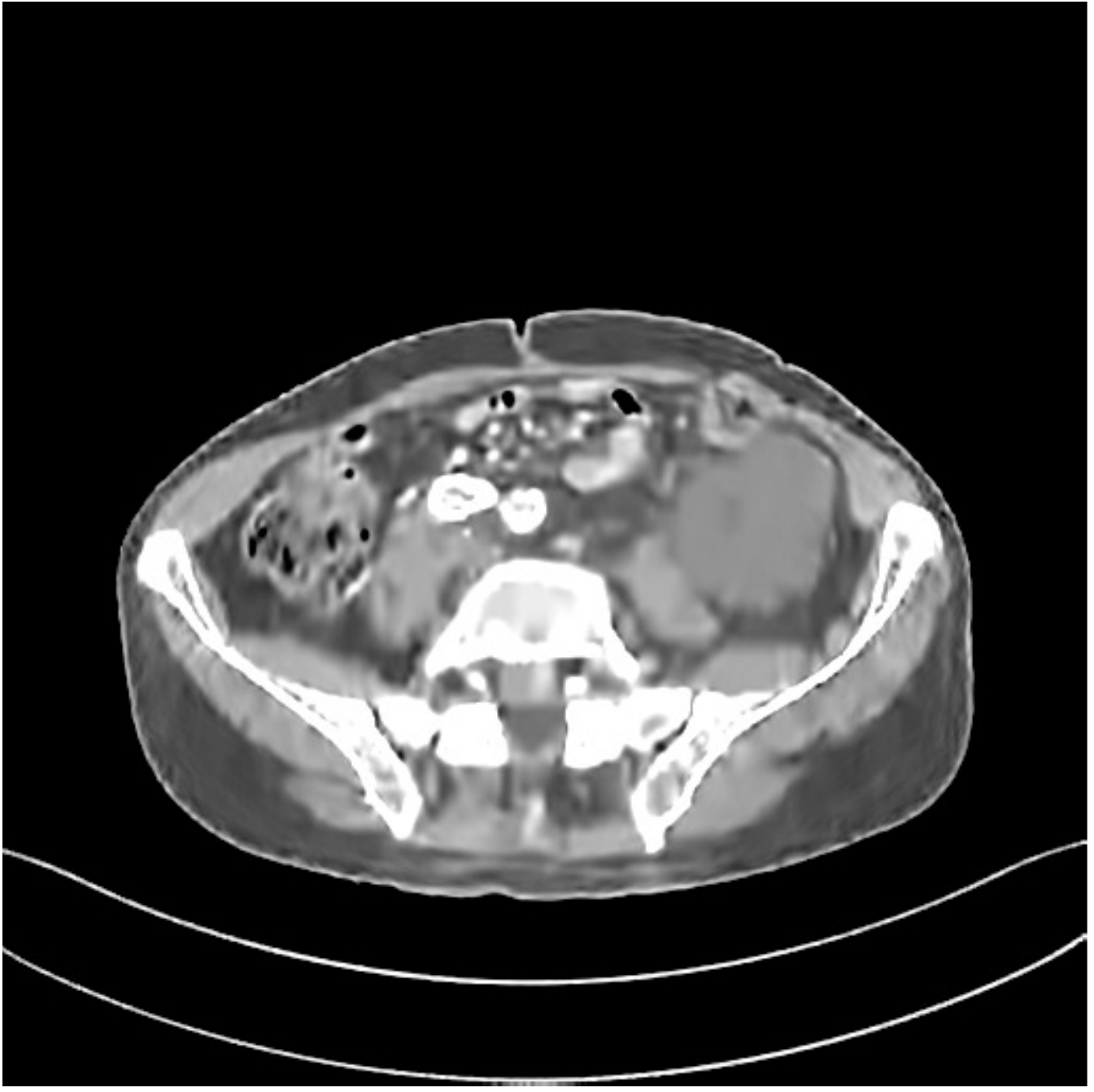}	};
	\spy [green, draw, height = 1.2cm, width = 1.2cm, magnification = 3.0,
	connect spies] on (-0.60,0.40) in node [left] at (-0.76,1.37);
	\spy [green, draw, height = 0.8cm, width = 0.8cm, magnification = 2.5,
	connect spies] on (-0.25,-0.00) in node [left] at (-1.16,-1.55);
	\spy [green, draw, height = 0.8cm, width = 0.8cm, magnification = 3.0,
	connect spies] on (0.30,0.10) in node [left] at (1.95,-1.56);	
	\spy [green, draw, height = 1.2cm, width = 1.2cm, magnification = 2.6,
	connect spies] on (-0.07,0.45) in node [left] at (1.95,1.38);
	\draw[line width=1pt, ultra thin, -latex, red] (-0.15,-0.10) -- node[xshift=-0.05cm,yshift=-0.09cm] {\tiny{}} (-0.25,-0.00);
	\end{tikzpicture}
	%	\put(-95,105){ \color{white}{\bf \small{RMSE:27.5}}}
	%	\put(-95,93){ \color{white}{\bf \small{SSIM:0.797}}}
	\put(-83,8){ \color{white}{\bf \small{MARS5}}}
	\hspace{-0.15in}
	\begin{tikzpicture}
	[spy using outlines={rectangle,green,magnification=2,size=7mm, connect spies}]
	\node {\includegraphics[width=0.24\textwidth]
		{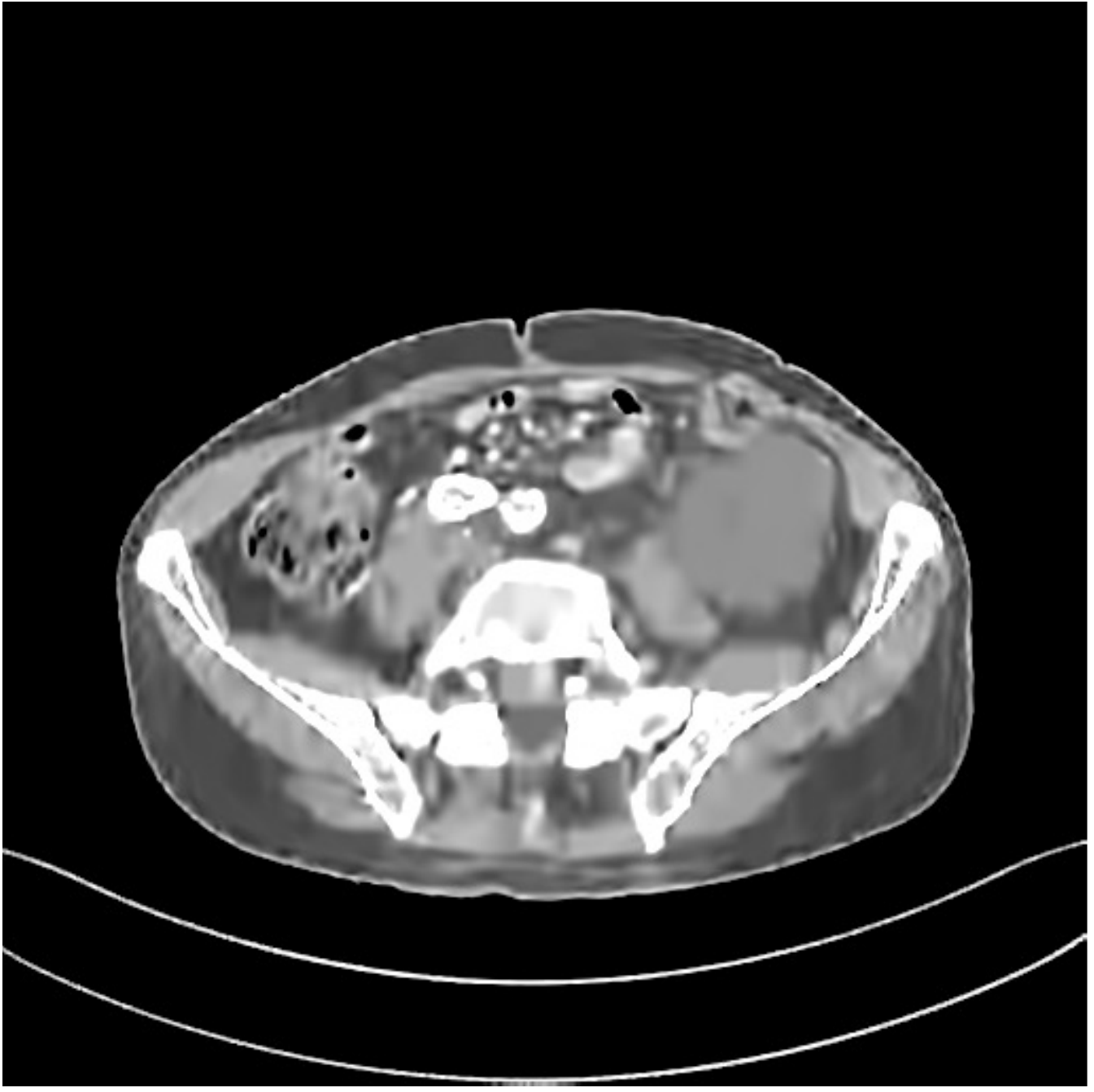}	};
	\spy [green, draw, height = 1.2cm, width = 1.2cm, magnification = 3.0,
	connect spies] on (-0.60,0.40) in node [left] at (-0.76,1.37);
	\spy [green, draw, height = 0.8cm, width = 0.8cm, magnification = 2.5,
	connect spies] on (-0.25,-0.00) in node [left] at (-1.16,-1.55);
	\spy [green, draw, height = 0.8cm, width = 0.8cm, magnification = 3.0,
	connect spies] on (0.30,0.10) in node [left] at (1.95,-1.56);	
	\spy [green, draw, height = 1.2cm, width = 1.2cm, magnification = 2.6,
	connect spies] on (-0.07,0.45) in node [left] at (1.95,1.38);
	\draw[line width=1pt, ultra thin, -latex, red] (-0.15,-0.10) -- node[xshift=-0.05cm,yshift=-0.09cm] {\tiny{}} (-0.25,-0.00);	
	\end{tikzpicture}
	%	\put(-95,105){ \color{white}{\bf \small{RMSE:28.7}}}
	%	\put(-95,93){ \color{white}{\bf \small{SSIM:0.802}}}
	\put(-83,8){ \color{white}{\bf \small{MARS7}}}
	\caption{Reconstructions of slice 100 of patient L506 at incident photon intensity $I_0=1\times 10^4$. The first row shows the reference image and reconstructions with FBP, PWLS-EP, and PWLS-ST, respectively, and the second row shows the results with MARS models with $2$, $3$, $5$, and $7$ layers, respectively. The display window is [800, 1200] HU.}
	\label{fig:recon_mayo_L506}
	\vspace{0.2in}
\end{figure}

Figs.~\ref{fig:recon_mayo_L109},~\ref{fig:recon_mayo_L192},~\ref{fig:recon_mayo_L333}, and~\ref{fig:recon_mayo_L506} show the reconstructions of the four independent slices using the FBP, PWLS-EP, PWLS-ST, PWLS-MARS2, PWLS-MARS3, PWLS-MARS5, and PWLS-MARS7 schemes, respectively. Additional Mayo Clinic experimental results of the parameter tuning case (Fig.~15) are shown in the supplementary document.
%at incident photon intensity $I_0=10^{4}$. 
Table~\ref{tab:MRST} lists the RMSE and SSIM values of reconstructions of the four independent test slices, with the best values bolded. 
Generally, the five and seven layer models provided the best RMSE and SSIM values. They outperform the single-layer model by $1.9$ HU in RMSE on average. However, the MARS5 and MARS7 models perform similarly.
In order to strengthen the benefits of the multi-layer model, Table~\ref{tab:MRST_ROI} lists the RMSE of the reconstructions in four different ROIs (shown in the reference of Fig.~\ref{fig:recon_mayo_L506}) with seven methods for slice 100 of patient L506.
By observing the reconstructed images, we see that although the ST model achieves a cleaner reconstruction result than FBP and PWLS-EP, it still sacrifices some sharpness of the central region and suffers from loss of details. The deeper models have a somewhat more positive effect in terms of maintaining subtle features, which is clearly more essential to clinical diagnosis. Furthermore, as we will discuss later, after considerable parameter tuning, we found that the information contained in residual maps is gradually decreased with the number of layers, eventually vanishing at some layer, which suggests that very deep unsupervised models might not offer significantly better image quality.
%Furthermore, as we will discuss later, after considerable parameter tuning, we have observed that the deeper models offer more stable image quality as $\beta$ is varied, i.e., they are more robust to oversmoothing.

\begin{table}[!h]	
	\vspace{0.2in}
	\centering
	\caption{RMSE in HU (first row) and SSIM (second row) of reconstructions with FBP, PWLS-EP, PWLS-ST, PWLS-MARS2, PWLS-MARS3, PWLS-MARS5, and PWLS-MARS7, for four slices of the Mayo Clinic data at incident photon intensity $I_0 = 1\times10^4$.}
	\label{tab:MRST}	 	
	%\vspace{-0.05in}
	\renewcommand\tabcolsep{5.0pt}
	\footnotesize{
		\begin{tabular}{c|ccccccc}		
			\toprule
			&FBP  & EP &PWLS-ST  &PWLS-MARS2    &PWLS-MARS3   & PWLS-MARS5  & PWLS-MARS7  \\
			\midrule
			%			\multirow{3}{*}{\shortstack{L067\\slice 120}} & 175.9  & 45.6  & 40.4  & 38.3  & 38.1   & \textbf{37.7}  & 38.2 \\
			%			\cmidrule{2-8}
			%			& 0.306  & 0.746  & 0.736   & 0.749  & 0.753   & 0.762  & \textbf{0.768}\\
			\multirow{3}{*}{\shortstack{L109\\slice90}}  & 107.1  & 33.5   &29.0    &28.1   & 27.8   &\textbf{27.6}   &28.1  \\
			\cmidrule{2-8}
			&0.343   &0.734  &0.716    &0.727   &0.731    &0.744    &\textbf{0.753} \\
			\midrule
			\multirow{3}{*}{\shortstack{L192\\slice90}}  & 93.7  & 31.5  & 26.3   & 25.3  & 24.9   & \textbf{24.6}  & 24.9 \\
			\cmidrule{2-8}
			& 0.350  & 0.747  & 0.737   & 0.744  & 0.750   & 0.765   & \textbf{0.781}\\
			\midrule
			\multirow{3}{*}{\shortstack{L333\\slice140}}  & 113.1  & 36.3   &29.7    &28.5   &28.3    &\textbf{28.1}   &28.4  \\
			\cmidrule{2-8}
			&0.358   &0.758  &0.739    &0.744   &0.750    &0.766    &\textbf{0.786} \\
			\midrule
			%			\multirow{3}{*}{\shortstack{L333\\slice 130}}  & 70.2  & 35.2  & 29.3   & 28.1  & 27.6   & \textbf{27.5}  & 27.7 \\
			%			\cmidrule{2-8}
			%			\cmidrule{2-8}
			%			& 0.409  & 0.757  & 0.736   & 0.741  & 0.748   & 0.764   & \textbf{0.782}\\
			%			\midrule
			\multirow{3}{*}{\shortstack{L506\\slice 100}}  & 65.3  & 34.3  & 27.5   & 26.2  & 25.6  & \textbf{25.3}  & 25.7 \\
			\cmidrule{2-8}
			& 0.461  & 0.778  & 0.760   & 0.766  & 0.773   & 0.790  & \textbf{0.809} \\
			\bottomrule
		\end{tabular}
	}
	\vspace{0.1in}
\end{table}

\begin{table}[!h]	
	\vspace{0.2in}
	\centering
	\caption{RMSE (HU) in four ROIs of reconstructions with FBP, PWLS-EP, PWLS-ST, PWLS-MARS2, PWLS-MARS3, PWLS-MARS5, and PWLS-MARS7, for slice 100 of patient L506 of the Mayo Clinic data at incident photon intensity $I_0 = 1\times10^4$.}
	\label{tab:MRST_ROI}	 	
	%\vspace{-0.05in}
	\renewcommand\tabcolsep{5.0pt}
	\footnotesize{
		\begin{tabular}{c|ccccccc}		
			\toprule
			&FBP  & EP &PWLS-ST  &PWLS-MARS2    &PWLS-MARS3   & PWLS-MARS5  & PWLS-MARS7  \\
			\midrule
			ROI-1 &1.05  &0.71   &0.68   &0.62   &0.60    &\textbf{0.59}   &0.59  \\
			\cmidrule{1-8}
			ROI-2 &0.90  &0.78   &0.69   &0.63   &0.62    &\textbf{0.61}   &0.63 \\
			\midrule
			ROI-3  &2.17  &1.88   &1.75   &1.57   &1.53    &\textbf{1.51}   &1.55 \\
			\cmidrule{1-8}
			ROI-4 &1.91  &0.96   &1.03   &0.91   &0.90    &\textbf{0.89}   &0.91 \\
			\bottomrule
		\end{tabular}
	}
	\vspace{0.1in}
\end{table}

\subsubsection{Analysis of Residual Maps}\label{sec:res_analysis}

\begin{figure}[!h]
	\centering  
	%\vspace{-0.1in}
	\begin{tikzpicture}
	[spy using outlines={rectangle,green,magnification=2,size=7mm, connect spies}]
	\node {\includegraphics[height=0.21\textwidth]{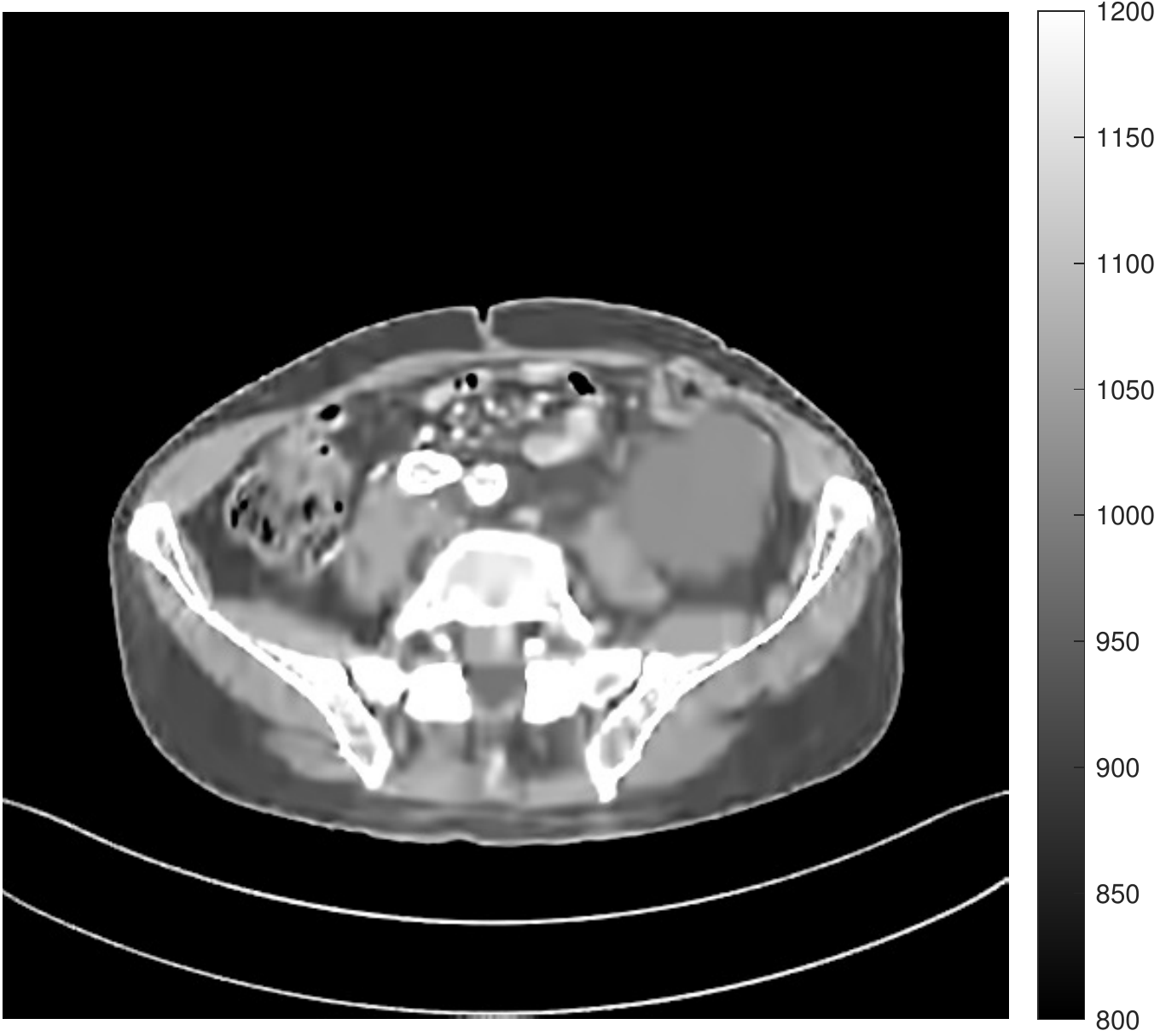}	};
	\end{tikzpicture}
	%	\put(-95,105){ \color{white}{\bf \small{RMSE:0.00}}}
	%	\put(-95,93){ \color{white}{\bf \small{SSIM:1.000}}}
	\put(-105,8){ \color{white}{\bf \small{Recon Image}}} 
	\hspace{-0.15in}
	\begin{tikzpicture}
	[spy using outlines={rectangle,green,magnification=2,size=7mm, connect spies}]
	\node {\includegraphics[height=0.21\textwidth]{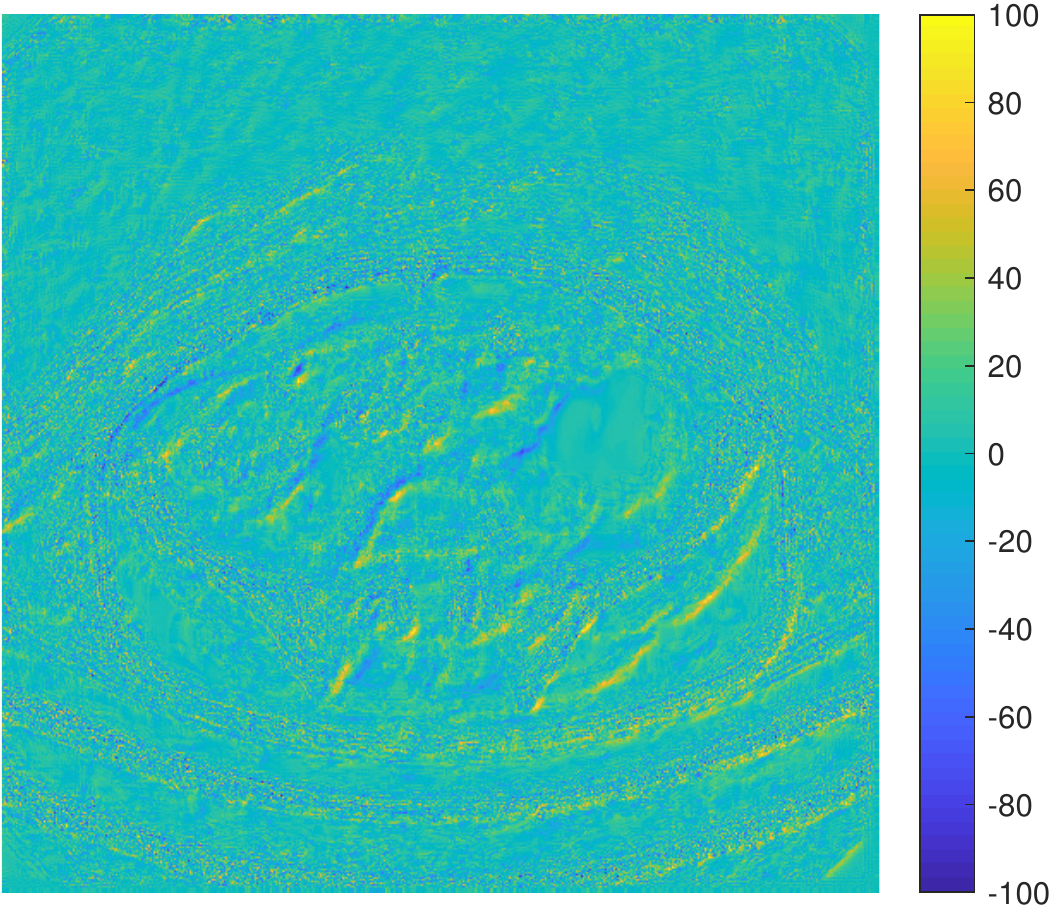}	};
	\end{tikzpicture}
	%	\put(-95,105){ \color{white}{\bf \small{RMSE:56.8}}}
	%	\put(-95,93){ \color{white}{\bf \small{SSIM:0.489}}}
	\put(-96,8){ \color{white}{\bf \small{1st layer}}} 
	\hspace{-0.15in}
	\begin{tikzpicture}
	[spy using outlines={rectangle,green,magnification=2,size=7mm, connect spies}]
	\node {\includegraphics[height=0.21\textwidth]{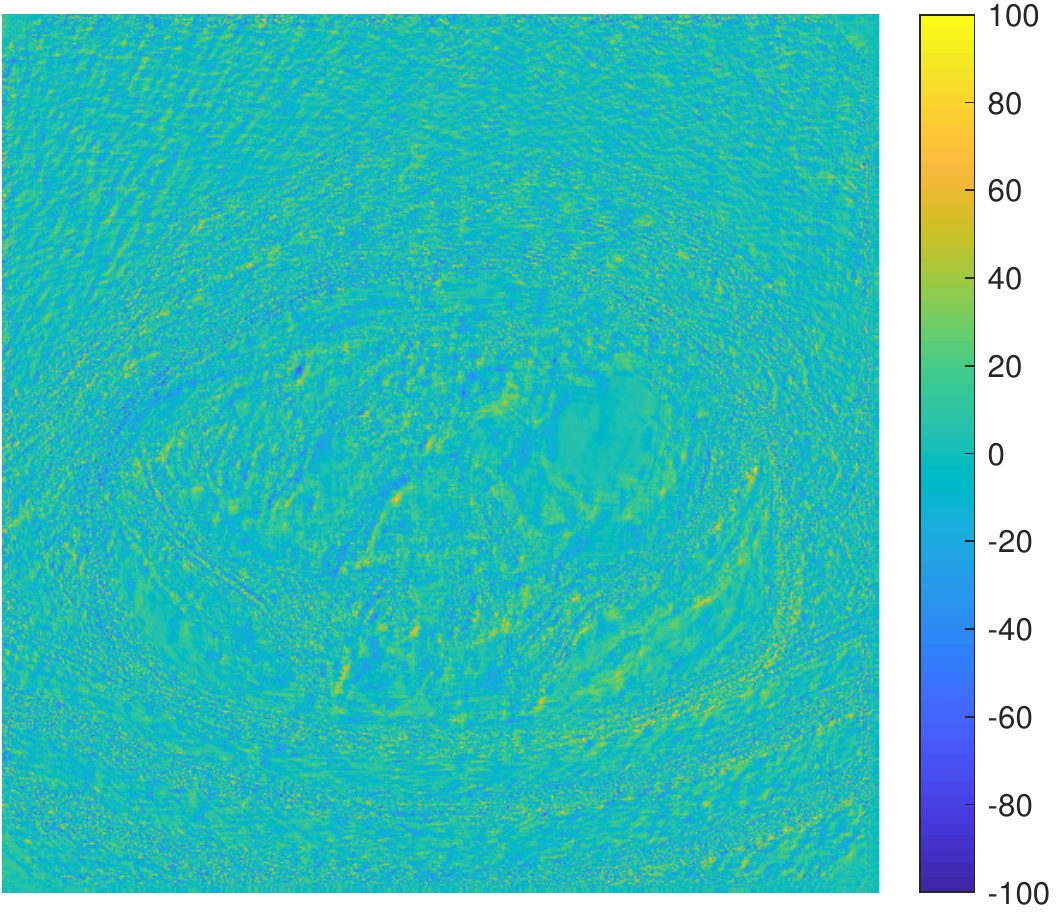}	};
	\end{tikzpicture}
	%	\put(-95,105){ \color{white}{\bf \small{RMSE:32.6}}}
	%	\put(-95,93){ \color{white}{\bf \small{SSIM:0.780}}}
	\put(-96,8){ \color{white}{\bf \small{2nd layer}}}
	\hspace{-0.15in}	
	\begin{tikzpicture}
	[spy using outlines={rectangle,green,magnification=2,size=7mm, connect spies}]
	\node {\includegraphics[height=0.21\textwidth]{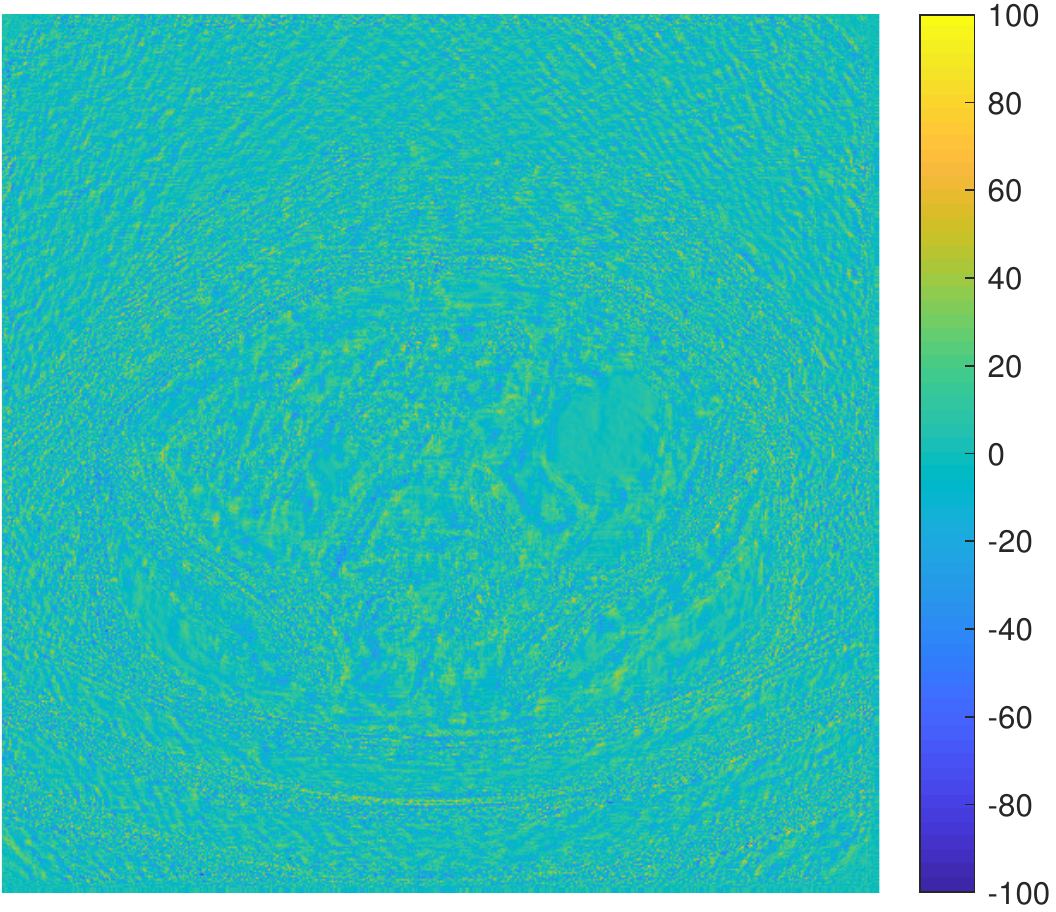}	};
	\end{tikzpicture}	
	%	\put(-95,105){ \color{white}{\bf \small{RMSE:27.1}}}
	%	\put(-95,93){ \color{white}{\bf \small{SSIM:0.790}}}
	\put(-96,8){ \color{white}{\bf \small{3rd layer}}}
	\\
	\begin{tikzpicture}
	[spy using outlines={rectangle,green,magnification=2,size=7mm, connect spies}]
	\node {\includegraphics[height=0.21\textwidth]{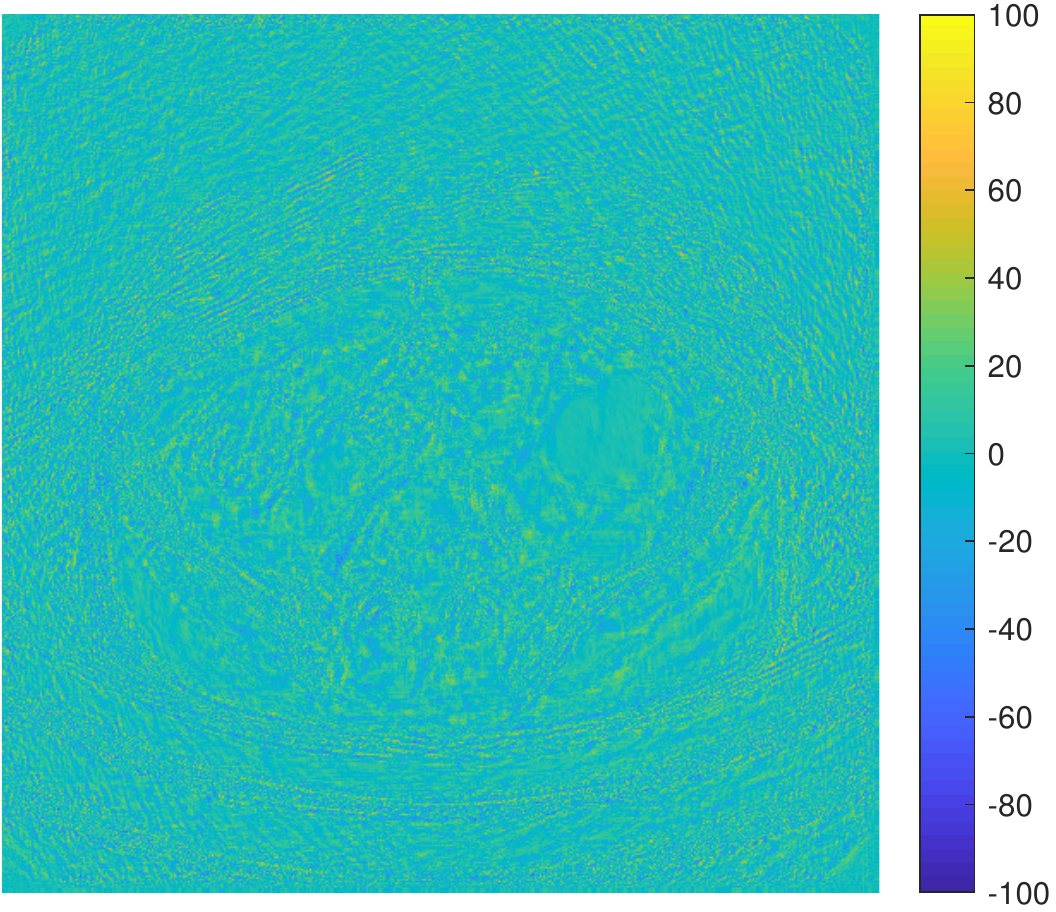}	};	
	\end{tikzpicture}
	%	\put(-95,105){ \color{white}{\bf \small{RMSE:26.0}}}
	%	\put(-95,93){ \color{white}{\bf \small{SSIM:0.791}}}
	\put(-96,8){ \color{white}{\bf \small{4th layer}}}
	\hspace{-0.15in}
	\begin{tikzpicture}
	[spy using outlines={rectangle,green,magnification=2,size=7mm, connect spies}]
	\node {\includegraphics[height=0.21\textwidth]
		{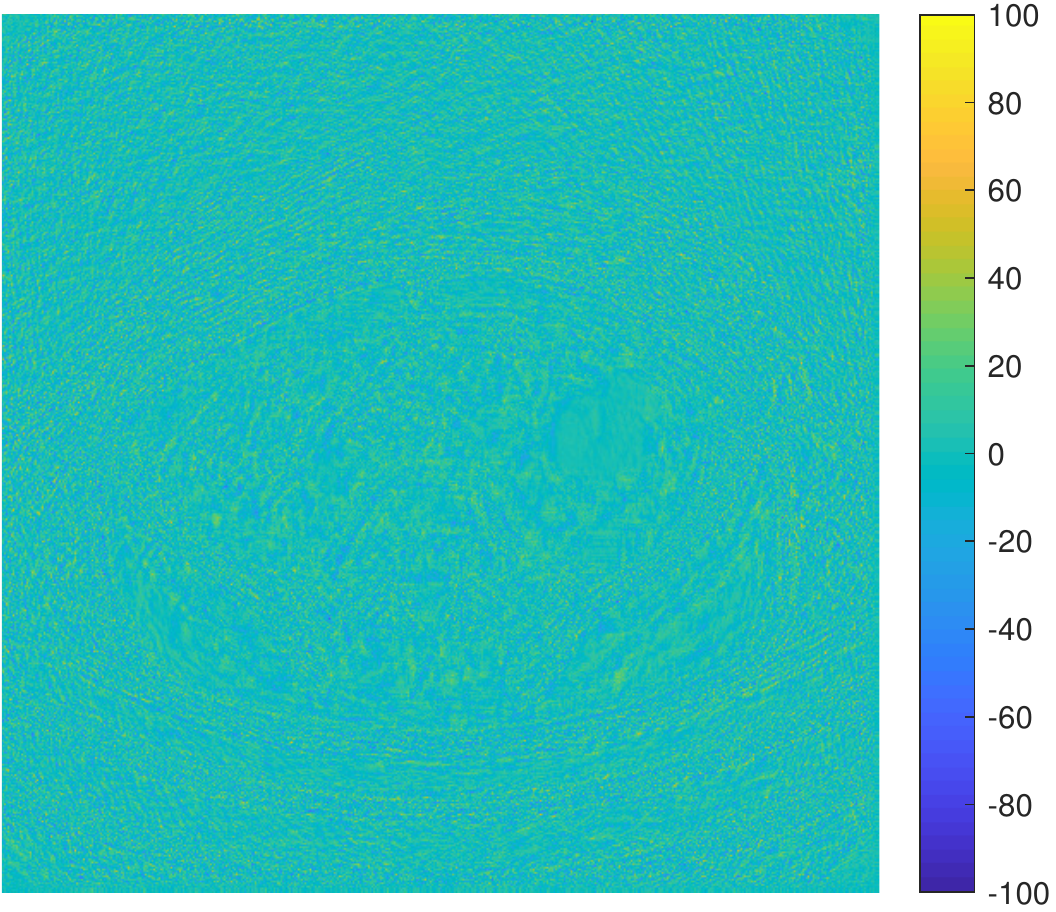}	};
	\end{tikzpicture}
	%	\put(-95,105){ \color{white}{\bf \small{RMSE:26.2}}}
	%	\put(-95,93){ \color{white}{\bf \small{SSIM:0.793}}}
	\put(-96,8){ \color{white}{\bf \small{5th layer}}}
	\hspace{-0.15in}
	\begin{tikzpicture}
	[spy using outlines={rectangle,green,magnification=2,size=7mm, connect spies}]
	\node {\includegraphics[height=0.21\textwidth]
		{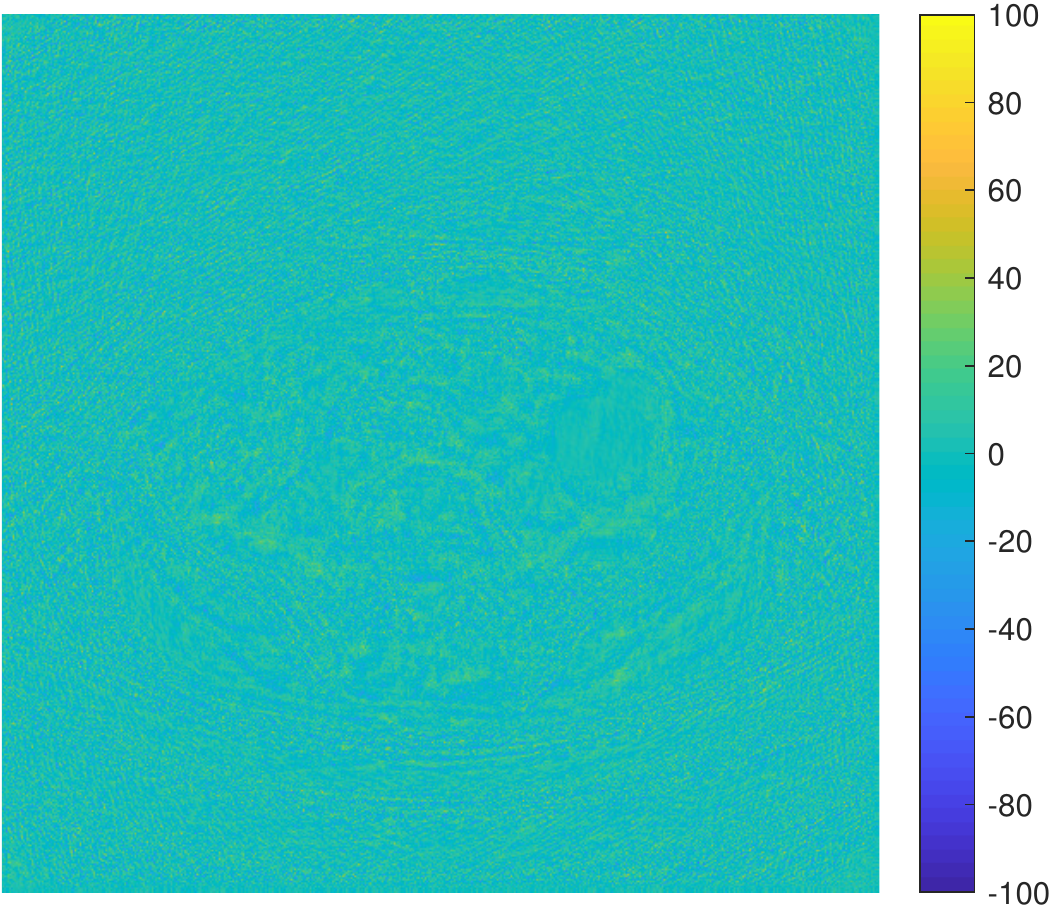}	};
	\end{tikzpicture}
	%	\put(-95,105){ \color{white}{\bf \small{RMSE:27.5}}}
	%	\put(-95,93){ \color{white}{\bf \small{SSIM:0.797}}}
	\put(-96,8){ \color{white}{\bf \small{6th layer}}}
	\hspace{-0.15in}
	\begin{tikzpicture}
	[spy using outlines={rectangle,green,magnification=2,size=7mm, connect spies}]
	\node {\includegraphics[height=0.21\textwidth]
		{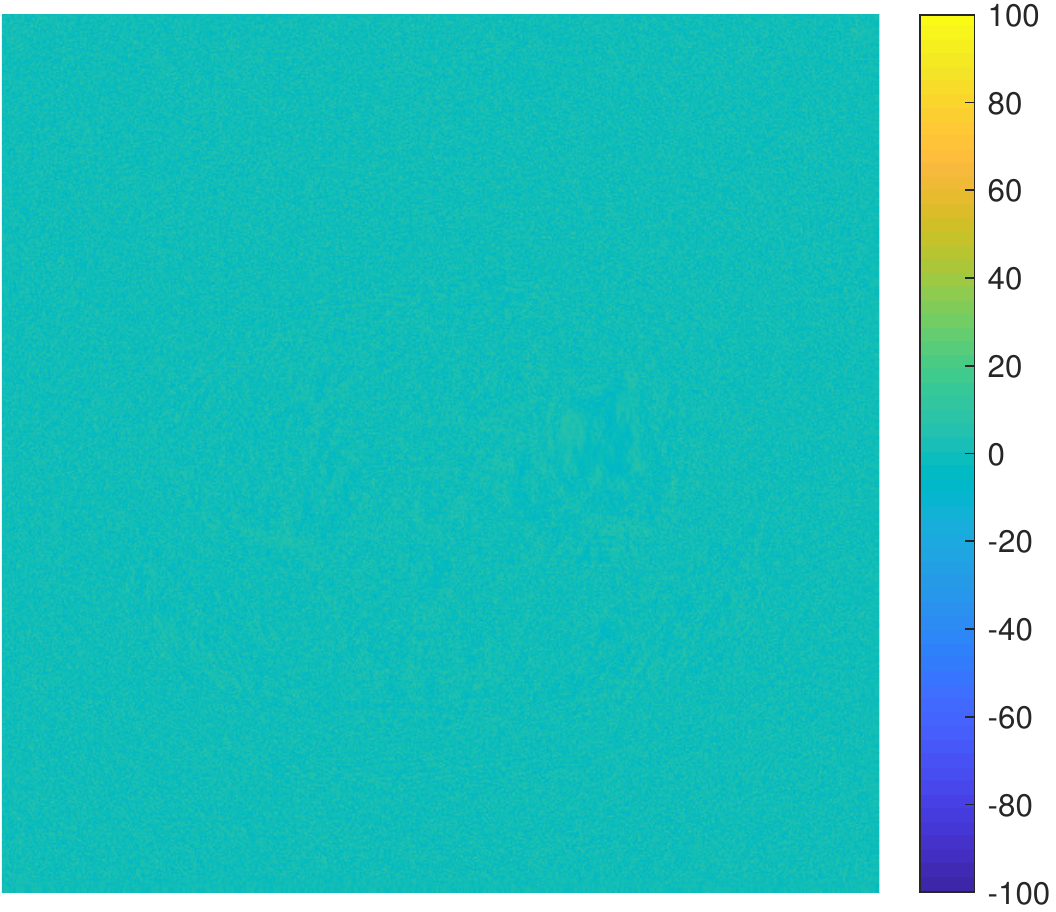}	};
	\end{tikzpicture}
	%	\put(-95,105){ \color{white}{\bf \small{RMSE:28.7}}}
	%	\put(-95,93){ \color{white}{\bf \small{SSIM:0.802}}}
	\put(-96,8){ \color{white}{\bf \small{7th layer}}}
	\caption{Reconstruction and transform-domain residual images for slice 100 of patient L506. The leftmost image on the first row is the reconstruction with PWLS-MARS7, while the other images are the residual maps in different layers. The display windows are [800, 1200] HU and [-100, 100] HU, respectively, for the reconstruction and the residual image, respectively.}
	\label{fig:residual_mayo_L506}
	\vspace{0.2in}
\end{figure}
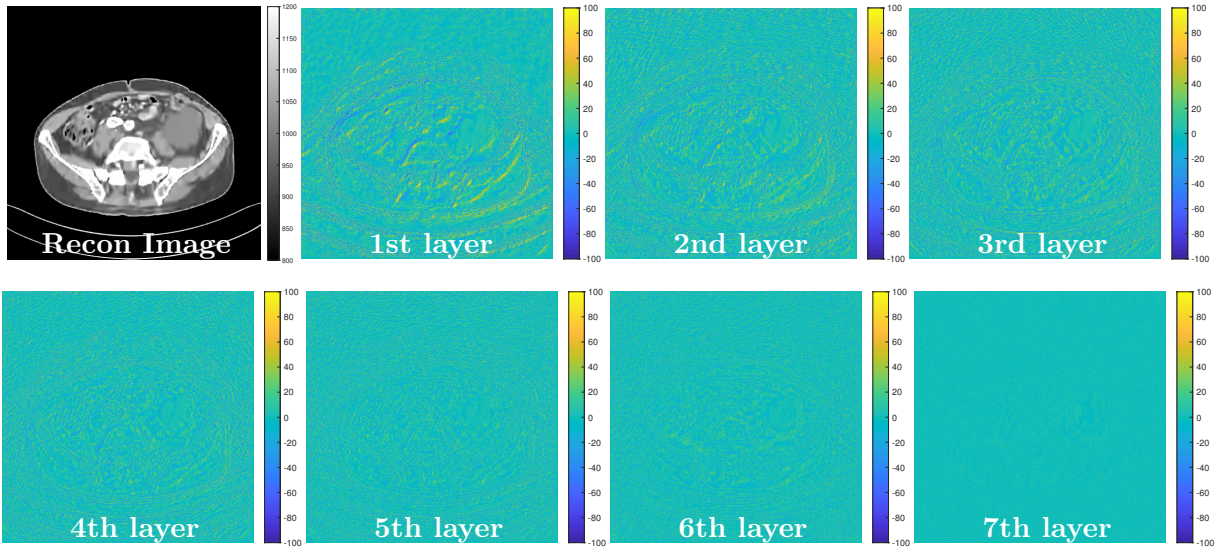

Here, we investigate the residual images over the layers of the MARS7 model. Fig.~\ref{fig:residual_mayo_L506} displays the image reconstructed with MARS7 along with the residual images in different layers. The residual images are generated by applying the restoring operation $(\P^j)^T$ to the corresponding columns of each residual matrix $\R_l, 1\leq l \leq L$, forming images $\sum_j (\P^j)^T \R_l^j$. Essentially, all the columns of $\R_l$ are transformed into $8 \times 8$ patches and accumulated back in the image to form the residual image in the $l$th layer. We can observe that the residual images in the first three layers contain explicit structural information and we still find some delicate details in the fourth and fifth layers. However, we hardly see any valuable features in the residual images for the following layers, which is consistent with the fact that the transform is overwhelmed by noise in quite deep layers. Therefore, the ceiling for the potential of multi-layer sparsifying transform model may be 5 or 7 layers. The quantitive result also implies the same conclusion. 
\vspace{0.8in}

\subsection{Runtimes for MARS}
We also discuss the runtimes for the proposed MARS model. 
%		The computational complexity of transform learning algorithm has been stated to be $O(p^2N)$ per iteration~\cite{ravishankar:15:lst}. Multi-layer extension requires addtional computation, while the cost is dominated by the calculation of backpropagation matrix $\B_p^q$. And the total computational cost of learning algorithm is supposed to be $O(\frac{L^3+6L}{6}p^2N+\frac{L^4-2L^3-L^2+2L}{24}p^3)$.
%		As illustrated in Section~II.B.2., the total computational complexity for image reconstruction is $O(N_p^3+5N_p^2+\frac{L^3-L}{6}p^3+\frac{L^2+L}{2}p^2N_p)$, where $N_p$ denotes the pixel number of reconstruction image. 
Table~\ref{tab:MRST_time} shows the average runtimes per iteration (MARS schemes were run for the same overall number of iterations) for various MARS models for both the XCAT phantom and Mayo Clinic data experiments. We ran the Matlab code on a machine with two 2.4GHz 14-core Intel Xeon E5-2680 v4 processors. We find that although training the deep models (which would be done once offline) takes several times as long as the shallow (single layer) model, the cost of the reconstruction/testing step is much more similar between deep and shallow models.
\vspace{0.2in}
\begin{table}[H]	
	\vspace{0.1in}
	\centering
	\caption{Average runtime per iteration of various MARS models with both XCAT phantom and Mayo Clinic data experiments. Each number displayed in this table is in seconds.}
	\label{tab:MRST_time}	 	
	%\vspace{-0.05in}
	\renewcommand\tabcolsep{5.0pt}
	\footnotesize{
		\begin{tabular}{c|c|ccccc}		
			\toprule
			&  &PWLS-ST  &PWLS-MARS2    &PWLS-MARS3   & PWLS-MARS5  & PWLS-MARS7  \\
			\midrule
			\multirow{3}{*}{\shortstack{XCAT\\phantom}}    & Training  & \textbf{0.8}   &  1.4  &  3.5  & 4.7   & 7.8 \\
			\cmidrule{2-7}
			& Testing  & \textbf{2.9}   & 3.2  &  3.6  & 4.4   & 5.1 \\
			\midrule
			\multirow{3}{*}{\shortstack{Mayo Clinic\\data}}    & Training  & \textbf{1.5}  & 2.8  & 7.4  & 9.3  & 15.2 \\
			\cmidrule{2-7}
			& Testing  & \textbf{3.1}  & 3.4   & 4.1  & 5.0  & 5.8 \\
			\bottomrule
		\end{tabular}
	}
	\vspace{0.1in}
\end{table}

\section{Discussion and Conclusion}

In this work, we presented a strategy for unsupervised learning of deep transform models from limited data and with nested network structure, where the input of each layer comprises of the sparsifiable residual map from the preceding layer. The learned Multi-lAyer Residual Sparsifying transform (MARS) model is used to form a data-driven regularizer in model-based image reconstruction and proves effective for low-dose CT image reconstruction. The proposed algorithms for learning MARS models and for image reconstruction use highly efficient updates and are scalable.

We trained models from patches of (regular-dose) slices of the XCAT phantom and Mayo Clinic data and tested the models for reconstructing other slices. The learned multi-layer models contain complex features and structures, which help enhance image reconstruction quality of MARS models over single layer models. Experiments with both simulated data from the XCAT phantom and with the synthesized clinical data reveal that PWLS-MARS provides better reconstruction metrics 
%and visual quality 
and image details compared to other methods such as FBP, PWLS-EP, and PWLS-ST. 
In Figs.~\ref{fig:recon_mayo_L109},~\ref{fig:recon_mayo_L192},~\ref{fig:recon_mayo_L333}, and~\ref{fig:recon_mayo_L506}, we observed that the reconstruction incorporating deep transform model prior presented more subtle details, especially for the central region, which normally suffers from severe artifacts in low-dose CT reconstruction.

We also investigated the potential limitation in terms of the model depth. By observing Tables~\ref{tab:MRST} and \ref{tab:MRST_ROI}, we found deep models such as MARS7 only offer little additional benefit of RMSE and SSIM. Such a phenomenon also appears in other related work~\cite{Singhal:19:MLDT} in which the author believes that limited training dataset leads to the deterioration of the performance of deep models. In order to seek the underlying reason, we increased the training dataset from 7 slices to 14 slices while the approximate number of patches to be fed into network has been risen to 3 million. Table~\ref{tab:MRST_incredata} lists the reconstruction results of slice 100 of patient L506 with respect to training dataset of 7 slices and 14 slices. The tiny improvement leads us to conjecture that the limitation of the deep model may not be due to the small set of training images. Section.~\ref{sec:res_analysis} provides an alternative explanation. 
We found that very deep residual layers may not contain much structures, thus resulting in somewhat noisy transforms there, which may offer little additional benefit.
\vspace{0.2in}
%\blue{Therefore, considering the computational cost to train a deep MARS model, we prefer to MARS2 or MARS3 model as the feasible tool for future application.} 
%Nevertheless, the deep model indeed performs well on robustness against parameter tunning. To some extent, it also prevents some over-smoothing regions and severely blurring reconstruction images resulting from inappropriate parameter. 
\begin{table}[H]	
	\centering
	\caption{Comparison of reconstruction of slice 100 of patient L506 between training dataset of 7 slices and 14 slices respectively.}
	\label{tab:MRST_incredata}	 	
	\vspace{-0.05in}
	\renewcommand\tabcolsep{5.0pt}
	\footnotesize{
		\begin{tabular}{c|c|ccccc}		
			\toprule
			&  &PWLS-ST  &PWLS-MARS2    &PWLS-MARS3   & PWLS-MARS5  & PWLS-MARS7  \\
			\midrule
			\multirow{3}{*}{\shortstack{dataset of \\7 slices}}    & RMSE  & 27.5   & 26.2  & 25.6  & \textbf{25.3}  & 25.7 \\
			\cmidrule{2-7}
			& SSIM  & 0.760   & 0.766  & 0.773   & 0.790  & \textbf{0.809} \\
			\midrule
			\multirow{3}{*}{\shortstack{dataset of \\14 slices}}    & RMSE  & 27.4   & 26.2  & 25.6  & \textbf{25.4}  & 25.6 \\
			\cmidrule{2-7}
			& SSIM  & 0.759  & 0.766   & 0.773  & 0.790  & \textbf{0.810} \\
			\bottomrule
		\end{tabular}
	}
	\vspace{0.1in}
\end{table}

As shown in Section~\ref{alg:MARS}, the block coordinate descent (BCD) method was applied to train a MARS model. Since the problem we address in this work is nonconvex, there might not be a unique minimizer in general. 
Despite that we use the BCD algorithm to ensure the monotone decrease over iterations of the nonnegative objective like \eqref{eq:P0} with a reasonable initialization (i.e., with PWLS-EP).
%However, the BCD algorithm provides a very efficient way to minimize the cost function, and is shown to empirically work well with appropriate initialization. 
%% Fortunately, some previous works \cite{ravishankar:15:lst, zheng:18:pua} have proposed an alternating algorithm to obtain the solution with acceleration. 
%Recent works involving transform learning \cite{ravishankar:15:lst, ye:19:sld} have shown that such efficient alternating minimization or BCD algorithms can provably converge to the critical points of the underlying problems. 
A more thorough analysis of convergence for our scheme is left for future work.

To conclude, we proposed a general framework for multi-layer residual sparsifying transform (MARS) learning, where the transform domain residual maps over several layers are jointly sparsified. Our work then applied learned MARS models to low-dose CT (LDCT) image reconstruction by using a PWLS approach with a learned MARS regularizer. Experimental results illustrate the promising performance of the multi-layer scheme over single-layer learned sparsifying transforms. Learned MARS models also offer image quality improvements over typical nonadaptive methods. Future work will consider other strategies for learning deep sparsifying models by exploiting pooling and other operations. In addition, more studies are required to validate the proposed method's clinical applicability. 
%Moreover, we also anticipate the \blue{incorporation of MARS models} with other frameworks~\cite{zheng:18:pua,Li:19:sup} to make our model \green{richer} and \green{more} effective.

\section{Acknowledgments}
Xikai Yang and Yong Long are supported in part by NSFC (Grant No. 61501292).

The authors thank Dr. Cynthia McCollough, the Mayo Clinic, the American Association of Physicists in Medicine, and the National Institute of Biomedical Imaging and Bioengineering for providing the Mayo Clinic data.

The authors thank Xuehang Zheng, Shanghai Jiaotong University, China for his helpful suggestions on the experiments. 

\section{Conflict of Interest}
The authors have no conflicts to disclose.

\section{Data Availability}
The data that support the findings of this study are openly available in the National Cancer Institute’s The Cancer Imaging Archive (TCIA) at \url{https://doi.org/10.7937/9npb-2637}, reference number\cite{Mayo:16:data}.

%\clearpage
%\input{unnecessary-v1}

% following only if there is an appendix
\section*{Appendix I: Solution of the Sparse Coding Problem \eqref{eq:sub_pro_Z}}
First, we can split this objective function and rewrite (2) as follows,
\begin{equation}\label{eq:derive_Z_1}
\min_{\Z_l}\|\Z_l-\omg_l\R_l\|_F^2 + \sum_{i=l+1}^L \|\Z_i-\omg_i\R_i\|_F^2 + \eta_l\|\Z_l\|_0. 
\end{equation}

Under the condition that $\omg_l^T\omg_l = \I, \forall l$, the following steps are based on
\begin{equation}\label{eq:derive_Z_2}
\|\omg_l\R_l-\Z_l\|^2_F = \|\omg_l^T\omg_l\R_l-\omg_l^T\Z_l\|^2_F = \|\R_l-\omg_l^T\Z_l\|^2_F. 
\end{equation}

We use \eqref{eq:derive_Z_2} within \eqref{eq:derive_Z_1} repetitively, which leads to the equivalent problem shown in \eqref{eq:derive_Z_3},
\begin{equation}\label{eq:derive_Z_3}
\min_{\Z_l}\|\Z_l-\omg_l\R_l\|_F^2 + \sum_{i=l+1}^L \|\Z_l+\B_l^i-\omg_l\R_l\|_F^2 + \eta_l^2\|\Z_l\|_0.
\end{equation}

Combining all the quadratic terms involving $\Z_l$ leads to the following optimization problem:
\begin{equation}\label{eq:derive_Z_4}
\min_{\Z_l}(L-l+1)\times\bigg\|\Z_l - \bigg( \omg_l\R_l - \frac{1}{L-l+1}\sum_{i=l+1}^L \B_l^i \bigg)\bigg\|_F^2 + \eta_l^2\|\Z_l\|_0.
\end{equation}

The solution to \eqref{eq:derive_Z_4} is similar to $\ell_0$ transform sparse coding~\cite{ravishankar:15:lst} and is given as follows when $1\leq l \leq L-1$

\begin{equation}
\hat{\Z}_l=H_{\eta_l/\sqrt{L-l+1}} \bigg( \omg_l\R_l - \frac{1}{L-l+1}\sum_{i=l+1}^L \B_l^i \bigg)
\end{equation}
and when $l=L$, it is given as
\begin{equation}
\hat{\Z}_L=H_{\eta_L} (\omg_L\R_L)
\end{equation}

\section*{Appendix II: Solution of the Transform Update Problem \eqref{eq:sub_pro_mOmega}}
Equation \eqref{eq:derive_Z_2} also works well for simplifying \eqref{eq:sub_pro_mOmega} as follows,
\begin{equation}\label{eq:derive_omg_1}
\min_{\omg_l: \omg_l^T \omg_l = \I}(L-l+1)\times\bigg\|\omg_l\R_l - \Z_l - \frac{1}{L-l+1}\sum_{i=l+1}^L \B_l^i \bigg\|_F^2.
\end{equation}
Problem \eqref{eq:derive_omg_1} can be equivalently written as
\begin{equation}\label{eq:derive_omg_2}
\begin{aligned}
\min_{\omg_l: \omg_l^T \omg_l = \I} tr(\R_l\R_l^T) - 2tr\bigg(\omg_l\R_l\bigg(\Z_l + \frac{1}{L-l+1}\sum_{i=l+1}^L \B_l^i\bigg)^T\bigg).
\end{aligned}
\end{equation}
Ignoring the constant first term, we get
\begin{equation}\label{eq:derive_omg_3}
\max_{\omg_l: \omg_l^T \omg_l = \I} tr\bigg(\omg_l\R_l\bigg(\Z_l + \frac{1}{L-l+1}\sum_{i=l+1}^L \B_l^i\bigg)^T\bigg).
\end{equation}
Subproblem \eqref{eq:derive_omg_3} is identical to the corresponding subproblem in single-layer sparsifying transform learning~\cite{ravishankar:15:lst}. We denote the full singular value decomposition of the matrix $\mathbf{G}_l$ as $\U_l\Sig_l\V_l^T$. The optimal solution to \eqref{eq:derive_omg_3} is then given as $\V_l\U_l^T$ (cf.~\cite{ravishankar:15:lst}).
% \clearpage
\section*{References}
\addcontentsline{toc}{section}{\numberline{}References}
\vspace*{-15mm}

% Following assumes you are using bibtex. However, for submission to the
% journal you MUST explicitly INCLUDE THE REFERENCES IN THE TEX FILE. 
% In that case you need the following

% \begin{thebibliography}{10}
% insert the .bbl file generated by bibtex here
	%This will be a series of entries from your .bib file formatted
	%something like
	%\bibitem{Me09}
        %{I.~Meijsing, B.~W.~Raaymakers, A.~J.~E.~Raaijmakers \it et al.},
        %\newblock {Dosimetry for the MRI accelerator: the impact of a 
	%magnetic field on the response of a Farmer NE2571 ionization chamber},
        %\newblock Phys. Med. Biol. {\bf 54}, 2993 -- 3002 (2009).

% \end{thebibliography}

% The following is when using bibtex and picks up the example.bib file

\bibliography{refs.bib}
%\bibliography{./example}      %example.bib is on the same directory
% above points to where we find the master reference list
% and also causes the bibliography to be printed

% When creating your bibliography you should run bibtex on your local
% computer after running pdflatex on your .tex file. bibtex will
% generate a .bbl file.
% Copy the contents of this .bbl file into your main latex document,
% replacing the "\bibliography" command which was pointing at your .bib file.

% following defines style of .bbl file 

%\bibliographystyle{explicit relative path to medphy.bst}
\bibliographystyle{./medphy.bst}    %if this is installed on your system,
				    %it is not essential to have the    ./

% Note that you need to typeset once, then run bibtex, then typeset another
% two times to get the references working properly.

\end{document}